\documentclass[fleqn,usenatbib]{mnras}

\usepackage{booktabs}
\usepackage{adjustbox}

\usepackage[T1]{fontenc}

\usepackage{siunitx}

\DeclareRobustCommand{\VAN}[3]{#2}
\let\VANthebibliography\thebibliography
\def\thebibliography{\DeclareRobustCommand{\VAN}[3]{##3}\VANthebibliography}

\usepackage{graphicx}	%
\usepackage{amsmath}	%
\usepackage{newtxtext,newtxmath}
\usepackage{fix-cm}
\usepackage{comment}

\title[ZTF insights into Fermi-Detected NLSy1s]{ZTF Monitoring of $\gamma-$ray emitting Narrow Line Seyfert 1 Galaxies}

\author[Aman et al.]{
Aman Kumar,$^{1,2}$
Suvas Chandra Chaudhary,$^{3}$\thanks{E-mail: 2029666035@ufs4life.ac.za}
Raj Prince,$^{4}$\thanks{E-mail: priraj@bhu.ac.in}
Brian van Soelen, $^{3}$
and I.P.\, van der Westhuizen$^{3}$
\\  
$^{1}$ Inter University Centre for Astronomy and Astrophysics (IUCAA), 411007, Pune, India.\\
$^{2}$ Department of Physics, Savitribai Phule Pune University, Ganeshkhind, 411007, Pune, India\\
$^{3}$Department of Physics, University of the Free State, 205 Nelson Mandela Dr., Bloemfontein, 9300, South Africa.\\
$^{4}$Department of Physics, Institute of Science, Banaras Hindu University, Varanasi, 221005, India.\\
}

\pubyear{\the\year{}}

\begin{document}
\label{firstpage}
\pagerange{\pageref{firstpage}--\pageref{lastpage}}
\maketitle

\begin{abstract}
The $\gamma$-ray-emitting narrow-line Seyfert-I ($\gamma$-NLSy1) are among the most interesting systems for studying disk-jet coupling. The soft X-ray properties of these systems suggest the presence of a disc component, which peaks in the optical/UV regime, in addition to the active jet. In this work, we investigate the optical emission from $\gamma$-NLSy1 using long-term Zwicky Transient Facility (ZTF) observations and discussed in the context of blazars. We have reported the long-term flux and color variability in the g- and r-bands. The fractional variability ($F_{\rm var}$) goes as high as 72\%, with a mean value of 23\%, while the amplitude of variability ($\psi$) values range from 0.24 to 3.20, which is consistent with the long-term Swift-UVOT variability studies. The color-magnitude diagrams exhibit an RWB or BWB trend similar to that of blazars. The $t_{\rm var}$ suggests an emitting region size of $10^{15-17}$ cm, aligned with emissions coming from the inner accretion disk or  base of the jet. The PSD analysis using both DRW and CARMA modeling exhibits a characteristic break timescale of a few days to hundreds of days, which is likely linked to fundamental physical timescales in the system, such as thermal or viscous timescales in the accretion disk or timescales for acceleration and energy dissipation in the jet. The existence of these timescales acts as another signature of the disc-jet connection. These time scales are correlated with black hole mass, and the relation is consistent with previous studies.

\end{abstract}


\section{Introduction}

The unification of Active Galactic Nuclei (AGN) has traditionally been viewed through the lens of black hole mass and accretion rate, typically segregating jet-dominated blazars from disk-dominated Seyfert galaxies. However, the emergence of $\gamma$-ray emitting Narrow Line Seyfert 1 ($\gamma$-NLSy1) galaxies has challenged this dichotomy. These sources act as a crucial "missing link" in the AGN family, providing a unique laboratory to study how relativistic jets can be launched from systems with relatively lower black hole masses but extremely high accretion rates. Narrow Line Seyfert 1 (NLSy1) galaxies are the highly accreting lower mass ($M_{BH}\sim10^{6-8}M_{\odot}$) tail of the Active Galactic Nuclei (AGN) family \citep{1992ApJS...80..109B, 2004AJ....127.3168B, 2008RMxAC..32...86K, 2010ApJS..187...64G, 2012AJ....143...83X, 2017ApJS..229...39R} identified by few permitted broad emission lines as narrow lines such as H$\beta$ with FWHM $< $ 2000 km s$^{-1}$ \citep{1989ApJ...342..224G}, [O III]$\lambda$5007 ([O III]/H$\beta < 3$) and $\mathrm{F\,II}$/H$\beta$ in their optical spectra \citep{1996A&A...305...53B, 1985ApJ...297..166O}. A small fraction (2-3\%) of these NLSy1 galaxies are radio loud \citep{2006ApJS..166..128Z}, which is very small in comparison with radio-loud AGNs ($\sim$15\%, \cite{2016ApJ...831..168K}).\\
With the launch of Fermi-LAT, a handful of NLSy1 galaxies have been detected in $\gamma$-rays \citep{2022Univ....8..587F}. These $\gamma$-NLSy1 galaxies exhibit complex timing and spectral features in the entire electromagnetic spectrum. In the Fermi-LAT energy band, the isotropic $\gamma$-ray luminosity ranges from $L_{\gamma} \sim 10^{44 - 48} erg s^{-1}$ \citep{2019Galax...7...87D}, which corresponds to Flat Spectrum Radio Quasars (FSRQs) type blazars. Similarly, the Fermi spectral index of several $\gamma$-NLSy1 sources also shows spectral curvature, similar to that of FSRQs. They exhibit rapid, high-amplitude flux variability, especially during flares. For example, the flare of 1H 0323+342 in 2025 showed a 20-fold flux enhancement \citep{2025ATel17407....1L} in $\gamma$-ray and 4 times flux enhancement in hard X-ray \citep{2026ApJ...996..118C} band. In another case, PKS 2004-447 experienced flux increases up to 50 times the average during its 2019 flare \citep{2019ATel13229....1G}. These systems are well-known for their distinctive X-ray signatures, which set them apart as a distinct class of AGNs rather than FSRQs or NLSy1s, which exhibit mixed features of both. Specifically, the X-ray spectra of $\gamma$-NLSy1 above 2 keV show a pure non-thermal flux contribution coming from the jet, also implying inverse Compton (IC) as a key mechanism at higher energies \citep{2019A&A...632A.120B, 2026ApJ...996..118C}, similar to blazars. In contrast, the soft X-ray spectra reveal more complex features, such as the soft-excess below 2 keV reported in \citep{2006MNRAS.370..245G}, similar to NLSy1 \citep{1996A&A...305...53B, 2006MNRAS.365.1067C, 2006ApJS..166..128Z}. Building on these findings, a recent study presented in \cite{2025arXiv251213569C} used XMM-Newton observations and found that six of $\gamma$-NLSy1 galaxies show soft excess below 2 keV, and their X-ray spectra were well fitted by powerlaw+bbody (PL+BB). Similar soft-excess features have also been reported in some FSRQs, such as 3C 273 \citep{2023ApJ...955..121D} and NLSy1s \citep[][]{2020ApJ...896...95O}, making these systems a more interesting AGN class to test disk-jet coupling.This spectral duality suggests that $\gamma$-NLSy1s are not merely low-mass blazar-like systems, but complex ecosystems where the thermal glow of the accretion disk and the non-thermal emission of the jet coexist and interact. By observing the "soft excess" alongside hard X-ray jet emission, we can directly probe the dynamical coupling between the inflow of matter and the subsequent launching of relativistic particles. Understanding this coupling requires moving beyond static spectral snapshots to analyse the dynamic "heartbeat" of these systems across different timescales. 

In the optical band, some of the $\gamma$-NLSy1 galaxies have been studied extensively with the intra-night observations. Studies done in \cite{10.1093/mnras/sty3288, 10.1093/mnras/sts217} showed that some of them have strong intra-night optical variability (INOV). \cite{10.1093/mnras/sty3288} also suggests that these galaxies can have large contamination by thermal optical emission from the disk because they are high Eddington rate accretors. \cite{10.1093/mnras/sts217} has obtained the high duty cycle of INOV and suggests that they are similar to BL Lac type blazars, where the jet is closely aligned with the observer's line of sight. \cite{2025arXiv250503902O} has studied the sample of 23 $\gamma$-NLSy1 galaxies using the optical and mid-infrared data from ZTF and WISE, and quantified the relative contribution of thermal and non-thermal components.
They have also observed a strong positive correlation between optical and MIR emission, which further suggests the reprocessing of thermal radiation in the dusty torus region. Their color-color variability revealed that almost 50$\%$ $\gamma$-NLSy1 galaxies in their sample show a stronger redder-when-brighter trend. \cite{2025ApJ...990...79S} has studied the extremely radio-loud NLSy1 galaxies using the optical and radio data from the 1.2m telescope located at the Mount Abu Observatory, India, and the 3.0 GHz Very Large Array Sky Survey (VLASS) data. In optical observations, they observed a strong INOV with a high duty cycle, similar to that of blazars. In radio, they found that these sources are very luminous, compact, and flat-spectrum, resembling the FSRQ class of blazars. They also plot the radio luminosity at 1.4 GHz with the supermassive black hole mass and observed that these extremely radio-loud NLSy1s are low-z and low-luminosity analogs of flat-spectrum radio quasars. Collectively, these studies highlight a complex scenario where both jet and disk photons contribute significantly to the overall optical emission. Consequently, distinguishing whether the emission arises from the turbulent churning of the accretion disk or internal shocks within the jet remains a significant challenge. To untangle the origin of this emission, whether it arises from the turbulent churning of the accretion disk or the internal shocks within the jet, we must decode the temporal fingerprints of their variability. The Power Spectral Density (PSD) serves as the primary diagnostic tool in this endeavour, allowing us to translate stochastic flux flickers into physical characteristic timescales. Breaks in the PSD could suggest different timescales of variability caused by various processes; however, the projected value may be influenced by several factors, including data gaps and distortions from aliasing and red noise \citep[e.g,][]{2002MNRAS.332..231U, 2013ApJ...770...60S, 2025ApJ...993...50C, 2025ApJS..279....3X}. %
Examining the damping timescale and its relationship with physical parameters such as black hole mass, jet power, disk luminosity, and accretion rate can offer important insights into the fundamental mechanisms of variability. For example, \cite{2021Sci...373..789B} applied Gaussian process (GPs) regression to adjust a DRW model to the optical light curves of non-jetted AGNs and reported a correlation between damping timescale and the mass of the black hole, with the damping timescale matching the anticipated thermal timescale. While studies conducted by \cite{2025ApJS..279....3X} indicate that the optical variability of jetted AGNs with effective accretion might stem from the standard accretion disk comparable to those of non-jetted AGNs, and is closely connected to the shock acceleration in the jet. In jetted AGNs with inefficient accretion, the intrinsic timescale aligns with the escape timescale for electrons.\\ 
In this work, we leverage the multi-year baseline of the Zwicky Transient Facility (ZTF) \citep{2019PASP..131a8003M, bellm2019zwicky, bellm2019scheduling} to perform a high-resolution "temporal dissection" of $\gamma$-NLSy1s. By applying DRW and CARMA models, we aim to move beyond simple variability detection and instead map the characteristic break timescales to specific physical parameters. This allows us to definitively test whether the optical variability is a thermal product of the disk or a non-thermal byproduct of the jet, thereby resolving the intricate disk-jet-corona interplay in the low-mass regime. We can examine both intraday and long-term variability in disk and jet emission from ZTF's multi-year observations. We utilize statistical measures of this variability, including power spectral density (PSD), flux distribution, color variability, variability amplitude ($\psi$), and fractional variability ($F_{\rm var}$). PSD describes stochastic and quasi-periodic variability in the frequency domain. However, its estimation is often complicated by irregular sampling and data gaps, particularly for non-periodic light curves.\\
The following outlines the paper's structure. The sample selection and ZTF data processing are covered in Section \ref{sample}. Section \ref{Analysis} discusses various analysis techniques used to quantify flux variability and present the results, while Section \ref{results} provides a discussion and interpretation of the findings. A summary of the work is provided in Section \ref{con}.

\section{SAMPLE SELECTION and ZTF DATA COLLECTION} \label{sample}

Fermi-LAT has identified about two dozen narrow-line Seyfert 1 (NLSy1) galaxies \cite[see, e.g.,][]{2018rnls.confE..15K, 2020yCat..22470033A, 2022Univ....8..587F}. Here, we use data from the Zwicky Transient Facility (ZTF) DR 23 to analyze the long-term optical variability of these $\gamma$-ray-emitting NLSy1 galaxies. Twenty-two sources with available photometric data were identified by cross-matching the ZTF collection with known $\gamma$-NLSy1 sources. Table \ref{tab:nls1_gamma_sample} summarises these sources and their fundamental characteristics, such as coordinates, black hole mass, and redshift, as documented in the literature. The ZTF database provided the $g$- and $r$-band magnitudes for these objects. We queried the ZTF light curve in the database\footnote{\url{https://irsa.ipac.caltech.edu/Missions/ztf.html}} using the ZTF Light Curve API, searching within a $1.5^{\prime\prime}$ radius around the reported coordinates of each $\gamma$-NLSy1 source. Out of the 22 objects in our cross-matched sample, we successfully retrieved both $g$- and $r$-band light curves for 15 sources, while the remaining objects had either no available data or light curves in only a single band. The retrieved light curves were then subjected to a quality filtering step. Specifically, we removed all photometric points for which the ZTF catalog flag, \texttt{catflags}, was nonzero to make sure we have only reliable data points. This flag encodes various data-quality issues, such as poor image subtraction, contamination from nearby bright stars, saturation, or artifacts from bad pixels \citep{bellm2019zwicky}. By discarding all points with \texttt{catflags} $\neq 0$, we ensured that only reliable photometric measurements were retained for further analysis. Finally, for analyses requiring simultaneous multi-band coverage, we constructed matched $g$- and $r$-band light curves by selecting observations taken within a one-hour tolerance of each other. This matching ensures that variability signatures are not biased by temporal offsets between the two bands, which is particularly important for color variability studies and for model comparisons that rely on contemporaneous flux measurements. The resulting paired light curves provide a consistent dataset for probing correlated variability across optical bands, complementing the single-band analyses performed on the full ZTF light curves.

\begin{table*}
    \centering
    \caption{Sample of the $\gamma$-ray emitting Narrow Line Seyfert 1 galaxies with their basic parameters.}
    \label{tab:nls1_gamma_sample}
\begin{tabular}{lrrrrrrrrl}
    \toprule
  Sr. No & Name & RA (deg) & DEC (deg) & $M_\mathrm{BH}$ & $z$ & $\log P_{\mathrm{jet}}$ & $\log L_{\mathrm{disk}}$ & $\delta$ & References \\
   & & & & [$M_\odot$] & & [erg s$^{-1}$] & [erg s$^{-1}$] & & \\
    \midrule
   1 & J032441+341045 &  51.1713 &  34.1794 & 7.30  & 0.06 & 45.82 & 45.30 & 13.6 & \citet{zhou2007narrow}, \citet{paliya2019general} \\
   2 &  J084957+510829 & 132.4917 &  51.1414 & 7.59  & 0.58 & 46.05 & 45.43 & 19.1 & \citet{rakshit2017optical}, \citet{paliya2019general} \\
   3 &  J093241+530633 & 143.1713 &  53.1092 & 8.00  & 0.60 & 46.54 & 45.70 & 14.7 & \citet{rakshit2017optical}, \citet{paliya2019general} \\
   4 &   J093712+500851 & 144.3012 &  50.1478 & 7.56  & 0.28 & 46.41 & 43.71 & 15.4 & \citet{rakshit2017optical}, \citet{paliya2019general} \\
   5 &  J094635+101706 & 146.6463 &  10.2850 & 8.20  & 1.00 & - & - & - & \citet{yao2019sdss} \\
   6 &  J094857+002226 & 147.2388 &   0.3739 & 8.18  & 0.58 & 47.11 & 45.70 & 15.7 & \citet{paliya2019general} \\
  7 &   J122222+041315 & 185.5938 &   4.2211 & 8.85  & 0.97 & 47.59 & 46.18 & 16.5 & \citet{yao2015identification}, \citet{paliya2019general} \\
   8 &  J142105+385522 & 215.2750 &  38.9231 & 8.48  & 0.49 & 46.77 & 45.34 & 13.6 & \citet{rakshit2017optical}, \citet{paliya2019general} \\
   9 &   J144318+472556 & 220.8275 &  47.4322 & 7.36  & 0.70 & -- & -- & -- & \citet{liao2015discovery} \\
  10  &   J150506+032631 & 226.2771 &   3.4419 & 7.6  & 0.41 & 46.09 & 44.78 & 17.2 & \citet{rakshit2017optical}, \citet{paliya2019general} \\
   11 &  J164442+261913 & 251.1771 &  26.3203 & 7.70  & 0.14 & 45.91 & 44.48 & 14.7 & \citet{rakshit2017optical}, \citet{paliya2019general} \\
   12 &  J211852$-$073228 & 319.7204 & $-$7.5411 & 7.20  & 0.26 & 45.93 & 44.00 & 17.2 & \citet{rakshit2017optical}, \citet{paliya2019general} \\
    13  & J133108+303032 & 202.7846 &  30.5089 & 8.11  & 0.85 & -- & -- & -- & \citet{yao2021spectroscopic} \\
   14  &  J003159+093618 &   8.0000 &   9.6050 & 6.71  & 0.22  & -- & -- & -- & \citet{mao2021search}\\
   15  &  J164100+345453 & 250.2504 &  34.9147 & 7.15  & 0.16 & -- & -- & -- & \citet{lahteenmaki2018radio} \\  
    \bottomrule
\end{tabular}

\end{table*}

\section{Analysis Techniques and Results} \label{Analysis}

\subsection{Flux Variability}

Fractional Variability is a widely used method to quantify the amplitude of variability in astronomical time series data \citep{vaughan2003characterizing}. This is defined by the equation

\begin{equation} \label{FVs}
    F_{\rm var} = \sqrt{\frac{S^{2} - \bar{\sigma}^{2}_{\rm err}}{\bar{X}^2}},
\end{equation}
Where $S^2$ is the light curve variance, $\bar {\sigma^2}$ is the flux mean square error, and $\bar X$ is the mean flux. The associated error in the $F_{var}$ is obtained using,
\begin{equation}
   \sigma_{F_{\mathrm{var}}} =  \sqrt{\left(\frac{1}{\sqrt{2N}}\frac{\bar{\sigma}^{2}_{err}}{F_{var}}\frac{1}{\bar{X}^2}\right)^2 + \left(\sqrt{\frac{\bar{\sigma}^{2}_{err}}{N}}\frac{1}{\bar{X}^2}\right)^2}. 
\end{equation}

Another important measure of flux variability is the variability amplitude, defined by \cite{romero1999optical}

\begin{equation}
    \psi = \sqrt{((A_{max} - A_{min})^2 - 2\sigma^2},
\end{equation}

\begin{figure}
    \centering
    \includegraphics[width=\linewidth]{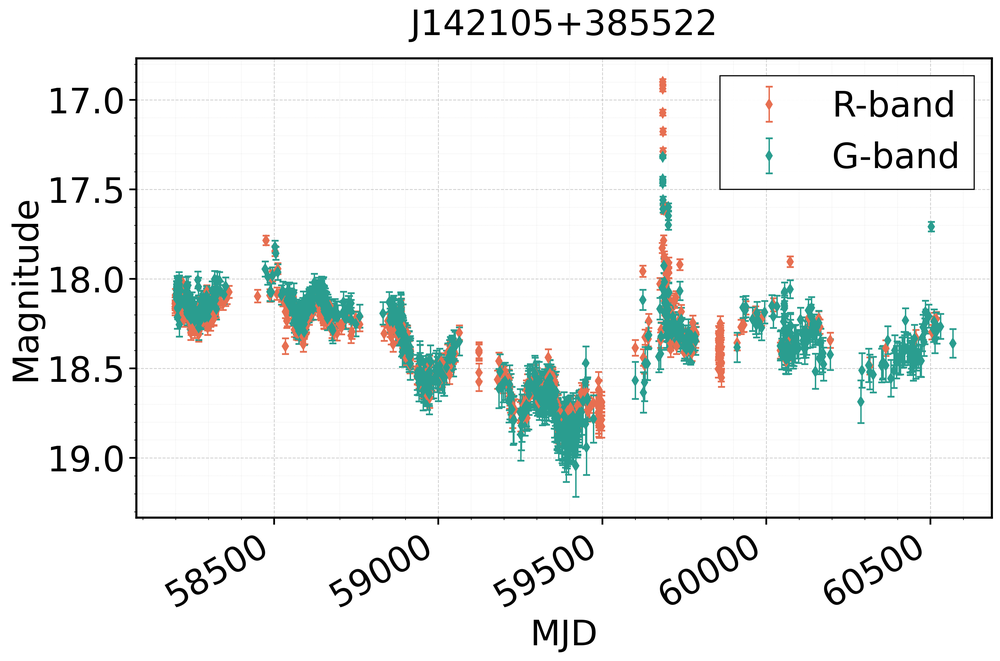}
    \caption{ZTF lightcurve for $\gamma$-NLSY1 J142105+385522 in $g$ and $r$ band. Other lightcurves are shown in Appendix \ref{sec:Ind_plots}.}
    \label{lc}
\end{figure}

An example light curve is shown in the Figure \ref{lc}, exhibiting a short \& long-term variability of the source J142105+385522 in both g- and r-bands. 
Table \ref{tab:variability} summarizes the optical variability properties of $\gamma$-NLSy1 galaxies in the g- and r-bands based on long-term ZTF data. The $F_{\rm var}$ varies from 1.75$\pm$0.11\% for J133108+303032 to 53.49$\pm$0.18\% for J084957+510829 in the r-band and from 1.03$\pm$0.12\% for J133108+303032 to 72.59$\pm$0.21\% for J093712+500851 in the g-band. Likewise, the $\psi$ ranges from 0.21$\pm$0.03 for J124634+023808 to 2.86$\pm$0.17 for J084957+510829 in the r-band and from 0.25$\pm$0.02 to 3.2$\pm$0.16 for J093712+500851 in the g-band. Like blazars, non-thermal synchrotron emission from relativistic jets is the primary source of optical fluctuations observed in $\gamma$-NLSy1 galaxies. Strong Doppler-boosted jet activity is indicated by the high levels of peak-to-peak variability ($\psi$) and $F_{var}$, especially in the g-band. Conversely, a higher contribution from thermal emission coming from the accretion disk is suggested by decreased variability in some sources. Variability within the sample reflects individual source variances in accretion characteristics, direction, and jet dominance. \cite{2022MNRAS.510.1791N} have studied a large blazar sample and found that most of the sources have higher g-band $F_{var}$ ($\sim$85\%, BL Lacs and $\sim$65\%, FSRQs), suggesting a frequency-dependent variability, however, the $\gamma$-NLSy1 doest follow this trend, while some of them having equal g- and r-band variability, indicating similar disc and jet contributions  \citep{2017ApJ...844..107I}.
The variability timescale ($\tau$) is another key parameter that can be inferred from light curves, providing valuable constraints on the physical mechanisms driving the observed flux variations. Following the approach of \citet{jorstad2013tight, burbidge1974physics}, we estimated $\tau$ for those sources in our sample that exhibit significant variability, using the definition

\begin{equation}
\tau \equiv \frac{\Delta t}{\ln(S_2/S_1)},
\end{equation}
where $S_1$ and $S_2$ are flux values at times values $t_1$ and $t_2$, respectively, and $\Delta t = |t_2 - t_1|$. To ensure robustness, we considered only those flux pairs that satisfy the following conditions: (i) $S_2 > S_1$, and (ii) $S_2 - S_1 > 3(\sigma_{S_1} + \sigma_{S_2})/2$, where $\sigma_{S_1}$ and $\sigma_{S_2}$ are the corresponding measurement uncertainties. The following equation defines the error.
\begin{equation}
\sigma_{\tau_{\rm var}} \approx 
\sqrt{ \frac{ S_1^2 \, \Delta S_2^2 + S_2^2 \, \Delta S_1^2 }
{ S_1^2 S_2^2 \left( \ln \left[ S_1 / S_2 \right] \right)^4 } } \, \cdot \Delta t ,
\end{equation}
The $\tau_{min}$ timescale in g- \& r-band throughout the sample ranges from 0.09 to 1.01 day \& 0.07 to 5.03 day, respectively.

\subsection{Color Variability}

\begin{figure}
    \centering
    \includegraphics[width=\linewidth]{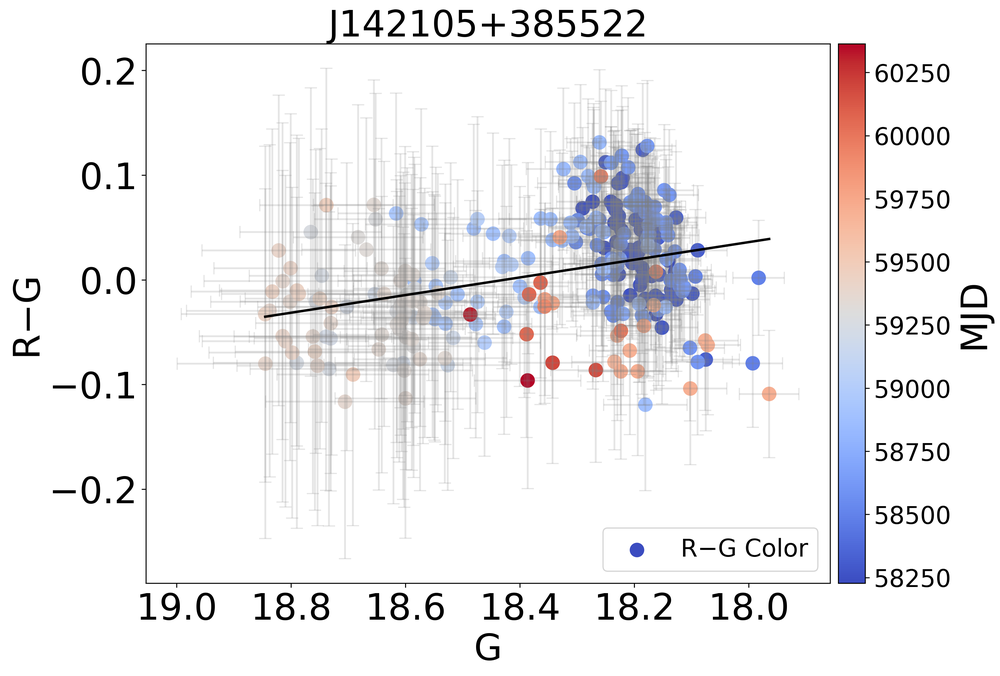}
    \caption{Colour Magnitude Plot of J142105+385522. Other colour-magnitude plots are shown in Appendix \ref{sec:Ind_plots}.}
    \label{cm}
\end{figure}

$\gamma$-NLSy1 exhibits rapid flux and spectral variations in the entire electromagnetic spectrum. Variations in jet, disk emission, or a combination of both account for the observed color changes. Hot-spots in accretion disks can also cause color variations \citep{2014ApJ...797...19R}. Examining the spectral variations of $\gamma$-NLSy1 can aid us in identifying the various components that contribute to the observed flux. The quasi-simultaneous $g$ and $r$-band observations in this work can provide clues to identifying any universal patterns in the spectral variations of the $\gamma$-NLSy1 class, similar to blazars over diverse timescales; the color variability mimics the spectral variability. This understanding could lead to more profound insights into the physical processes governing these enigmatic objects. The ZTF color–magnitude diagrams ($r-g$ vs. $g$) for our sample sources indicate that most show a clear redder-when-brighter (RWB) or bluer-when-brighter (BWB) trend, though a few display more complex patterns. Five sources show an RWB; the optical emission shifts to redder wavelengths as the flux increases, consistent with a scenario in which the jet's redder synchrotron emission increasingly overshadows the bluer thermal contribution from the accretion disk during bright states. Such features are well known in flat-spectrum radio quasars (FSRQs), when the jet emission outshines the disk \citep{2004A&A...421...83R, 2012ApJ...756...13B, 2022MNRAS.510.1791N, 2023MNRAS.519.5263Z}. While nine sources in our sample exhibit a bluer-when-brighter (BWB) trend, this aligns with the behavior observed in BL Lac-type blazars. In these, flux enhancements are linked to a shift of the synchrotron emission peak toward higher energies, producing a bluer optical spectrum \citep{1998MNRAS.299...47M, 2003MmSAI..74..963V, 2011PASJ...63..639I, 2015MNRAS.450..541A, 2022MNRAS.510.1791N}. This pattern is typically attributed to enhanced particle acceleration or the presence of high-energy electrons in the jet. An opposite trend of BWB in a sample of FSRQs was also seen in \cite{2011JApA...32...87G}. Figure \ref{cm} depicts a representative case of RWB, and the color–magnitude diagrams for the other sources can be found in the Appendix. Several AGNs also exhibit a mixed color variability trend reported in \citep{2017ApJ...844..107I, 2026ApJ...996..118C}, suggesting a jet-disk connection. A long-term study of 3C 279 conducted from 2005 to 2016 by \cite{2017ApJS..229...21X} again reveals a variable RWB trend. The source PMN J0948+0022 was monitored by \cite{2022RAA....22g5001X} throughout a somewhat narrow magnitude range (18.2–17.2), where a RWB trend was evident. However, our comprehensive ZTF monitoring spans a wider range (19–17.5), where a BWB trend governs the overall behavior. This suggests that the source flux state has a significant influence on the observed color variation, with conflicting tendencies emerging within various magnitude ranges.

\subsection{Discrete Correlation Function}

The correlation between two unevenly sampled light curves was estimated using the Discrete Correlation Function (DCF; \citealt{1988ApJ...333..646E}), which avoids interpolation. For measurements $(x_i, y_j)$ with time separation $\tau_{ij}$, the unbinned correlation is
\begin{equation}
u_{ij} = \frac{(x_i - \bar{x})(y_j - \bar{y})}{s_x s_y},
\end{equation}
where $\bar{x}, \bar{y}$ are sample means and $s_x, s_y$ the standard deviations. The DCF at lag $\tau$ is obtained by averaging all $u_{ij}$ within the bin centered on $\tau$:
\begin{equation}
r_{\mathrm{DCF}}(\tau) = \frac{1}{n} \sum_{\tau_{ij} \in \mathrm{bin}} u_{ij},
\end{equation}
with uncertainties from the scatter of $u_{ij}$ in each bin.

To mitigate small-sample biases, we adopt the $z$-transformed DCF (ZDCF; \citealt{1997ASSL..218..163A}), which applies Fisher’s $z$-transform,
\begin{equation}
z = \tfrac{1}{2}\ln\left(\frac{1+r}{1-r}\right), \quad r = \tanh(z),
\end{equation}
yielding approximately normal errors with mean $\bar{z}$ and variance $s_z^2$. Equal-population binning is enforced with a minimum of $n_{\min}=11$ statistically independent pairs per bin. Noise effects are assessed via Monte Carlo methods, including observational errors, following the implementation of \citet{2014ascl.soft04002A}.

\begin{figure}
    \centering
    \includegraphics[width=\linewidth]{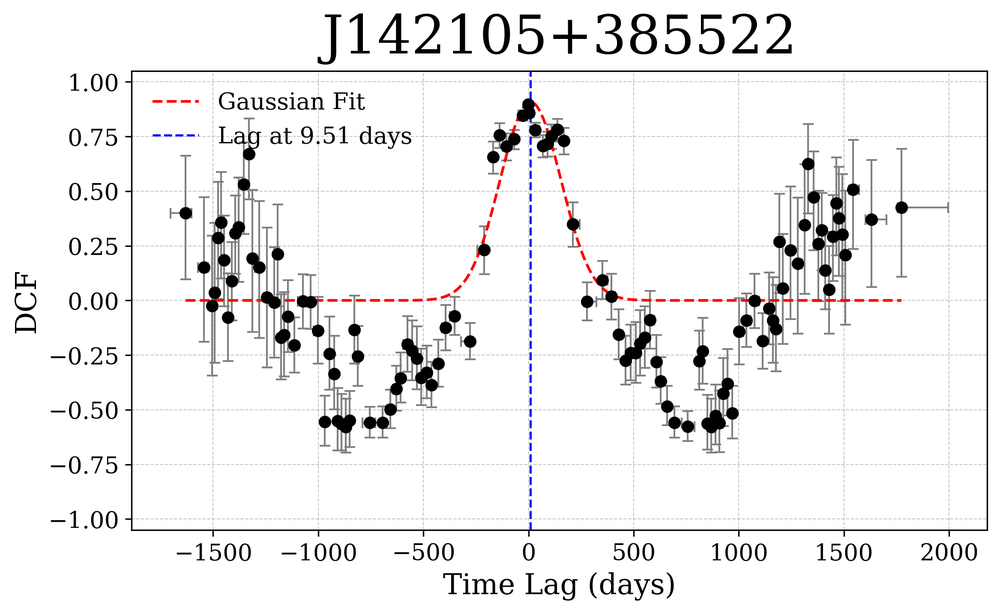}
     \includegraphics[width=.99\linewidth]{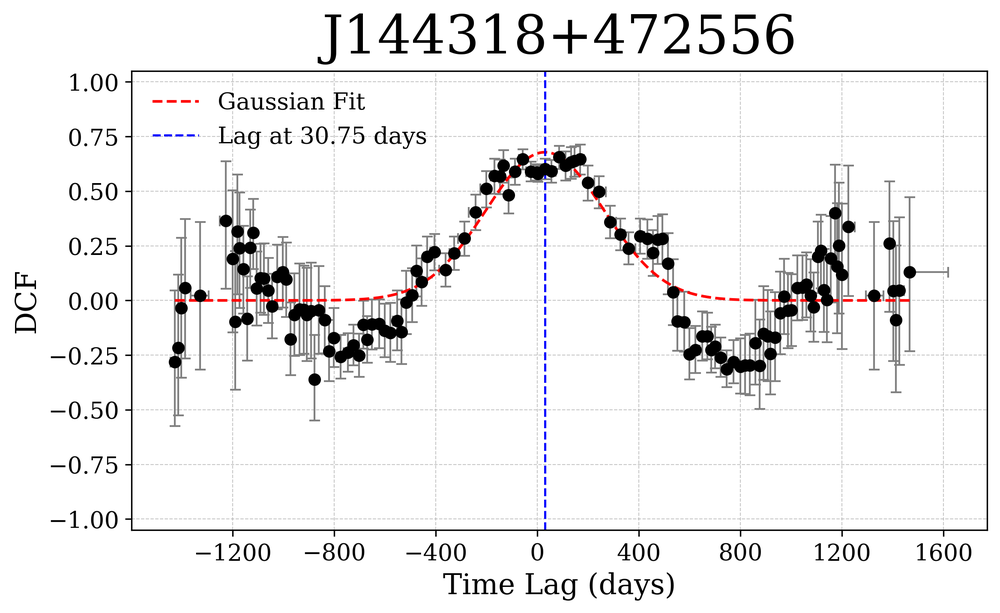}
    \caption{An example DCF for two sources is shown: Other DCFs are shown in Appendix \ref{sec:Ind_plots}.}
    \label{dcf}
\end{figure}
The DCF analysis of the $r$- and $g$-band optical lightcurves suggests an asymmetry in lag distributions. With a mean of -34.59 days, a median of -2.18 and a skewness of -2.04, the negative lags indicate that the $g$-band may lag significantly behind the $r$-band in certain sources. While a few sources indicate that the r-band typically leads the g-band, as evidenced by the positive lags' greater mean value of 33.40 and skewness of 0.89. Two sources, J144318+472556 and J142105+385522, exhibit a strong correlation (see Figure \ref{dcf}) with lags of -3.56 days and 9.51 days, respectively. This indicates that variability in both bands likely comes from a single zone, most likely the jet, where synchrotron emission dominates. The other sources, however, display several weak, jitter-like patterns instead of a clear DCF peak as shown in the appendix \ref{sec:Ind_plots}. These features may represent low-amplitude variability or blended signals from the host galaxy, the accretion disk, and the jet, which weaken the inter-band correlation. Conversely, the synchrotron origin of blazar variability is linked to similarly high, near-zero-lag optical band correlations, especially for FSRQs and BL Lacs \citep{2012ApJ...756...13B, 2013arXiv1303.1898P}. In contrast, Radio galaxies, AGNs, and Seyfert galaxies typically exhibit measurable time lags between $\gamma$-ray/X-ray/optical/UV/radio bands over periods ranging from hours to days \citep{2025ApJ...985...39S, 2010MNRAS.403..605B}. These lags result from the reprocessing of changing disk and coronal emission throughout the accretion disk regions.

\subsection{Flux Distribution}
Insights into the variability characteristics of astrophysical sources—such as emission states and the fundamental physical processes that control their overall emission—can be obtained by analyzing their flux distribution. Probability distribution functions (PDFs), which describe the shape of the flux distribution, are believed to reflect the underlying physical mechanisms involved. If the total emission results from additive processes, a normal flux distribution is usually expected \cite[e.g.,][]{2005MNRAS.359..345U, 2010A&A...524A..48T, 2018Galax...6..135R}. Conversely, a lognormal flux distribution often suggests that multiplicative processes are occurring \citep[e.g.,][]{2004ApJ...612L..21G, 2010A&A...524A..48T, 2009A&A...503..797G}. For each source in our sample, we constructed histograms of flux (in magnitudes) for both the $r$- and $g$-bands. We then fitted the distributions using three functional forms: (i) a Gaussian (normal) distribution, (ii) a lognormal distribution, and (iii) a two-component Gaussian mixture model, in cases where the data exhibited evidence of bimodality. The functional forms of the normal and lognormal distributions are given below:

\begin{equation}
f_{\mathrm{Normal}}(x; \mu, \sigma) =
\frac{1}{\sigma \sqrt{2\pi}}
\exp\left( -\frac{(x - \mu)^2}{2\sigma^2} \right),
\end{equation}

\begin{equation}
f_{\mathrm{LogNormal}}(x; \mu, \sigma) =
\frac{1}{x  \sigma \sqrt{2\pi}}
\exp\left( -\frac{(\ln x - \mu)^2}{2\sigma^2} \right),
\quad x > 0 .
\end{equation}

\begin{figure}
    \centering
    \includegraphics[width=\linewidth]{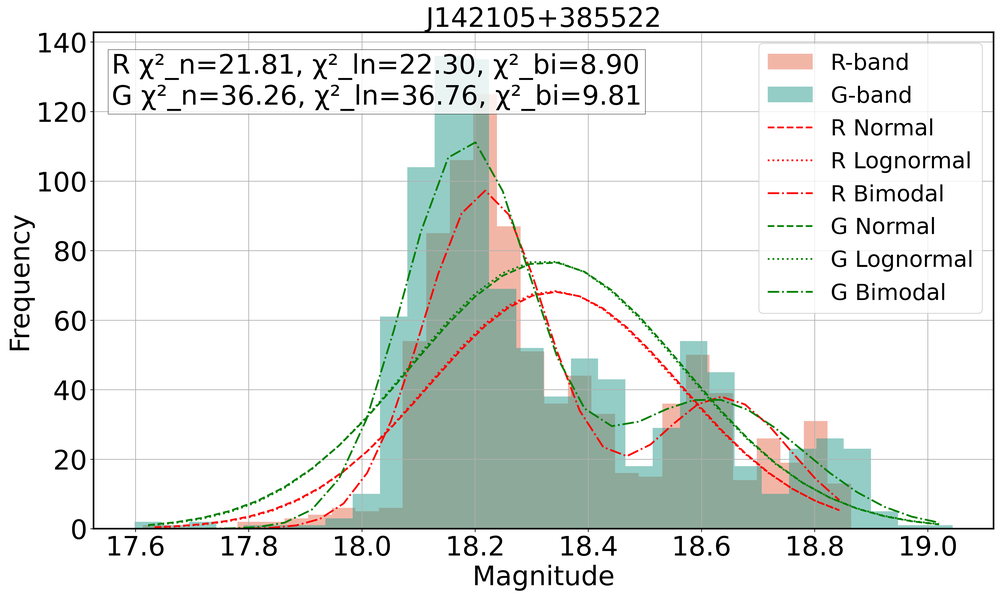}
    \caption{Flux Distribution of object J142105+385522. Other distributions are shown in Appendix \ref{sec:Ind_plots}.}
    \label{fd}
\end{figure}

In most cases, we observed a Gaussian distribution with a hint of bimodality. To test for bimodality, we employed multiple statistical criteria. First, we compared Bayesian Information Criterion (BIC) values between single- and double-component Gaussian mixture models (GMM). A significant reduction in BIC favoured a bimodal interpretation. Second, we estimated the bimodality coefficient (BC) \citep{sas1990sas, joanes1998comparing, pfister2013good}, with BC values $> 0.55$ suggesting a departure from unimodality. Finally, we required the two fitted Gaussian modes to be well separated in mean and reasonably balanced in weight, to avoid spurious bimodality due to noise. To evaluate goodness of fit, we computed reduced $\chi^2$ values for all three models (normal, lognormal, and bimodal). An example flux histogram, with fitted distributions overlaid, is shown in Figure~\ref{fd}. Out of 15 sources, four exhibit a bi-modal flux distribution as shown in Figure~\ref{fd}, while the remaining show unimodality (See Appendix \ref{sec:Ind_plots}). In the case of blazars and other AGNs, bimodality is particularly interesting. Bi-modal lognormal distributions are often very good at describing flux distributions of several blazars, such as RBS 2070, Mrk 501, OJ 287, PKS 0235-618, PKS 0035-252, and PKS 2155-304 \citep{2025ApJ...981..118B, 2024ApJ...973...10D, 2022MNRAS.510.5280M} and are typically associated with multiplicative variable processes in the jet, such as turbulence or oscillations in particle acceleration. 

\subsection{Flux-RMS Relationship}
\begin{figure}
    \centering
    \includegraphics[width=\linewidth]{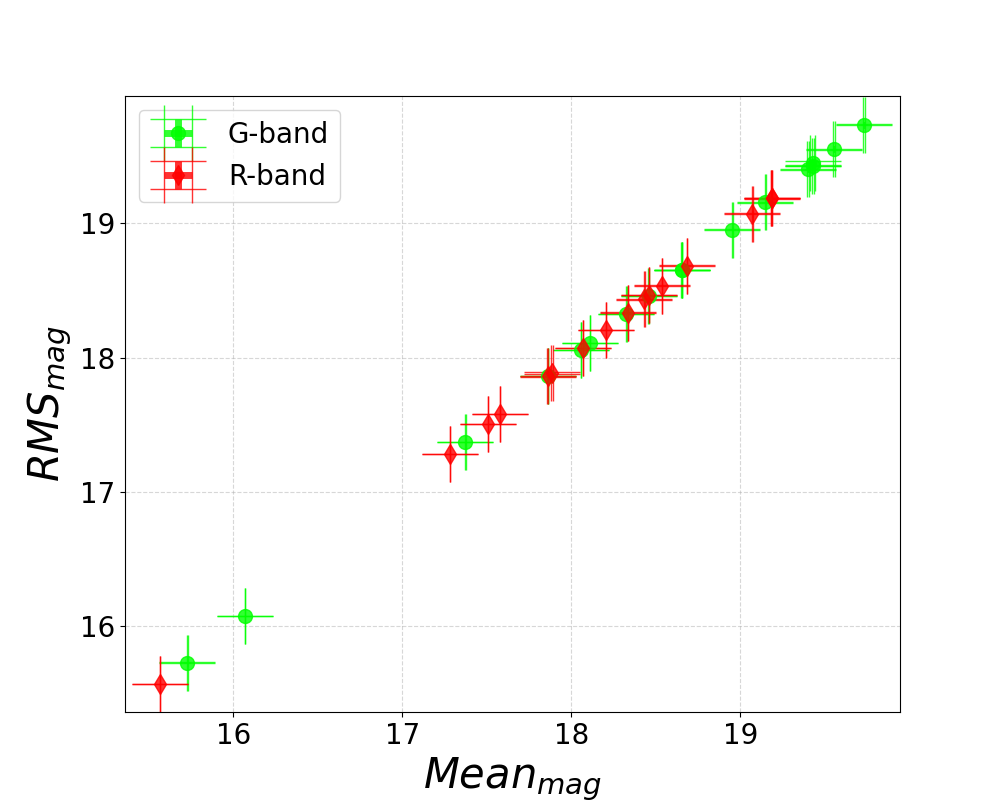}
    \caption{RMS vs Mean Magnitude relation for all the sources in R and G band.}
    \label{fig:rms_mean}
\end{figure}

The flux-rms relation \citep{vaughan2003characterizing} is a powerful tool for linking flux variability across different timescales. This places strict limits on models that explain flux variability in accreting compact objects. Similar to the propagating fluctuation model \citep{1997MNRAS.292..679L}, the preferred theories involve longer-term fluctuations originating in the outer disk that move inward with the accretion flow and affect the shorter-timescale variability generated at smaller radii. Our $\gamma$-NLSy1 sample shows that as the source brightness increases, the rms variability also increases, shown in the Figure \ref{fig:rms_mean}, suggesting multiplicative variability. This pattern resembles blazars, where jet emissions cause correlated flux and variability due to relativistic outflows \citep{2020ApJ...897...25B,2024ApJ...973...10D}. Similar flux–rms trends are seen in other accreting systems like Seyferts and X-ray binaries, typically linked to inward accretion disk fluctuations \cite{2004A&A...414.1091G, 2012A&A...548A.123B}. In $\gamma$-NLSy1s, though, both jet and disk variability seem to play a role, highlighting their hybrid nature between blazars and non-jetted AGNs. 

\subsection{CARMA Modeling}

For our study, we use continuous-time autoregressive moving-average (CARMA) models to derive the PSDs \citep{Kelly_2014}.
A zero-mean CARMA($p, q$) process $y(t)$ is defined to be the solution of a stochastic differential equation,

\begin{multline}
\frac{d^p y(t)}{dt^p} + \alpha_{p-1} \frac{d^{p-1} y(t)}{dt^{p-1}} 
+ \cdots + \alpha_0 y(t) \\
= \beta_q \frac{d^q \epsilon(t)}{dt^q} 
+ \beta_{q-1} \frac{d^{q-1} \epsilon(t)}{dt^{q-1}} 
+ \cdots + \epsilon(t)\,.
\end{multline}

where $\alpha_{p} = 1$, $\alpha_{p-1}$, ..., $\alpha_{0}$ and $\beta_{q}$, $\beta_{q-1}$, ..., $\beta_{0}=1$ are the auto-regressive coefficients and moving average coefficients respectively, and
$\epsilon(t)$ is zero-mean continuous-time white-noise process. 

The CARMA($p,q$) family provides a flexible framework for modeling stochastic time series, as it generalizes both continuous-time autoregressive (CAR) and continuous-time moving-average (CMA) processes. The power spectral density (PSD) corresponding to a CARMA($p,q$) process can be written analytically as
\begin{equation}
\label{eq:PSD}
P(\omega) = \sigma^2 , \frac{\left| \sum_{j=0}^{q} \beta_j (i\omega)^j \right|^2}{\left| \sum_{k=0}^{p} \alpha_k (i\omega)^k \right|^2},
\end{equation}
where $\omega$ is the angular frequency and $\sigma^2$ denotes the variance of the driving white-noise process $\epsilon(t)$.

The roots of the auto-regressive polynomial
\begin{equation}
A(z) = z^p + \alpha_{p-1} z^{p-1} + \cdots + \alpha_0
\end{equation}
determine the characteristic timescales of the process, while the moving average polynomial
\begin{equation}
B(z) = \beta_q z^q + \beta_{q-1} z^{q-1} + \cdots + 1
\end{equation}
modulates the short-timescale correlations. Stability requires that all roots of $A(z)$ have negative real parts, ensuring the process is stationary. In practice, low-order models such as CAR(1), CARMA(1,0), and CARMA(2,1) are often sufficient to capture the variability properties of astrophysical light curves \citep[e.g.,][]{Kelly_2014, moreno2019stochastic}. In particular, the CAR(1) process (also known as the Ornstein–Uhlenbeck process) has been widely used to describe the damped random walk (DRW) behavior of AGN variability, whereas higher-order CARMA models allow for more complex PSD shapes, including multiple characteristic frequencies and quasi-periodic behavior. To model light curves under the assumption of a damped random walk (DRW; CAR(1)) process, we use the EZTao package \citep{Yu2022}, which provides an efficient implementation for fitting irregularly sampled time series. The DRW is the simplest member of the CARMA family and is mathematically equivalent to an Ornstein–Uhlenbeck (OU) process. Its dynamics are governed by the stochastic differential equation
\begin{equation}
\label{eq:drw_sde}
dX(t) = -\frac{1}{\tau}\big[X(t)-\mu\big]\,dt + \sigma\, dW_t,
\end{equation}
where $\mu$ is the long-term mean, $\sigma$ is the short-timescale variability amplitude, and $dW_t$ denotes a Wiener increment. The coefficient $1/\tau$ represents the linear restoring force, so the damping (characteristic) timescale $\tau$ specifies the exponential memory of the process or relaxation time ("damping") scale. It represents the time required for the light curve to become decorrelated from its previous state.

In the CARMA notation, the DRW corresponds to the CAR(1) model with an autoregressive polynomial
\begin{equation}
A(z) = z + \alpha_0,
\end{equation}
whose root $z=-\alpha_0$ determines the de-correlation timescale. Identifying the OU form $dX(t) = -a\,[X(t)-\mu]\,dt + \sigma\,dW_t$ gives $a=\alpha_0$ and therefore
\begin{equation}
\tau = \frac{1}{a} = -\frac{1}{z}.
\end{equation}
With only two parameters, $\tau$ and $\sigma$, the DRW provides a robust and interpretable baseline model for quasar and AGN variability.
\\ \\
However, the DRW assumption can be overly restrictive when the observed variability exhibits more complex temporal structure. To address this, we employ the \texttt{carma\_pack} software developed by \citet{Kelly_2014}, which allows us to fit higher-order CARMA($p,q$) models to the same light curves. The package employs a state-space representation of the CARMA process, combined with a Kalman filter, to efficiently evaluate the likelihood, making it well-suited for irregularly sampled astronomical time series. Model selection is performed by comparing the Akaike Information Criterion (AIC) values across different $(p,q)$ combinations. Specifically, for each candidate CARMA model, we compute
\begin{equation}
\mathrm{AIC} = 2k - 2 \ln \hat{\mathcal{L}},
\end{equation}
where $k$ is the number of free parameters while $\hat{\mathcal{L}}$ is the maximum likelihood of the model. The preferred model is the one that minimizes the AIC, balancing model complexity against goodness of fit. In our implementation, we restrict the moving average order to a maximum of $q=6$, to ensure a tractable and physically interpretable model space, and for each $q$ we set the auto-regressive order as $p_{\max} = q-1$, following the prescription of \citet{Kelly_2014}, for a stable, physically plausible process. The model then selects the optimal $(p,q)$ pair based on the AIC criterion, allowing us to determine whether a simple DRW process is sufficient or if a higher-order CARMA model provides a statistically significant improvement in describing the observed variability. Table \ref{table:CARMA_fititng} provides the list of (p,q) orders selected using the AIC value and the corresponding AIC values. Once the CARMA fitting is complete, we obtain the corresponding analytical and MCMC-based uncertainties on the PSD of the given light curve by taking 50,000 MCMC samples and 20,000 burn-in samples. The analytical PSD is computed directly from the maximum-likelihood estimates of the CARMA coefficients using Equation \ref{eq:PSD}. To account for parameter uncertainties, we also use the posterior samples generated by the MCMC chains in \texttt{carma\_pack}, which allow us to propagate parameter uncertainties to the PSD. This provides credible intervals on the PSD at each frequency, offering a more robust characterisation of the variability power spectrum.

Having established the modeling framework, we applied the DRW model to the light curves of our sample.  One example of the modeled light curve is shown in Figure 7, and in Figures 8 \& 9 we show the obtained PSD. A clear break is observed in both the $r$ and $g$-band PSD. Modeled light curves and the corresponding PSD for all other objects are presented in the Appendix, and their breaking time scale is summarized in Table \ref{TABDRW}.
The results for the $\gamma$-NLSy1s, presented in Table \ref{TABDRW}, suggests that the $g$- and $r$-band show a significant difference in the damping timescales ($\tau_{damping}$), for the $r$-band, it varies from a few minutes to 1100 days, with a mean value of 131 days, while in the $g$-band, it changes from an hour to a maximum of 838 days, with a mean value of 83 days. The differences in the $\tau_{\text{damping}}$ indicate the wavelength dependence of their optical variability and may reveal the physical processes regulating their accretion and jet activity. The longer variability timescales in the redder band likely trace slower processes in the outer disk or jet-dominated regions. On the other hand, shorter $\tau_{\text{damping}}$ in the g-band suggests that higher-energy emission regions, which are generally produced closer to the central black hole and inner accretion disk, show faster variability, with a mean of approximately 83 days. These estimates are consistent with the findings of \cite{2025arXiv251105268Z, 2025ApJS..279....3X} for blazars and Seyfert galaxies, where the $F_{var}$ and timescales are frequency dependent. For Seyferts, there is comparatively slow variability driven by the disk, whereas blazars show fast and high $F_{var}$ dominated by relativistic jets. The findings favor that the $\gamma$-NLSy1s as low-mass, high-accretion analogs to the blazar population, which could fill the gap between radio-quiet Seyferts and mighty jet-dominated AGN.

Beyond the baseline DRW description, the results of the higher-order CARMA modeling presented in Table \ref{table:CARMA_fititng} reveal a more complex temporal structure in the $\gamma$-NLSy1 light curves. Based on the minimization of the AIC, we find that the simple CAR(1) process is rarely the preferred model; instead, the majority of the sources are best described by higher-order processes, typically with auto-regressive orders of $p \geq 2$ and moving average orders up to $q = 3$. This preference for models such as CARMA(4,2) or (4,3) implies the presence of multiple characteristic timescales or deviations from a single thermal relaxation process, features that a standard damped random walk cannot capture.
\\
PSD derived from CARMA fitting shows multiple breaks, a feature showcasing the complex nature of temporal variability. To obtain a characteristic break timescale from these complex PSDs, we fit a double power-law to the CARMA PSDs and determine the corresponding break frequency. The characteristic break timescales derived from these higher-order fits exhibit a broad dynamic range, spanning from extremely rapid variations of around $0.03$ days (e.g., J150506+032631 in the r-band) to long-term evolution exceeding $250$ days (e.g., J144318+472556 in the r-band). Notably, for several sources, the optimal (p,q) combination differs between the g- and r-bands. This discrepancy suggests that while the bands are correlated, they may be sensitive to different physical components, such as the disk-jet coupling, which dominate the variability at different frequencies and break timescales. These timescales are consistent with \cite{lefkir2025variability} for bend timescales in AGNs. The detection of these short break timescales in the higher-order models further supports the hypothesis of compact emission regions in $\gamma$-NLSy1s, consistent with their rapid variability and high-energy nature.

\begin{table*}
\caption{Summary of variability properties for the sample.
$\langle F_r \rangle$ and $\langle F_g \rangle$ are the mean magnitudes in the $r$ and $g$ bands, respectively. $\mathrm{RMS}_r$ and $\mathrm{RMS}_g$ denote the root-mean-square variability in the two bands. $F_{\mathrm{var}, r}$ and $F_{\mathrm{var}, g}$ are the corresponding fractional variabilities. $A_r$ and $A_g$ represent the variability amplitudes (in magnitudes), while $\tau_{var, r}$ and $\tau_{var, g}$ give the minimum variability timescales (in days). PSD Slopes $\beta_P,r$ and $\beta_P,g$ refer to the slopes of the Lomb-Scargle power spectral density in the $r$ and $g$ bands.}
\adjustbox{width=\textwidth}{
\begin{tabular}{lcccccccccccc}
\toprule
Name & 
        $\langle F_r \rangle$ & $\langle F_g \rangle$ & 
        $\mathrm{RMS}_r$ & $\mathrm{RMS}_g$ & 
        $F_{\mathrm{var}, r}$ & $F_{\mathrm{var}, g}$ & 
        $\psi_r$ & $\psi_g$ & 
        $\tau_{var, r}$ & $\tau_{var, g}$ & 
        $\beta_P,r$ & $\beta_P,g$ \\
\midrule
J133108+303032 & 17.28  & 17.37  & 17.28  & 17.37  & 1.75 $\pm$ 0.11 & 1.03 $\pm$ 0.12 & 0.24 $\pm$ 0.03 & 0.29 $\pm$ 0.03 & 0.95 $\pm$ 0.04 & 1.71 $\pm$ 0.13 & -0.83 & -0.91 \\
J142105+385522 & 18.33  & 18.32  & 18.34  & 18.33  & 24.86 $\pm$ 0.13 & 20.91 $\pm$ 0.22 & 1.96 $\pm$ 0.07 & 1.72 $\pm$ 0.17 & 0.37 $\pm$ 0.04 & 5.03 $\pm$ 0.47 & -0.71 & -0.69 \\
J144318+472556 & 18.07  & 18.06  & 18.07  & 18.06  & 4.26 $\pm$ 0.09 & 4.85 $\pm$ 0.08 & 0.34 $\pm$ 0.04 & 0.42 $\pm$ 0.04 & 0.58 $\pm$ 0.05 & 0.22 $\pm$ 0.02 & -0.48 & -0.44 \\
J150506+032631 & 18.43  & 18.95  & 18.43  & 18.95  & 38.25 $\pm$ 0.13 & 44.27 $\pm$ 0.24 & 2.09 $\pm$ 0.06 & 2.32 $\pm$ 0.09 & 0.54 $\pm$ 0.03 & 158.90 $\pm$ 3.49 & -0.50 & -0.66 \\
J164100+345453 & 17.58  & 18.66  & 17.58  & 18.66  & 6.68 $\pm$ 0.05 & 7.90 $\pm$ 0.10 & 0.64 $\pm$ 0.03 & 0.86 $\pm$ 0.06 & 0.43 $\pm$ 0.02 & 0.22 $\pm$ 0.02 & -0.87 & -0.71 \\
J164442+261913 & 17.51  & 17.86  & 17.51  & 17.86  & 21.26 $\pm$ 0.05 & 20.75 $\pm$ 0.07 & 1.56 $\pm$ 0.03 & 1.63 $\pm$ 0.04 & 1.00 $\pm$ 0.11 & 0.82 $\pm$ 0.05 & -0.81 & -0.66 \\
J211852-073228 & 18.68  & 19.40  & 18.69  & 19.40  & 29.42 $\pm$ 0.20 & 44.19 $\pm$ 0.36 & 2.18 $\pm$ 0.07 & 2.49 $\pm$ 0.11 & 0.09 $\pm$ 0.00  & 0.27 $\pm$ 0.03 & -0.67 & -0.84 \\
J003159+093618 & 19.19  & 19.55  & 19.19  & 19.55  & 14.74 $\pm$ 0.27 & 19.64 $\pm$ 0.37 & 0.98 $\pm$ 0.10 & 1.17 $\pm$ 0.13 & 0.14 $\pm$ 0.01 & 2.24 $\pm$ 0.24 & -0.96 & -0.94 \\
J032441+341045 & 15.57  & 16.07  & 15.57  & 16.07  & 13.16 $\pm$ 0.03 & 14.85 $\pm$ 0.05 & 0.81 $\pm$ 0.02 & 0.84 $\pm$ 0.02 & 0.28 $\pm$ 0.01 & 0.35 $\pm$ 0.02 & -0.76 & -0.06 \\
J084957+510829 & 19.18  & 19.73  & 19.19  & 19.74  & 53.49 $\pm$ 0.18 & 59.48 $\pm$ 0.34 & 2.86 $\pm$ 0.17 & 3.08 $\pm$ 0.23 & 0.09 $\pm$ 0.01 & 0.11 $\pm$ 0.01 & -0.35 & -0.039 \\
J093241+530633 & 18.54  & 18.65  & 18.54  & 18.65  & 25.93 $\pm$ 0.14 & 18.68 $\pm$ 0.17 & 2.04 $\pm$ 0.07 & 1.52 $\pm$ 0.07 & 0.37 $\pm$ 0.04 & 0.07 $\pm$ 0.00  & -1.16 & -0.45 \\
J093712+500851 & 18.46  & 19.15  & 18.47  & 19.16  & 52.00 $\pm$ 0.13 & 72.59 $\pm$ 0.21 & 2.72 $\pm$ 0.11 & 3.20 $\pm$ 0.16 & 0.12 $\pm$ 0.01 & 1.39 $\pm$ 0.12 & -1.01 & -0.40 \\
J094635+101706 & 19.07  & 19.43 & 19.07  & 19.43 & 11.45 $\pm$ 0.27 & 11.40 $\pm$ 0.54 & 0.87 $\pm$ 0.11 & 0.83 $\pm$ 0.12 & 0.74 $\pm$ 0.07 & 204.20 $\pm$ 7.85 & -0.26 & -0.17 \\
J094857+002226 & 18.20  & 18.46  & 18.21  & 18.46  & 29.49 $\pm$ 0.17 & 17.70 $\pm$ 0.25 & 1.94 $\pm$ 0.06 & 1.24 $\pm$ 0.07 & 0.25 $\pm$ 0.02 & 0.47 $\pm$ 0.02 & -0.59 & -0.82 \\
J122222+041315 & 17.86  & 18.11  & 17.86  & 18.11  & 14.79 $\pm$ 0.14 & 10.62 $\pm$ 0.18 & 1.12 $\pm$ 0.05 & 1.16 $\pm$ 0.07 & 1.01 $\pm$ 0.07 & 342.87 $\pm$ 9.95 & -0.51 & -0.50 \\

\bottomrule
\end{tabular}}

\label{tab:variability}
\end{table*}

\begin{table*}
    \caption{Comparison of intrinsic variability timescales with theoretical accretion disk timescales. The table lists the source name, band, best-fit CARMA $(p,q)$ order, and AIC value. The Break Timescale corresponds to the characteristic variability time ($1/\nu_{break}$) derived from the broken power-law fit to the CARMA PSD, corrected for cosmological redshift ($z$) and relativistic beaming (Lorentz factor). These are compared with the theoretical thermal ($t_{th}$) and viscous ($t_{vis}$) timescales.}
    \label{table:CARMA_fititng}
    \centering
    \begin{tabular}{lllrrrr}
    \toprule
    Name & Band & (p,q) & AIC & Break Timescale & $t_{th}$ & $t_{vis}$ \\
     & & & & (days) & (days) & (days) \\
    \midrule
    J133108+303032 & R & (4,1) & -3862.260 & 1.301 & 0.298 & 29.82 \\
     & G & (3,0) & -4147.910 & 76.653 & & \\
    J142105+385522 & R & (4,2) & -3972.790 & 0.374 & 0.699 & 69.91 \\
     & G & (4,1) & -3821.080 & 0.806 & & \\
    J144318+472556 & R & (4,2) & -6625.820 & 258.843 & 0.053 & 5.30 \\
     & G & (4,3) & -6949.510 & 14.404 & & \\
    J150506+032631 & R & (4,3) & -1999.030 & 0.034 & 0.092 & 9.22 \\
     & G & (4,1) & -980.540 & 7.484 & & \\
    J164100+345453 & R & (4,2) & -6557.780 & 1.393 & 0.033 & 3.27 \\
     & G & (4,3) & -5790.280 & 0.059 & & \\
    J164442+261913 & R & (3,2) & -5082.790 & 0.356 & 0.116 & 11.60 \\
     & G & (3,2) & -4344.840 & 0.171 & & \\
    J211852-073228 & R & (3,2) & -803.110 & 6.903 & 0.037 & 3.67 \\
     & G & (3,2) & -480.720 & 6.214 & & \\
    J003159+093618 & R & (4,1) & -2170.100 & 65.226 & 0.012 & 1.19 \\
     & G & (4,2) & -1449.570 & 201.383 & & \\
    J032441+341045 & R & (4,3) & -9785.370 & 0.111 & 0.046 & 4.62 \\
     & G & (4,3) & -4778.960 & 6.063 & & \\
    J084957+510829 & R & (4,2) & -4321.680 & 3.564 & 0.090 & 9.01 \\
     & G & (4,2) & -2029.790 & 2.168 & & \\
    J093241+530633 & R & (4,1) & -3501.920 & 17.432 & 0.231 & 23.15 \\
     & G & (3,2) & -2330.800 & 26.277 & & \\
    J093712+500851 & R & (2,1) & -3292.250 & 4.297 & 0.084 & 8.40 \\
     & G & (4,1) & -1189.110 & 4.090 & & \\
    J094635+101706 & R & (3,1) & -2246.990 & 64.436 & 0.367 & 36.69 \\
     & G & (3,1) & -735.160 & 36.795 & & \\
    J094857+002226 & R & (2,1) & -1319.710 & 25.392 & 0.350 & 35.04 \\
     & G & (4,3) & -687.960 & 23.630 & & \\
    J122222+041315 & R & (3,1) & -2342.150 & 2.763 & 1.639 & 163.88 \\
     & G & (3,1) & -1839.070 & 3.814 & & \\
    \bottomrule
\end{tabular}
\end{table*}

\begin{table}
\centering
\caption{Results from the Damped Random Walk (DRW) modeling of the ZTF light curves. The table lists the best-fit parameters for both R and G bands: $\sigma$ represents the asymptotic variability amplitude, and $\tau$ denotes the characteristic damping timescale in days (source rest frame).}
\label{TABDRW}
\begin{tabular}{@{}lcccc@{}}
\toprule
Name & $\sigma_r$ & $\tau_r$ (days) & $\sigma_g$ & $\tau_g$ (days) \\
\midrule
J133108+303032 & 0.67 & 671     & 0.87 & 118   \\
J142105+385522 & 1.33 & 11.1    & 1.18 & 18.5  \\
J144318+472556 & 1.02 & 1100    & 0.96 & 838   \\
J150506+032631 & 2.55 & 7.17    & 1.87 & 14.01    \\
J164100+345453 & 1.06 & 0.004 & 1.21 & 0.04  \\
J164442+261913 & 1.92 & 1.43    & 1.34 & 0.86  \\
J211852-073228 & 1.98 & 0.04    & 2.38 & 2.52  \\
J003159+093618 & 1.01 & 116.03     & 1.22 & 142.02   \\
J032441+341045 & 1.12 & 5.93    & 1.29 & 63.4  \\
J084957+510829 & 2.08 & 9.2     & 1.94 & 6.75  \\
J093241+530633 & 2.47 & 2.85    & 1.89 & 4.63  \\
J093712+500851 & 1.74 & 8.75    & 1.70 & 6.33  \\
J094635+101706 & 1.31 & 25.6    & 1.06 & 26.1  \\
J094857+002226 & 2.97 & 10.2    & 1.41 & 2.79  \\
J122222+041315 & 1.19 & 8.77    & 0.95 & 10.7  \\
\bottomrule
\end{tabular}
\end{table}

\begin{figure}
    \centering
    \includegraphics[width=\linewidth]{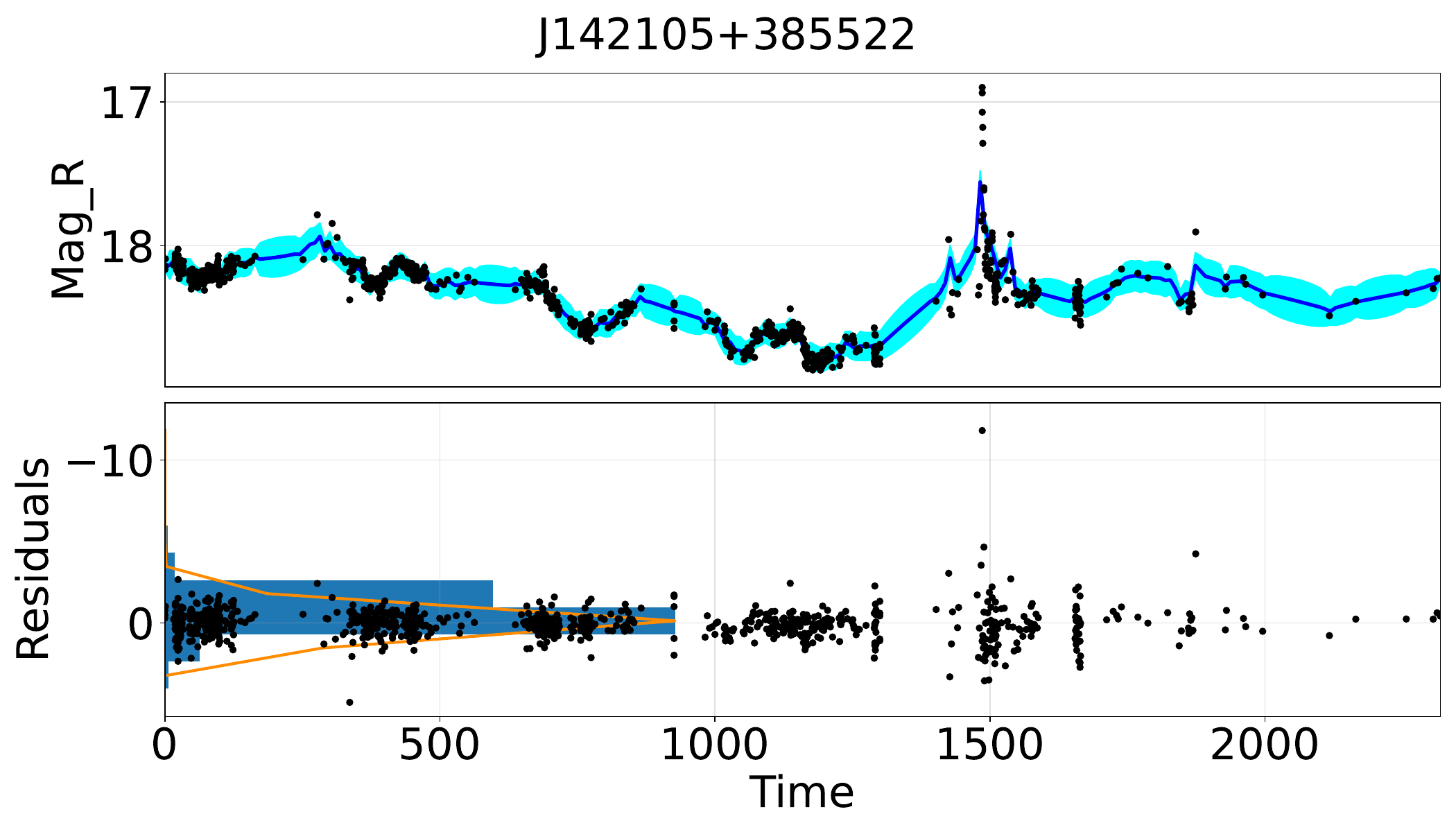}
    \caption{CARMA fitting of J142105+385522, R-band lightcurve. CARMA fitting for the remaining sources is shown in the Appendix \ref{sec:Ind_plots}.}
    \label{fig:enter-label1}
\end{figure}

\begin{figure}
    \centering
    \includegraphics[width=.99\linewidth]{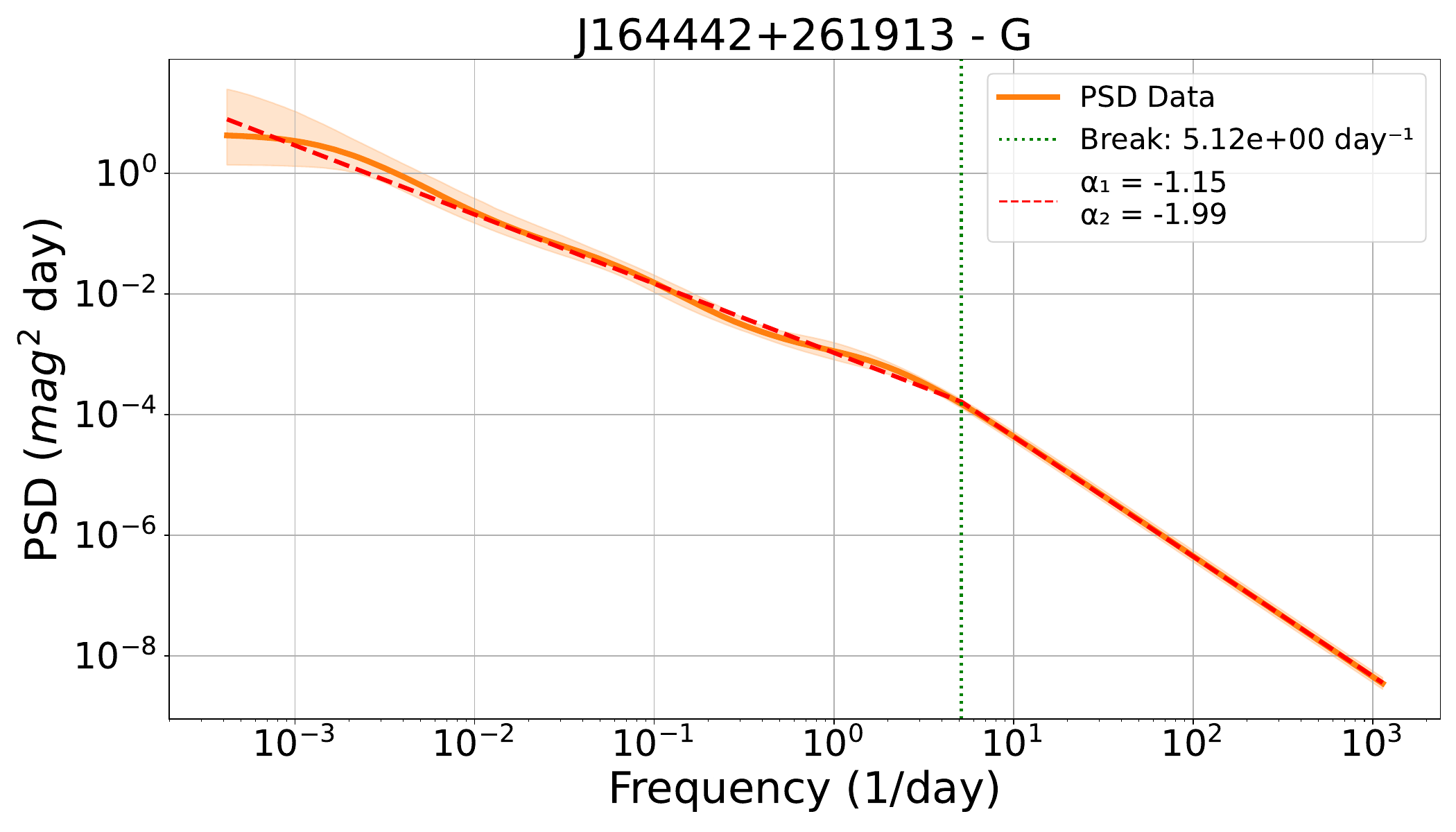}
    \includegraphics[width=.99\linewidth]{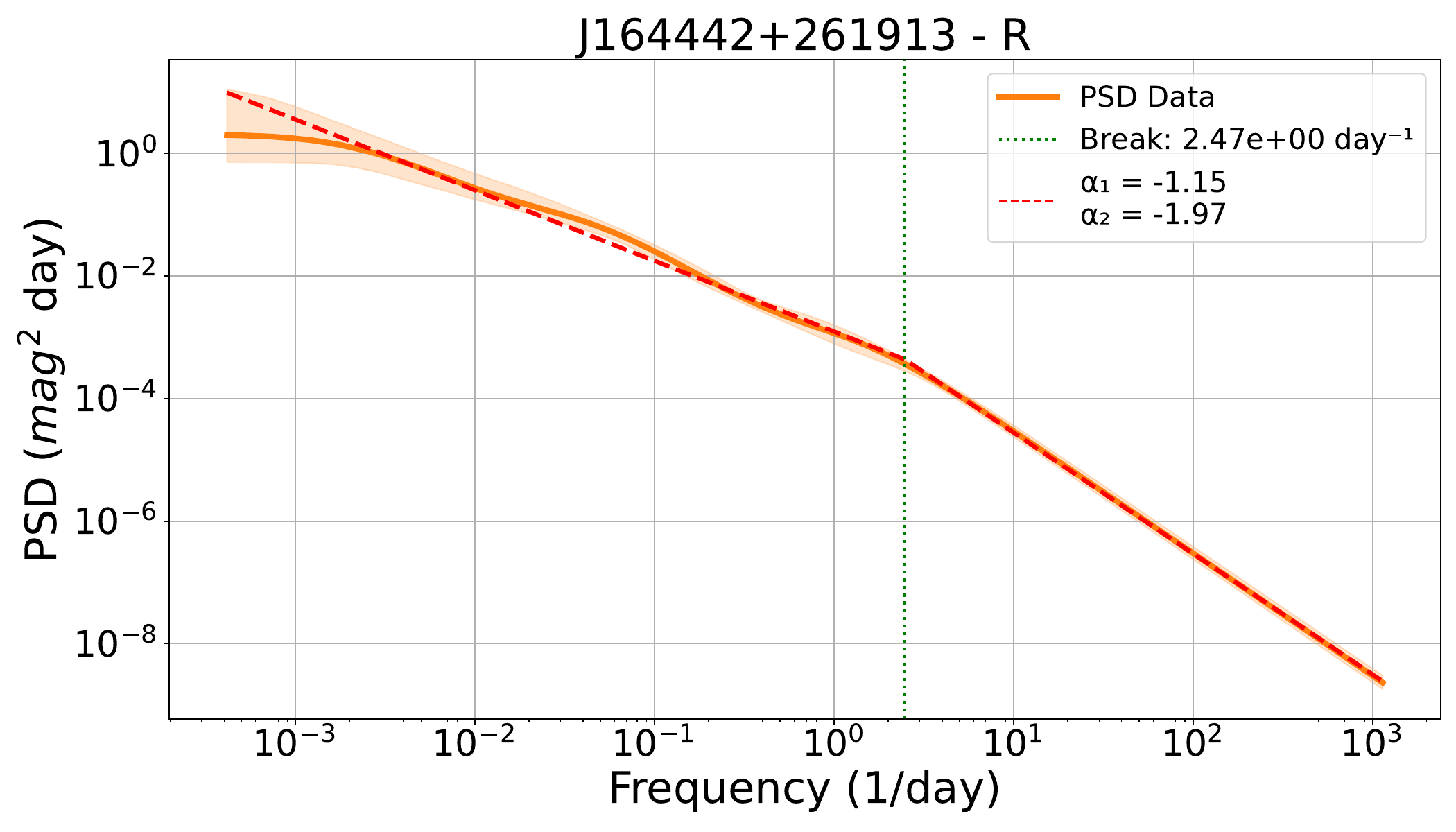}
    \caption{PSD plots of J164442+261913; PSD plots for the remaining sources are shown in the Appendix \ref{sec:Ind_plots}.}
    \label{fig:enter-label3}
\end{figure}

\begin{figure}
    \centering
    \includegraphics[width=\linewidth]{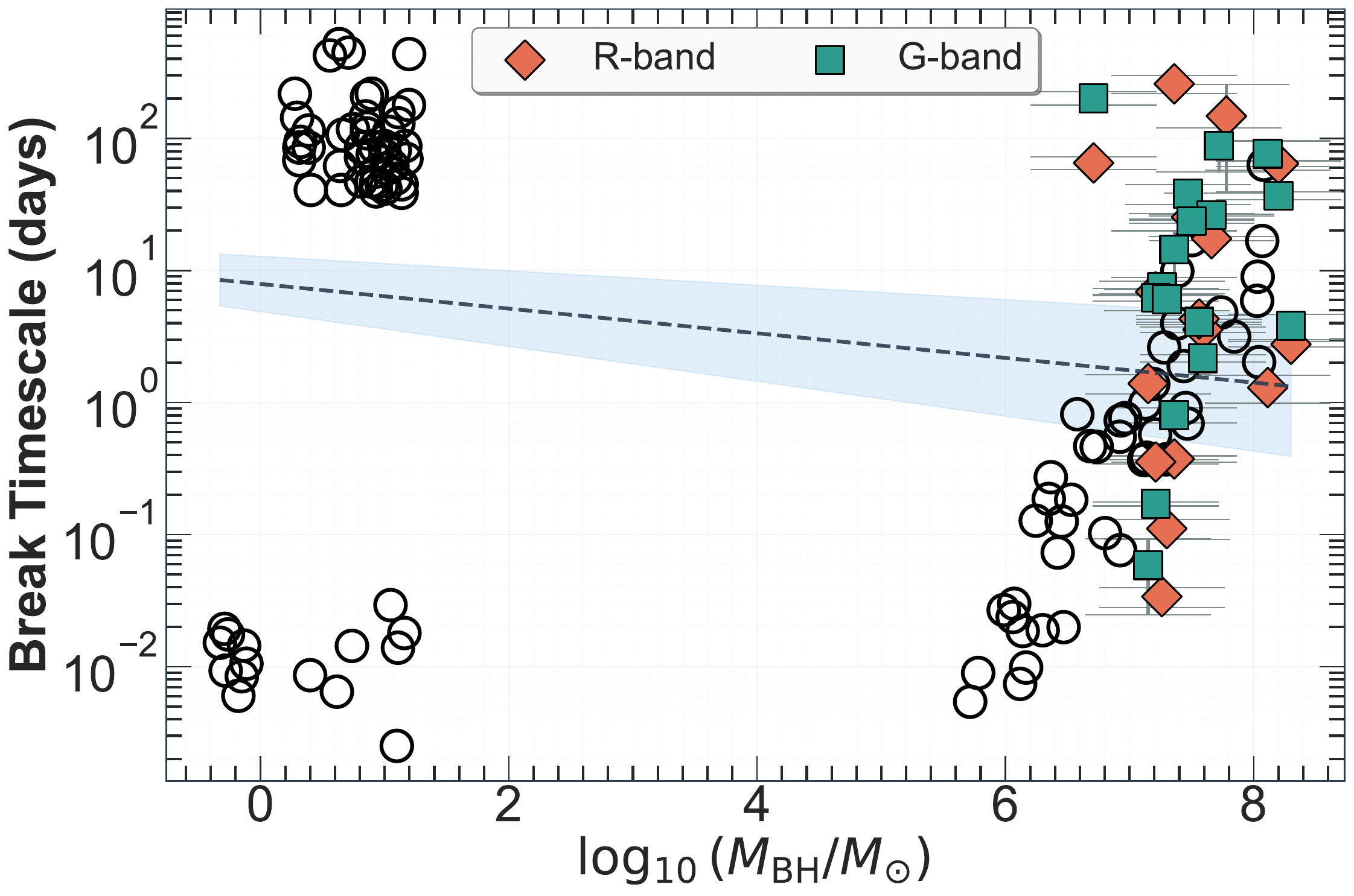}
    \caption{SMBH Mass vs Break Time scale obtained from CARMA PSD. Background white points are the AGN bend timescales from \citep{lefkir2025variability} and Stellar mass XRB, WD and Microquasars from \citep{zhang2025mass,burke2021characteristic,zhang2024discovering}.}
    \label{fig:carma_with_XRBs}
\end{figure}

\begin{figure}
    \centering
     \includegraphics[width=\linewidth]{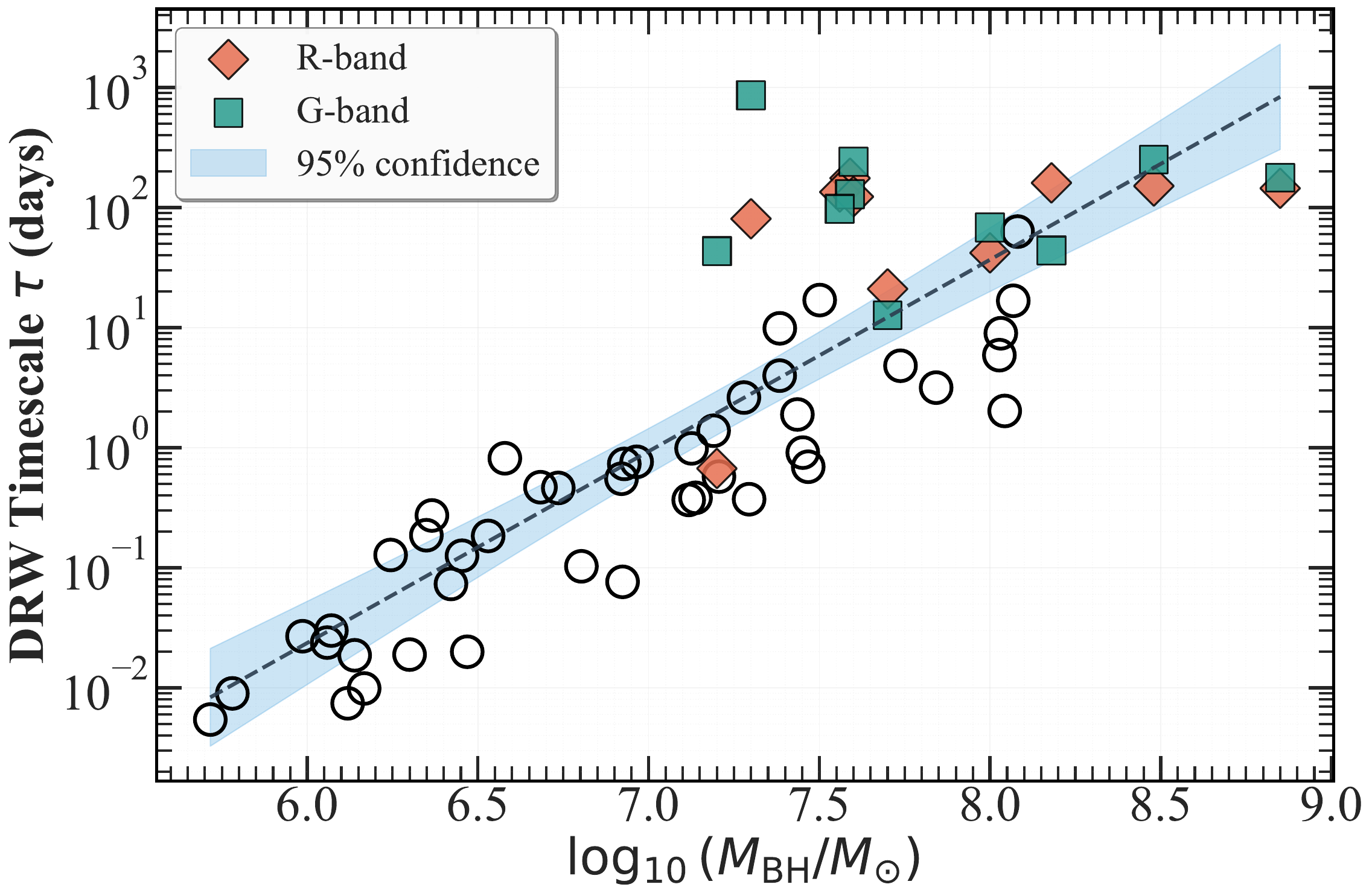}
    \includegraphics[width=\linewidth]{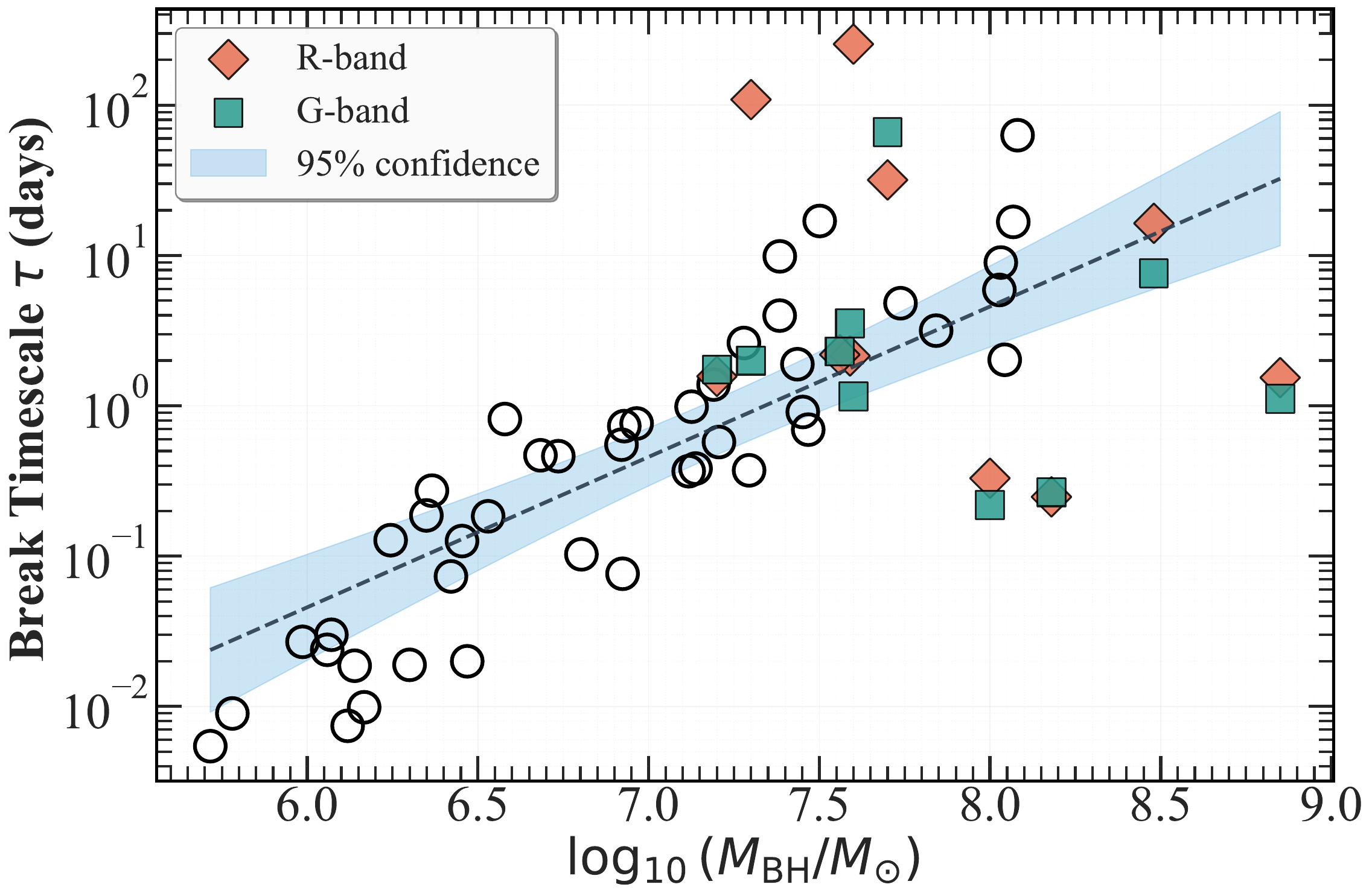}
    \caption{Top panel: SMBH Mass vs Damping Time scale ($\tau \propto M_{BH}^{1.34}$); Bottom panel: SMBH Mass vs Break Time scale obtained from CARMA PSD ($\tau \propto M_{BH}^{1.41}$). Background white points are the bend timescales from \citep{lefkir2025variability}.}
    \label{fig:carma_timescale}
\end{figure}

\begin{figure}
    \centering
     \includegraphics[width=\linewidth]{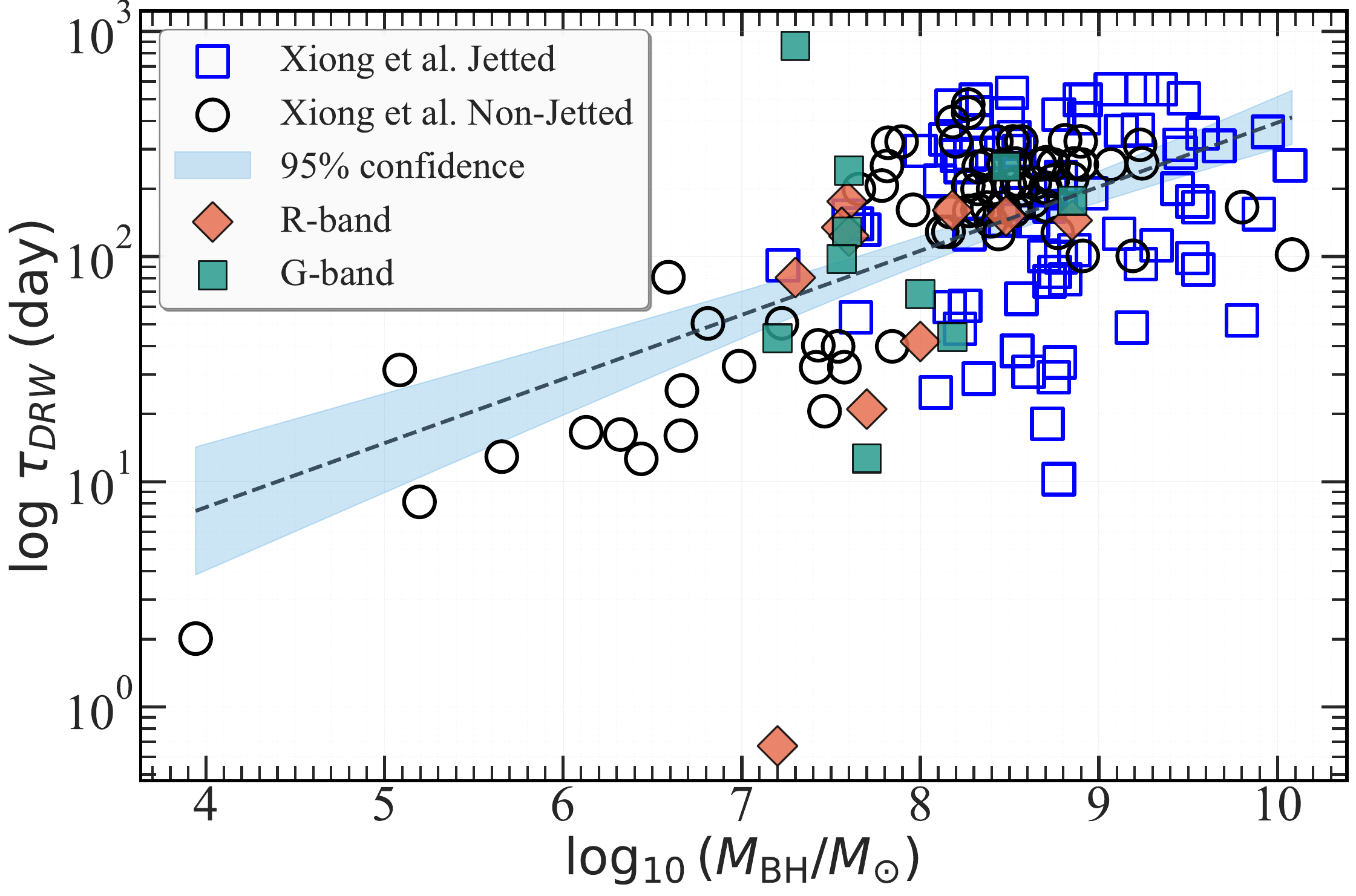}
    \includegraphics[width=\linewidth]{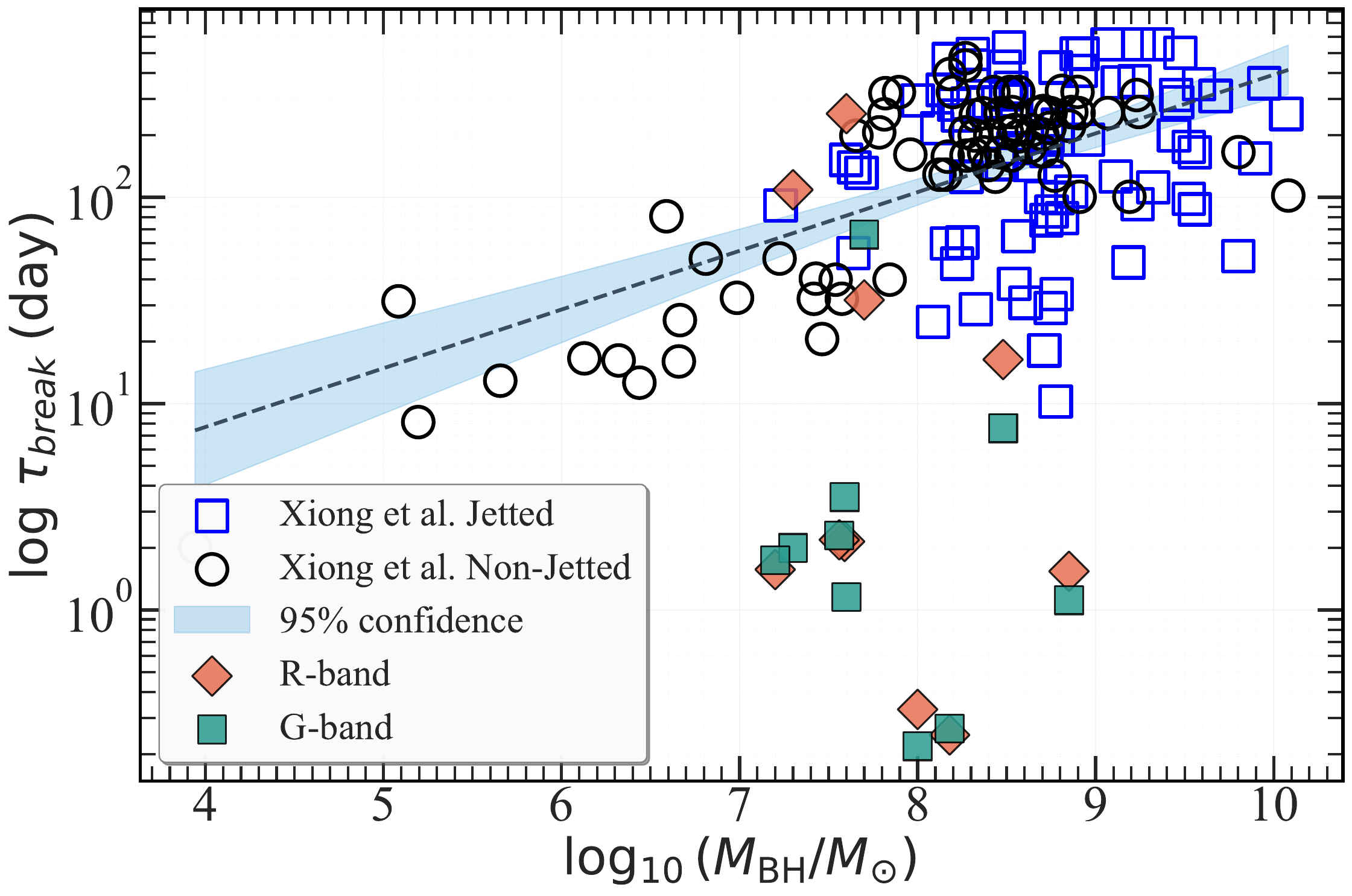}
    \caption{Top panel: SMBH Mass vs Damping Time scale; Bottom panel: SMBH Mass vs Break Time scale obtained from CARMA PSD. Background white points are DRW timescale from \citep{xiong2025characteristic}.}
    \label{fig:jetted_nonjetted_comparison}
\end{figure}

\section{Discussion and Interpretation} \label{results}
High cadence and long-term monitoring with ZTF provides a unique opportunity in the optical waveband to look at both short-term and long-term variability, as well as the nature of variability and the processes that produce it. We selected all the $\gamma-ray$-detected NLSy1 from Fermi-LAT and cross-matched them in the ZTF. Out of the total 23 objects, we identified 22 objects in the ZTF catalog, and of these, 15 have well-sampled ZTF light curves and were selected for further study. We retrieved both $g$ and $r$-band light curves for these objects. An exemplary light curve is shown in Figure \ref{lc}.
The list of all the objects, along with their basic properties, is quoted in Table \ref{tab:nls1_gamma_sample}. The fractional variability estimated in both bands for all the objects was found to be very high. In the $r$-band, it ranges from 4 to nearly 50$\%$, and in the $g$-band, it goes from even higher up to 70$\%$ in one of the objects (Table \ref{tab:variability}). Similarly, the variability amplitude is also observed to be very high for these objects. Our results are consistent with the results obtained in \cite{2025arXiv250503902O}. The high cadence light curve helps us to determine the fastest/shortest variability time present in the ZTF light curves. The time scale in both bands ranges from a fraction of a day (0.07 days) to a few days (5.03 days) and is tabulated in Table \ref{tab:variability}. We can estimate the size of the emission region producing this variability. timescale using the expression,$R$ is given by $R = c\,\tau_{\mathrm{var}}\,\delta / (1 + z)$. We take the value of $\delta$ = 15, an average from the \cite{2019ApJ...872..169P} and the mean redshift, z = 0.7 from \cite{2024MNRAS.527.7055P}, and estimated the size of the emission region in the range of 1.60$\times$10$^{15}$ to 1.15$\times$10$^{17}$ cm. This is also the range most commonly observed in blazars from gamma-ray studies.

Using the ZTF $g$- and $ r$-band light curves, we have also investigated the color-color variability. Most of the objects (9 out of 15) in the sample exhibit the bluer-when-brighter trend, while 5 objects show the redder-when-brighter (RWB) trend, and in one object, the trend is inconclusive (J164442+261913). All the colour plots are shown in the Appendix \ref{sec:Ind_plots}. In some of the objects, a strong trend is observed; this can be easily seen in the appendix plots. However, our results differ from the study done in \cite{2025arXiv250503902O}, where they show that $g-r$ color in most of the $\gamma$NLSy1 (almost 47$\%$ in their sample) lacks clear correlation. In the literature, the studies have shown that RWB trends are mostly seen by FSRQs, whereas BWB trend is seen in BL Lac objects \cite{Bonning_2012, 10.1093/pasj/63.3.327}. The RWB behaviour implies a redder jet emission, and the BWB trend is believed to be produced by the disc emission. In this context, our results show that the radiation of some objects is dominated by the jet, while in other cases, it is dominated by the disk, once again demonstrating that these objects are ideal for studying the disk-jet coupling. We have found that a strong flux-rms relation exists in both ZTF bands, implying that a multiplicative process (i.e., shocks and turbulence) driven by jets dominates over accretion-driven fluctuations, causing rapid optical variability. 

PSD analysis derived from DRW and CARMA modeling reveals a spectral break, providing estimates of the damping timescales present in the system. To correlate the characteristic time scale estimated here with the physical time scale present in the disk, we compared the damping time scale with the dynamical, thermal, and viscous time scales. This comparison can directly tell us about the origin of variability in these objects and put possible constraints on the disk-jet coupling. The dynamical time scale in the accretion disk can be defined as, $t_{dyn} = 2\times10^3~R^{3/2}~M_8$ seconds, where $R$ = $r/r_s$ is the distance of the hot flow in units of Schwarschild’s radius ($r_s$), and in the case of jet or hot corona it should be much closer to the central SMBH, and $M_8$ represents the BH mass in units of 10$^8$ M$_{\odot}$. 
For a highly spinning BH, we can consider $r = r_s$ and the $t_{dyn}$ for mass distribution 10$^{7}$ to 10$^{8}$ is estimated between a few minutes and a few hours.
It is also believed that the local temperature of the accretion disk can vary over a certain timescale, which is defined as the ratio of the disk's internal energy to the heating or cooling rate. Considering that NLSy1 hosts the standard accretion disk, the thermal time scale can be defined as $t_{th} = t_{dyn}/\alpha$, where $\alpha$ is the viscosity parameter, which is taken to be 0.1 \citep{2006ASPC..350..183W}. Using the value of $t_{dyn}$, the $t_{th}$ is estimated between a few tens of minutes to a few tens of hours. 

On the other hand, the viscous time scale can be defined as the time taken by a perturbation originating at the outer disk to travel to the inner part of the accretion disk. In the standard accretion disk scenario, it is defined as $t_{vis} = t_{th}~ (r/h)^2$, where $h$ represents the height of the accretion disk. In most cases, $h/r$ is taken to be 0.1 or less \citep{2006ASPC..350..183W}. The viscous timescale is estimated to be between a few days and a few hundred days. Comparing these time scales with the break time scale calculated from the CARMA modeling (see Table \ref{table:CARMA_fititng}), we conclude that in some cases, viscous time scales dominate, and in the rest of the objects, thermal time scales match the CARMA break time scale. Since these theoretical timescales are estimated for the accretion disk, and the measured CARMA timescales are from the jet emission, we conclude that the accretion disk variability plays a role in modulating the jet variability. The possible jet-disk coupling has been argued in some of the blazar objects as well. This would not be so preposterous as to propose that NLSy1s are the best candidate to probe the accretion disk jet coupling. As the \cite{1982MNRAS.199..883B} mechanism argues, the magnetic field lines and the ionized plasma in the accretion flow are frozen in the field, and the ionized plasma follows the field lines and gets coupled with the SMBH rotation to launch a jet. Therefore, it is highly possible that the variability in the disk and the jet are also coupled.

Figure~\ref{fig:carma_with_XRBs} illustrates the dependence of the CARMA-derived optical break timescale on black hole mass over a broad dynamical range. The upper-left region is populated by stellar-mass XRBs, which exhibit short characteristic masses but comparatively longer break timescales, while the lower-left corner corresponds to white dwarfs with lower masses and shorter timescales. The high-mass end is dominated by AGNs, which occupy the regime of longer characteristic timescales. Overall, a positive correlation between compact object mass and break timescale is evident, consistent with the expected mass scaling of accretion-driven variability where the break frequency decreases with increasing mass. The dashed trend in log–log space aligns with the framework of scale-invariant accretion physics. The moderate scatter likely reflects differences in accretion rate, disk structure, and emission region (R- versus G-band). Taken together, the distribution across XRBs, white dwarfs, and AGNs strengthens the case for a universal variability mechanism operating across compact accretors spanning many orders of magnitude in mass.

To investigate the physical origin of the optical variability in $\gamma$-NLSy1s, we examine the dependence of the characteristic timescales on the supermassive black hole mass ($M_{BH}$).  Theoretically, if variability arises from accretion disk instabilities, the characteristic timescales, whether thermal, viscous, or dynamical, should scale positively with the mass of the compact object. Figures. \ref{fig:carma_timescale} presents the CARMA-derived break timescales and the DRW damping timescales as a function of $M_{BH}$. For comparison, we overplot the extensive sample of AGN from \citet{lefkir2025variability}, which follows a steep power-law relation with $\tau \propto M_{BH}^{1.34}$ and $\tau \propto M_{BH}^{1.41}$ for the above two cases. The derived timescales for our $\gamma$-NLSy1 sample align closely with established global scaling relations, populating the same parameter space as the background sample within the bounds of intrinsic scatter. This agreement is physically significant; it suggests that despite the presence of relativistic jets in $\gamma$-NLSy1s, the fundamental "optical variability clock" remains primarily regulated by the characteristic dimensions and physics of the accretion flow, mirroring the behavior observed in radio-quiet Seyferts.

\cite{2025ApJS..279....3X} investigated characteristic variability timescale for $\sim$1700 jetted AGNs using ZTF data and updated the scaling relation and revealed that the $\tau$ has dependency on both blackhole mass and jet power ($P_{jet}$). Additionally, they have also quoted how the scaling relation differs with the accretion rates ($\tau_{\rm in} \propto M_{\rm BH}^{0.29} P_{\rm jet}^{-0.30}$, for high accretion rate, and $\tau_{\rm in} \propto M_{\rm BH}^{0.06} P_{\rm jet}^{0.37}$, for low accretion rate), highlighting disk and jet driven activities.
We also included the sample of \cite{2025ApJS..279....3X} and plotted them along with our sample in Fig~\ref{fig:jetted_nonjetted_comparison}. The samples overlap with our sample very much, and the scaling relation of Fig~\ref{fig:carma_timescale} is plotted here to show the relation. The scaling relation shows that the characteristic time scales in non-jetted AGN with low black hole mass are extended to high black hole mass AGNs, and mostly their variability is dominated by the disk rather than the jets. \\ 
The alignment of our observed DRW/CARMA timescales with theoretical viscous and thermal scales ($t_{vis}$ and $t_{th}$) provides empirical weight to the Blandford-Znajek framework in the low-mass regime. If the jet variability were entirely decoupled, we would expect timescales to be dominated by purely relativistic effects or shock-crossing times independent of the disc's thermal structure. Instead, our results suggest a \emph{top-down} modulation, where instabilities in the accretion flow act as the primary driver, regulating the energy injection into the relativistic jet.

\section{Conclusions} \label{con}
In this work, we have investigated long-term optical monitoring of 15 $\gamma$-NLSy1 galaxies using ZTF g- \& r-band observations as a tracer of different physical timescales present in these systems to understand the disk and jet-driven activities. 
\begin{itemize}
    \item These sources are extremely active. We observed high fractional variability, reaching as high as 72\% in the g-band and approximately 53\% in the r-band. This intense variability points to strong activity powered by Doppler-boosted jets, similar to what is seen in blazars.
    \item Our color-magnitude analysis reveals that these galaxies show a mixed class behaviour. Most sources (9 out of 15) become bluer when they get brighter (BWB), resembling BL Lac objects. However, five sources become redder when brighter (RWB), similar to Flat Spectrum Radio Quasars (FSRQs).
    \item The inter-band correlation analysis displays a range of behaviours. While sources like J144318+472556 and J142105+385522 show strong near-zero lag correlations indicative of a single emitting zone (likely the jet), others exhibit weak or "jitter-like" correlations. These weaker correlations suggest a complex blending of signals from the host galaxy, accretion disk, and jet, rather than a solitary dominant component.
    \item The flux histograms for these sources are primarily Gaussian, with four sources specifically exhibiting a bimodal distribution. This bimodality indicates transitions between different activity states, likely representing shifts between disk-dominated and jet-dominated emission regimes or changes in jet power.
    \item We identify a linear correlation between the RMS variability and the mean flux across the sample. This confirms that the variability is driven by multiplicative mechanisms, such as cascade processes or turbulence within the jet/disk system, rather than by independent, additive random fluctuations.
    \item Through CARMA modelling, we extracted characteristic break timescales ranging from minutes to hundreds of days. Crucially, these timescales scale with black hole mass in a manner consistent with theoretical thermal and viscous timescales of the accretion disk. This indicates that despite the presence of a powerful jet, the fundamental temporal behaviour of $\gamma$-NLSy1s is still governed by the physical properties of the accretion flow.
\end{itemize}

\section*{Acknowledgements}
BvS acknowledges this work is based on the research supported in part by the National Research Foundation of South Africa (Ref Numbers 119430 and CSRP23041894484)

\section*{Data Availability}
All the light curves are publicly available on the ZTF webpage (\url{https://irsa.ipac.caltech.edu/Missions/ztf.html}) and can be easily accessed. The estimated results and plots are presented in the paper and can be made available on request.

\bibliographystyle{mnras}
\bibliography{Refs} %

@ARTICLE{2019PASP..131a8003M,
       author = {{Masci}, Frank J. and {Laher}, Russ R. and {Rusholme}, Ben and {Shupe}, David L. and {Groom}, Steven and {Surace}, Jason and {Jackson}, Edward and {Monkewitz}, Serge and {Beck}, Ron and {Flynn}, David and {Terek}, Scott and {Landry}, Walter and {Hacopians}, Eugean and {Desai}, Vandana and {Howell}, Justin and {Brooke}, Tim and {Imel}, David and {Wachter}, Stefanie and {Ye}, Quan-Zhi and {Lin}, Hsing-Wen and {Cenko}, S. Bradley and {Cunningham}, Virginia and {Rebbapragada}, Umaa and {Bue}, Brian and {Miller}, Adam A. and {Mahabal}, Ashish and {Bellm}, Eric C. and {Patterson}, Maria T. and {Juri{\'c}}, Mario and {Golkhou}, V. Zach and {Ofek}, Eran O. and {Walters}, Richard and {Graham}, Matthew and {Kasliwal}, Mansi M. and {Dekany}, Richard G. and {Kupfer}, Thomas and {Burdge}, Kevin and {Cannella}, Christopher B. and {Barlow}, Tom and {Van Sistine}, Angela and {Giomi}, Matteo and {Fremling}, Christoffer and {Blagorodnova}, Nadejda and {Levitan}, David and {Riddle}, Reed and {Smith}, Roger M. and {Helou}, George and {Prince}, Thomas A. and {Kulkarni}, Shrinivas R.},
        title = "{The Zwicky Transient Facility: Data Processing, Products, and Archive}",
      journal = {\pasp},
         year = 2019,
        month = jan,
       volume = {131},
       number = {995},
        pages = {018003},
          doi = {10.1088/1538-3873/aae8ac},
archivePrefix = {arXiv},
       eprint = {1902.01872},
 primaryClass = {astro-ph.IM},
       adsurl = {https://ui.adsabs.harvard.edu/abs/2019PASP..131a8003M}
}

@ARTICLE{2021Sci...373..789B,
       author = {{Burke}, Colin J. and {Shen}, Yue and {Blaes}, Omer and {Gammie}, Charles F. and {Horne}, Keith and {Jiang}, Yan-Fei and {Liu}, Xin and {McHardy}, Ian M. and {Morgan}, Christopher W. and {Scaringi}, Simone and {Yang}, Qian},
        title = "{A characteristic optical variability time scale in astrophysical accretion disks}",
      journal = {Science},
         year = 2021,
        month = aug,
       volume = {373},
       number = {6556},
        pages = {789-792},
          doi = {10.1126/science.abg9933},
archivePrefix = {arXiv},
       eprint = {2108.05389},
 primaryClass = {astro-ph.GA},
       adsurl = {https://ui.adsabs.harvard.edu/abs/2021Sci...373..789B}
}

@ARTICLE{2025ApJ...993...50C,
       author = {{Chen}, Yongyun and {Gu}, Qiusheng and {Fan}, Junhui and {Xiong}, Dingrong and {Yu}, Xiaoling and {Guo}, Xiaotong and {Ding}, Nan and {Yi}, Ting-Feng},
        title = "{The Relation between the Optical Variability Timescale, Magnetic Field of Jets, and Black Hole Spin in Active Galactic Nuclei}",
      journal = {\apj},
         year = 2025,
        month = nov,
       volume = {993},
       number = {1},
          eid = {50},
        pages = {50},
          doi = {10.3847/1538-4357/ae08ac},
archivePrefix = {arXiv},
       eprint = {2510.21263},
 primaryClass = {astro-ph.HE},
       adsurl = {https://ui.adsabs.harvard.edu/abs/2025ApJ...993...50C}
}

@ARTICLE{2013ApJ...770...60S,
       author = {{Shimizu}, T. Taro and {Mushotzky}, Richard F.},
        title = "{The First Hard X-Ray Power Spectral Density Functions of Active Galactic Nucleus}",
      journal = {\apj},
         year = 2013,
        month = jun,
       volume = {770},
       number = {1},
          eid = {60},
        pages = {60},
          doi = {10.1088/0004-637X/770/1/60},
archivePrefix = {arXiv},
       eprint = {1304.7002},
 primaryClass = {astro-ph.HE},
       adsurl = {https://ui.adsabs.harvard.edu/abs/2013ApJ...770...60S}
}

@ARTICLE{2002MNRAS.332..231U,
       author = {{Uttley}, P. and {McHardy}, I.~M. and {Papadakis}, I.~E.},
        title = "{Measuring the broad-band power spectra of active galactic nuclei with RXTE}",
      journal = {\mnras},
         year = 2002,
        month = may,
       volume = {332},
       number = {1},
        pages = {231-250},
          doi = {10.1046/j.1365-8711.2002.05298.x},
archivePrefix = {arXiv},
       eprint = {astro-ph/0201134},
 primaryClass = {astro-ph},
       adsurl = {https://ui.adsabs.harvard.edu/abs/2002MNRAS.332..231U}}

@book{sas1990sas,
  title={SAS/STAT User's Guide: GLM-VARCOMP},
  author={SAS Institute},
  isbn={9781555443764},
  series={SAS/STAT User's Guide: Version 6},
  url={https://books.google.co.za/books?id=PIRXAAAAMAAJ},
  year={1990},
  publisher={SAS Institute Incorporated}
}

@article{joanes1998comparing,
  title={Comparing measures of sample skewness and kurtosis},
  author={Joanes, Derrick N and Gill, Christine A},
  journal={Journal of the Royal Statistical Society: Series D (The Statistician)},
  volume={47},
  number={1},
  pages={183--189},
  year={1998},
  publisher={Wiley Online Library}
}

@ARTICLE{2026ApJ...996..118C,
       author = {{Chaudhary}, Suvas Chandra and {Prince}, Raj and {van Soelen}, Brian and {Meintjes}, Pieter},
        title = "{Exploring Hard X-Ray Properties of Gamma-Ray Emitting Narrow-line Seyfert I Galaxies through NuSTAR Observations}",
      journal = {\apj},
         year = 2026,
        month = jan,
       volume = {996},
       number = {2},
          eid = {118},
        pages = {118},
          doi = {10.3847/1538-4357/ae18cc},
archivePrefix = {arXiv},
       eprint = {2504.04492},
 primaryClass = {astro-ph.HE},
       adsurl = {https://ui.adsabs.harvard.edu/abs/2026ApJ...996..118C}}

@ARTICLE{1982MNRAS.199..883B,
       author = {{Blandford}, R.~D. and {Payne}, D.~G.},
        title = "{Hydromagnetic flows from accretion disks and the production of radio jets.}",
      journal = {\mnras},
         year = 1982,
        month = jun,
       volume = {199},
        pages = {883-903},
          doi = {10.1093/mnras/199.4.883},
       adsurl = {https://ui.adsabs.harvard.edu/abs/1982MNRAS.199..883B}
}

@INPROCEEDINGS{2006ASPC..350..183W,
       author = {{Wiita}, P.~J.},
        title = "{Accretion Disks, Jets and Blazar Variability}",
    booktitle = {Blazar Variability Workshop II: Entering the GLAST Era},
         year = 2006,
       editor = {{Miller}, H.~R. and {Marshall}, K. and {Webb}, J.~R. and {Aller}, M.~F.},
       series = {Astronomical Society of the Pacific Conference Series},
       volume = {350},
        month = jul,
        pages = {183},
          doi = {10.48550/arXiv.astro-ph/0507141},
archivePrefix = {arXiv},
       eprint = {astro-ph/0507141},
 primaryClass = {astro-ph},
       adsurl = {https://ui.adsabs.harvard.edu/abs/2006ASPC..350..183W}
}

@article{xiong2025characteristic,
  title={A Characteristic Optical Variability Timescale in Jetted Active Galactic Nuclei: A Large Gamma-Ray Emission Sample},
  author={Xiong, Dingrong and Sun, Mouyuan and Wang, Jun-Xian and Fan, Junhui and Xue, Yongquan and Gu, Minfeng and Chen, Liang and Chen, Yongyun and Ding, Nan and Guo, Fei and others},
  journal={The Astrophysical Journal Supplement Series},
  volume={279},
  number={1},
  pages={3},
  year={2025},
  publisher={IOP Publishing}
}

@inproceedings{bellm2019scheduling,
  title={Scheduling the Zwicky transient facility surveys},
  author={Bellm, Eric and Kulkarni, Shrinivas and Graham, Matthew},
  booktitle={American Astronomical Society Meeting Abstracts\# 233},
  volume={233},
  pages={363--08},
  year={2019}
}

@ARTICLE{2023ApJ...955..121D,
       author = {{Dinesh}, Adithiya and {Bhatta}, Gopal and {Adhikari}, Tek P. and {Mohorian}, Maksym and {Dhital}, Niraj and {Chaudhary}, Suvas C. and {P{\'a}nis}, Radim and {G{\'o}ra}, Dariusz},
        title = "{Constraining X-Ray Variability of the Blazar 3C 273 Using XMM-Newton Observations over Two Decades}",
      journal = {\apj},
         year = 2023,
        month = oct,
       volume = {955},
       number = {2},
          eid = {121},
        pages = {121},
          doi = {10.3847/1538-4357/acf316},
archivePrefix = {arXiv},
       eprint = {2309.00406},
 primaryClass = {astro-ph.HE},
       adsurl = {https://ui.adsabs.harvard.edu/abs/2023ApJ...955..121D}
}

@ARTICLE{2020ApJ...896...95O,
       author = {{Ojha}, Vineet and {Chand}, Hum and {Dewangan}, Gulab Chand and {Rakshit}, Suvendu},
        title = "{A Comparison of X-Ray Photon Indices among the Narrow- and Broad-line Seyfert 1 Galaxies}",
      journal = {\apj},
         year = 2020,
        month = jun,
       volume = {896},
       number = {2},
          eid = {95},
        pages = {95},
          doi = {10.3847/1538-4357/ab94ac},
archivePrefix = {arXiv},
       eprint = {2005.08352},
 primaryClass = {astro-ph.HE},
       adsurl = {https://ui.adsabs.harvard.edu/abs/2020ApJ...896...95O}
}

@ARTICLE{2025arXiv251213569C,
       author = {{Chaudhary}, Suvas Chandra and {Prince}, Raj and {van Soelen}, Brian and {van der Westhuizen}, I.~P.},
        title = "{Unveiling the X-ray Secrets of Fermi-detected Narrow-Line Seyfert 1 Galaxies with XMM-Newton Observations}",
      journal = {arXiv e-prints},
         year = 2025,
        month = dec,
          eid = {arXiv:2512.13569},
        pages = {arXiv:2512.13569},
archivePrefix = {arXiv},
       eprint = {2512.13569},
 primaryClass = {astro-ph.HE},
       adsurl = {https://ui.adsabs.harvard.edu/abs/2025arXiv251213569C}
}

@ARTICLE{2025ApJS..279....3X,
       author = {{Xiong}, Dingrong and {Sun}, Mouyuan and {Wang}, Jun-Xian and {Fan}, Junhui and {Xue}, Yongquan and {Gu}, Minfeng and {Chen}, Liang and {Chen}, Yongyun and {Ding}, Nan and {Guo}, Fei and {Mao}, Jirong and {Ren}, Guowei and {Xue}, Rui and {Yan}, Dahai and {Yang}, Shenbang and {Zhang}, Haiyun and {Bai}, Jinming},
        title = "{A Characteristic Optical Variability Timescale in Jetted Active Galactic Nuclei: A Large Gamma-Ray Emission Sample}",
      journal = {\apjs},
         year = 2025,
        month = jul,
       volume = {279},
       number = {1},
          eid = {3},
        pages = {3},
          doi = {10.3847/1538-4365/add481},
archivePrefix = {arXiv},
       eprint = {2504.15638},
 primaryClass = {astro-ph.GA},
       adsurl = {https://ui.adsabs.harvard.edu/abs/2025ApJS..279....3X}
}

@ARTICLE{2025arXiv251105268Z,
       author = {{Zhang}, Haoyang and {Yang}, Shenbang and {Zhang}, Li and {Dai}, Benzhong},
        title = "{A Mass-Independent Damping Timescale in Black Hole Accretion Systems}",
      journal = {arXiv e-prints},
         year = 2025,
        month = nov,
          eid = {arXiv:2511.05268},
        pages = {arXiv:2511.05268},
archivePrefix = {arXiv},
       eprint = {2511.05268},
 primaryClass = {astro-ph.HE},
       adsurl = {https://ui.adsabs.harvard.edu/abs/2025arXiv251105268Z}
}

@ARTICLE{2025ATel17407....1L,
       author = {{Longo}, F. and {Holzmann Airasca}, A. and {La Mura}, G.},
        title = "{Fermi-LAT detection of renewed gamma-ray activity from the Radio-Loud Narrow-Line Seyfert 1 1H 0323+342}",
      journal = {The Astronomer's Telegram},
         year = 2025,
        month = sep,
       volume = {17407},
        pages = {1},
       adsurl = {https://ui.adsabs.harvard.edu/abs/2025ATel17407....1L}
}

@article{10.1093/mnras/sty3288,
    author = {Ojha, Vineet and Krishna, Gopal and Chand, Hum},
    title = {Intra-night optical monitoring of three γ-ray detected narrow-line Seyfert 1 galaxies},
    journal = {Monthly Notices of the Royal Astronomical Society},
    volume = {483},
    number = {3},
    pages = {3036-3047},
    year = {2018},
    month = {12},
    issn = {0035-8711},
    doi = {10.1093/mnras/sty3288},
    url = {https://doi.org/10.1093/mnras/sty3288},
    eprint = {https://academic.oup.com/mnras/article-pdf/483/3/3036/27212341/sty3288.pdf},
}

@article{10.1093/mnras/sts217,
    author = {Paliya, Vaidehi S. and Stalin, C. S. and Kumar, Brijesh and Kumar, Brajesh and Bhatt, V. K. and Pandey, S. B. and Yadav, R. K. S.},
    title = {Intranight optical variability of γ-ray-loud narrow-line Seyfert 1 galaxies},
    journal = {Monthly Notices of the Royal Astronomical Society},
    volume = {428},
    number = {3},
    pages = {2450-2458},
    year = {2012},
    month = {11},
    issn = {0035-8711},
    doi = {10.1093/mnras/sts217},
    url = {https://doi.org/10.1093/mnras/sts217},
    eprint = {https://academic.oup.com/mnras/article-pdf/428/3/2450/3710238/sts217.pdf},
}

@article{Bonning_2012,
doi = {10.1088/0004-637X/756/1/13},
url = {https://doi.org/10.1088/0004-637X/756/1/13},
year = {2012},
month = {aug},
publisher = {The American Astronomical Society},
volume = {756},
number = {1},
pages = {13},
author = {Bonning, Erin and Megan Urry, C. and Bailyn, Charles and Buxton, Michelle and Chatterjee, Ritaban and Coppi, Paolo and Fossati, Giovanni and Isler, Jedidah and Maraschi, Laura},
title = {SMARTS OPTICAL AND INFRARED MONITORING OF 12 GAMMA-RAY BRIGHT BLAZARS},
journal = {The Astrophysical Journal}
}

@article{10.1093/pasj/63.3.327,
    author = {Ikejiri, Yuki and Uemura, Makoto and Sasada, Mahito and Ito, Ryosuke and Yamanaka, Masayuki and Sakimoto, Kiyoshi and Arai, Akira and Fukazawa, Yasushi and Ohsugi, Takashi and Kawabata, Koji S. and Yoshida, Michitoshi and Sato, Shuji and Kino, Masaru},
    title = {Photopolarimetric Monitoring of Blazars in the Optical and Near-Infrared Bands with the Kanata Telescope. I. Correlations between Flux, Color, and Polarization},
    journal = {Publications of the Astronomical Society of Japan},
    volume = {63},
    number = {3},
    pages = {639-675},
    year = {2011},
    month = {06},
    issn = {0004-6264},
    doi = {10.1093/pasj/63.3.327},
    url = {https://doi.org/10.1093/pasj/63.3.327},
    eprint = {https://academic.oup.com/pasj/article-pdf/63/3/639/54692259/pasj_63_3_639.pdf},
}

@ARTICLE{2025arXiv250503902O,
       author = {{Ojha}, Vineet and {Wu}, Xue-Bing and {Ho}, Luis C.},
        title = "{The Relative Contributions of Accretion Disk versus Jet to the Optical and Mid-infrared Variability of Seyfert Galaxies}",
      journal = {arXiv e-prints},
         year = 2025,
        month = may,
          eid = {arXiv:2505.03902},
        pages = {arXiv:2505.03902},
          doi = {10.48550/arXiv.2505.03902},
archivePrefix = {arXiv},
       eprint = {2505.03902},
 primaryClass = {astro-ph.HE},
       adsurl = {https://ui.adsabs.harvard.edu/abs/2025arXiv250503902O}
}

@ARTICLE{2025ApJ...990...79S,
       author = {{Singh}, Veeresh and {Kumar}, Parveen and {Das}, Avik Kumar and {Ojha}, Vineet},
        title = "{Intra-night Optical Variability and Radio Characteristics of Extremely Radio-loud Narrow-line Seyfert 1 Galaxies}",
      journal = {\apj},
         year = 2025,
        month = sep,
       volume = {990},
       number = {1},
          eid = {79},
        pages = {79},
          doi = {10.3847/1538-4357/adf210},
archivePrefix = {arXiv},
       eprint = {2507.16486},
 primaryClass = {astro-ph.GA},
       adsurl = {https://ui.adsabs.harvard.edu/abs/2025ApJ...990...79S}
}

@ARTICLE{2006MNRAS.365.1067C,
       author = {{Crummy}, J. and {Fabian}, A.~C. and {Gallo}, L. and {Ross}, R.~R.},
        title = "{An explanation for the soft X-ray excess in active galactic nuclei}",
      journal = {\mnras},
         year = 2006,
        month = feb,
       volume = {365},
       number = {4},
        pages = {1067-1081},
          doi = {10.1111/j.1365-2966.2005.09844.x},
archivePrefix = {arXiv},
       eprint = {astro-ph/0511457},
 primaryClass = {astro-ph},
       adsurl = {https://ui.adsabs.harvard.edu/abs/2006MNRAS.365.1067C}
}

@ARTICLE{2017ApJS..229...39R,
       author = {{Rakshit}, Suvendu and {Stalin}, C.~S. and {Chand}, Hum and {Zhang}, Xue-Guang},
        title = "{A Catalog of Narrow Line Seyfert 1 Galaxies from the Sloan Digital Sky Survey Data Release 12}",
      journal = {\apjs},
         year = 2017,
        month = apr,
       volume = {229},
       number = {2},
          eid = {39},
        pages = {39},
          doi = {10.3847/1538-4365/aa6971},
archivePrefix = {arXiv},
       eprint = {1704.07700},
 primaryClass = {astro-ph.GA},
       adsurl = {https://ui.adsabs.harvard.edu/abs/2017ApJS..229...39R}
}

@ARTICLE{2012AJ....143...83X,
       author = {{Xu}, Dawei and {Komossa}, S. and {Zhou}, Hongyan and {Lu}, Honglin and {Li}, Cheng and {Grupe}, Dirk and {Wang}, Jing and {Yuan}, Weimin},
        title = "{Correlation Analysis of a Large Sample of Narrow-line Seyfert 1 Galaxies: Linking Central Engine and Host Properties}",
      journal = {\aj},
         year = 2012,
        month = apr,
       volume = {143},
       number = {4},
          eid = {83},
        pages = {83},
          doi = {10.1088/0004-6256/143/4/83},
archivePrefix = {arXiv},
       eprint = {1201.2810},
 primaryClass = {astro-ph.CO},
       adsurl = {https://ui.adsabs.harvard.edu/abs/2012AJ....143...83X}
}

@ARTICLE{2024MNRAS.527.7055P,
       author = {{Paliya}, Vaidehi S. and {Stalin}, C.~S. and {Dom{\'\i}nguez}, Alberto and {Saikia}, D.~J.},
        title = "{Narrow-line Seyfert 1 galaxies in Sloan Digital Sky Survey: a new optical spectroscopic catalogue}",
      journal = {\mnras},
         year = 2024,
        month = jan,
       volume = {527},
       number = {3},
        pages = {7055-7069},
          doi = {10.1093/mnras/stad3650},
archivePrefix = {arXiv},
       eprint = {2311.13818},
 primaryClass = {astro-ph.GA},
       adsurl = {https://ui.adsabs.harvard.edu/abs/2024MNRAS.527.7055P}
}

@ARTICLE{2019ApJ...872..169P,
       author = {{Paliya}, Vaidehi S. and {Parker}, M.~L. and {Jiang}, J. and {Fabian}, A.~C. and {Brenneman}, L. and {Ajello}, M. and {Hartmann}, D.},
        title = "{General Physical Properties of Gamma-Ray-emitting Narrow-line Seyfert 1 Galaxies}",
      journal = {\apj},
         year = 2019,
        month = feb,
       volume = {872},
       number = {2},
          eid = {169},
        pages = {169},
          doi = {10.3847/1538-4357/ab01ce},
archivePrefix = {arXiv},
       eprint = {1901.07613},
 primaryClass = {astro-ph.HE},
       adsurl = {https://ui.adsabs.harvard.edu/abs/2019ApJ...872..169P}
}

@ARTICLE{2006MNRAS.370..245G,
       author = {{Gallo}, L.~C. and {Edwards}, P.~G. and {Ferrero}, E. and {Kataoka}, J. and {Lewis}, D.~R. and {Ellingsen}, S.~P. and {Misanovic}, Z. and {Welsh}, W.~F. and {Whiting}, M. and {Boller}, Th. and {Brinkmann}, W. and {Greenhill}, J. and {Oshlack}, A.},
        title = "{The spectral energy distribution of PKS 2004-447: a compact steep-spectrum source and possible radio-loud narrow-line Seyfert 1 galaxy}",
      journal = {\mnras},
         year = 2006,
        month = jul,
       volume = {370},
       number = {1},
        pages = {245-254},
          doi = {10.1111/j.1365-2966.2006.10482.x},
archivePrefix = {arXiv},
       eprint = {astro-ph/0604480},
 primaryClass = {astro-ph},
       adsurl = {https://ui.adsabs.harvard.edu/abs/2006MNRAS.370..245G}
}

@ARTICLE{2019A&A...632A.120B,
       author = {{Berton}, M. and {Braito}, V. and {Mathur}, S. and {Foschini}, L. and {Piconcelli}, E. and {Chen}, S. and {Pogge}, R.~W.},
        title = "{Broadband X-ray observations of four gamma-ray narrow-line Seyfert 1 galaxies}",
      journal = {\aap},
         year = 2019,
        month = dec,
       volume = {632},
          eid = {A120},
        pages = {A120},
          doi = {10.1051/0004-6361/201935929},
archivePrefix = {arXiv},
       eprint = {1910.10925},
 primaryClass = {astro-ph.GA},
       adsurl = {https://ui.adsabs.harvard.edu/abs/2019A&A...632A.120B}
}

@ARTICLE{2019ATel13229....1G,
       author = {{Gokus}, Andrea},
        title = "{Fermi LAT detection of a GeV flare from the radio-loud narrow-line Seyfert 1 Galaxy PKS 2004-447}",
      journal = {The Astronomer's Telegram},
         year = 2019,
        month = oct,
       volume = {13229},
        pages = {1},
       adsurl = {https://ui.adsabs.harvard.edu/abs/2019ATel13229....1G}
}

@ARTICLE{2019Galax...7...87D,
       author = {{D'Ammando}, Filippo},
        title = "{Relativistic Jets in Gamma-Ray-Emitting Narrow-Line Seyfert 1 Galaxies}",
      journal = {Galaxies},
         year = 2019,
        month = nov,
       volume = {7},
       number = {4},
          eid = {87},
        pages = {87},
          doi = {10.3390/galaxies7040087},
archivePrefix = {arXiv},
       eprint = {1911.03500},
 primaryClass = {astro-ph.HE},
       adsurl = {https://ui.adsabs.harvard.edu/abs/2019Galax...7...87D}
}

@ARTICLE{2016ApJ...831..168K,
       author = {{Kellermann}, K.~I. and {Condon}, J.~J. and {Kimball}, A.~E. and {Perley}, R.~A. and {Ivezi{\'c}}, {\v{Z}}eljko},
        title = "{Radio-loud and Radio-quiet QSOs}",
      journal = {\apj},
         year = 2016,
        month = nov,
       volume = {831},
       number = {2},
          eid = {168},
        pages = {168},
          doi = {10.3847/0004-637X/831/2/168},
archivePrefix = {arXiv},
       eprint = {1608.04586},
 primaryClass = {astro-ph.GA},
       adsurl = {https://ui.adsabs.harvard.edu/abs/2016ApJ...831..168K}
}

@ARTICLE{2006ApJS..166..128Z,
       author = {{Zhou}, Hongyan and {Wang}, Tinggui and {Yuan}, Weimin and {Lu}, Honglin and {Dong}, Xiaobo and {Wang}, Junxian and {Lu}, Youjun},
        title = "{A Comprehensive Study of 2000 Narrow Line Seyfert 1 Galaxies from the Sloan Digital Sky Survey. I. The Sample}",
      journal = {\apjs},
         year = 2006,
        month = sep,
       volume = {166},
       number = {1},
        pages = {128-153},
          doi = {10.1086/504869},
archivePrefix = {arXiv},
       eprint = {astro-ph/0603759},
 primaryClass = {astro-ph},
       adsurl = {https://ui.adsabs.harvard.edu/abs/2006ApJS..166..128Z}
}

@ARTICLE{1985ApJ...297..166O,
       author = {{Osterbrock}, D.~E. and {Pogge}, R.~W.},
        title = "{The spectra of narrow-line Seyfert 1 galaxies.}",
      journal = {\apj},
         year = 1985,
        month = oct,
       volume = {297},
        pages = {166-176},
          doi = {10.1086/163513},
       adsurl = {https://ui.adsabs.harvard.edu/abs/1985ApJ...297..166O}
}

@ARTICLE{1996A&A...305...53B,
       author = {{Boller}, T. and {Brandt}, W.~N. and {Fink}, H.},
        title = "{Soft X-ray properties of narrow-line Seyfert 1 galaxies.}",
      journal = {\aap},
         year = 1996,
        month = jan,
       volume = {305},
        pages = {53},
          doi = {10.48550/arXiv.astro-ph/9504093},
archivePrefix = {arXiv},
       eprint = {astro-ph/9504093},
 primaryClass = {astro-ph},
       adsurl = {https://ui.adsabs.harvard.edu/abs/1996A&A...305...53B}
}

@ARTICLE{1989ApJ...342..224G,
       author = {{Goodrich}, Robert W.},
        title = "{Spectropolarimetry of ``Narrow-Line'' Seyfert 1 Galaxies}",
      journal = {\apj},
         year = 1989,
        month = jul,
       volume = {342},
        pages = {224},
          doi = {10.1086/167586},
       adsurl = {https://ui.adsabs.harvard.edu/abs/1989ApJ...342..224G}
}

@INPROCEEDINGS{2008RMxAC..32...86K,
       author = {{Komossa}, S.},
        title = "{Narrow-line Seyfert 1 Galaxies}",
    booktitle = {Revista Mexicana de Astronomia y Astrofisica Conference Series},
         year = 2008,
       series = {Revista Mexicana de Astronomia y Astrofisica Conference Series},
       volume = {32},
        month = apr,
        pages = {86-92},
          doi = {10.48550/arXiv.0710.3326},
archivePrefix = {arXiv},
       eprint = {0710.3326},
 primaryClass = {astro-ph},
       adsurl = {https://ui.adsabs.harvard.edu/abs/2008RMxAC..32...86K}
}

@ARTICLE{2004AJ....127.3168B,
       author = {{Botte}, V. and {Ciroi}, S. and {Rafanelli}, P. and {Di Mille}, F.},
        title = "{Exploring Narrow-Line Seyfert 1 Galaxies through the Physical Properties of Their Hosts}",
      journal = {\aj},
         year = 2004,
        month = jun,
       volume = {127},
       number = {6},
        pages = {3168-3179},
          doi = {10.1086/420803},
archivePrefix = {arXiv},
       eprint = {astro-ph/0402627},
 primaryClass = {astro-ph},
       adsurl = {https://ui.adsabs.harvard.edu/abs/2004AJ....127.3168B}
}

@ARTICLE{1992ApJS...80..109B,
       author = {{Boroson}, Todd A. and {Green}, Richard F.},
        title = "{The Emission-Line Properties of Low-Redshift Quasi-stellar Objects}",
      journal = {\apjs},
         year = 1992,
        month = may,
       volume = {80},
        pages = {109},
          doi = {10.1086/191661},
       adsurl = {https://ui.adsabs.harvard.edu/abs/1992ApJS...80..109B}
}

@ARTICLE{2010ApJS..187...64G,
       author = {{Grupe}, Dirk and {Komossa}, Stefanie and {Leighly}, Karen M. and {Page}, Kim L.},
        title = "{The Simultaneous Optical-to-X-Ray Spectral Energy Distribution of Soft X-Ray Selected Active Galactic Nuclei Observed by Swift}",
      journal = {\apjs},
         year = 2010,
        month = mar,
       volume = {187},
       number = {1},
        pages = {64-106},
          doi = {10.1088/0067-0049/187/1/64},
archivePrefix = {arXiv},
       eprint = {1001.3140},
 primaryClass = {astro-ph.CO},
       adsurl = {https://ui.adsabs.harvard.edu/abs/2010ApJS..187...64G}
}

@ARTICLE{2010MNRAS.403..605B,
       author = {{Breedt}, E. and {McHardy}, I.~M. and {Ar{\'e}valo}, P. and {Uttley}, P. and {Sergeev}, S.~G. and {Minezaki}, T. and {Yoshii}, Y. and {Sakata}, Y. and {Lira}, P. and {Chesnok}, N.~G.},
        title = "{Twelve years of X-ray and optical variability in the Seyfert galaxy NGC 4051}",
      journal = {\mnras},
         year = 2010,
        month = apr,
       volume = {403},
       number = {2},
        pages = {605-619},
          doi = {10.1111/j.1365-2966.2009.16146.x},
archivePrefix = {arXiv},
       eprint = {0912.0544},
 primaryClass = {astro-ph.GA},
       adsurl = {https://ui.adsabs.harvard.edu/abs/2010MNRAS.403..605B}
}

@ARTICLE{2024ApJ...973...10D,
       author = {{Dingler}, Ryne and {Smith}, Krista Lynne},
        title = "{Optical Variability Properties of Southern TESS Blazars}",
      journal = {\apj},
         year = 2024,
        month = sep,
       volume = {973},
       number = {1},
          eid = {10},
        pages = {10},
          doi = {10.3847/1538-4357/ad4f87},
archivePrefix = {arXiv},
       eprint = {2406.10346},
 primaryClass = {astro-ph.HE},
       adsurl = {https://ui.adsabs.harvard.edu/abs/2024ApJ...973...10D}
}

@ARTICLE{2025ApJ...985...39S,
       author = {{Sinitsyna}, Vera G. and {Sinitsyna}, Vera Y.},
        title = "{Multiwavelength Long-term Studies of Radio Galaxy NGC 1275}",
      journal = {\apj},
         year = 2025,
        month = may,
       volume = {985},
       number = {1},
          eid = {39},
        pages = {39},
          doi = {10.3847/1538-4357/adc112},
       adsurl = {https://ui.adsabs.harvard.edu/abs/2025ApJ...985...39S}
}

@ARTICLE{1988ApJ...333..646E,
       author = {{Edelson}, R.~A. and {Krolik}, J.~H.},
        title = "{The Discrete Correlation Function: A New Method for Analyzing Unevenly Sampled Variability Data}",
      journal = {\apj},
         year = 1988,
        month = oct,
       volume = {333},
        pages = {646},
          doi = {10.1086/166773},
       adsurl = {https://ui.adsabs.harvard.edu/abs/1988ApJ...333..646E}
}

@ARTICLE{2013arXiv1303.1898P,
       author = {{Pati{\~n}o-{\'A}lvarez}, V. and {Carrami{\~n}ana}, A. and {Carrasco}, L. and {Chavushyan}, V.},
        title = "{A Multiwavelength Cross-Correlation Variability Study of Fermi-LAT Blazars}",
      journal = {arXiv e-prints},
         year = 2013,
        month = mar,
          eid = {arXiv:1303.1898},
        pages = {arXiv:1303.1898},
          doi = {10.48550/arXiv.1303.1898},
archivePrefix = {arXiv},
       eprint = {1303.1898},
 primaryClass = {astro-ph.HE},
       adsurl = {https://ui.adsabs.harvard.edu/abs/2013arXiv1303.1898P}
}

@ARTICLE{2022MNRAS.510.5280M,
       author = {{Mohorian}, Maksym and {Bhatta}, Gopal and {Adhikari}, Tek P. and {Dhital}, Niraj and {P{\'a}nis}, Radim and {Dinesh}, Adithiya and {Chaudhary}, Suvas C. and {Bachchan}, Rajesh K. and {Stuchl{\'\i}k}, Zden{\v{e}}k},
        title = "{X-ray timing and spectral variability properties of blazars S5 0716 + 714, OJ 287, Mrk 501, and RBS 2070}",
      journal = {\mnras},
         year = 2022,
        month = mar,
       volume = {510},
       number = {4},
        pages = {5280-5301},
          doi = {10.1093/mnras/stab3738},
archivePrefix = {arXiv},
       eprint = {2112.11272},
 primaryClass = {astro-ph.HE},
       adsurl = {https://ui.adsabs.harvard.edu/abs/2022MNRAS.510.5280M}
}

@ARTICLE{2025ApJ...981..118B,
       author = {{Bhatta}, Gopal and {Chaudhary}, Suvas Chandra and {Dhital}, Niraj and {Adhikari}, Tek P. and {Mohorian}, Maksym and {Dinesh}, Adithiya and {P{\'a}nis}, Radim and {Neupane}, Raghav and {Maharjan}, Yogesh Singh},
        title = "{Probing X-Ray Timing and Spectral Variability in the Blazar PKS 2155{\textendash}304 over a Decade of XMM-Newton Observations}",
      journal = {\apj},
         year = 2025,
        month = mar,
       volume = {981},
       number = {2},
          eid = {118},
        pages = {118},
          doi = {10.3847/1538-4357/adb0c9},
archivePrefix = {arXiv},
       eprint = {2410.01278},
 primaryClass = {astro-ph.HE},
       adsurl = {https://ui.adsabs.harvard.edu/abs/2025ApJ...981..118B}}

@article{pfister2013good,
  title={Good things peak in pairs: a note on the bimodality coefficient},
  author={Pfister, Roland and Schwarz, Katharina A and Janczyk, Markus and Dale, Rick and Freeman, Jonathan B},
  journal={Frontiers in psychology},
  volume={4},
  pages={700},
  year={2013},
  publisher={Frontiers Media SA}
}

@article{burbidge1974physics,
  title={Physics of compact nonthermal sources. III-Energetic considerations},
  author={Burbidge, G Ri and Jones, TW and O'dell, SL},
  journal={Astrophysical Journal, vol. 193, Oct. 1, 1974, pt. 1. p. 43-54. NSF-supported research;},
  volume={193},
  pages={43--54},
  year={1974}
}

@article{jorstad2013tight,
  title={A tight connection between gamma-ray outbursts and parsec-scale jet activity in the quasar 3c 454.3},
  author={Jorstad, Svetlana G and Marscher, Alan P and Smith, Paul S and Larionov, Valeri M and Agudo, Iv{\'a}n and Gurwell, Mark and Wehrle, Ann E and L{\"a}hteenm{\"a}ki, Anne and Nikolashvili, Maria G and Schmidt, Gary D and others},
  journal={The Astrophysical Journal},
  volume={773},
  number={2},
  pages={147},
  year={2013},
  publisher={IOP Publishing}
}

@article{bellm2019zwicky,
  title={The zwicky transient facility: Surveys and scheduler},
  author={Bellm, Eric C and Kulkarni, Shrinivas R and Barlow, Tom and Feindt, Ulrich and Graham, Matthew J and Goobar, Ariel and Kupfer, Thomas and Ngeow, Chow-Choong and Nugent, Peter and Ofek, Eran and others},
  journal={Publications of the Astronomical Society of the Pacific},
  volume={131},
  number={1000},
  pages={068003},
  year={2019},
  publisher={IOP Publishing}
}

@MISC{Yu2022,
       author = {{Yu}, Weixiang and {Richards}, Gordon T.},
        title = "{EzTao: Easier CARMA Modeling}",
 howpublished = {Astrophysics Source Code Library, record ascl:2201.001},
         year = 2022,
        month = jan,
          eid = {ascl:2201.001},
        pages = {ascl:2201.001},
archivePrefix = {ascl},
       eprint = {2201.001},
       adsurl = {https://ui.adsabs.harvard.edu/abs/2022ascl.soft01001Y}
}

@article{moreno2019stochastic,
  title={Stochastic modeling handbook for optical AGN variability},
  author={Moreno, Jackeline and Vogeley, Michael S and Richards, Gordon T and Yu, Weixiang},
  journal={Publications of the Astronomical Society of the Pacific},
  volume={131},
  number={1000},
  pages={063001},
  year={2019},
  publisher={IOP Publishing}
}

@ARTICLE{1997MNRAS.292..679L,
       author = {{Lyubarskii}, Yu. E.},
        title = "{Flicker noise in accretion discs}",
      journal = {\mnras},
         year = 1997,
        month = dec,
       volume = {292},
       number = {3},
        pages = {679-685},
          doi = {10.1093/mnras/292.3.679},
       adsurl = {https://ui.adsabs.harvard.edu/abs/1997MNRAS.292..679L}
}

@ARTICLE{2020ApJ...897...25B,
       author = {{Bhattacharyya}, Joy and {Ghosh}, Ritesh and {Chatterjee}, Ritaban and {Das}, Nabanita},
        title = "{Blazar Variability: A Study of Nonstationarity and the Flux-Rms Relation}",
      journal = {\apj},
         year = 2020,
        month = jul,
       volume = {897},
       number = {1},
          eid = {25},
        pages = {25},
          doi = {10.3847/1538-4357/ab91a8},
archivePrefix = {arXiv},
       eprint = {2005.05230},
 primaryClass = {astro-ph.HE},
       adsurl = {https://ui.adsabs.harvard.edu/abs/2020ApJ...897...25B}
}

@ARTICLE{2012A&A...548A.123B,
       author = {{Biteau}, J. and {Giebels}, B.},
        title = "{The minijets-in-a-jet statistical model and the rms-flux correlation}",
      journal = {\aap},
         year = 2012,
        month = dec,
       volume = {548},
          eid = {A123},
        pages = {A123},
          doi = {10.1051/0004-6361/201220056},
archivePrefix = {arXiv},
       eprint = {1210.2045},
 primaryClass = {astro-ph.HE},
       adsurl = {https://ui.adsabs.harvard.edu/abs/2012A&A...548A.123B}
}

@ARTICLE{2004A&A...414.1091G,
       author = {{Gleissner}, T. and {Wilms}, J. and {Pottschmidt}, K. and {Uttley}, P. and {Nowak}, M.~A. and {Staubert}, R.},
        title = "{Long term variability of Cyg X-1. II. The rms-flux relation}",
      journal = {\aap},
         year = 2004,
        month = feb,
       volume = {414},
        pages = {1091-1104},
          doi = {10.1051/0004-6361:20031684},
archivePrefix = {arXiv},
       eprint = {astro-ph/0311039},
 primaryClass = {astro-ph},
       adsurl = {https://ui.adsabs.harvard.edu/abs/2004A&A...414.1091G}
}

@article{romero1999optical,
  title={Optical microvariability of southern AGNs},
  author={Romero, GE and Cellone, SA and Combi, JA},
  journal={Astronomy and Astrophysics Supplement Series},
  volume={135},
  number={3},
  pages={477--486},
  year={1999},
  publisher={EDP Sciences}
}

@article{vaughan2003characterizing,
  title={On characterizing the variability properties of X-ray light curves from active galaxies},
  author={Vaughan, Simon and Edelson, R and Warwick, RS and Uttley, P},
  journal={Monthly Notices of the Royal Astronomical Society},
  volume={345},
  number={4},
  pages={1271--1284},
  year={2003},
  publisher={Blackwell Science Ltd Oxford, UK}
}

@ARTICLE{2010A&A...524A..48T,
       author = {{Tluczykont}, M. and {Bernardini}, E. and {Satalecka}, K. and {Clavero}, R. and {Shayduk}, M. and {Kalekin}, O.},
        title = "{Long-term lightcurves from combined unified very high energy {\ensuremath{\gamma}}-ray data}",
      journal = {\aap},
         year = 2010,
        month = dec,
       volume = {524},
          eid = {A48},
        pages = {A48},
          doi = {10.1051/0004-6361/201015193},
archivePrefix = {arXiv},
       eprint = {1010.5659},
 primaryClass = {astro-ph.HE},
       adsurl = {https://ui.adsabs.harvard.edu/abs/2010A&A...524A..48T}
}

@ARTICLE{2018Galax...6..135R,
       author = {{Romoli}, Carlo and {Chakraborty}, Nachiketa and {Dorner}, Daniela and {Taylor}, Andrew M. and {Blank}, Michael},
        title = "{Flux Distribution of Gamma-Ray Emission in Blazars: The Example of Mrk 501}",
      journal = {Galaxies},
         year = 2018,
        month = dec,
       volume = {6},
       number = {4},
          eid = {135},
        pages = {135},
          doi = {10.3390/galaxies6040135},
archivePrefix = {arXiv},
       eprint = {1812.06204},
 primaryClass = {astro-ph.HE},
       adsurl = {https://ui.adsabs.harvard.edu/abs/2018Galax...6..135R}
}

@ARTICLE{2005MNRAS.359..345U,
       author = {{Uttley}, P. and {McHardy}, I.~M. and {Vaughan}, S.},
        title = "{Non-linear X-ray variability in X-ray binaries and active galaxies}",
      journal = {\mnras},
         year = 2005,
        month = may,
       volume = {359},
       number = {1},
        pages = {345-362},
          doi = {10.1111/j.1365-2966.2005.08886.x},
archivePrefix = {arXiv},
       eprint = {astro-ph/0502112},
 primaryClass = {astro-ph},
       adsurl = {https://ui.adsabs.harvard.edu/abs/2005MNRAS.359..345U}
}

@ARTICLE{2004ApJ...612L..21G,
       author = {{Gaskell}, C. Martin},
        title = "{Lognormal X-Ray Flux Variations in an Extreme Narrow-Line Seyfert 1 Galaxy}",
      journal = {\apjl},
         year = 2004,
        month = sep,
       volume = {612},
       number = {1},
        pages = {L21-L24},
          doi = {10.1086/424565},
       adsurl = {https://ui.adsabs.harvard.edu/abs/2004ApJ...612L..21G}
}

@ARTICLE{2009A&A...503..797G,
       author = {{Giebels}, B. and {Degrange}, B.},
        title = "{Lognormal variability in BL Lacertae}",
      journal = {\aap},
         year = 2009,
        month = sep,
       volume = {503},
       number = {3},
        pages = {797-799},
          doi = {10.1051/0004-6361/200912303},
archivePrefix = {arXiv},
       eprint = {0907.2425},
 primaryClass = {astro-ph.CO},
       adsurl = {https://ui.adsabs.harvard.edu/abs/2009A&A...503..797G}
}

@ARTICLE{2015MNRAS.450..541A,
       author = {{Agarwal}, Aditi and {Gupta}, Alok C.},
        title = "{Multiband optical variability studies of BL Lacertae}",
      journal = {\mnras},
         year = 2015,
        month = jun,
       volume = {450},
       number = {1},
        pages = {541-551},
          doi = {10.1093/mnras/stv625},
archivePrefix = {arXiv},
       eprint = {1506.00574},
 primaryClass = {astro-ph.GA},
       adsurl = {https://ui.adsabs.harvard.edu/abs/2015MNRAS.450..541A}
}

@ARTICLE{1998MNRAS.299...47M,
       author = {{Massaro}, E. and {Nesci}, R. and {Maesano}, M. and {Montagni}, F. and {D'Alessio}, F.},
        title = "{Fast variability of BL Lacertae at 1mum}",
      journal = {\mnras},
         year = 1998,
        month = aug,
       volume = {299},
       number = {1},
        pages = {47-50},
          doi = {10.1046/j.1365-8711.1998.01696.x},
       adsurl = {https://ui.adsabs.harvard.edu/abs/1998MNRAS.299...47M}
}

@ARTICLE{2012ApJ...756...13B,
       author = {{Bonning}, Erin and {Urry}, C. Megan and {Bailyn}, Charles and {Buxton}, Michelle and {Chatterjee}, Ritaban and {Coppi}, Paolo and {Fossati}, Giovanni and {Isler}, Jedidah and {Maraschi}, Laura},
        title = "{SMARTS Optical and Infrared Monitoring of 12 Gamma-Ray Bright Blazars}",
      journal = {\apj},
         year = 2012,
        month = sep,
       volume = {756},
       number = {1},
          eid = {13},
        pages = {13},
          doi = {10.1088/0004-637X/756/1/13},
archivePrefix = {arXiv},
       eprint = {1201.4380},
 primaryClass = {astro-ph.HE},
       adsurl = {https://ui.adsabs.harvard.edu/abs/2012ApJ...756...13B}
}

@ARTICLE{2011JApA...32...87G,
       author = {{Gu}, Minfeng and {Ai}, Y.~L.},
        title = "{Spectral Variability of FSRQs}",
      journal = {Journal of Astrophysics and Astronomy},
         year = 2011,
        month = jun,
       volume = {32},
       number = {1-2},
        pages = {87-90},
          doi = {10.1007/s12036-011-9051-2},
archivePrefix = {arXiv},
       eprint = {1101.2258},
 primaryClass = {astro-ph.CO},
       adsurl = {https://ui.adsabs.harvard.edu/abs/2011JApA...32...87G}
}

@ARTICLE{2004A&A...421...83R,
       author = {{Ram{\'\i}rez}, A. and {de Diego}, J.~A. and {Dultzin-Hacyan}, D. and {Gonz{\'a}lez-P{\'e}rez}, J.~N.},
        title = "{Optical variability of <ASTROBJ>PKS 0736+017</ASTROBJ>}",
      journal = {\aap},
         year = 2004,
        month = jul,
       volume = {421},
        pages = {83-89},
          doi = {10.1051/0004-6361:20034449},
archivePrefix = {arXiv},
       eprint = {astro-ph/0401626},
 primaryClass = {astro-ph},
       adsurl = {https://ui.adsabs.harvard.edu/abs/2004A&A...421...83R}
}

@ARTICLE{2023MNRAS.519.5263Z,
       author = {{Zhang}, Bing-Kai and {Tang}, Wei-Feng and {Wang}, Chun-Xiao and {Wu}, Qi and {Jin}, Min and {Dai}, Ben-Zhong and {Zhu}, Feng-Rong},
        title = "{The optical spectral features of 27 Fermi blazars}",
      journal = {\mnras},
         year = 2023,
        month = mar,
       volume = {519},
       number = {4},
        pages = {5263-5270},
          doi = {10.1093/mnras/stac3795},
archivePrefix = {arXiv},
       eprint = {2212.12331},
 primaryClass = {astro-ph.HE},
       adsurl = {https://ui.adsabs.harvard.edu/abs/2023MNRAS.519.5263Z}
}

@ARTICLE{2014ApJ...797...19R,
       author = {{Ruan}, J.~J. and {Anderson}, S.~F. and {Plotkin}, R.~M. and {Brandt}, W.~N. and {Burnett}, T.~H. and {Myers}, A.~D. and {Schneider}, D.~P.},
        title = "{The Nature of Transition Blazars}",
      journal = {\apj},
         year = 2014,
        month = dec,
       volume = {797},
       number = {1},
          eid = {19},
        pages = {19},
          doi = {10.1088/0004-637X/797/1/19},
archivePrefix = {arXiv},
       eprint = {1410.1539},
 primaryClass = {astro-ph.HE},
       adsurl = {https://ui.adsabs.harvard.edu/abs/2014ApJ...797...19R}
}

@ARTICLE{2017ApJS..229...21X,
       author = {{Xiong}, Dingrong and {Bai}, Jinming and {Zhang}, Haojing and {Fan}, Junhui and {Gu}, Minfeng and {Yi}, Tingfeng and {Zhang}, Xiong},
        title = "{Multicolor Optical Monitoring of the Quasar 3C 273 from 2005 to 2016}",
      journal = {\apjs},
         year = 2017,
        month = apr,
       volume = {229},
       number = {2},
          eid = {21},
        pages = {21},
          doi = {10.3847/1538-4365/aa64d2},
archivePrefix = {arXiv},
       eprint = {1703.01645},
 primaryClass = {astro-ph.HE},
       adsurl = {https://ui.adsabs.harvard.edu/abs/2017ApJS..229...21X}
}

@ARTICLE{2022RAA....22g5001X,
       author = {{Xin}, Yu-Xin and {Xiong}, Ding-Rong and {Bai}, Jin-Ming and {Liu}, Hong-Tao and {Lu}, Kai-Xing and {Mao}, Ji-Rong},
        title = "{Multicolor Optical Monitoring of the {\ensuremath{\gamma}} -Ray Emitting Narrow-line Seyfert 1 Galaxy PMN J0948+0022 from 2020 to 2021}",
      journal = {Research in Astronomy and Astrophysics},
         year = 2022,
        month = jul,
       volume = {22},
       number = {7},
          eid = {075001},
        pages = {075001},
          doi = {10.1088/1674-4527/ac684e},
       adsurl = {https://ui.adsabs.harvard.edu/abs/2022RAA....22g5001X}
}

@ARTICLE{2011PASJ...63..639I,
       author = {{Ikejiri}, Yuki and {Uemura}, Makoto and {Sasada}, Mahito and {Ito}, Ryosuke and {Yamanaka}, Masayuki and {Sakimoto}, Kiyoshi and {Arai}, Akira and {Fukazawa}, Yasushi and {Ohsugi}, Takashi and {Kawabata}, Koji S. and {Yoshida}, Michitoshi and {Sato}, Shuji and {Kino}, Masaru},
        title = "{Photopolarimetric Monitoring of Blazars in the Optical and Near-Infrared Bands with the Kanata Telescope. I. Correlations between Flux, Color, and Polarization}",
      journal = {\pasj},
         year = 2011,
        month = jun,
       volume = {63},
        pages = {639},
          doi = {10.1093/pasj/63.3.327},
archivePrefix = {arXiv},
       eprint = {1105.0255},
 primaryClass = {astro-ph.HE},
       adsurl = {https://ui.adsabs.harvard.edu/abs/2011PASJ...63..639I}
}

@ARTICLE{2017ApJ...844..107I,
       author = {{Isler}, Jedidah C. and {Urry}, C.~M. and {Coppi}, P. and {Bailyn}, C. and {Brady}, M. and {MacPherson}, E. and {Buxton}, M. and {Hasan}, I.},
        title = "{A Consolidated Framework of the Color Variability in Blazars: Long-term Optical/Near-infrared Observations of 3C 279}",
      journal = {\apj},
         year = 2017,
        month = aug,
       volume = {844},
       number = {2},
          eid = {107},
        pages = {107},
          doi = {10.3847/1538-4357/aa79fc},
archivePrefix = {arXiv},
       eprint = {1706.09891},
 primaryClass = {astro-ph.HE},
       adsurl = {https://ui.adsabs.harvard.edu/abs/2017ApJ...844..107I}
}

@ARTICLE{2022MNRAS.510.1791N,
       author = {{Negi}, Vibhore and {Joshi}, Ravi and {Chand}, Krishan and {Chand}, Hum and {Wiita}, Paul and {Ho}, Luis C. and {Singh}, Ravi S.},
        title = "{Optical flux and colour variability of blazars in the ZTF survey}",
      journal = {\mnras},
         year = 2022,
        month = feb,
       volume = {510},
       number = {2},
        pages = {1791-1800},
          doi = {10.1093/mnras/stab3591},
archivePrefix = {arXiv},
       eprint = {2112.00790},
 primaryClass = {astro-ph.GA},
       adsurl = {https://ui.adsabs.harvard.edu/abs/2022MNRAS.510.1791N}
}

@ARTICLE{2003MmSAI..74..963V,
       author = {{Vagnetti}, F. and {Trevese}, D.},
        title = "{Color Variability of AGNs}",
      journal = {\memsai},
         year = 2003,
        month = jan,
       volume = {74},
        pages = {963},
          doi = {10.48550/arXiv.astro-ph/0212474},
archivePrefix = {arXiv},
       eprint = {astro-ph/0212474},
 primaryClass = {astro-ph},
       adsurl = {https://ui.adsabs.harvard.edu/abs/2003MmSAI..74..963V}
}

@INPROCEEDINGS{1997ASSL..218..163A,
       author = {{Alexander}, Tal},
        title = "{Is AGN Variability Correlated with Other AGN Properties? ZDCF Analysis of Small Samples of Sparse Light Curves}",
    booktitle = {Astronomical Time Series},
         year = 1997,
       editor = {{Maoz}, D. and {Sternberg}, A. and {Leibowitz}, E.~M.},
       series = {Astrophysics and Space Science Library},
       volume = {218},
        month = jan,
        pages = {163},
          doi = {10.1007/978-94-015-8941-3_14},
       adsurl = {https://ui.adsabs.harvard.edu/abs/1997ASSL..218..163A}
}

@misc{2014ascl.soft04002A,
       author = {{Alexander}, Tal},
        title = "{ZDCF: Z-Transformed Discrete Correlation Function}",
 howpublished = {Astrophysics Source Code Library, record ascl:1404.002},
         year = 2014,
        month = apr,
          eid = {ascl:1404.002},
       adsurl = {https://ui.adsabs.harvard.edu/abs/2014ascl.soft04002A}
}

@article{zhang2025mass,
  title={A Mass-Independent Damping Timescale in Black Hole Accretion Systems},
  author={Zhang, Haoyang and Yang, Shenbang and Zhang, Li and Dai, Benzhong},
  journal={arXiv preprint arXiv:2511.05268},
  year={2025}
}

@article{Kelly_2014,
	doi = {10.1088/0004-637x/788/1/33},
	url = {https://doi.org/10.1088/0004-637x/788/1/33},
	year = 2014,
	month = {may},
	publisher = {American Astronomical Society},
	volume = {788},
	number = {1},
	pages = {33},
	author = {Brandon C. Kelly and Andrew C. Becker and Malgosia Sobolewska and Aneta Siemiginowska and Phil Uttley},
	title = {{FLEXIBLE} {AND} {SCALABLE} {METHODS} {FOR} {QUANTIFYING} {STOCHASTIC} {VARIABILITY} {IN} {THE} {ERA} {OF} {MASSIVE} {TIME}-{DOMAIN} {ASTRONOMICAL} {DATA} {SETS}},
	journal = {The Astrophysical Journal}}

@INPROCEEDINGS{2018rnls.confE..15K,
       author = {{Komossa}, S.},
        title = "{Multi-wavelength properties of radio-loud Narrow-line Seyfert 1 galaxies}",
    booktitle = {Revisiting Narrow-Line Seyfert 1 Galaxies and their Place in the Universe},
         year = 2018,
        month = apr,
          eid = {15},
        pages = {15},
          doi = {10.22323/1.328.0015},
archivePrefix = {arXiv},
       eprint = {1807.03666},
 primaryClass = {astro-ph.GA},
       adsurl = {https://ui.adsabs.harvard.edu/abs/2018rnls.confE..15K}
}

@misc{2020yCat..22470033A,
       author = {{Abdollahi}, S. and {Acero}, F. and {Ackermann}, M. and {Ajello}, M. and {Atwood}, W.~B. and {Axelsson}, M. and {Baldini}, L. and {Ballet}, J. and {Barbiellini}, G. and {Bastieri}, D. and {Becerra Gonzalez}, J. and {Bellazzini}, R. and {Berretta}, A. and {Bissaldi}, E. and {Blandford}, R.~D. and {Bloom}, E.~D. and {Bonino}, R. and {Bottacini}, E. and {Brandt}, T.~J. and {Bregeon}, J. and {Bruel}, P. and {Buehler}, R. and {Burnett}, T.~H. and {Buson}, S. and {Cameron}, R.~A. and {Caputo}, R. and {Caraveo}, P.~A. and {Casandjian}, J.~M. and {Castro}, D. and {Cavazzuti}, E. and {Charles}, E. and {Chaty}, S. and {Chen}, S. and {Cheung}, C.~C. and {Chiaro}, G. and {Ciprini}, S. and {Cohen-Tanugi}, J. and {Cominsky}, L.~R. and {Coronado-Blazquez}, J. and {Costantin}, D. and {Cuoco}, A. and {Cutini}, S. and {D'Ammando}, F. and {Deklotz}, M. and {de La Torre Luque}, P. and {de Palma}, F. and {Desai}, A. and {Digel}, S.~W. and {di Lalla}, N. and {di Mauro}, M. and {di Venere}, L. and {Dominguez}, A. and {Dumora}, D. and {Fana Dirirsa}, F. and {Fegan}, S.~J. and {Ferrara}, E.~C. and {Franckowiak}, A. and {Fukazawa}, Y. and {Funk}, S. and {Fusco}, P. and {Gargano}, F. and {Gasparrini}, D. and {Giglietto}, N. and {Giommi}, P. and {Giordano}, F. and {Giroletti}, M. and {Glanzman}, T. and {Green}, D. and {Grenier}, I.~A. and {Griffin}, S. and {Grondin}, M. -H. and {Grove}, J.~E. and {Guiriec}, S. and {Harding}, A.~K. and {Hayashi}, K. and {Hays}, E. and {Hewitt}, J.~W. and {Horan}, D. and {Johannesson}, G. and {Johnson}, T.~J. and {Kamae}, T. and {Kerr}, M. and {Kocevski}, D. and {Kovac'evic'}, M. and {Kuss}, M. and {Landriu}, D. and {Larsson}, S. and {Latronico}, L. and {Lemoine-Goumard}, M. and {Li}, J. and {Liodakis}, I. and {Longo}, F. and {Loparco}, F. and {Lott}, B. and {Lovellette}, M.~N. and {Lubrano}, P. and {Madejski}, G.~M. and {Maldera}, S. and {Malyshev}, D. and {Manfreda}, A. and {Marchesini}, E.~J. and {Marcotulli}, L. and {Marti-Devesa}, G. and {Martin}, P. and {Massaro}, F. and {Mazziotta}, M.~N. and {McEnery}, J.~E. and {Mereu}, I. and {Meyer}, M. and {Michelson}, P.~F. and {Mirabal}, N. and {Mizuno}, T. and {Monzani}, M.~E. and {Morselli}, A. and {Moskalenko}, I.~V. and {Negro}, M. and {Nuss}, E. and {Ojha}, R. and {Omodei}, N. and {Orienti}, M. and {Orlando}, E. and {Ormes}, J.~F. and {Palatiello}, M. and {Paliya}, V.~S. and {Paneque}, D. and {Pei}, Z. and {Pena-Herazo}, H. and {Perkins}, J.~S. and {Persic}, M. and {Pesce-Rollins}, M. and {Petrosian}, V. and {Petrov}, L. and {Piron}, F. and {Poon}, H. and {Porter}, T.~A. and {Principe}, G. and {Raino}, S. and {Rando}, R. and {Razzano}, M. and {Razzaque}, S. and {Reimer}, A. and {Reimer}, O. and {Remy}, Q. and {Reposeur}, T. and {Romani}, R.~W. and {Saz Parkinson}, P.~M. and {Schinzel}, F.~K. and {Serini}, D. and {Sgro}, C. and {Siskind}, E.~J. and {Smith}, D.~A. and {Spandre}, G. and {Spinelli}, P. and {Strong}, A.~W. and {Suson}, D.~J. and {Tajima}, H. and {Takahashi}, M.~N. and {Tak}, D. and {Thayer}, J.~B. and {Thompson}, D.~J. and {Tibaldo}, L. and {Torres}, D.~F. and {Torresi}, E. and {Valverde}, J. and {van Klaveren}, B. and {van Zyl}, P. and {Wood}, K. and {Yassine}, M. and {Zaharijas}, G.},
        title = "{VizieR Online Data Catalog: The Fermi LAT fourth source catalog (4FGL) (Abdollahi+, 2020)}",
 howpublished = {VizieR On-line Data Catalog: J/ApJS/247/33. Originally published in: 2020ApJS..247...33A},
         year = 2020,
        month = jun,
          eid = {J/ApJS/247/33},
          doi = {10.26093/cds/vizier.22470033},
       adsurl = {https://ui.adsabs.harvard.edu/abs/2020yCat..22470033A}
}

@ARTICLE{2022Univ....8..587F,
       author = {{Foschini}, Luigi and {Lister}, Matthew L. and {Andernach}, Heinz and {Ciroi}, Stefano and {Marziani}, Paola and {Ant{\'o}n}, Sonia and {Berton}, Marco and {Dalla Bont{\`a}}, Elena and {J{\"a}rvel{\"a}}, Emilia and {March{\~a}}, Maria J.~M. and {Romano}, Patrizia and {Tornikoski}, Merja and {Vercellone}, Stefano and {Vietri}, Amelia},
        title = "{A New Sample of Gamma-Ray Emitting Jetted Active Galactic Nuclei}",
      journal = {Universe},
         year = 2022,
        month = nov,
       volume = {8},
       number = {11},
          eid = {587},
        pages = {587},
          doi = {10.3390/universe8110587},
archivePrefix = {arXiv},
       eprint = {2211.03400},
 primaryClass = {astro-ph.HE},
       adsurl = {https://ui.adsabs.harvard.edu/abs/2022Univ....8..587F}
}

@article{paliya2019general,
  title={General physical properties of gamma-ray-emitting narrow-line Seyfert 1 galaxies},
  author={Paliya, Vaidehi S and Parker, ML and Jiang, J and Fabian, AC and Brenneman, L and Ajello, M and Hartmann, D},
  journal={The Astrophysical Journal},
  volume={872},
  number={2},
  pages={169},
  year={2019},
  publisher={IOP Publishing}
}

@article{yao2019sdss,
  title={SDSS J094635. 06+ 101706.1: a redshift one, very radio-loud, $\gamma$-ray emitting narrow-line Seyfert 1 galaxy},
  author={Yao, Su and Komossa, S and Liu, Wen-Juan and Yi, Weimin and Yuan, Weimin and Zhou, Hongyan and Wu, Xue-Bing},
  journal={Monthly Notices of the Royal Astronomical Society: Letters},
  volume={487},
  number={1},
  pages={L40--L45},
  year={2019},
  publisher={Oxford University Press}
}

@article{liao2015discovery,
  title={Discovery of $$\backslash$gamma $-ray emission from steep radio spectrum NLS1s},
  author={Liao, Neng-Hui and Liang, Yun-Feng and Weng, Shan-Shan and Berton, Marco and Gu, Min-Feng and Fan, Yi-Zhong},
  journal={arXiv preprint arXiv:1510.05584},
  year={2015}
}

@article{yao2021spectroscopic,
  title={Spectroscopic classification, variability, and SED of the Fermi-detected CSS 3C 286: the radio-loudest NLS1 galaxy?},
  author={Yao, Su and Komossa, S},
  journal={Monthly Notices of the Royal Astronomical Society},
  volume={501},
  number={1},
  pages={1384--1393},
  year={2021},
  publisher={Oxford University Press}
}

@article{mao2021search,
  title={A Search for Rapid Mid-infrared Variability in Gamma-Ray-emitting Narrow-line Seyfert 1 Galaxies},
  author={Mao, Lisheng and Yi, Tingfeng},
  journal={The Astrophysical Journal Supplement Series},
  volume={255},
  number={1},
  pages={10},
  year={2021},
  publisher={IOP Publishing}
}

@article{lahteenmaki2018radio,
  title={Radio jets and gamma-ray emission in radio-silent narrow-line Seyfert 1 galaxies},
  author={L{\"a}hteenm{\"a}ki, A and J{\"a}rvel{\"a}, E and Ramakrishnan, V and Tornikoski, M and Tammi, J and Vera, RJC and Chamani, Wara},
  journal={Astronomy \& Astrophysics},
  volume={614},
  pages={L1},
  year={2018},
  publisher={EDP Sciences}
}

@article{zhou2007narrow,
  title={A narrow-line Seyfert 1-blazar composite nucleus in 2MASX J0324+ 3410},
  author={Zhou, Hongyan and Wang, Tinggui and Yuan, Weimin and Shan, Hongguang and Komossa, Stefanie and Lu, Honglin and Liu, Yi and Xu, Dawei and Bai, JM and Jiang, DR},
  journal={The Astrophysical Journal},
  volume={658},
  number={1},
  pages={L13},
  year={2007},
  publisher={IOP Publishing}
}

@article{rakshit2017optical,
  title={Optical variability of narrow-line and broad-line Seyfert 1 Galaxies},
  author={Rakshit, Suvendu and Stalin, CS},
  journal={The Astrophysical Journal},
  volume={842},
  number={2},
  pages={96},
  year={2017},
  publisher={IOP Publishing}
}

@article{yao2015identification,
  title={Identification of a new $\gamma$-ray-emitting narrow-line Seyfert 1 galaxy, at redshift~ 1},
  author={Yao, Su and Yuan, Weimin and Zhou, Hongyan and Komossa, S and Zhang, Jin and Qiao, Erlin and Liu, Bifang},
  journal={Monthly Notices of the Royal Astronomical Society: Letters},
  volume={454},
  number={1},
  pages={L16--L20},
  year={2015},
  publisher={Oxford University Press}
}

@article{lefkir2025variability,
  title={The variability of active galaxies--I. Broad-band noise X-ray power spectra from XMM--Newton and Swift},
  author={Lefkir, Mehdy and Vaughan, Simon and Goad, Mike and Huppenkothen, Daniela and Uttley, Phil},
  journal={Monthly Notices of the Royal Astronomical Society},
  volume={544},
  number={4},
  pages={3260--3279},
  year={2025},
  publisher={Oxford University Press}
}

@article{burke2021characteristic,
  title={A characteristic optical variability time scale in astrophysical accretion disks},
  author={Burke, Colin J and Shen, Yue and Blaes, Omer and Gammie, Charles F and Horne, Keith and Jiang, Yan-Fei and Liu, Xin and McHardy, Ian M and Morgan, Christopher W and Scaringi, Simone and others},
  journal={Science},
  volume={373},
  number={6556},
  pages={789--792},
  year={2021},
  publisher={American Association for the Advancement of Science}
}

@article{zhang2024discovering,
  title={Discovering the mass-scaled damping timescale from microquasars to blazars},
  author={Zhang, Haoyang and Yang, Shenbang and Dai, Benzhong},
  journal={The Astrophysical Journal Letters},
  volume={967},
  number={1},
  pages={L18},
  year={2024},
  publisher={IOP Publishing}
}

\appendix
\section{Individual Plots}
\label{sec:Ind_plots}

\subsection{Lightcurves, Color Magnitude Diagrams, Flux histograms, PSD, Simulated LCs.}

\begin{figure*}\label{J003159093618}
\centering
\caption{J003159+093618}
    \begin{minipage}{.3\textwidth}
        \centering
        \includegraphics[width=.99\linewidth]{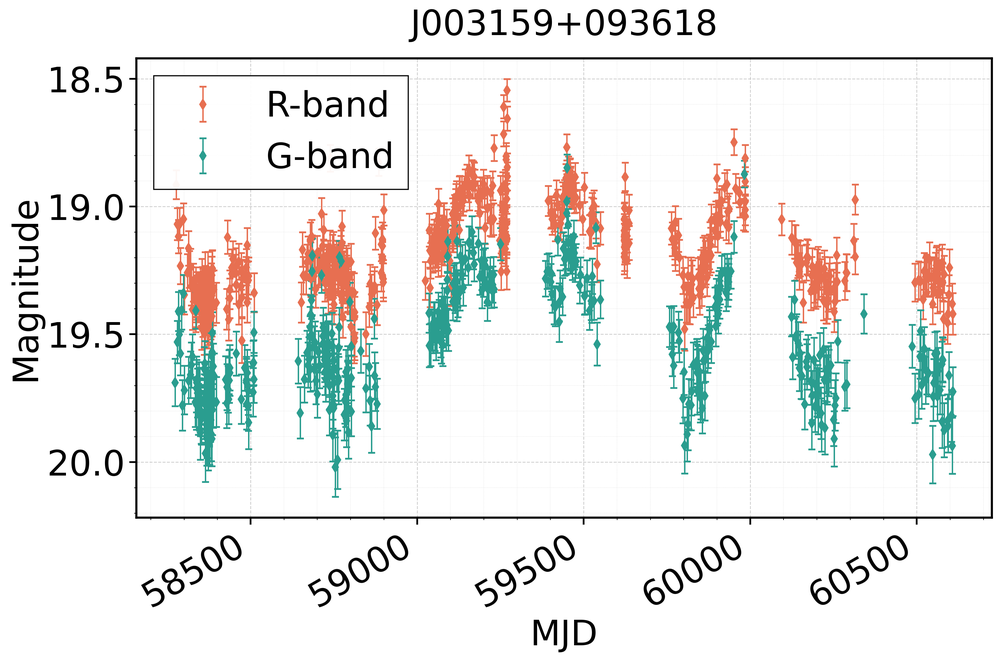}
    \end{minipage}
    \begin{minipage}{.3\textwidth}
        \centering
        \includegraphics[width=.99\linewidth]{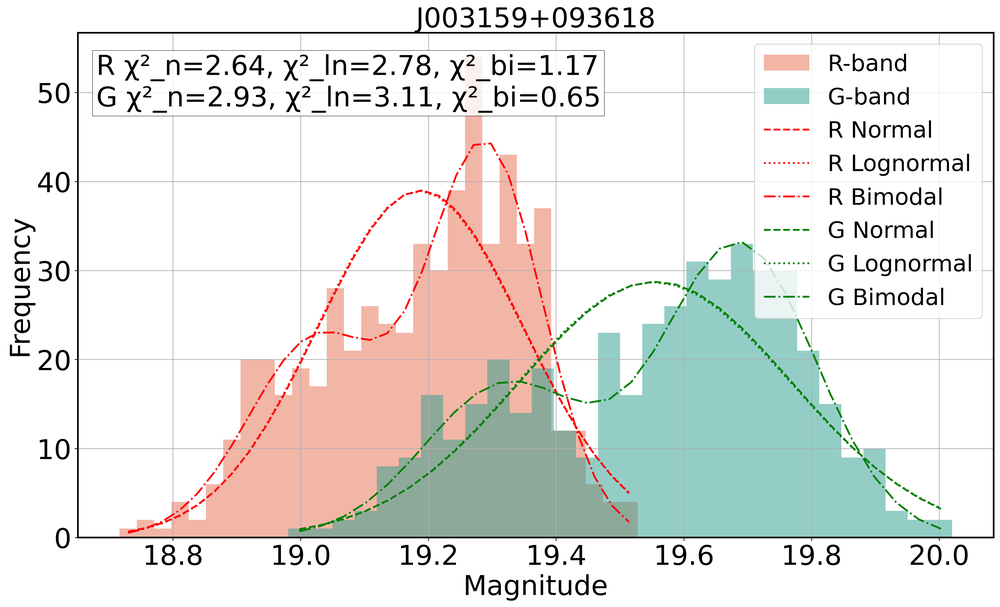}
    \end{minipage}
    \begin{minipage}{.3\textwidth}
        \centering
        \includegraphics[width=.99\linewidth]{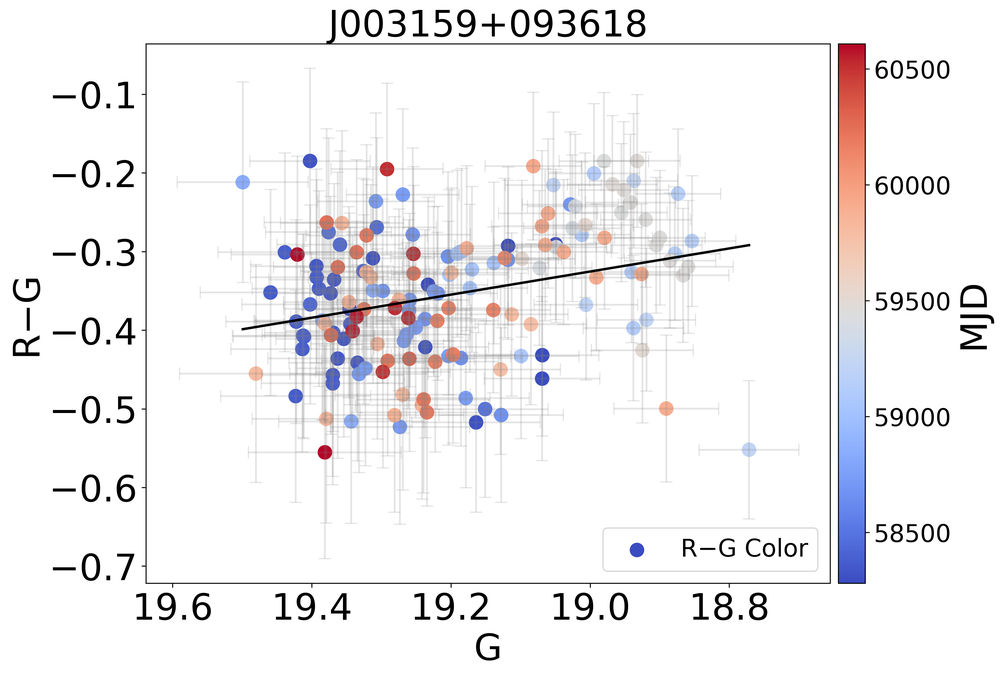}
    \end{minipage}
    \\
    \begin{minipage}{.3\textwidth}
        \centering
        \includegraphics[width=.99\linewidth]{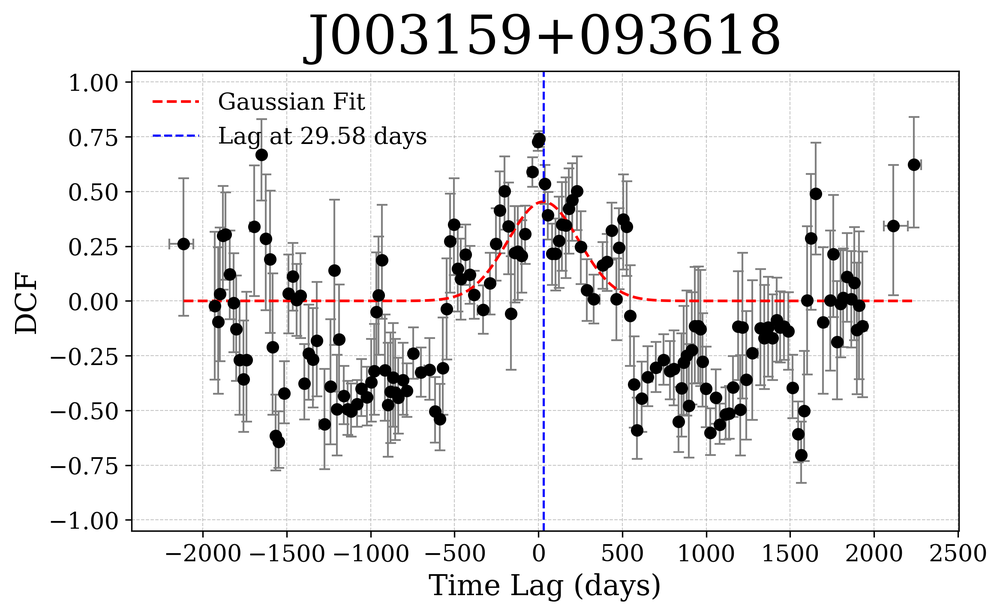}
    \end{minipage}
    \begin{minipage}{.3\textwidth}
        \centering
        \includegraphics[width=.99\linewidth]{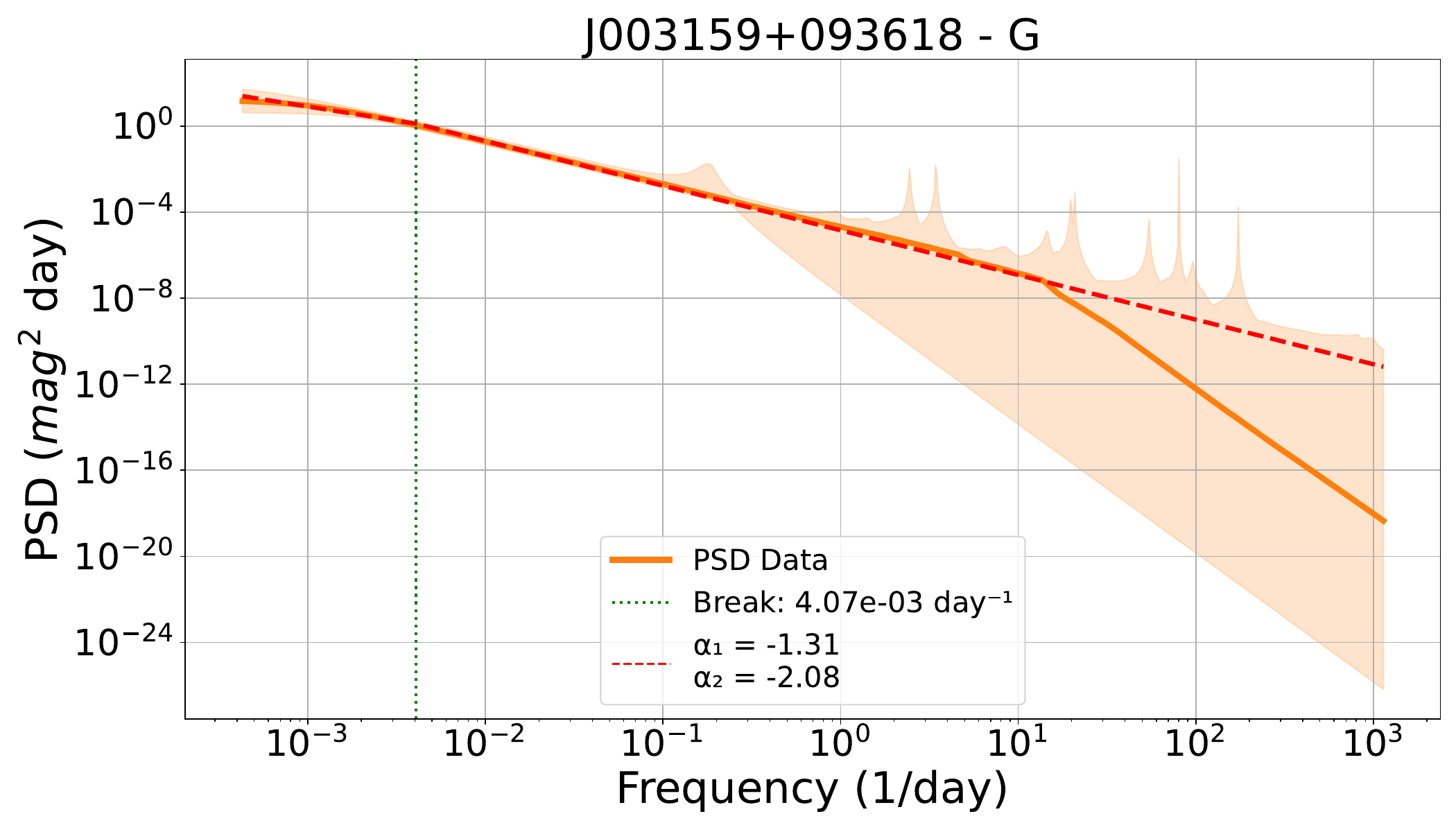}
    \end{minipage}
    \begin{minipage}{.3\textwidth}
        \centering
        \includegraphics[width=.99\linewidth]{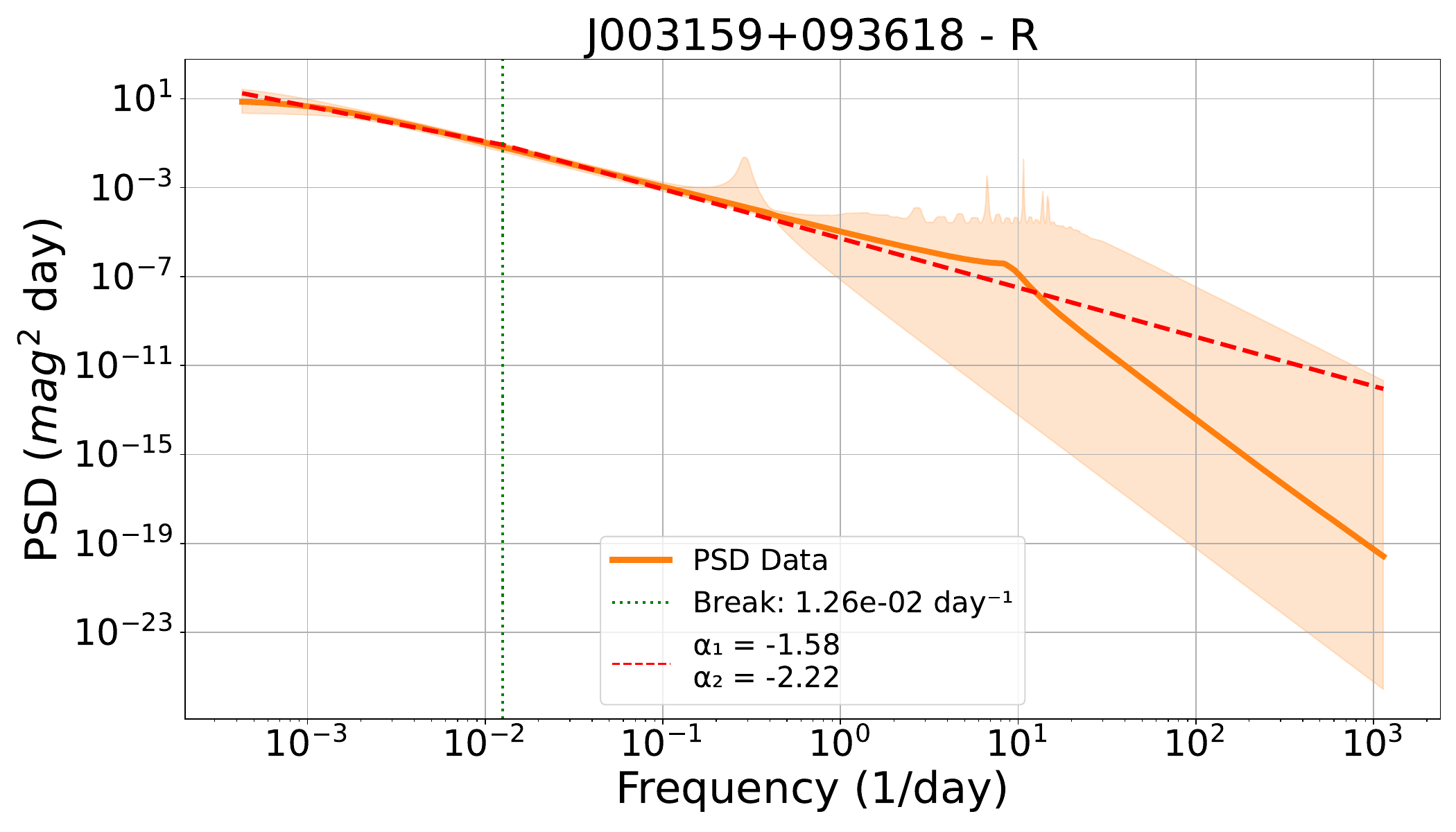}
    \end{minipage}
    \\
    \begin{minipage}{.3\textwidth}
        \centering
        \includegraphics[width=.99\linewidth]{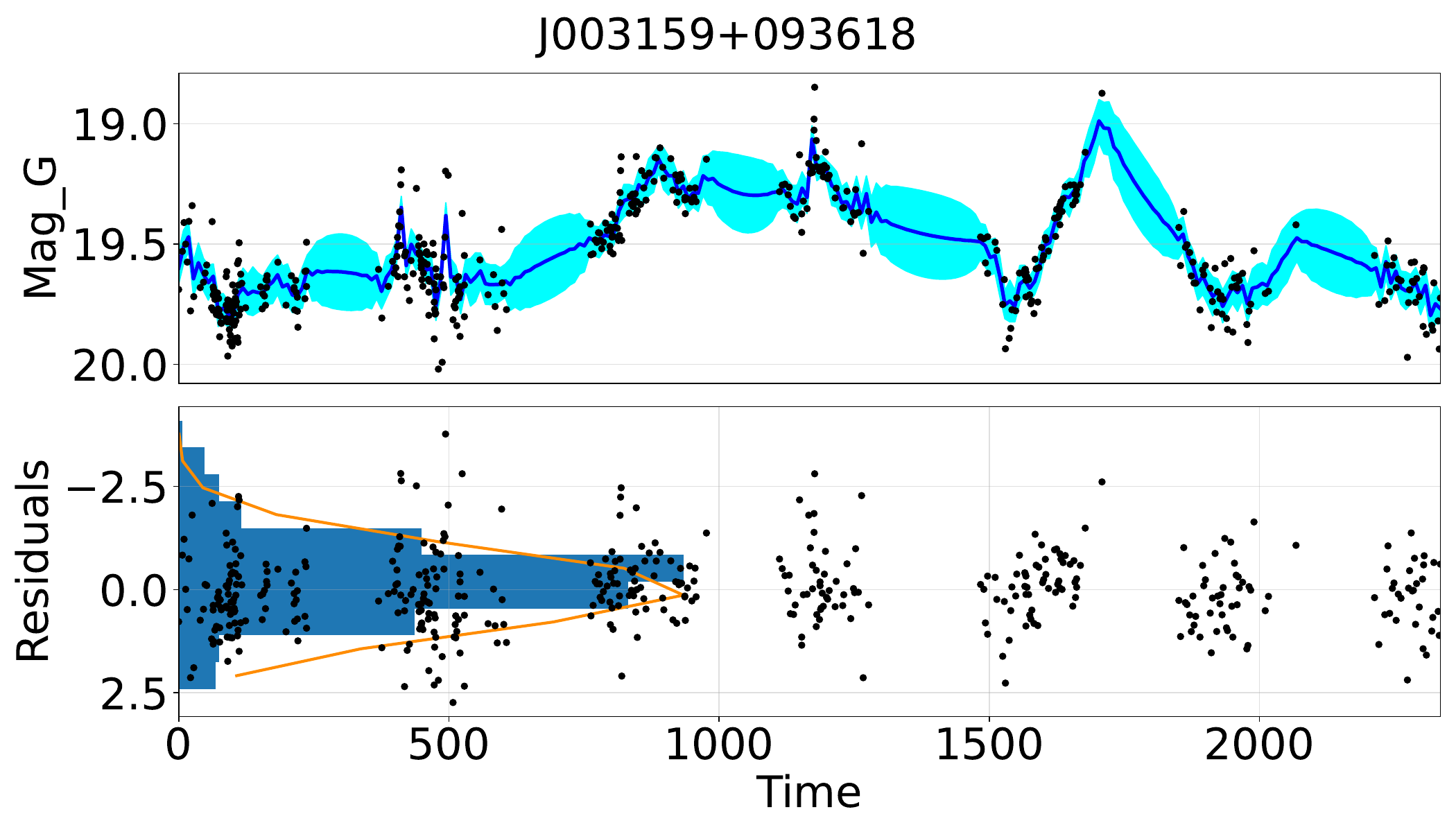}
    \end{minipage}
    \begin{minipage}{.3\textwidth}
        \centering
        \includegraphics[width=.99\linewidth]{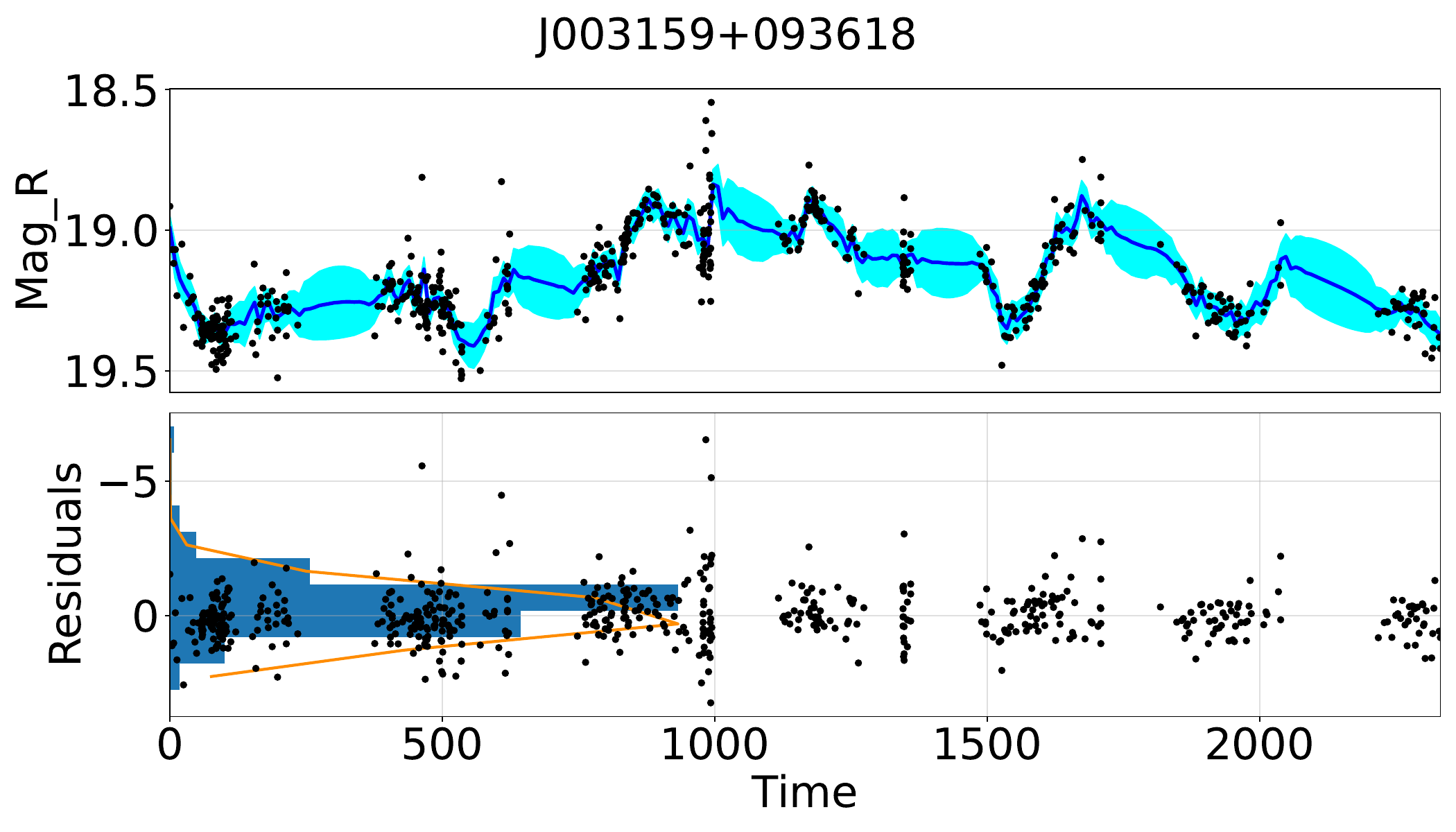}
    \end{minipage}
\end{figure*}

\begin{figure*}\label{J032441341045}
\centering
\caption{J032441+341045}
    \begin{minipage}{.3\textwidth}
        \centering
        \includegraphics[width=.99\linewidth]{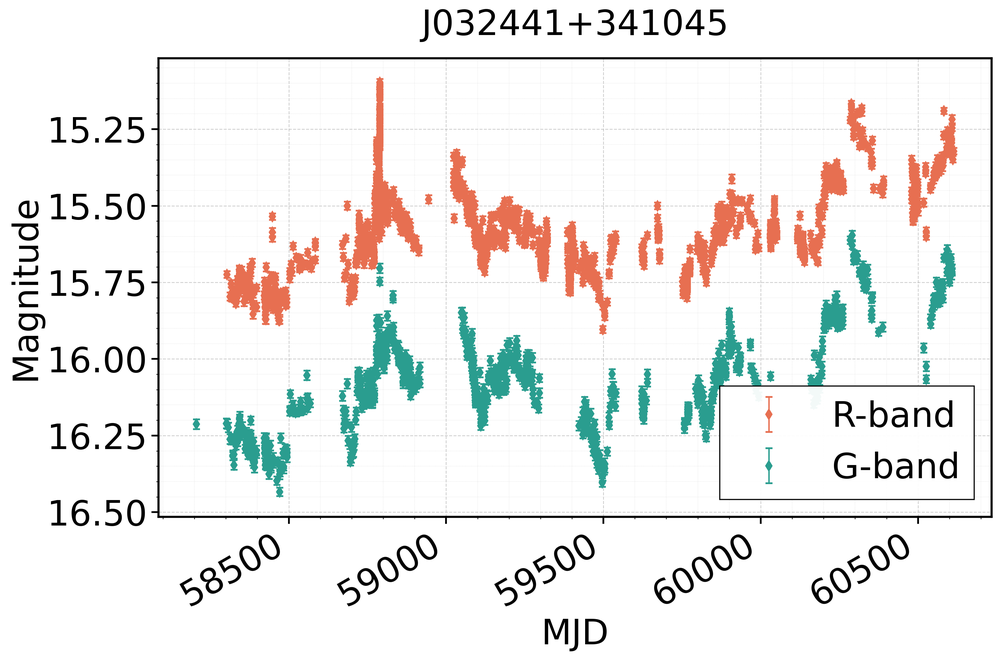}
    \end{minipage}
    \begin{minipage}{.3\textwidth}
        \centering
        \includegraphics[width=.99\linewidth]{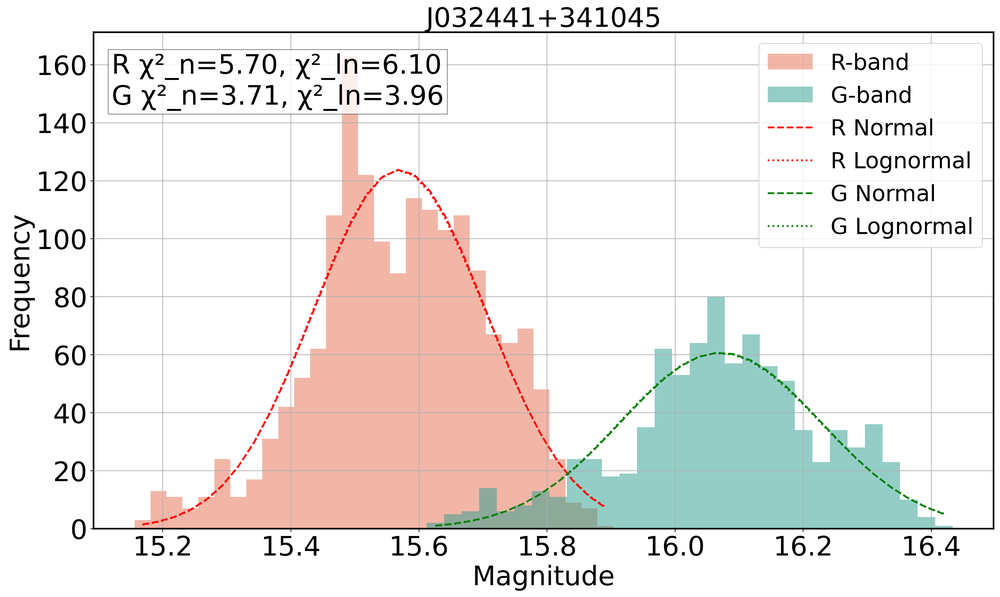}
    \end{minipage}
    \begin{minipage}{.3\textwidth}
        \centering
        \includegraphics[width=.99\linewidth]{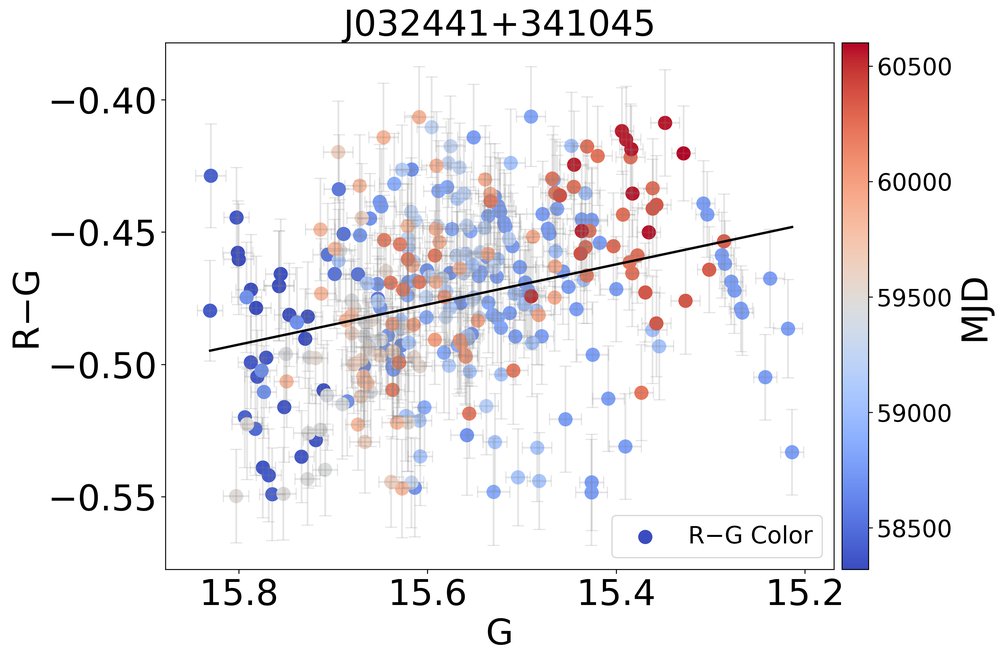}
    \end{minipage}
    \\
    \begin{minipage}{.3\textwidth}
        \centering
        \includegraphics[width=.99\linewidth]{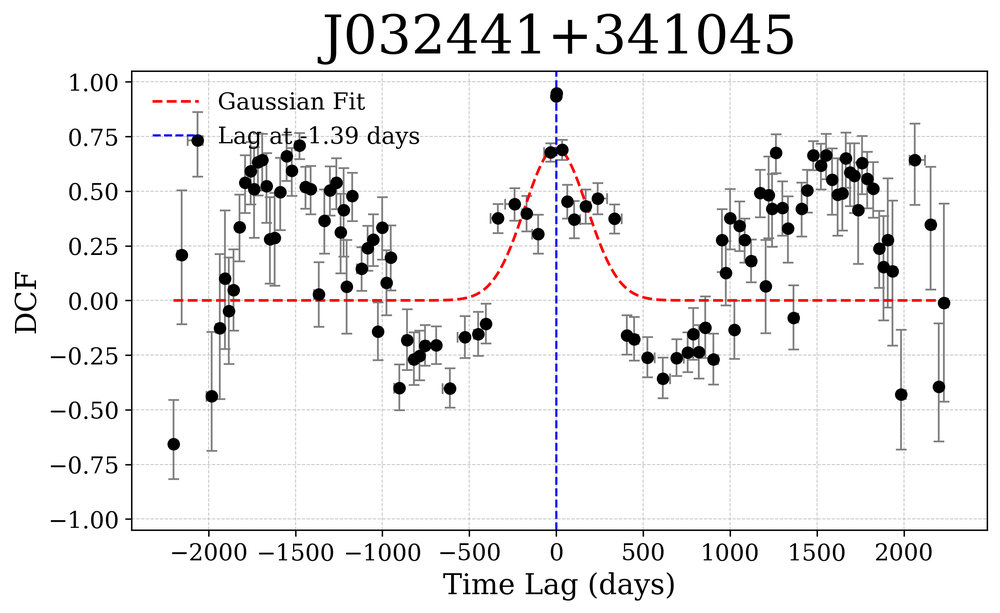}
    \end{minipage}
    \begin{minipage}{.3\textwidth}
        \centering
        \includegraphics[width=.99\linewidth]{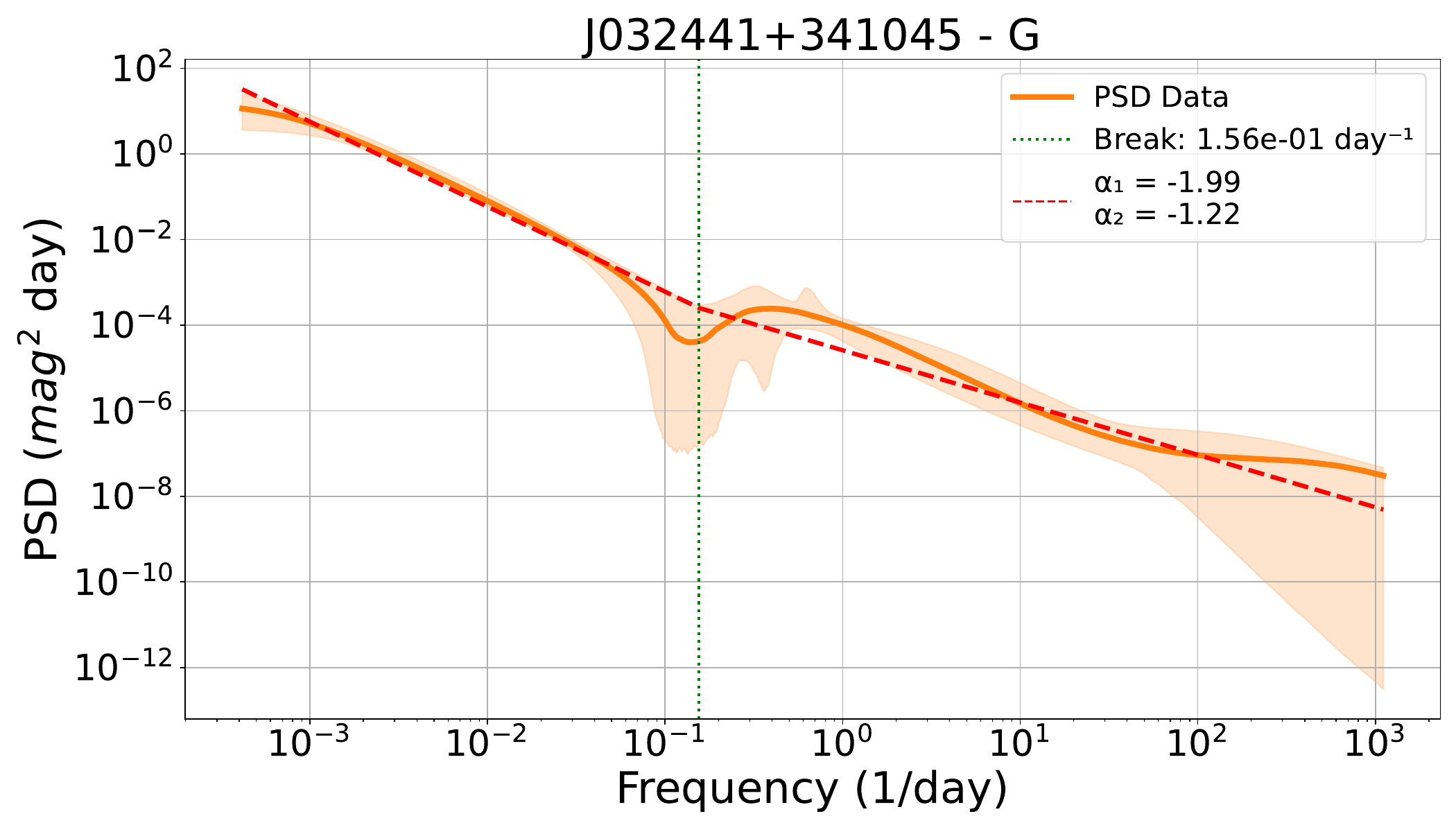}
    \end{minipage}
    \begin{minipage}{.3\textwidth}
        \centering
        \includegraphics[width=.99\linewidth]{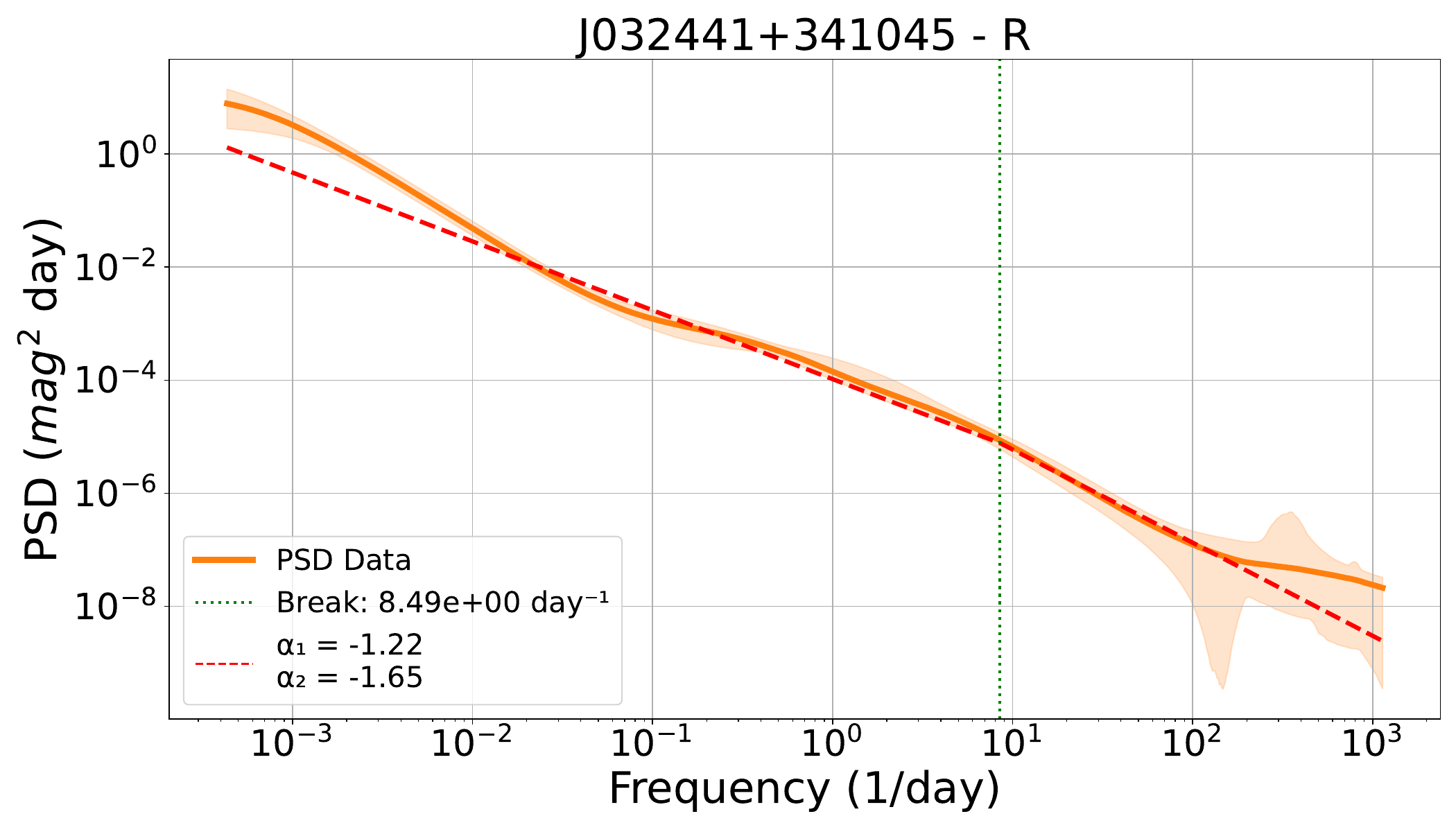}
    \end{minipage}
    \\
    \begin{minipage}{.3\textwidth}
        \centering
        \includegraphics[width=.99\linewidth]{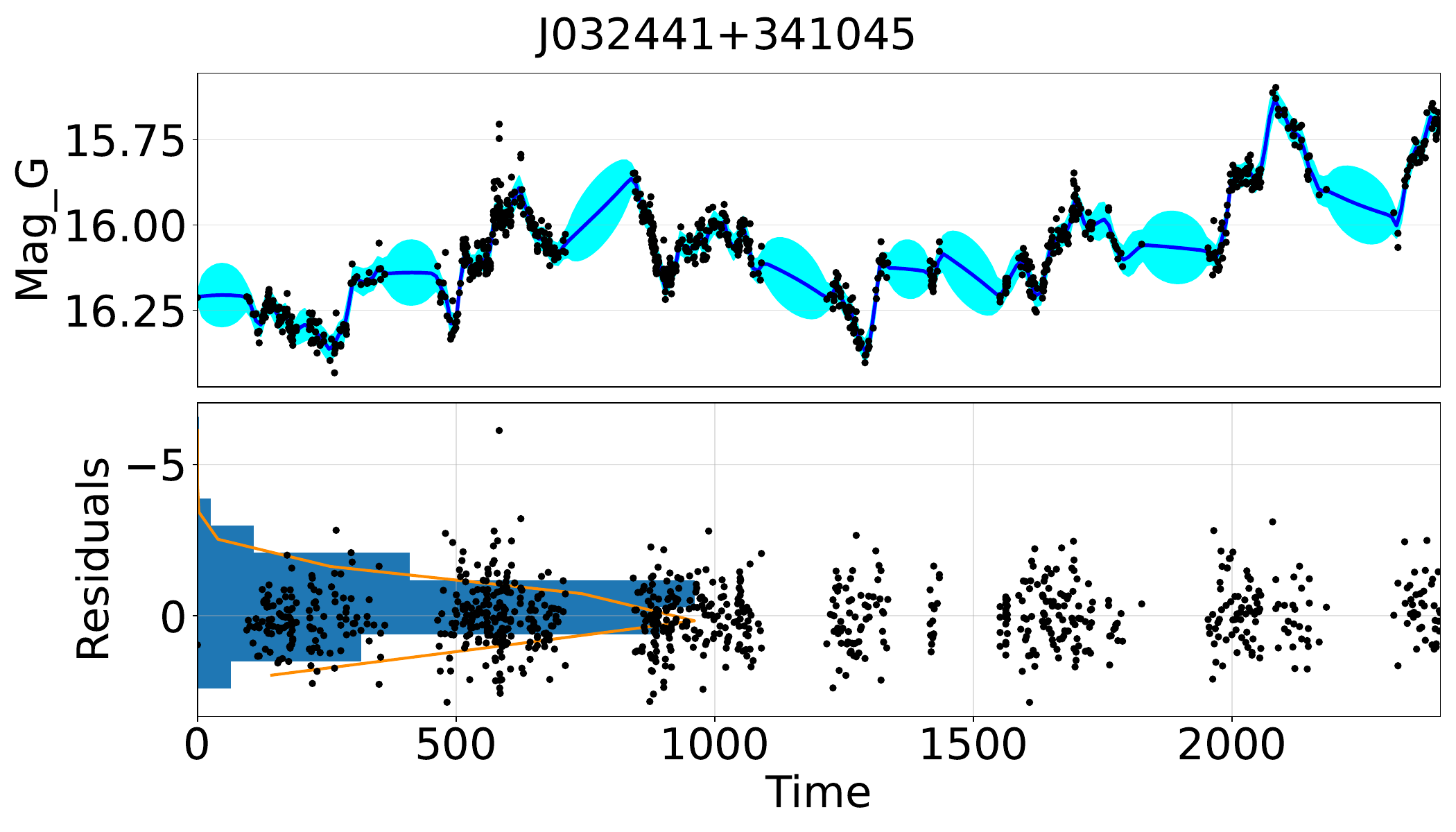}
    \end{minipage}
    \begin{minipage}{.3\textwidth}
        \centering
        \includegraphics[width=.99\linewidth]{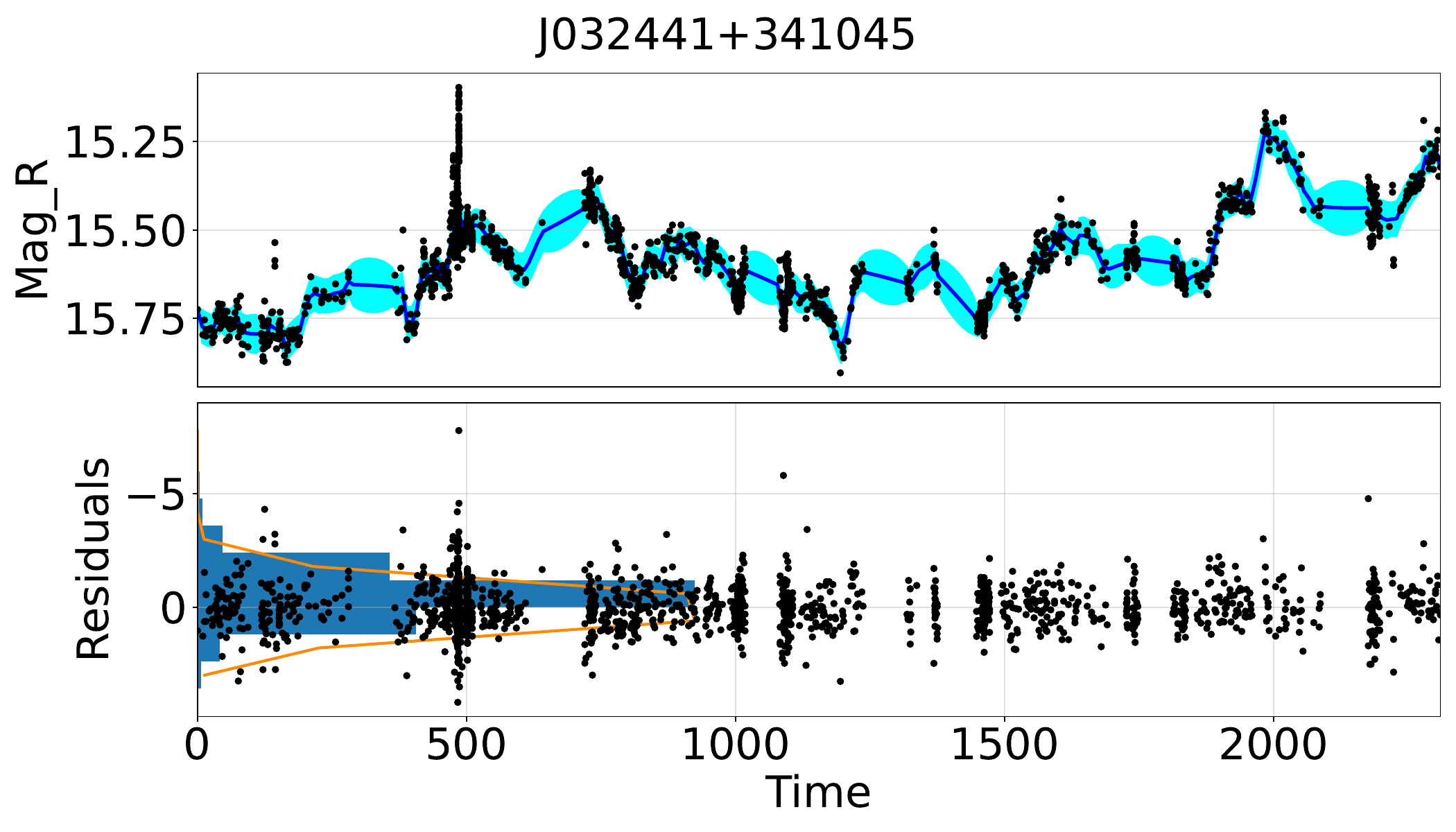}
    \end{minipage}
\end{figure*}

\begin{figure*}\label{J084957510829}
\centering
\caption{J084957+510829}
    \begin{minipage}{.3\textwidth}
        \centering
        \includegraphics[width=.99\linewidth]{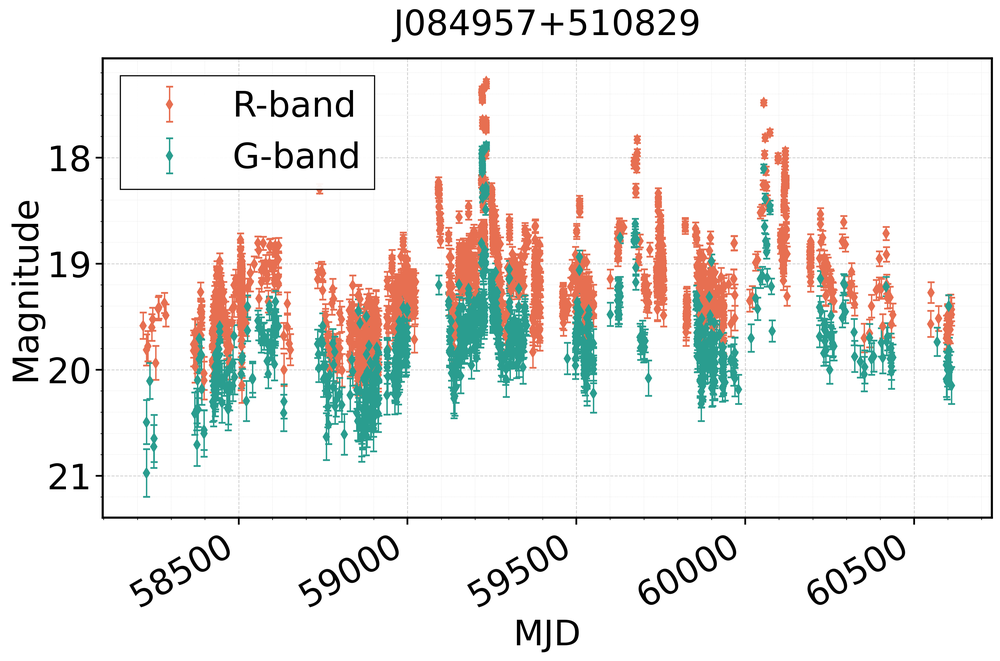}
    \end{minipage}
    \begin{minipage}{.3\textwidth}
        \centering
        \includegraphics[width=.99\linewidth]{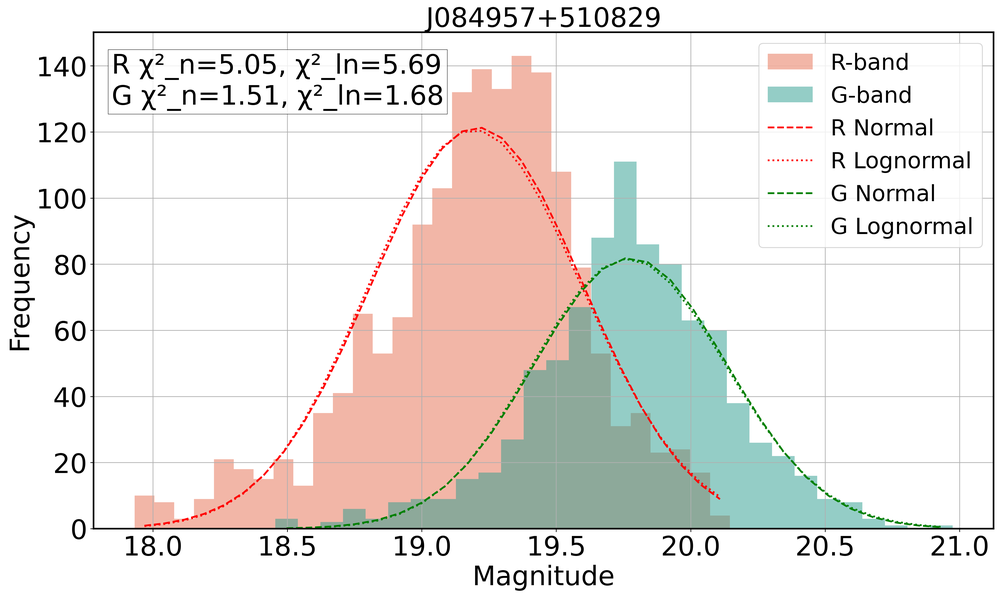}
    \end{minipage}
    \begin{minipage}{.3\textwidth}
        \centering
        \includegraphics[width=.99\linewidth]{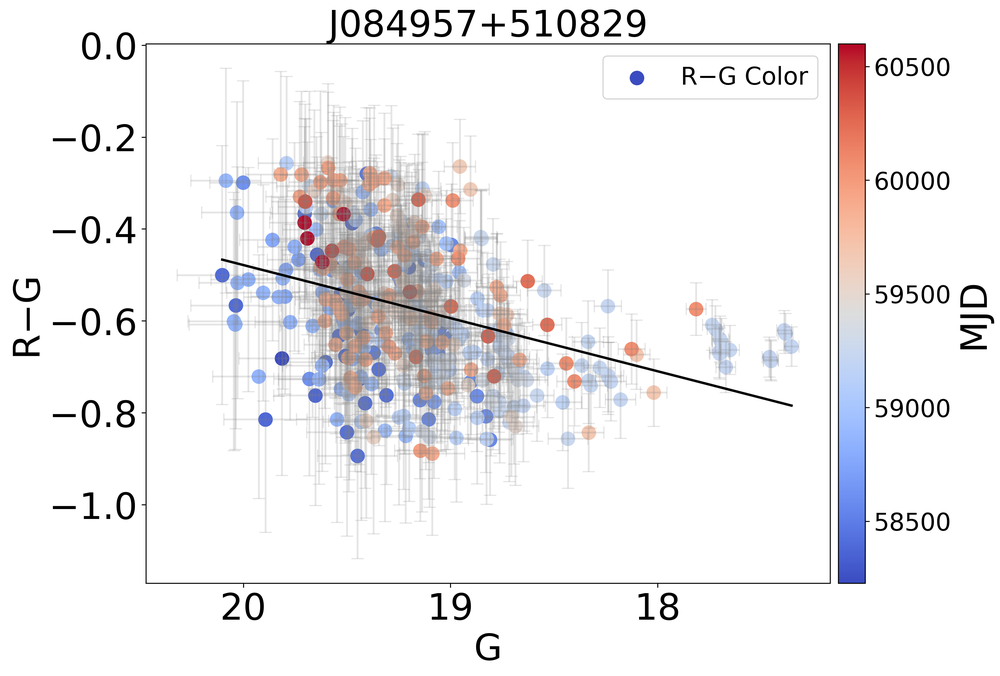}
    \end{minipage}
    \\
    \begin{minipage}{.3\textwidth}
        \centering
        \includegraphics[width=.99\linewidth]{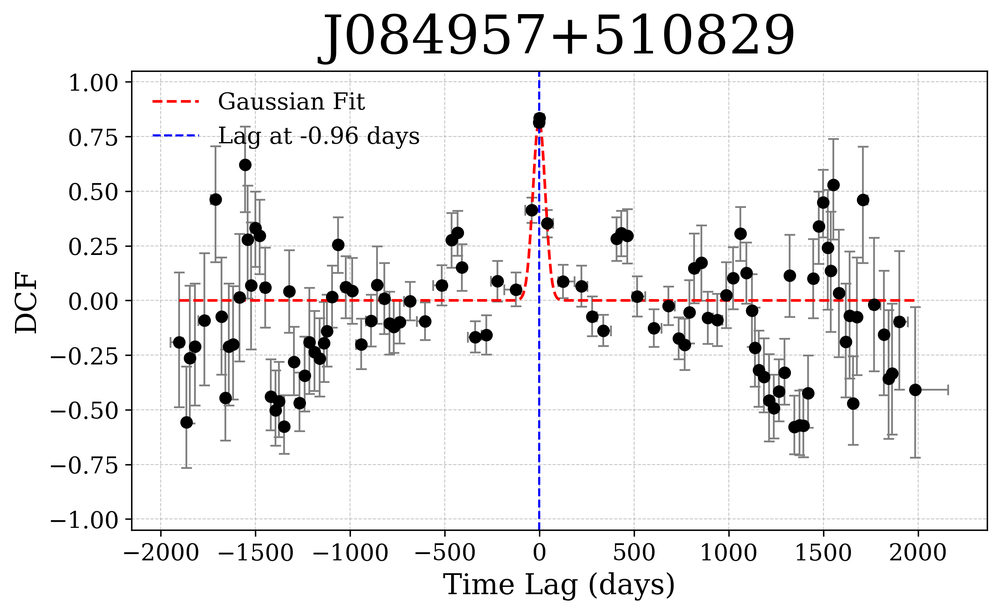}
    \end{minipage}
    \begin{minipage}{.3\textwidth}
        \centering
        \includegraphics[width=.99\linewidth]{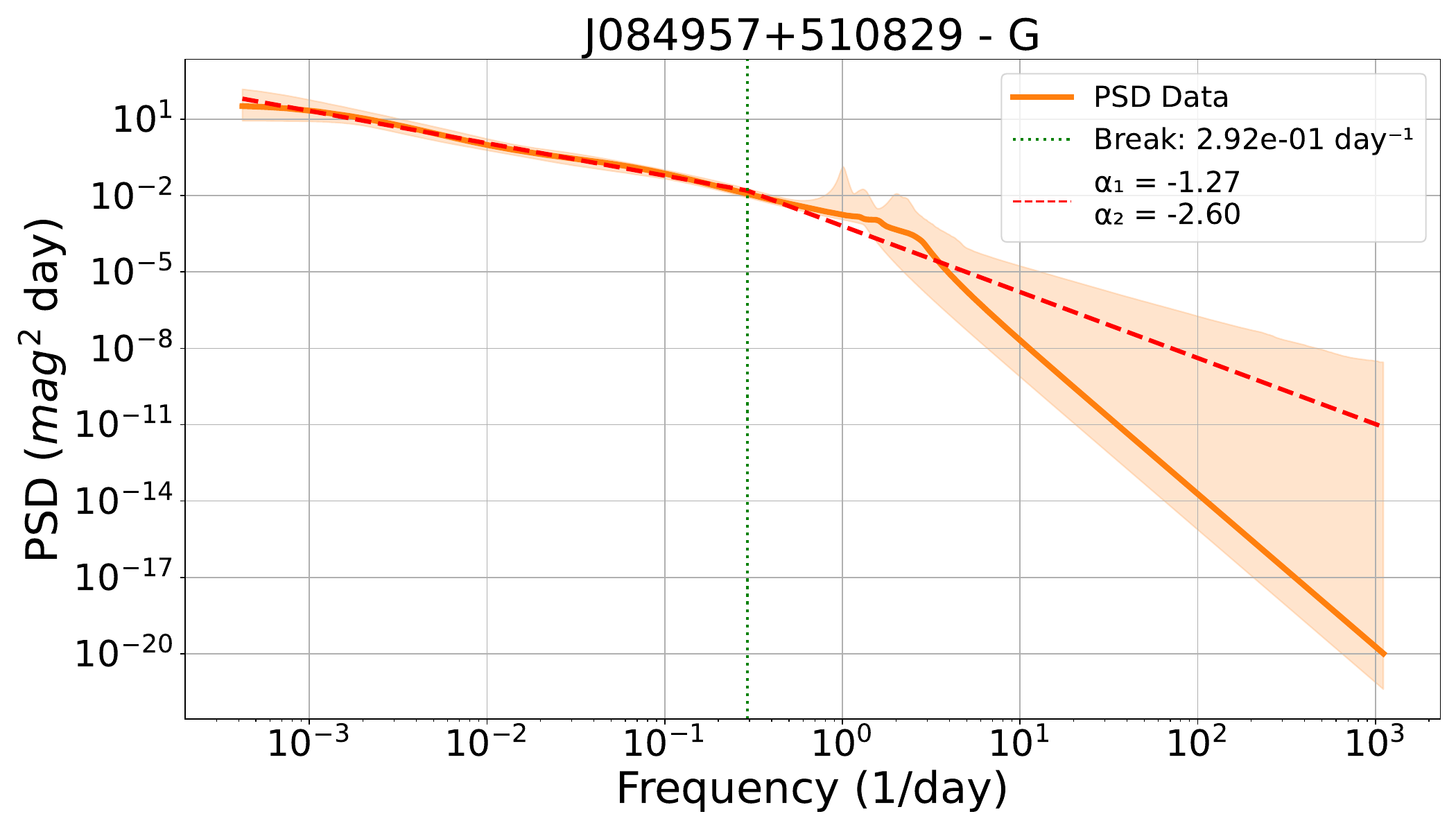}
    \end{minipage}
    \begin{minipage}{.3\textwidth}
        \centering
        \includegraphics[width=.99\linewidth]{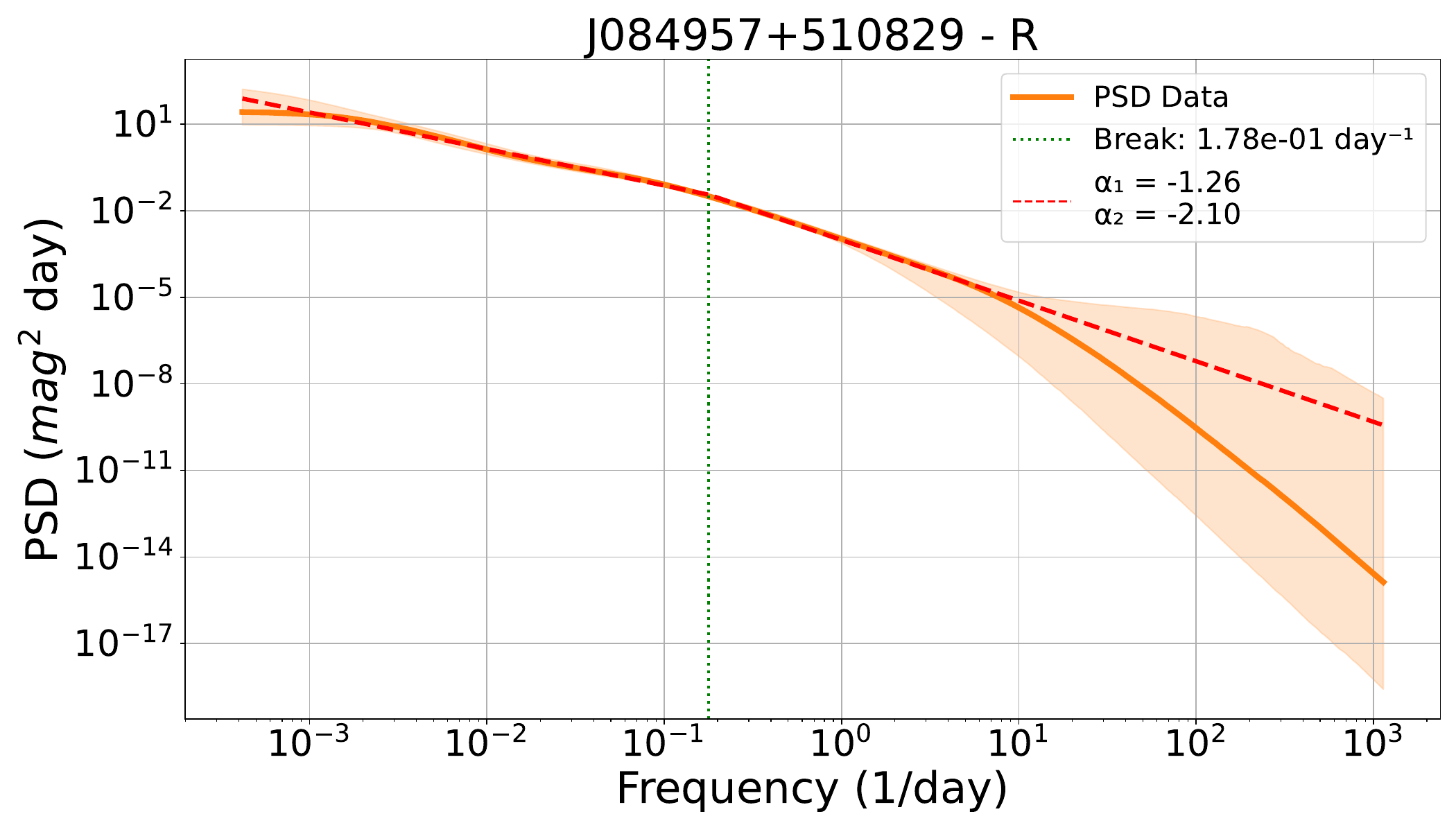}
    \end{minipage}
    \\
    \begin{minipage}{.3\textwidth}
        \centering
        \includegraphics[width=.99\linewidth]{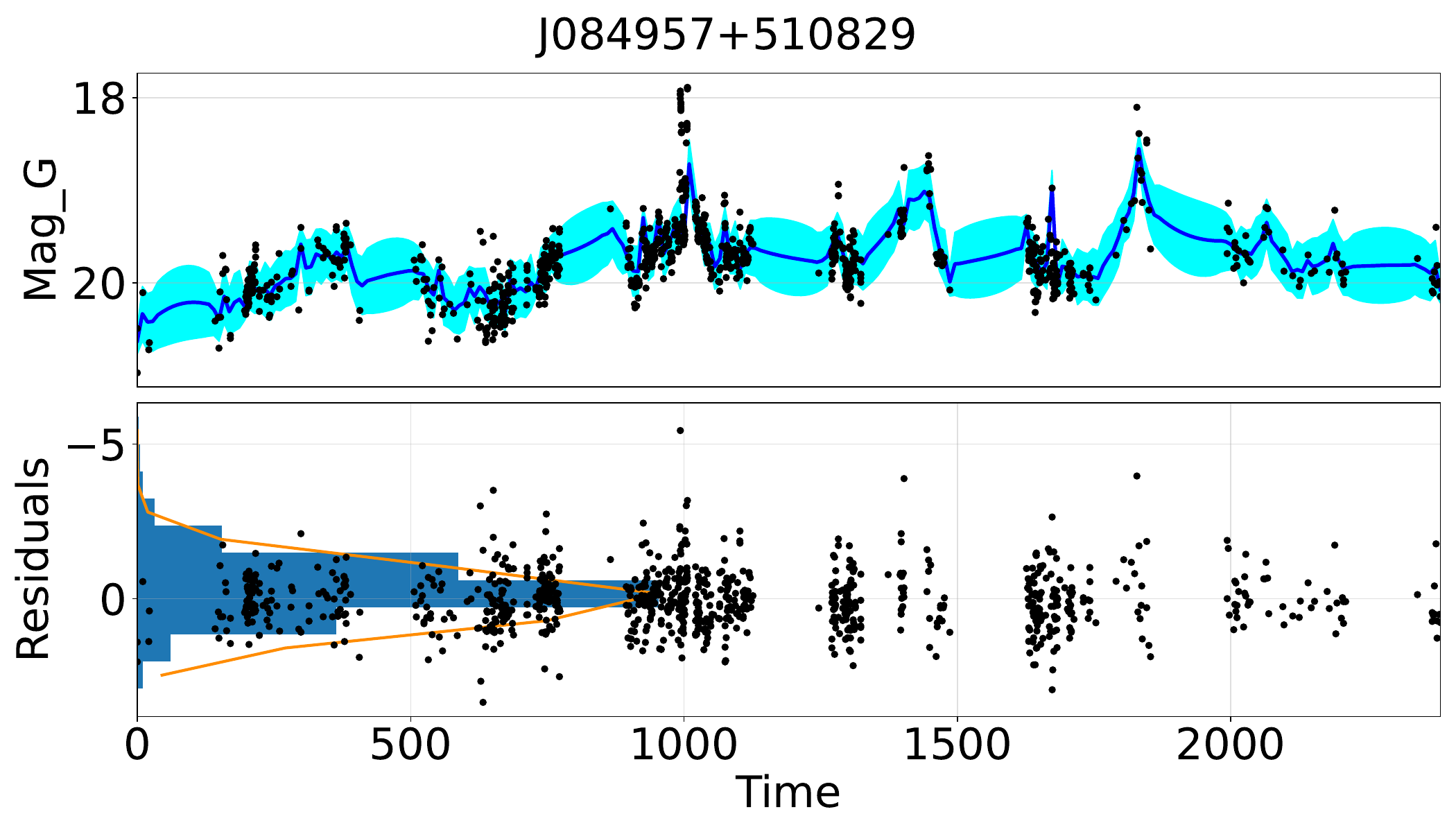}
    \end{minipage}
    \begin{minipage}{.3\textwidth}
        \centering
        \includegraphics[width=.99\linewidth]{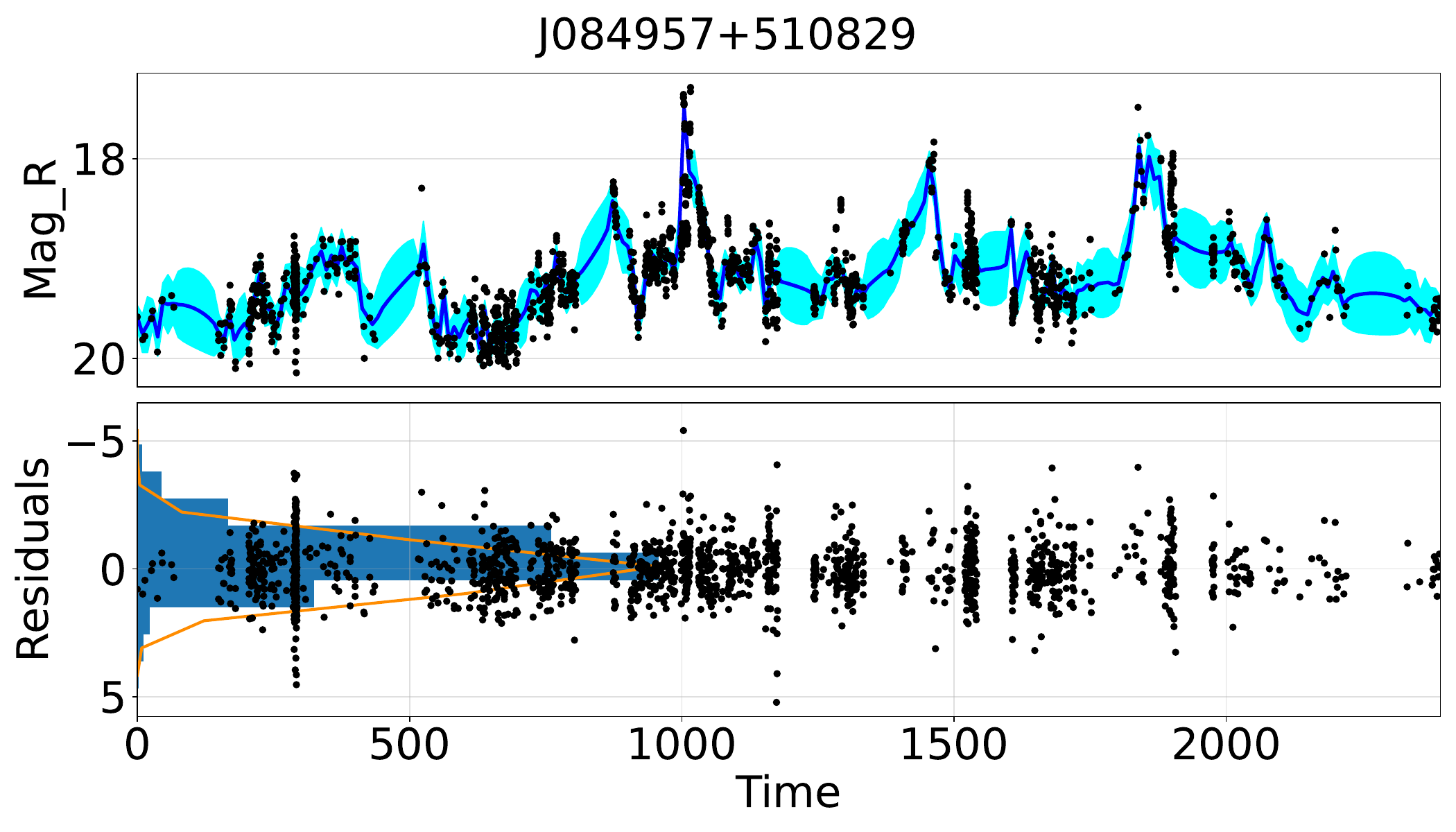}
    \end{minipage}
\end{figure*}

\begin{figure*}\label{J093241530633}
\centering
\caption{J093241+530633}
    \begin{minipage}{.3\textwidth}
        \centering
        \includegraphics[width=.99\linewidth]{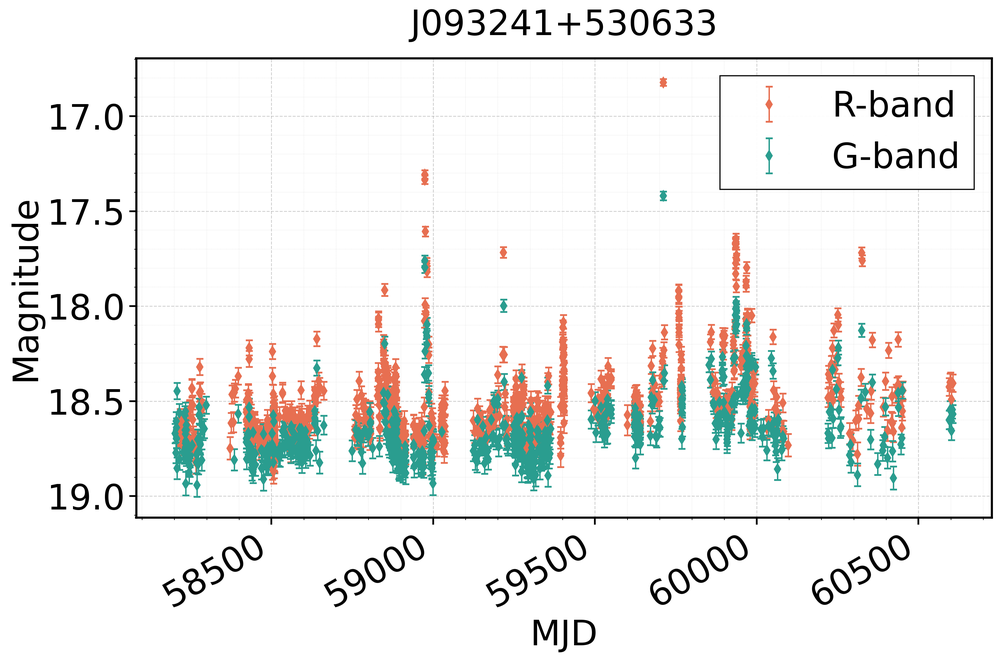}
    \end{minipage}
    \begin{minipage}{.3\textwidth}
        \centering
        \includegraphics[width=.99\linewidth]{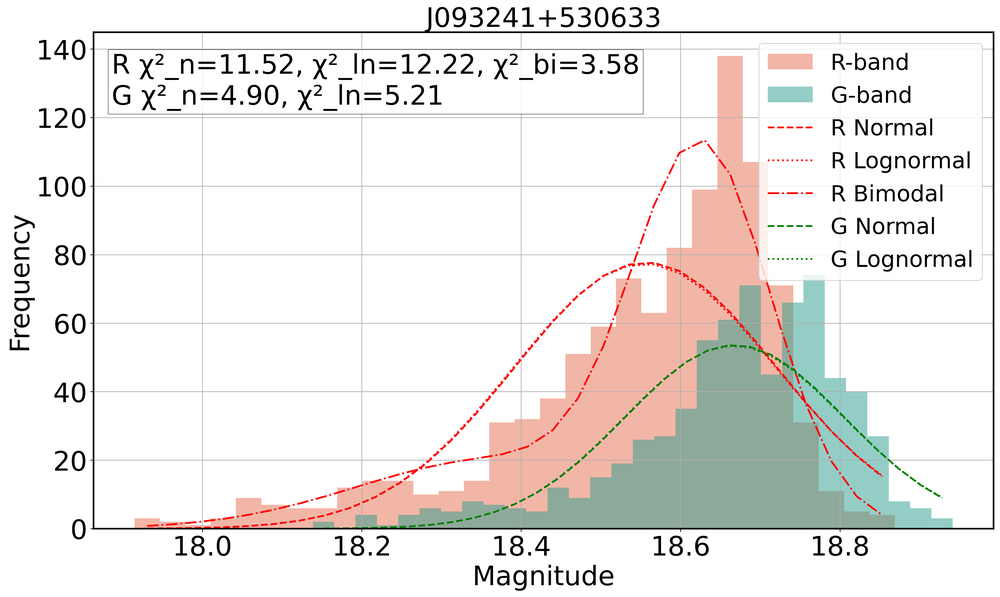}
    \end{minipage}
    \begin{minipage}{.3\textwidth}
        \centering
        \includegraphics[width=.99\linewidth]{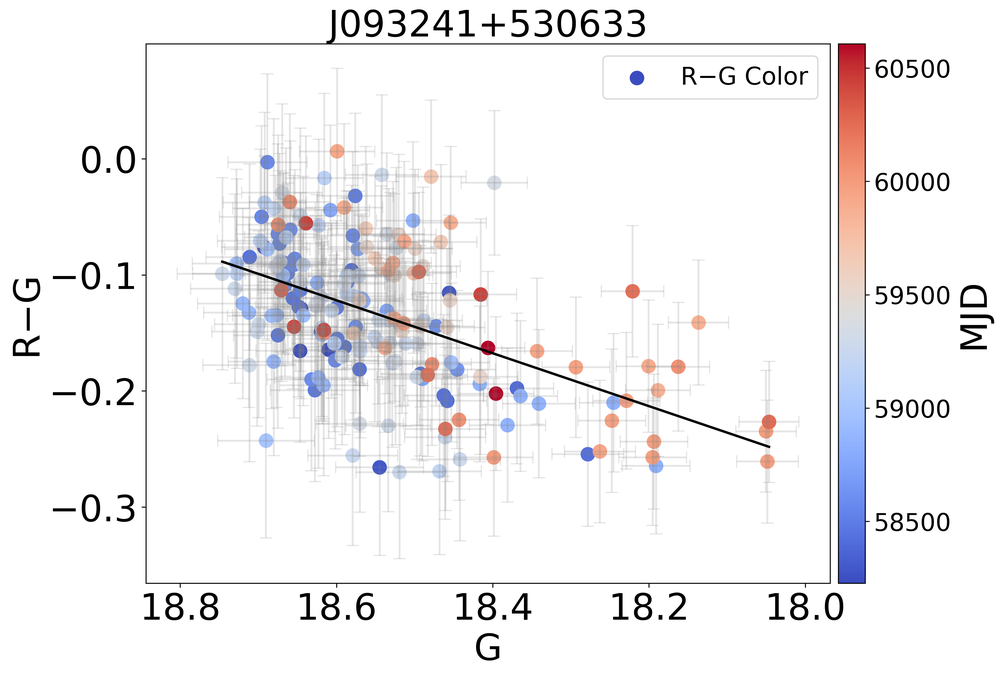}
    \end{minipage}
    \\
    \begin{minipage}{.3\textwidth}
        \centering
        \includegraphics[width=.99\linewidth]{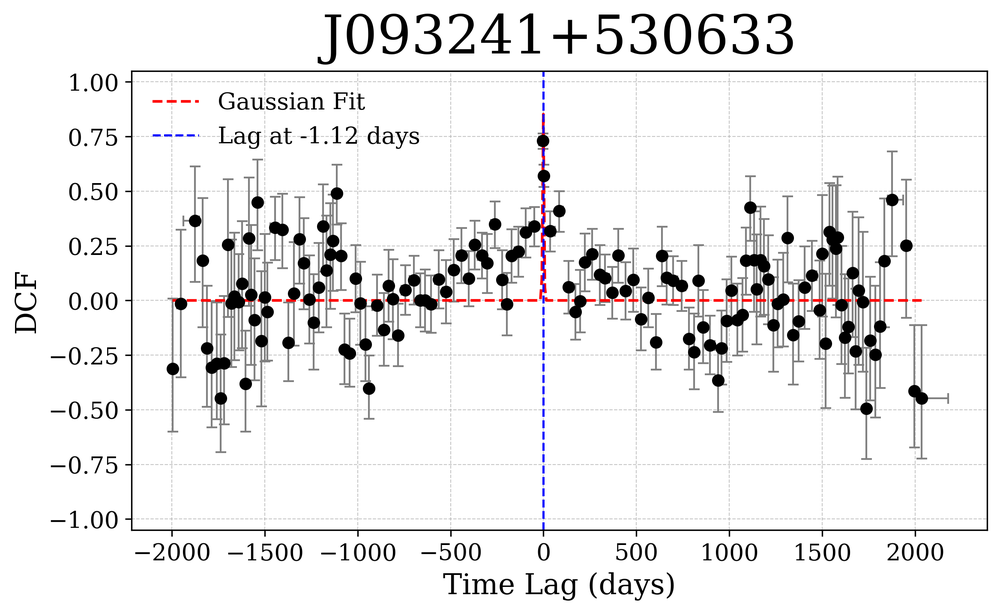}
    \end{minipage}
    \begin{minipage}{.3\textwidth}
        \centering
        \includegraphics[width=.99\linewidth]{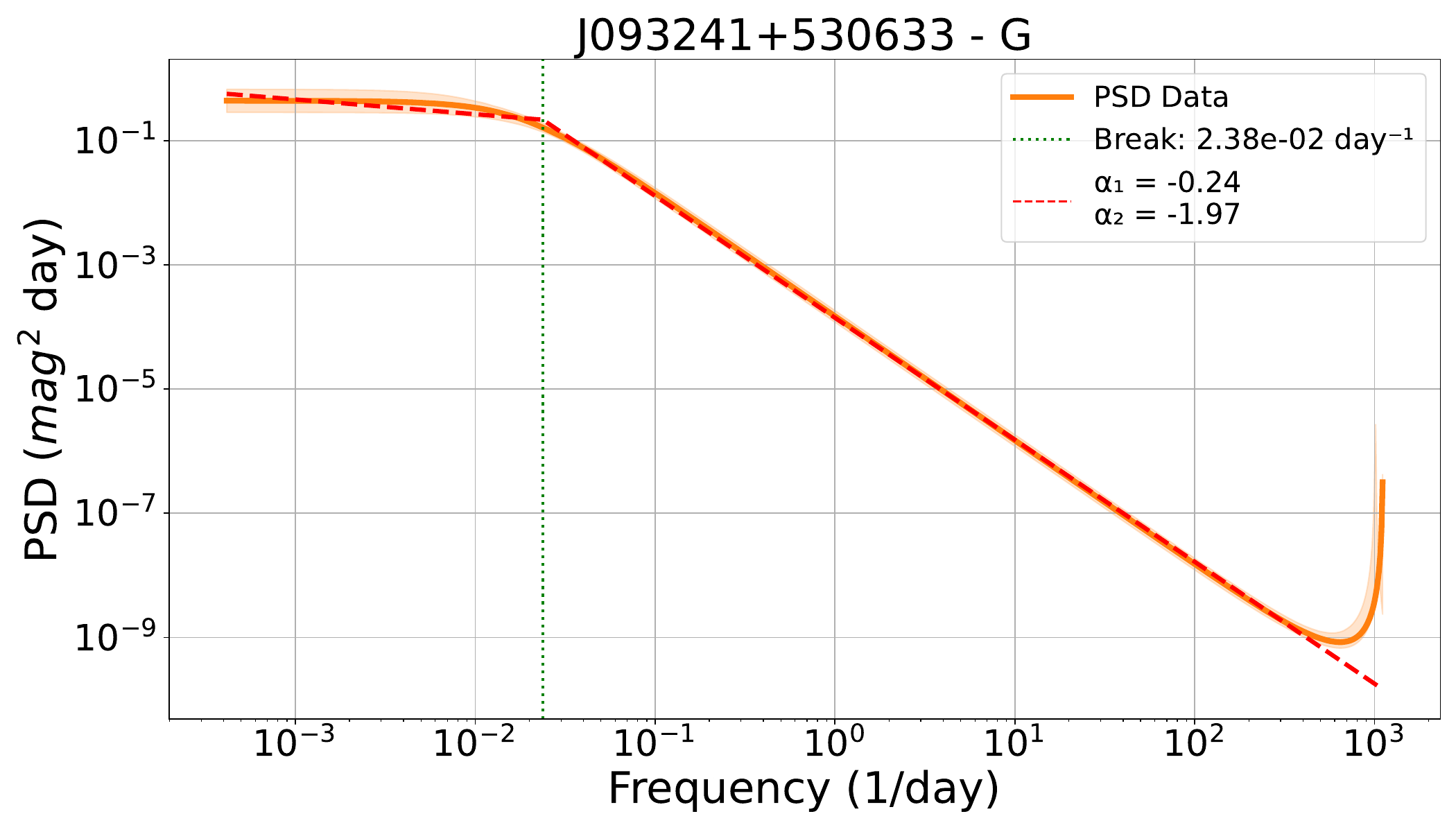}
    \end{minipage}
    \begin{minipage}{.3\textwidth}
        \centering
        \includegraphics[width=.99\linewidth]{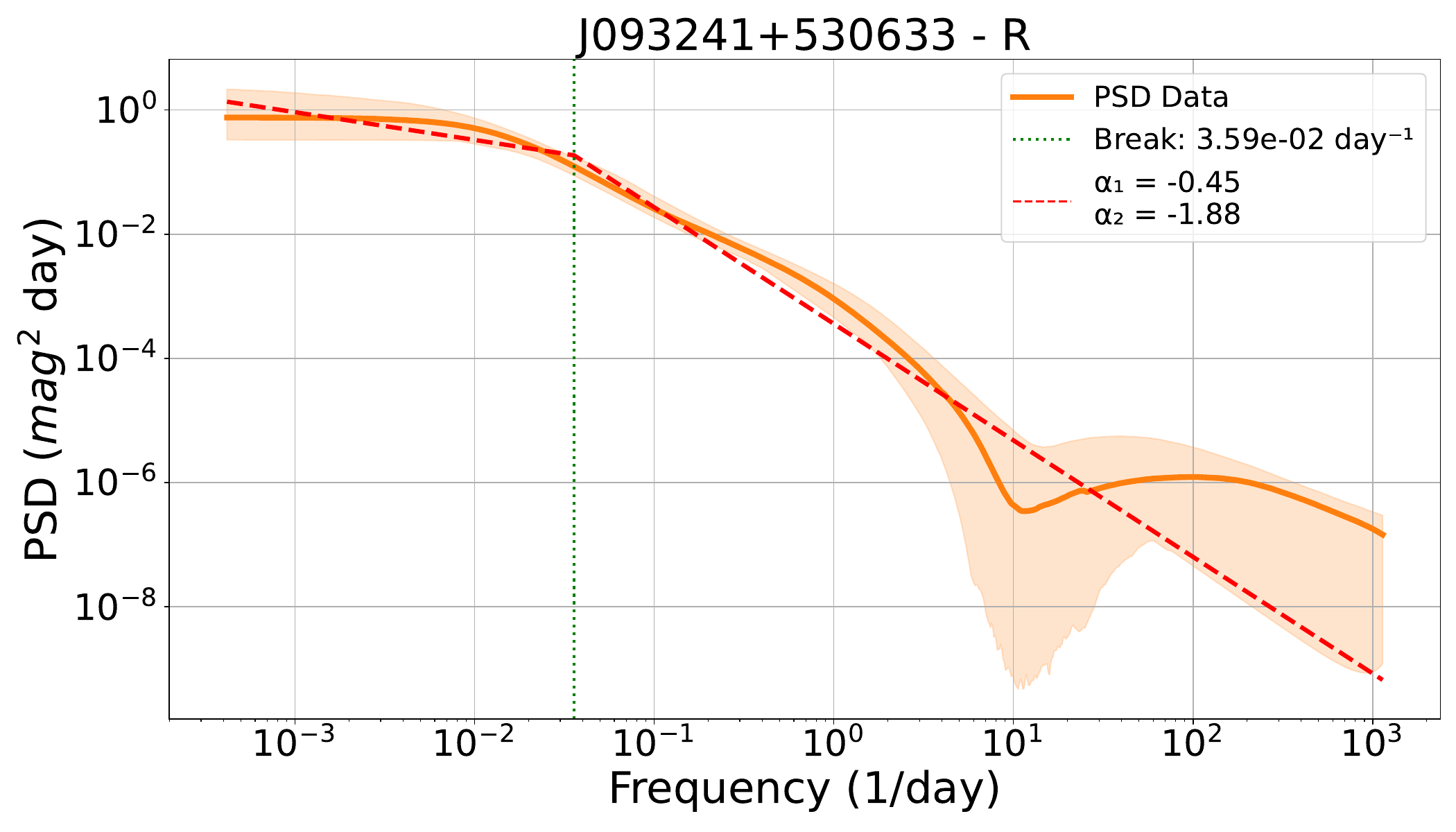}
    \end{minipage}
    \\
    \begin{minipage}{.3\textwidth}
        \centering
        \includegraphics[width=.99\linewidth]{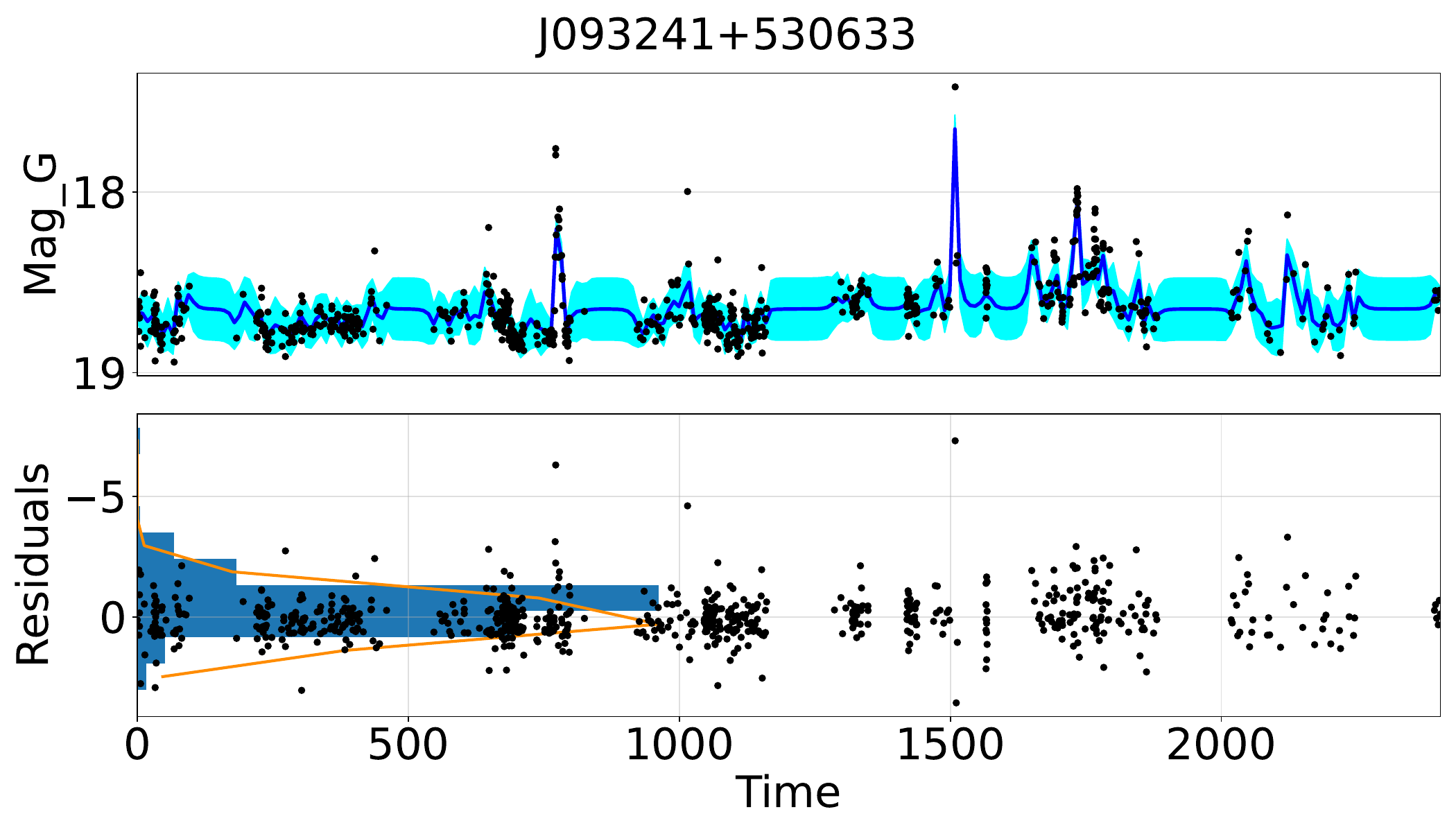}
    \end{minipage}
    \begin{minipage}{.3\textwidth}
        \centering
        \includegraphics[width=.99\linewidth]{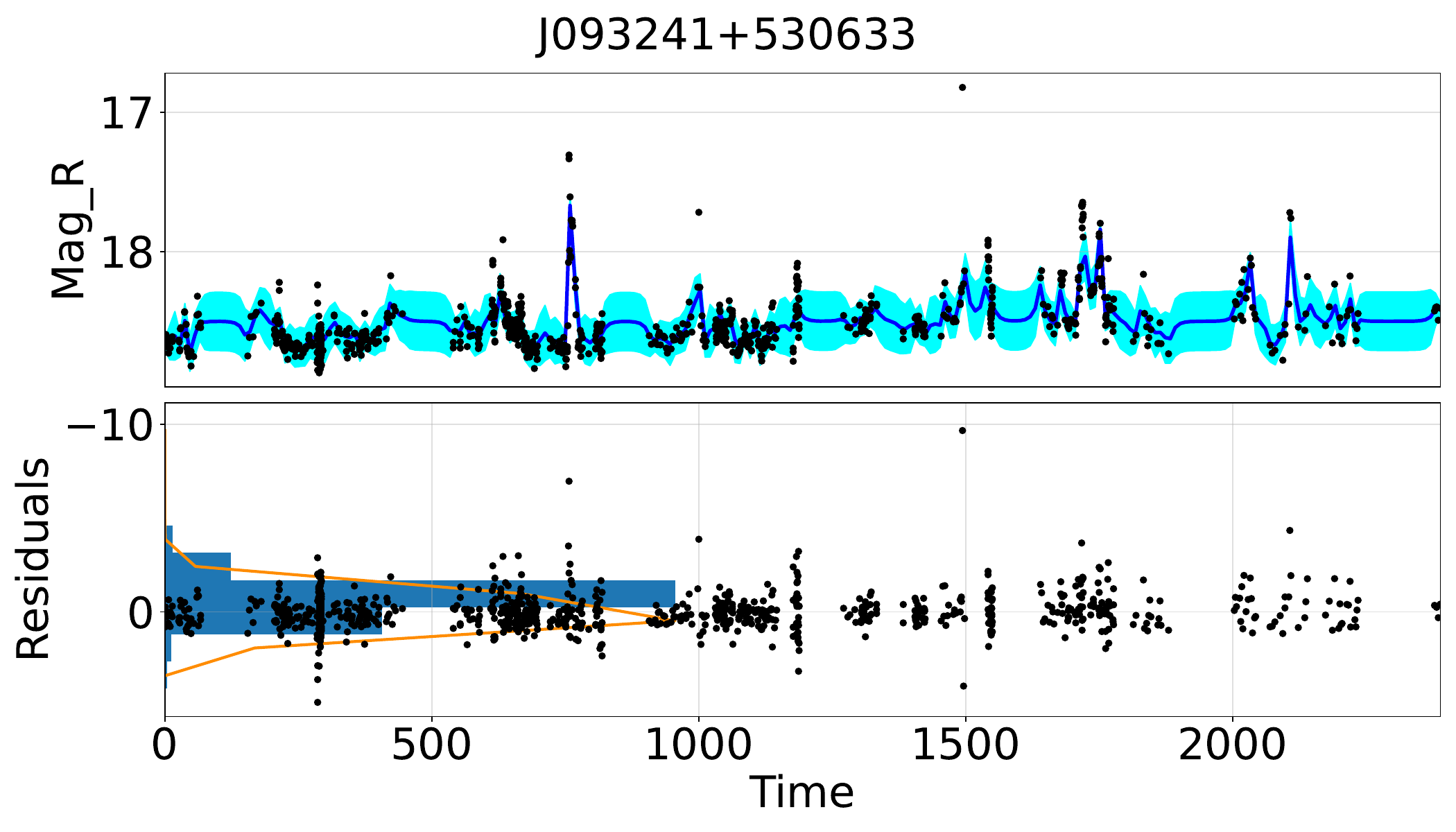}
    \end{minipage}
\end{figure*}

\begin{figure*}\label{J093712500851}
\centering
\caption{J093712+500851}
    \begin{minipage}{.3\textwidth}
        \centering
        \includegraphics[width=.99\linewidth]{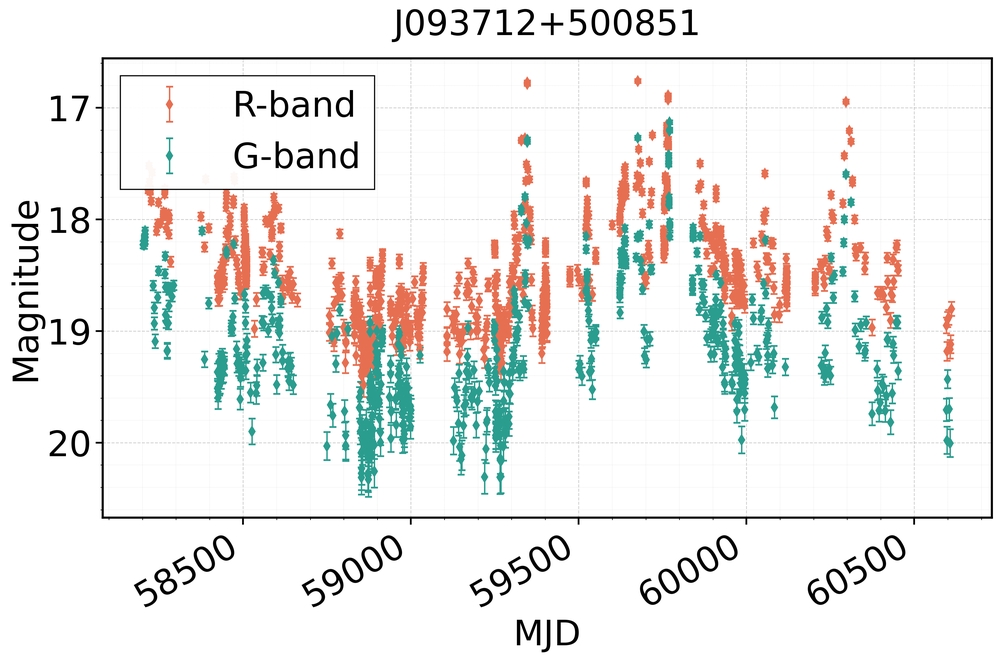}
    \end{minipage}
    \begin{minipage}{.3\textwidth}
        \centering
        \includegraphics[width=.99\linewidth]{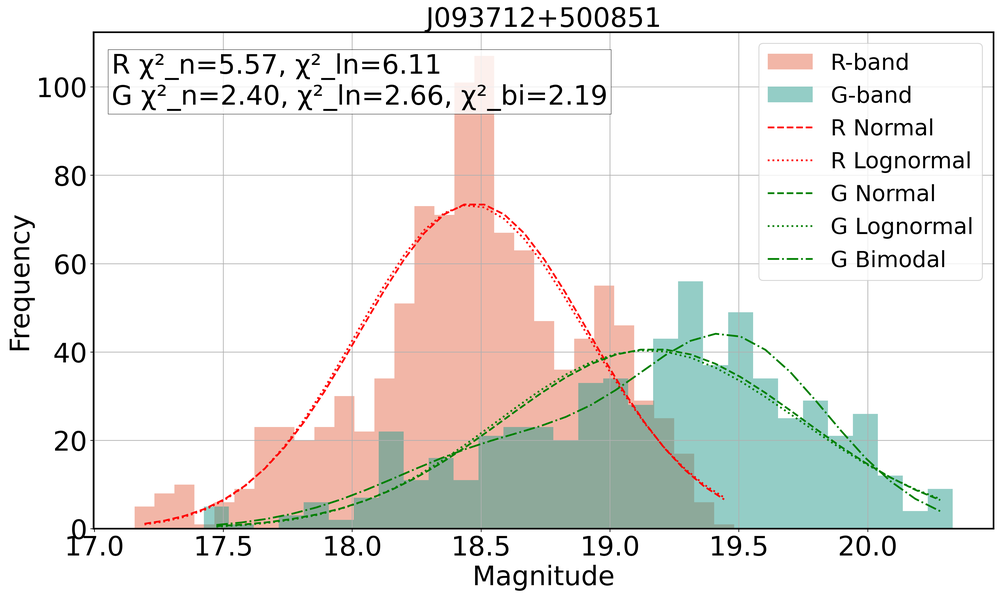}
    \end{minipage}
    \begin{minipage}{.3\textwidth}
        \centering
        \includegraphics[width=.99\linewidth]{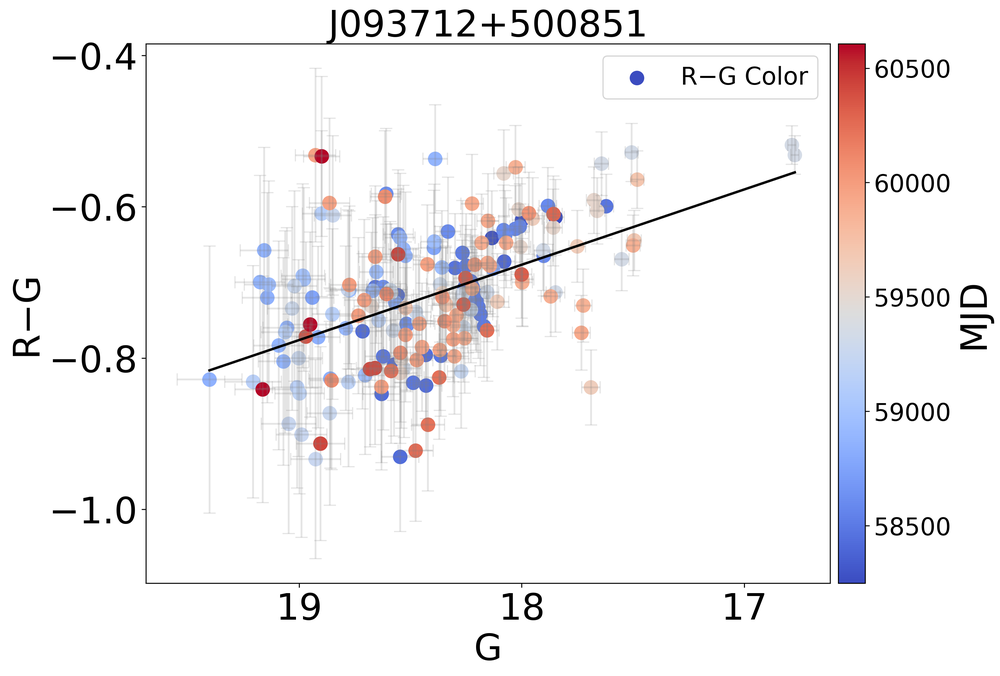}
    \end{minipage}
    \\
    \begin{minipage}{.3\textwidth}
        \centering
        \includegraphics[width=.99\linewidth]{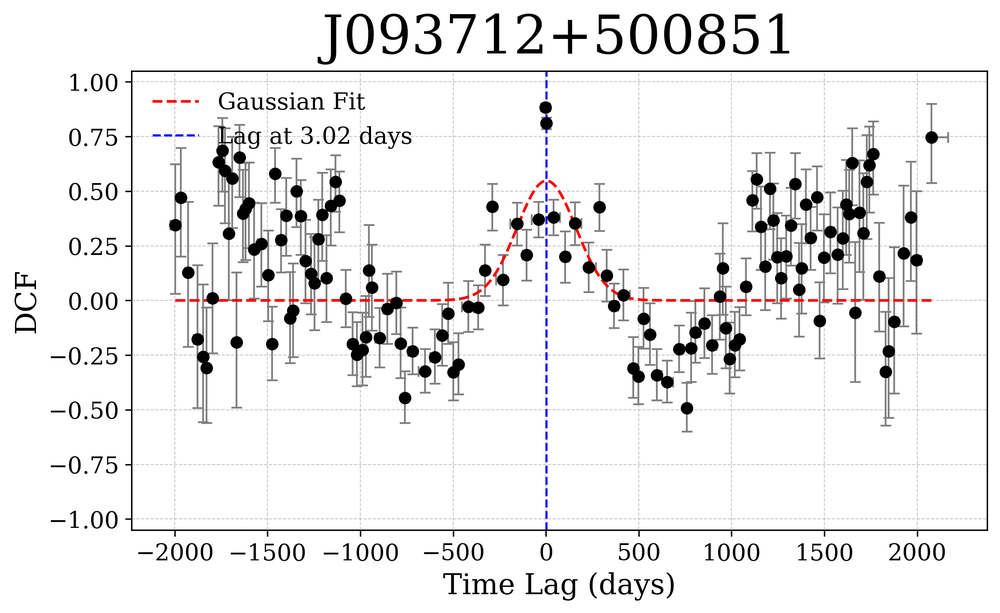}
    \end{minipage}
    \begin{minipage}{.3\textwidth}
        \centering
        \includegraphics[width=.99\linewidth]{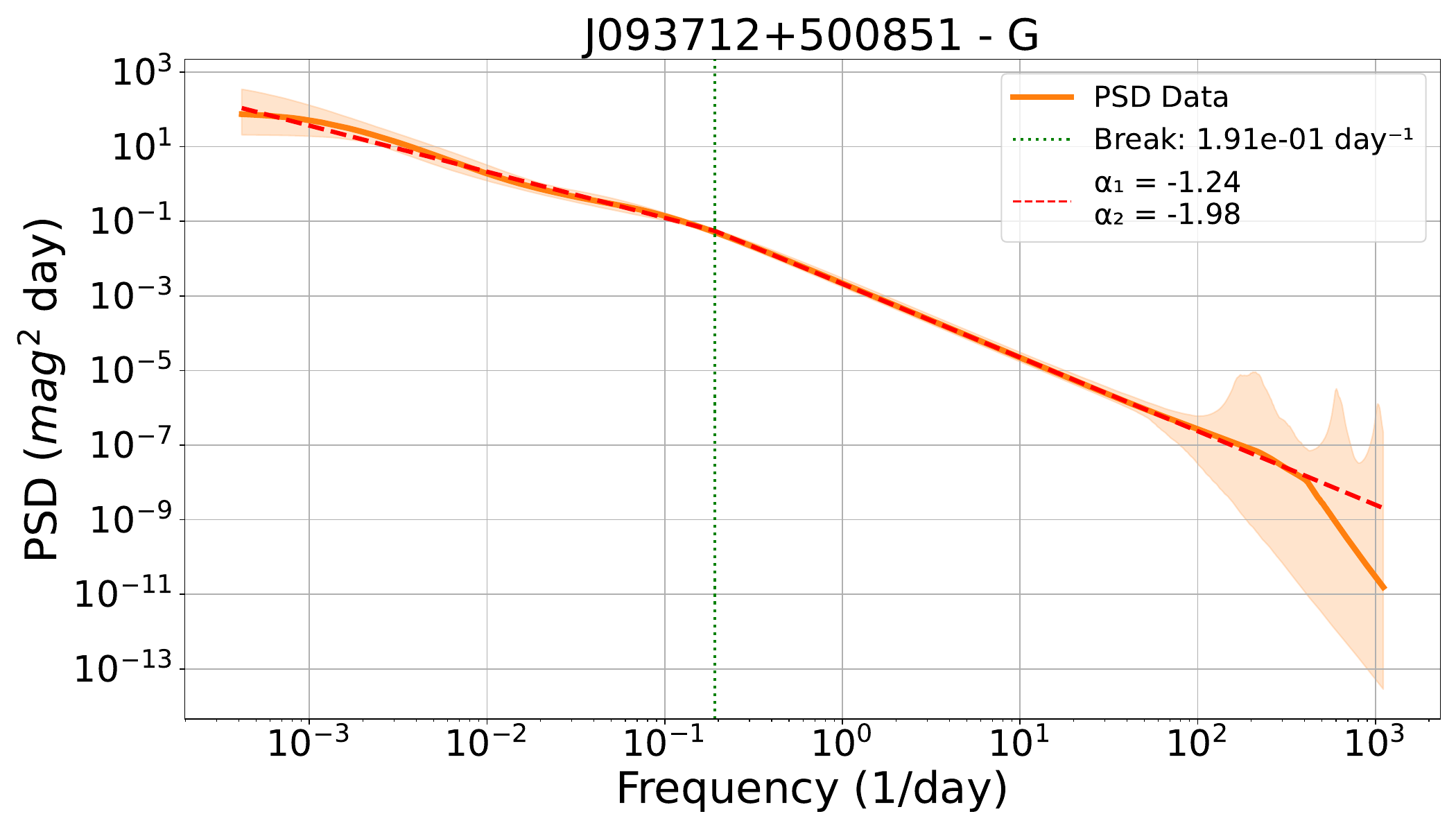}
    \end{minipage}
    \begin{minipage}{.3\textwidth}
        \centering
        \includegraphics[width=.99\linewidth]{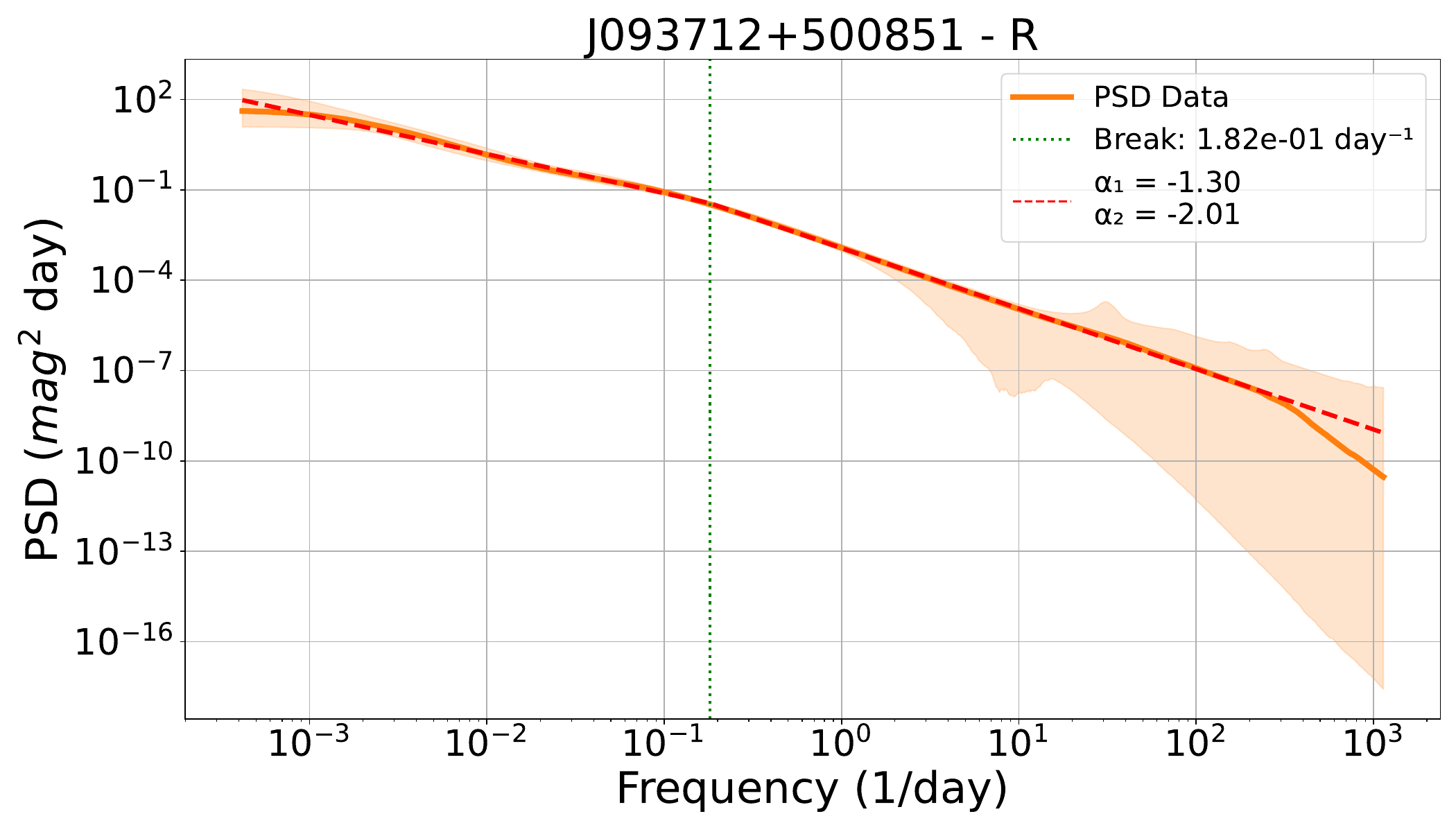}
    \end{minipage}
    \\
    \begin{minipage}{.3\textwidth}
        \centering
        \includegraphics[width=.99\linewidth]{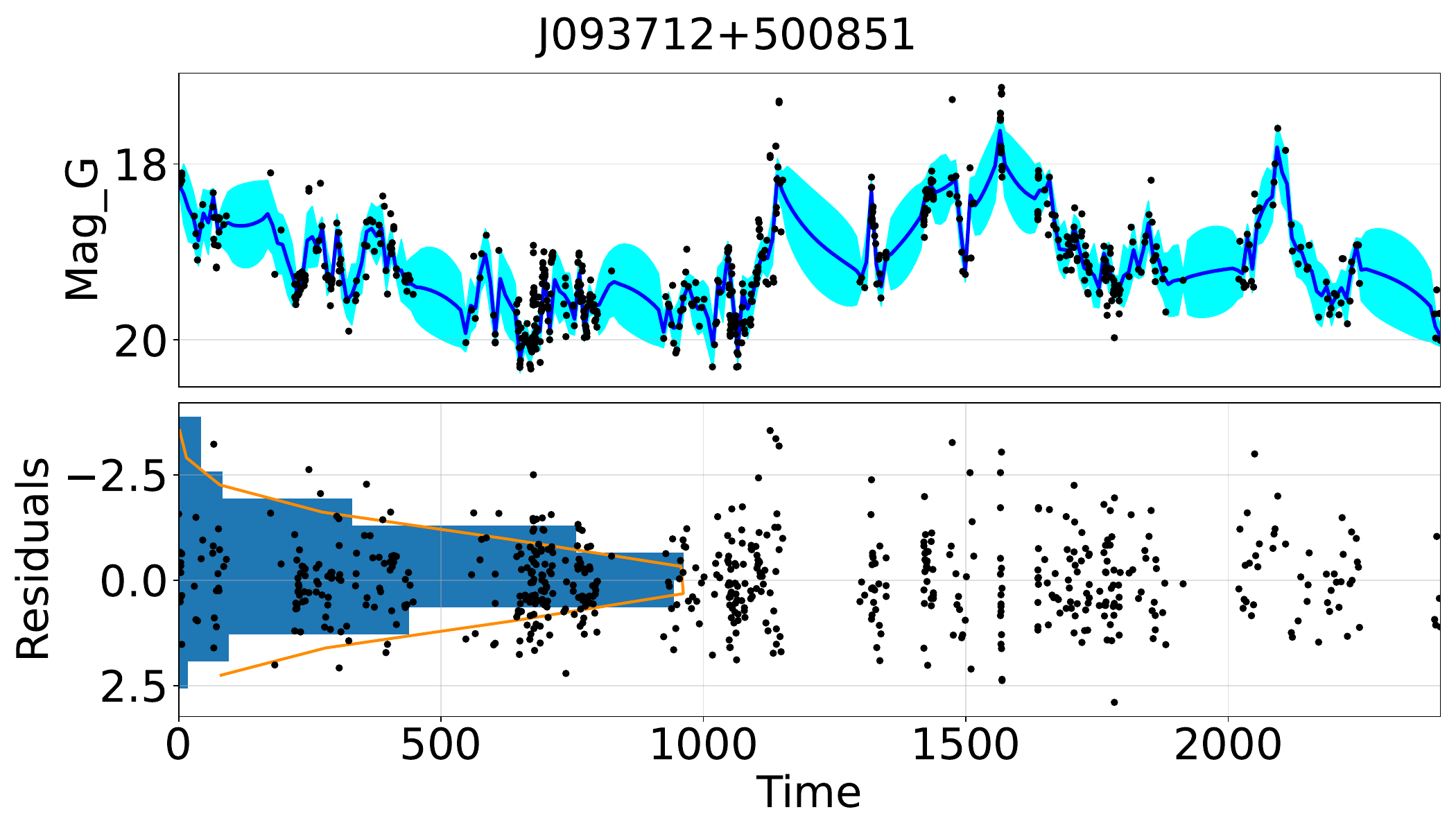}
    \end{minipage}
    \begin{minipage}{.3\textwidth}
        \centering
        \includegraphics[width=.99\linewidth]{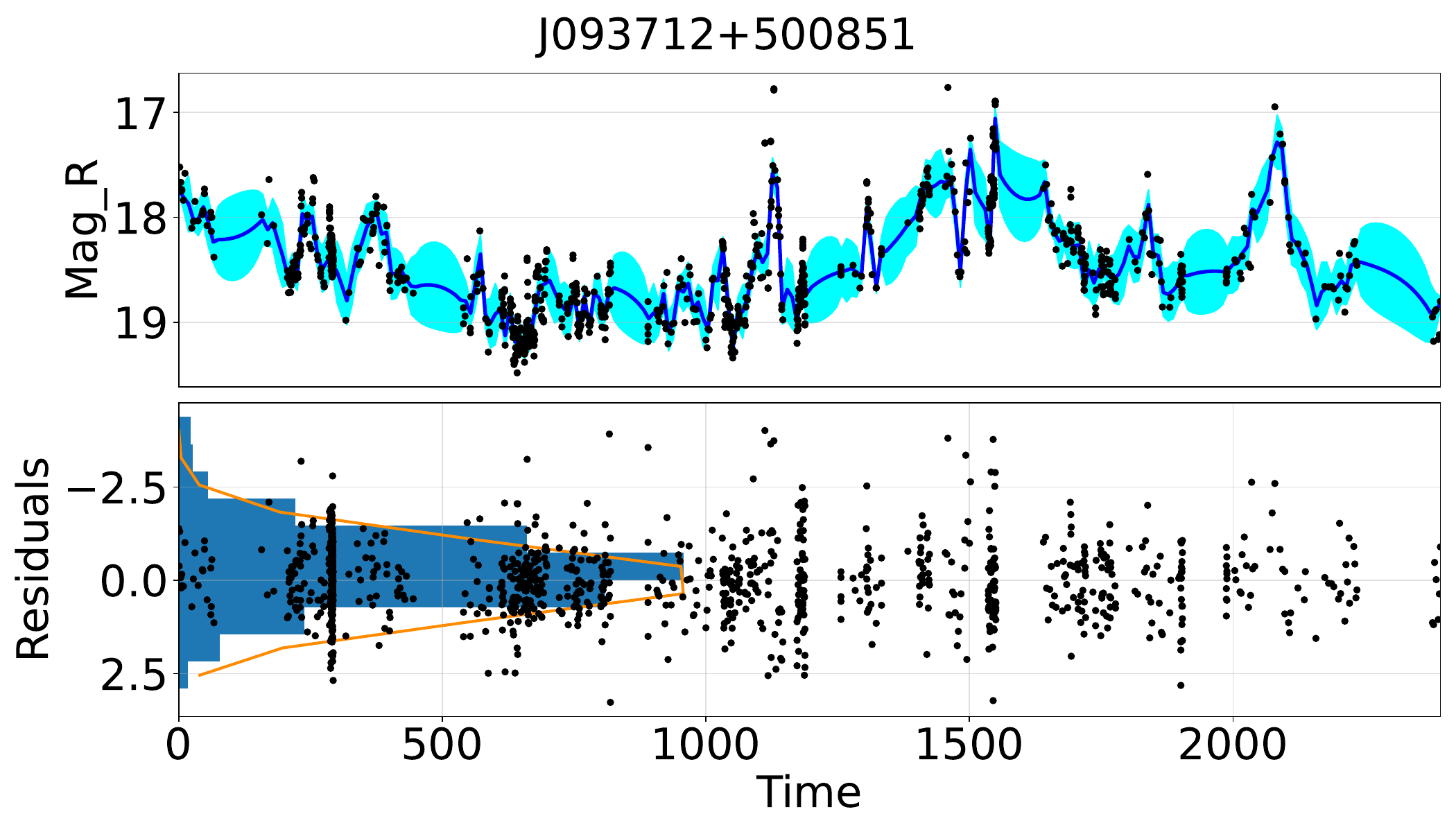}
    \end{minipage}
\end{figure*}

\begin{figure*}\label{J094635101706}
\centering
\caption{J094635+101706}
    \begin{minipage}{.3\textwidth}
        \centering
        \includegraphics[width=.99\linewidth]{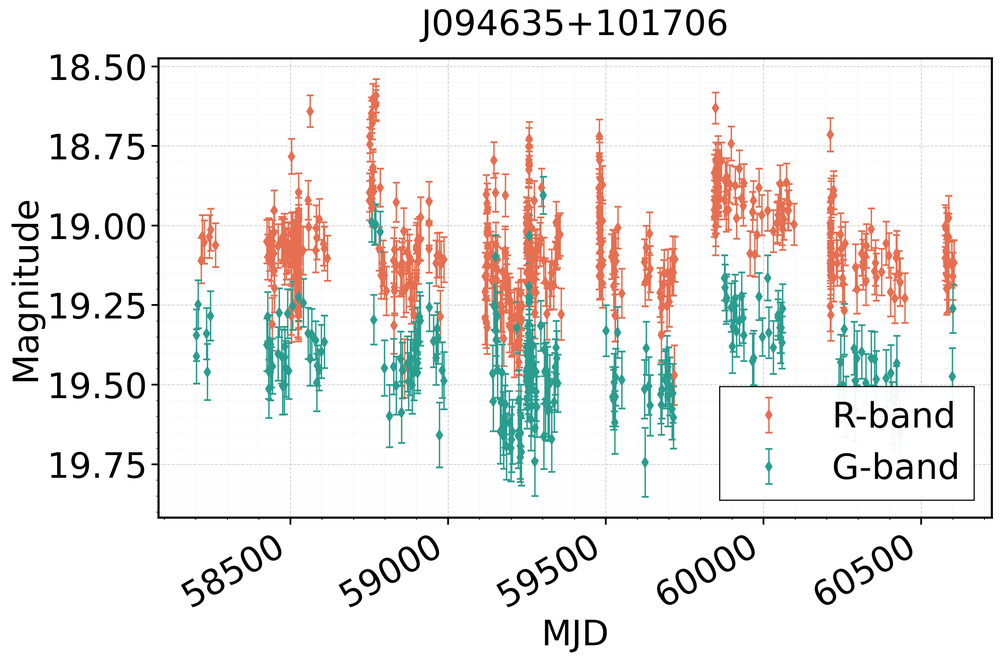}
    \end{minipage}
    \begin{minipage}{.3\textwidth}
        \centering
        \includegraphics[width=.99\linewidth]{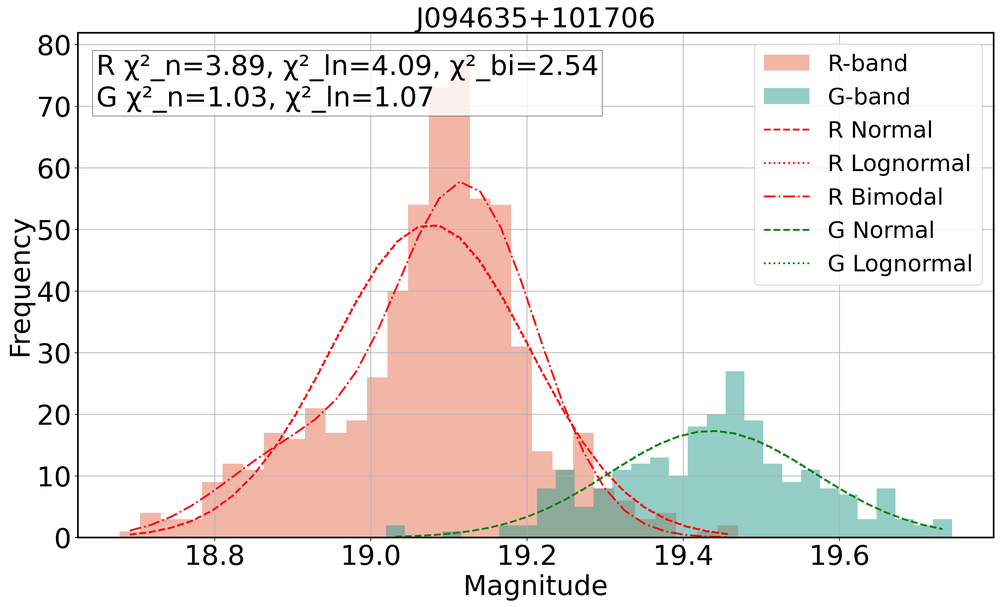}
    \end{minipage}
    \begin{minipage}{.3\textwidth}
        \centering
        \includegraphics[width=.99\linewidth]{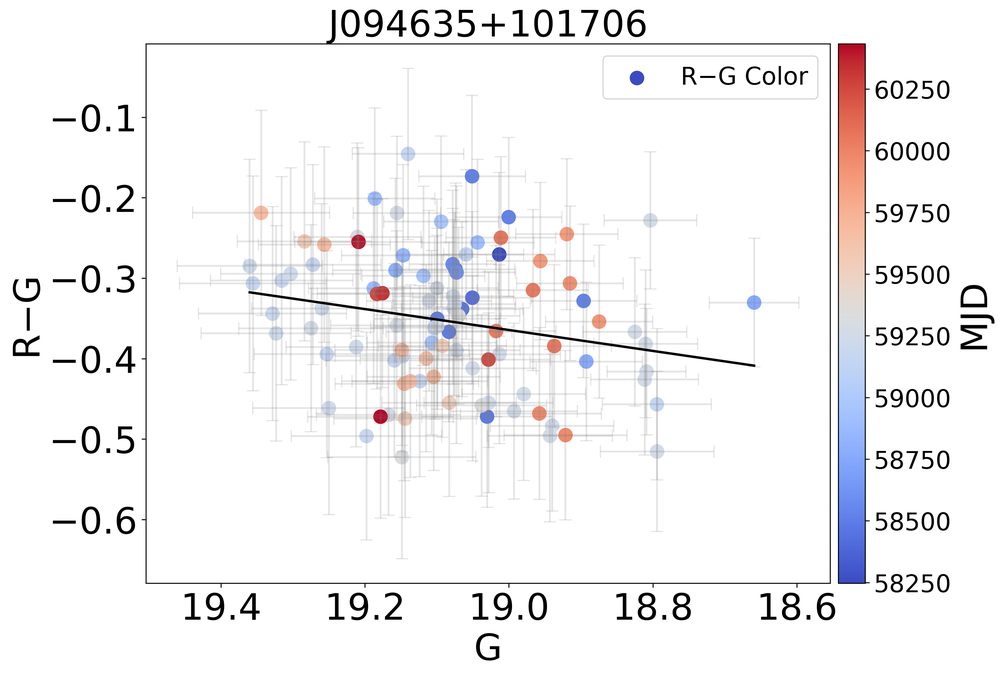}
    \end{minipage}
    \\
    \begin{minipage}{.3\textwidth}
        \centering
        \includegraphics[width=.99\linewidth]{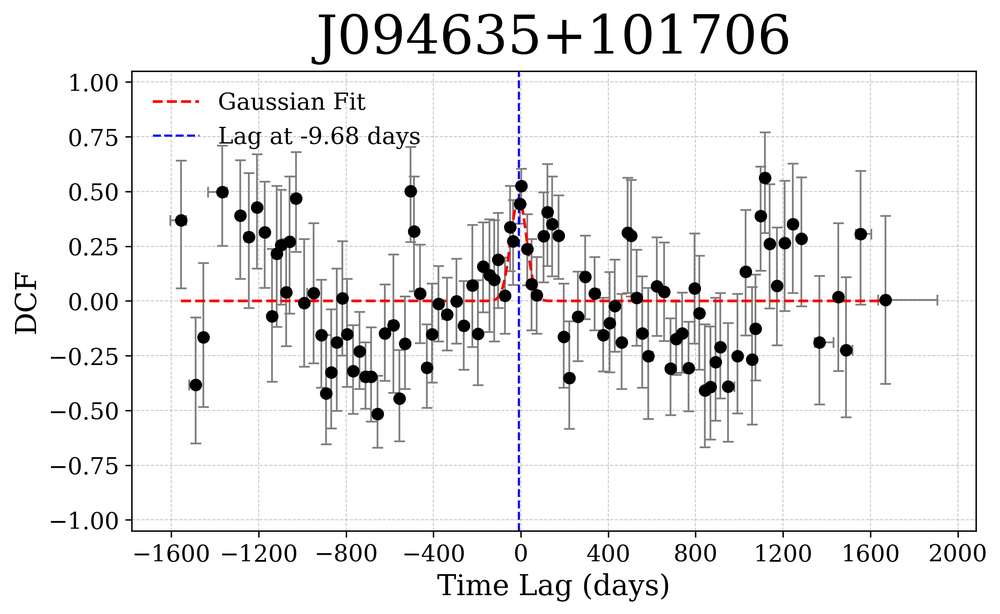}
    \end{minipage}
    \begin{minipage}{.3\textwidth}
        \centering
        \includegraphics[width=.99\linewidth]{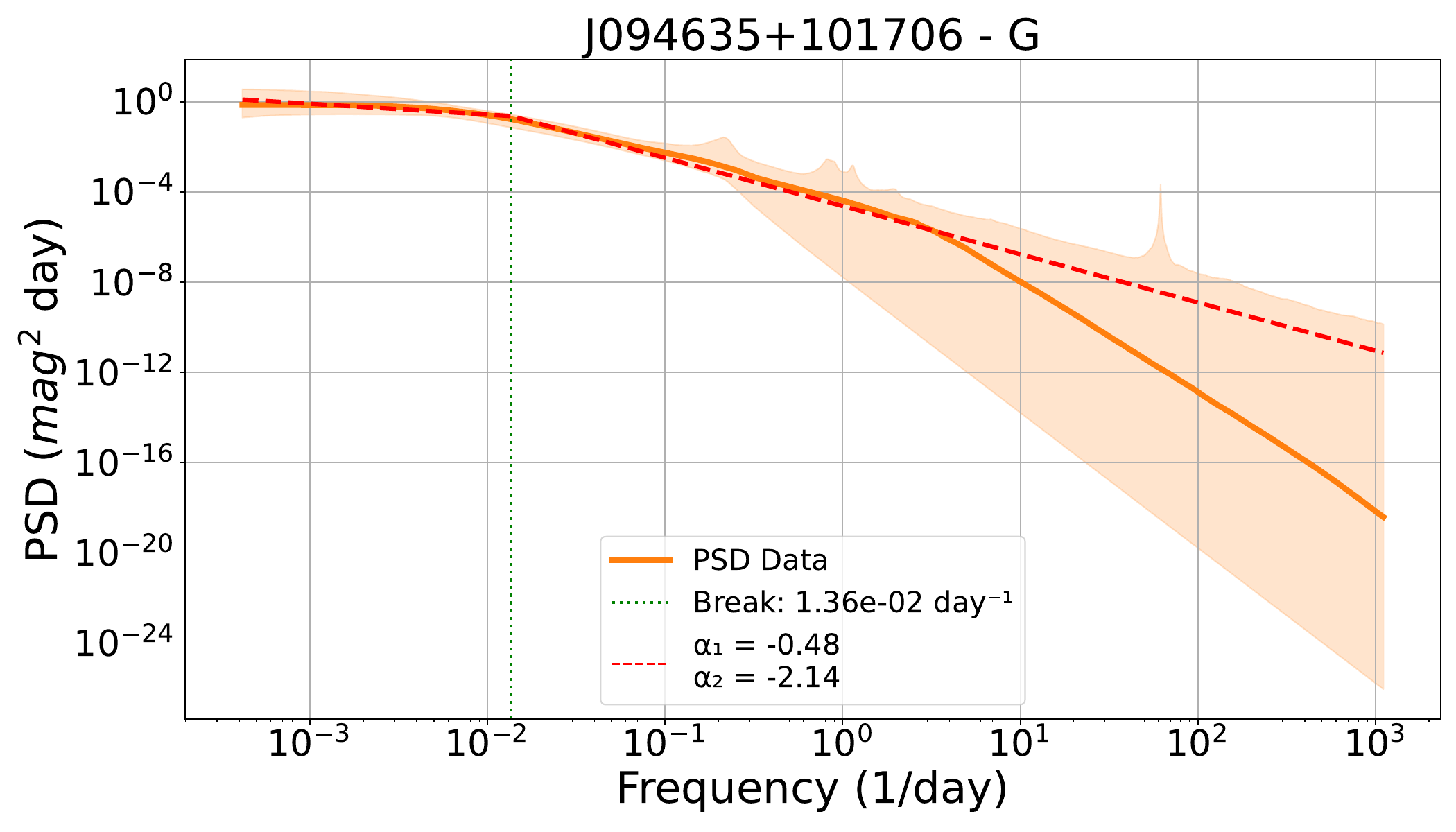}
    \end{minipage}
    \begin{minipage}{.3\textwidth}
        \centering
        \includegraphics[width=.99\linewidth]{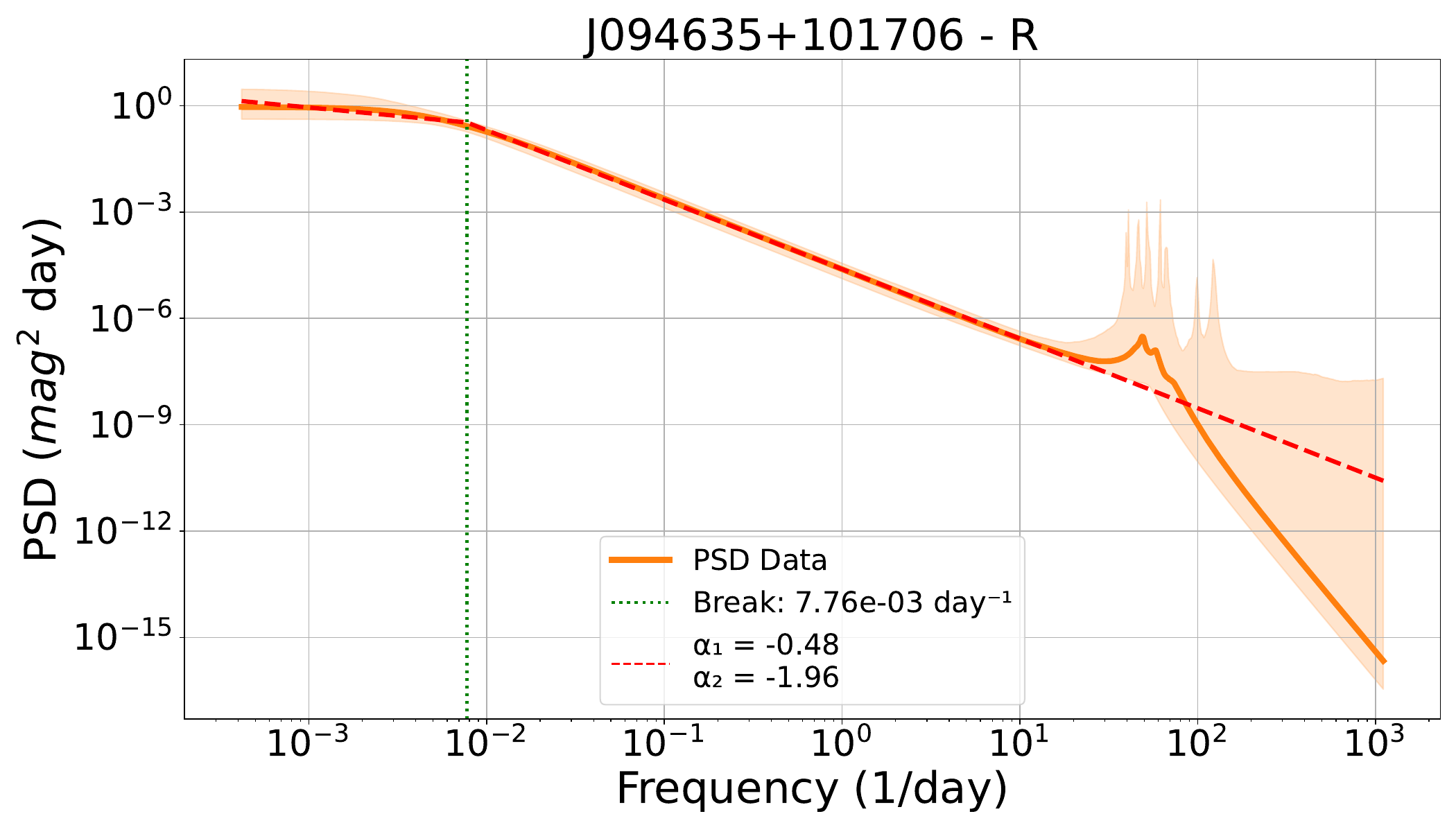}
    \end{minipage}
    \\
    \begin{minipage}{.3\textwidth}
        \centering
        \includegraphics[width=.99\linewidth]{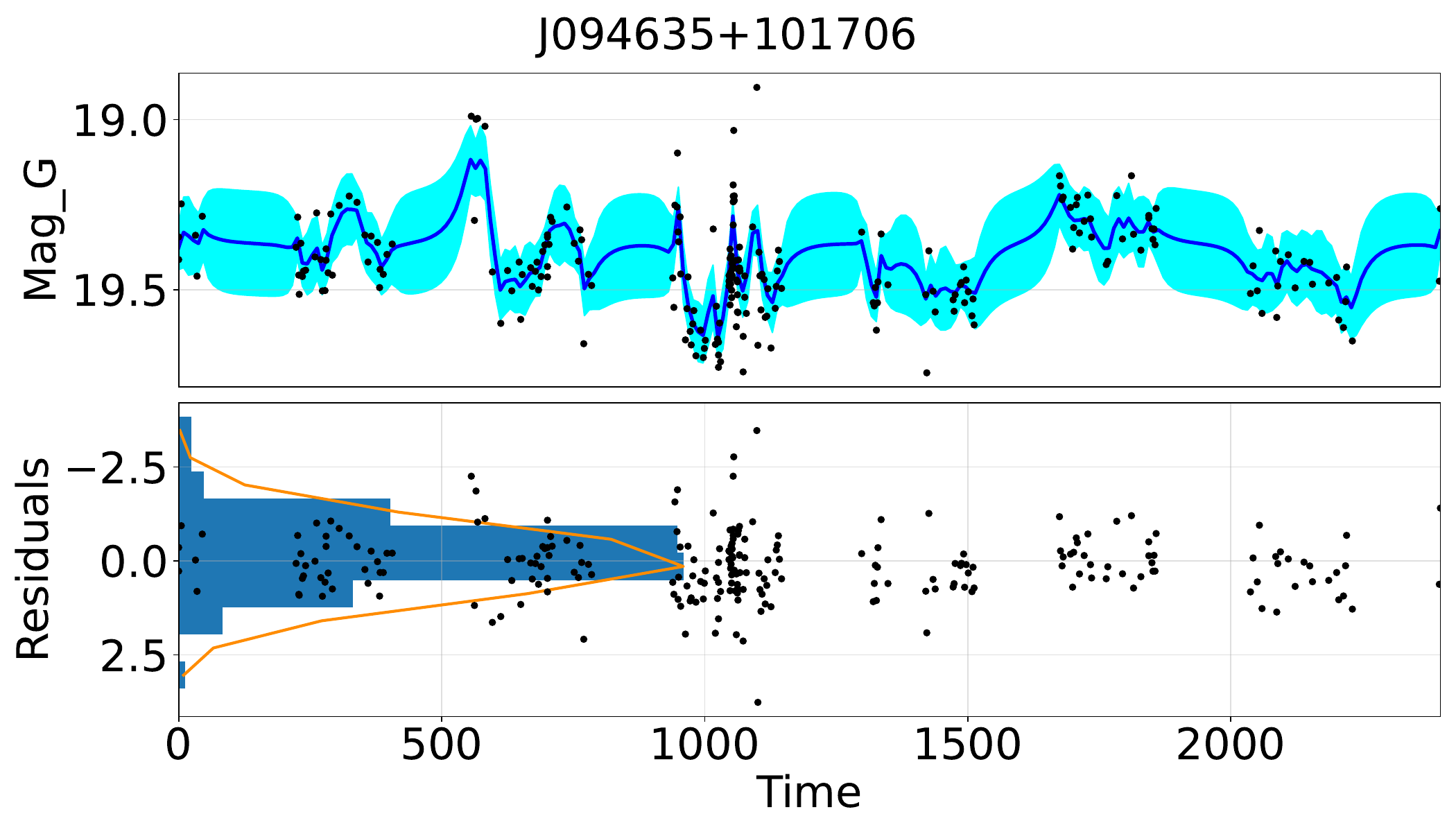}
    \end{minipage}
    \begin{minipage}{.3\textwidth}
        \centering
        \includegraphics[width=.99\linewidth]{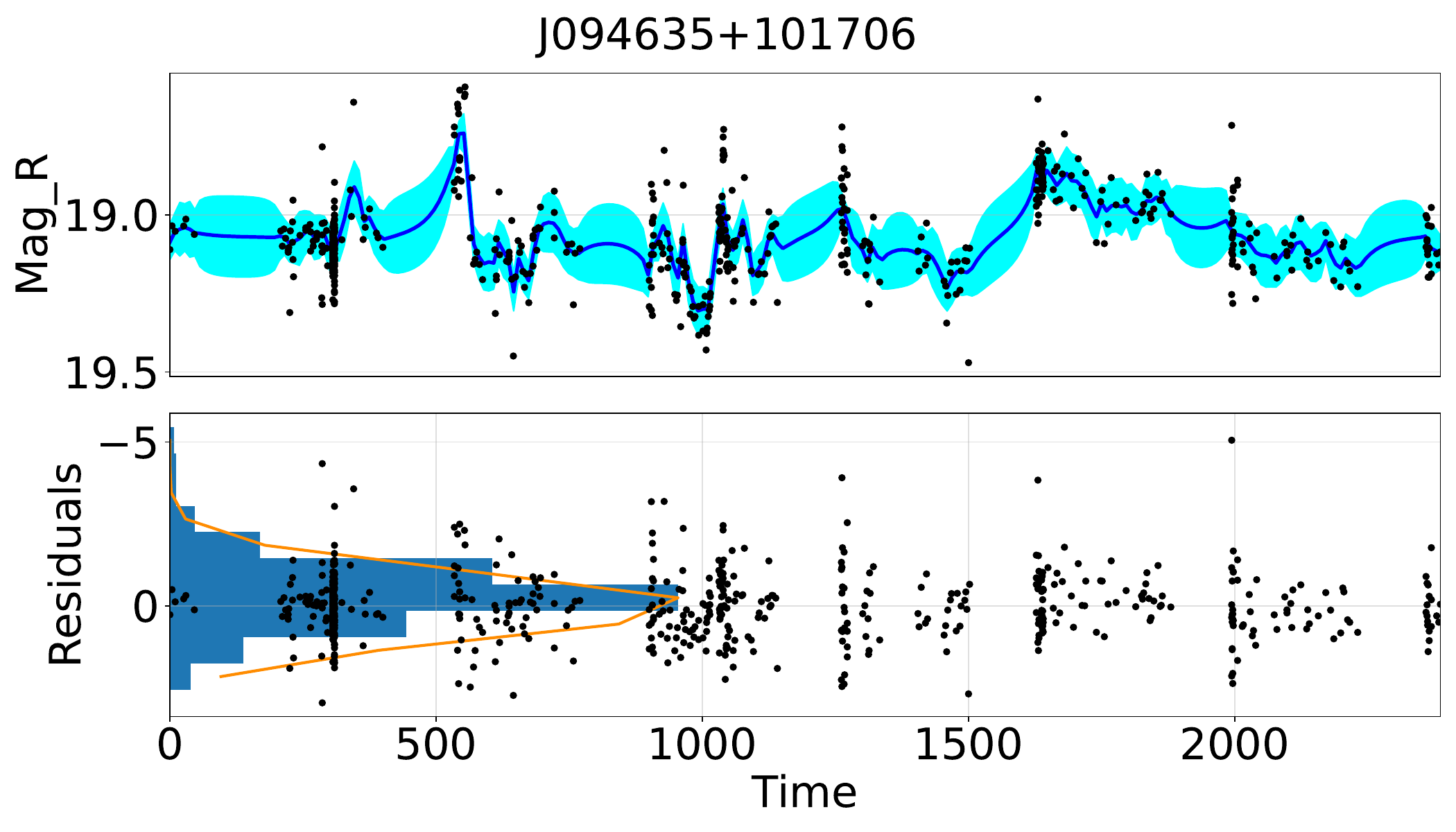}
    \end{minipage}
\end{figure*}

\begin{figure*}\label{J094857002226}
\centering
\caption{J094857+002226}
    \begin{minipage}{.3\textwidth}
        \centering
        \includegraphics[width=.99\linewidth]{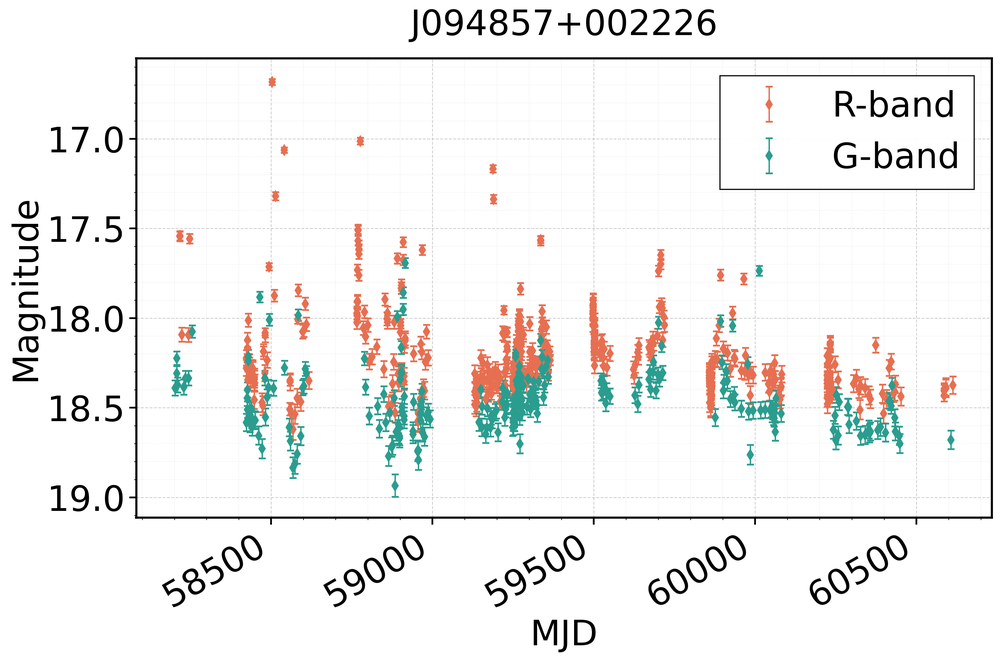}
    \end{minipage}
    \begin{minipage}{.3\textwidth}
        \centering
        \includegraphics[width=.99\linewidth]{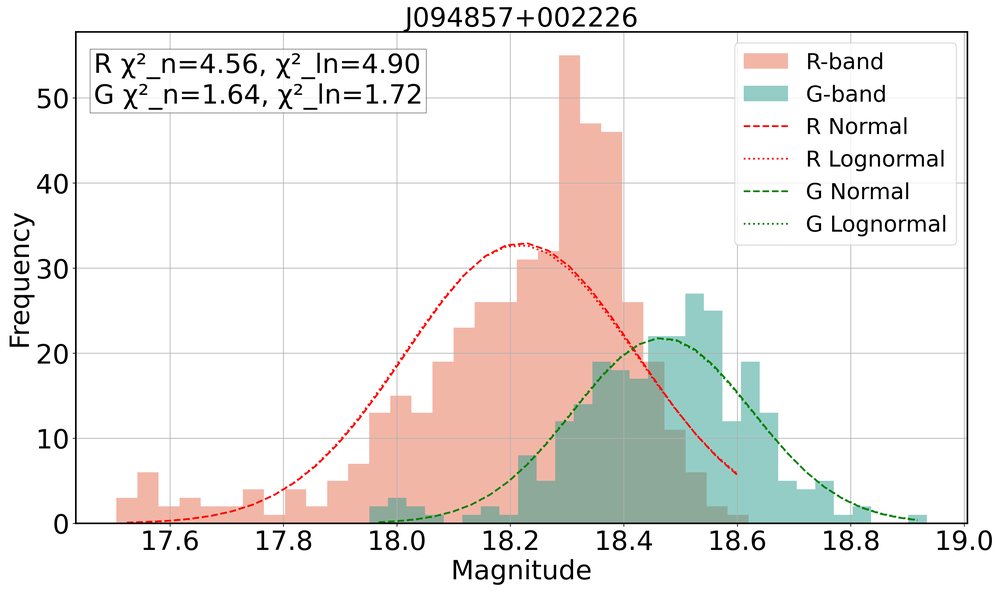}
    \end{minipage}
    \begin{minipage}{.3\textwidth}
        \centering
        \includegraphics[width=.99\linewidth]{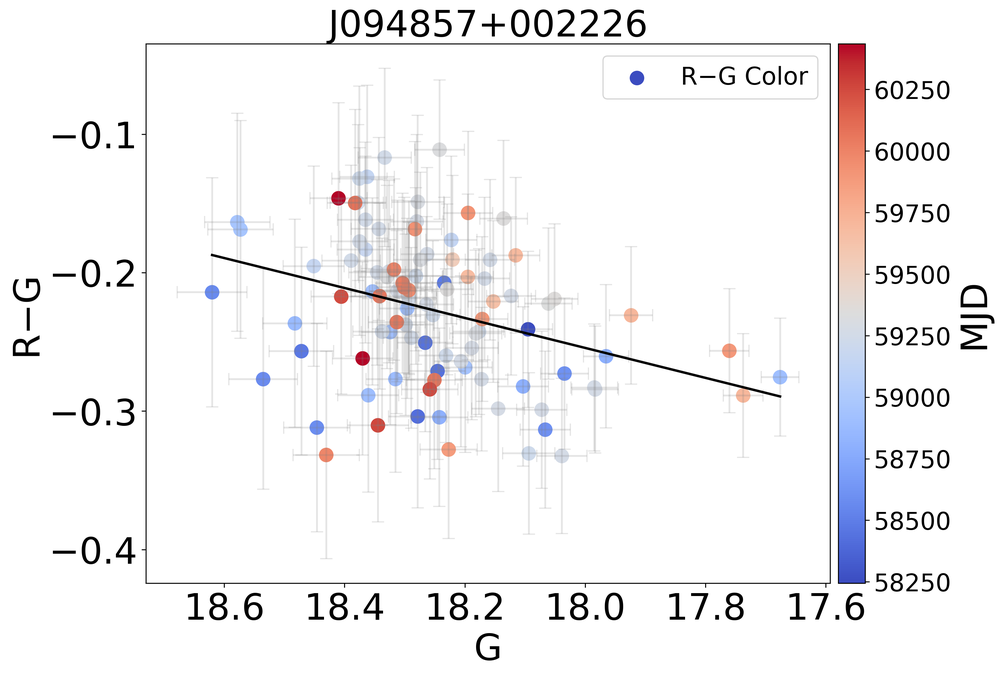}
    \end{minipage}
    \\
    \begin{minipage}{.3\textwidth}
        \centering
        \includegraphics[width=.99\linewidth]{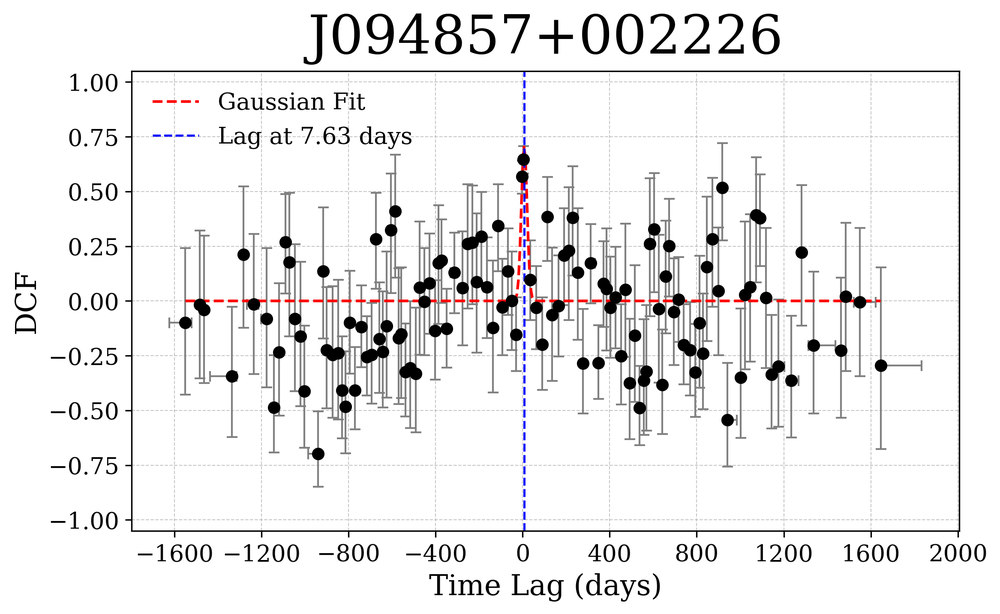}
    \end{minipage}
    \begin{minipage}{.3\textwidth}
        \centering
        \includegraphics[width=.99\linewidth]{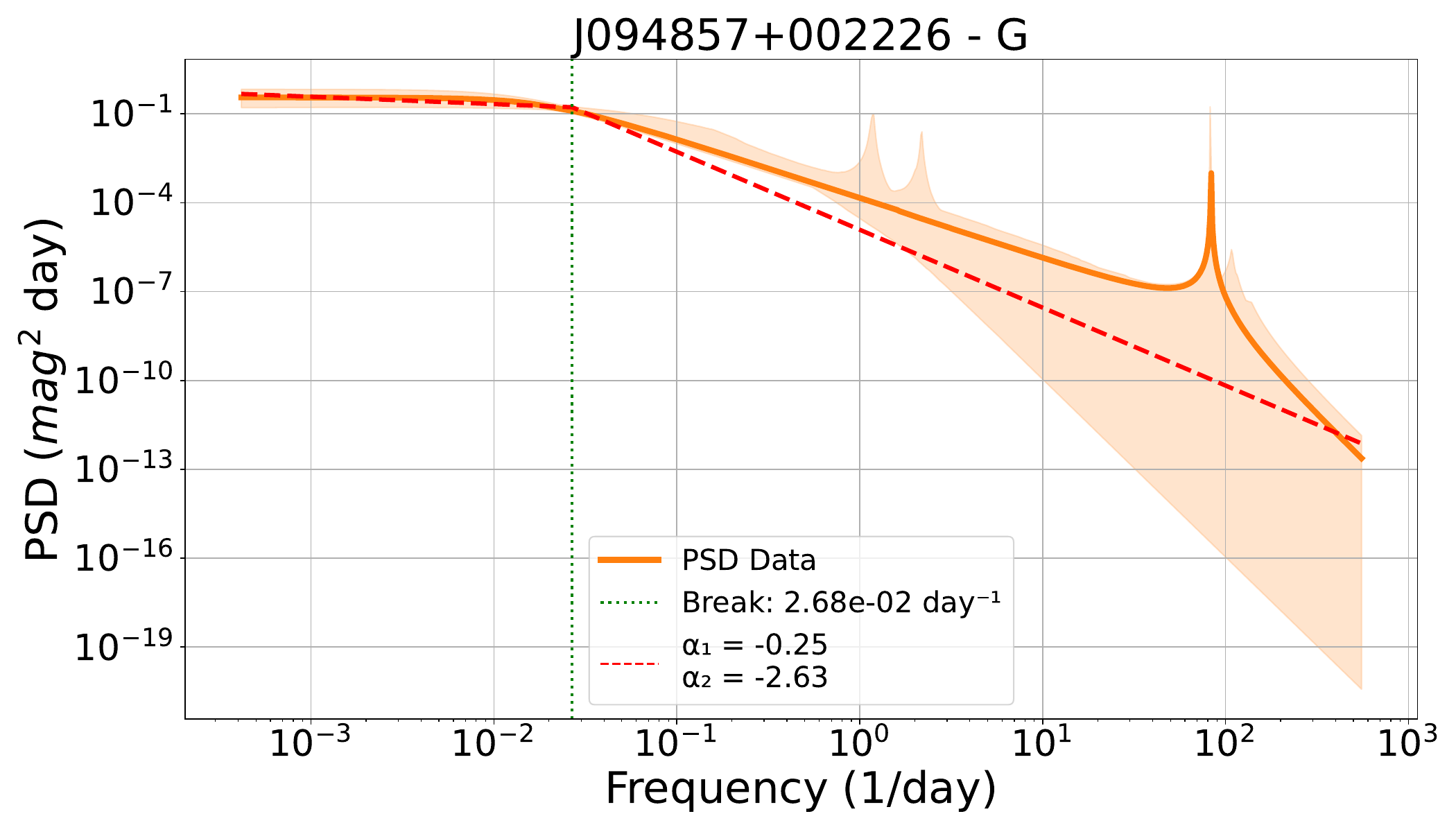}
    \end{minipage}
    \begin{minipage}{.3\textwidth}
        \centering
        \includegraphics[width=.99\linewidth]{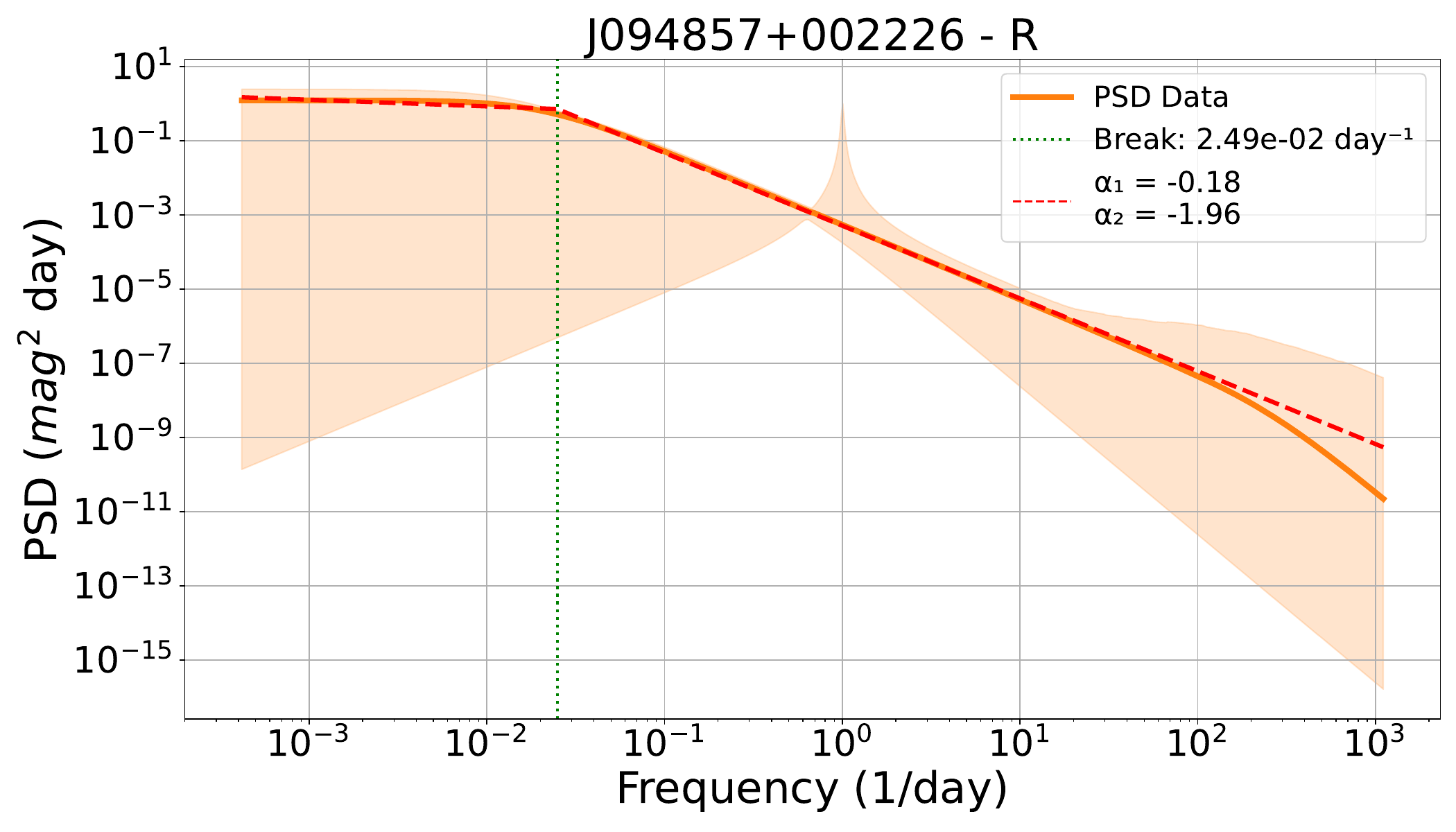}
    \end{minipage}
    \\
    \begin{minipage}{.3\textwidth}
        \centering
        \includegraphics[width=.99\linewidth]{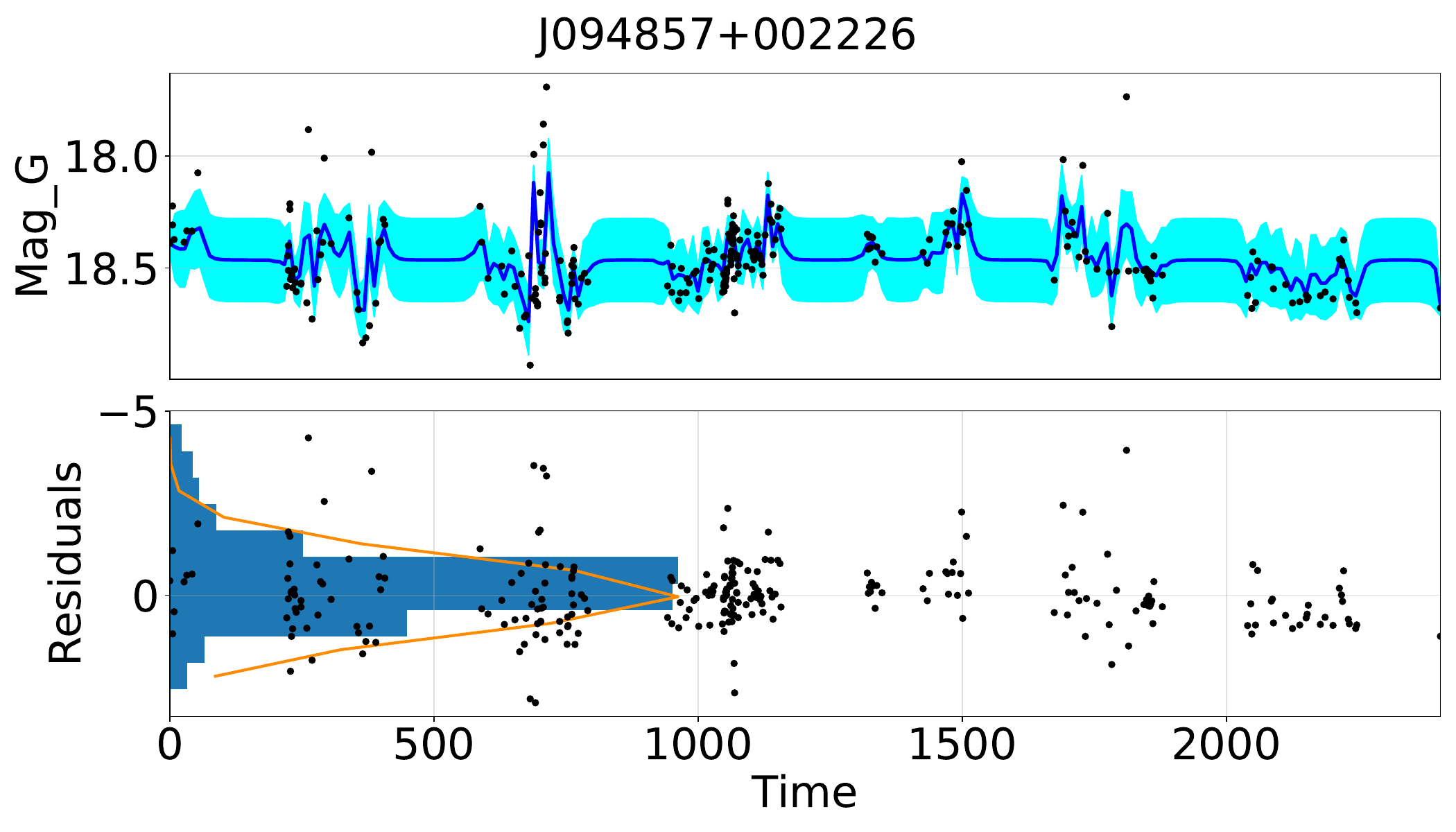}
    \end{minipage}
    \begin{minipage}{.3\textwidth}
        \centering
        \includegraphics[width=.99\linewidth]{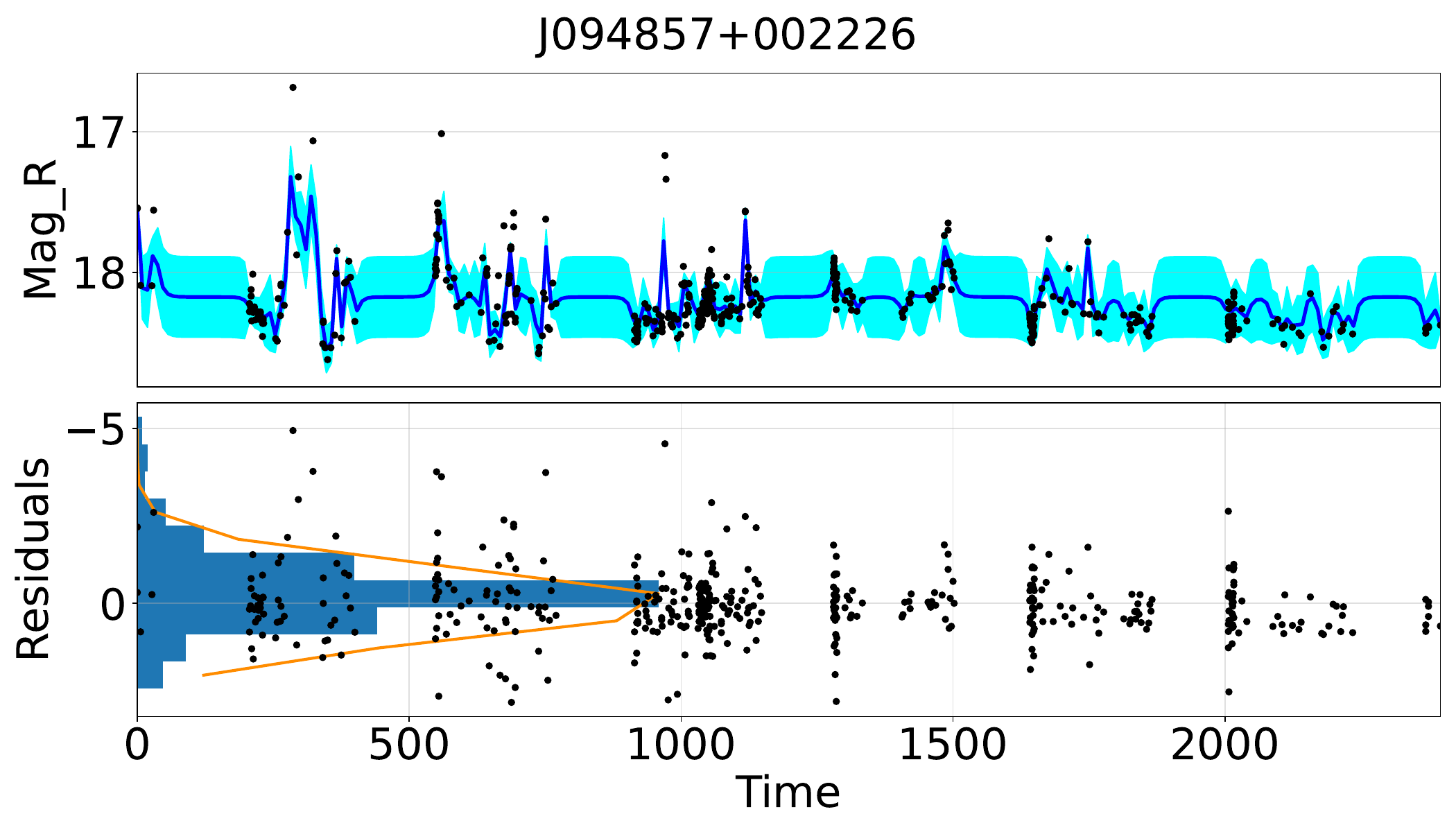}
    \end{minipage}
\end{figure*}

\begin{figure*}\label{J122222041315}
\centering
\caption{J122222+041315}
    \begin{minipage}{.3\textwidth}
        \centering
        \includegraphics[width=.99\linewidth]{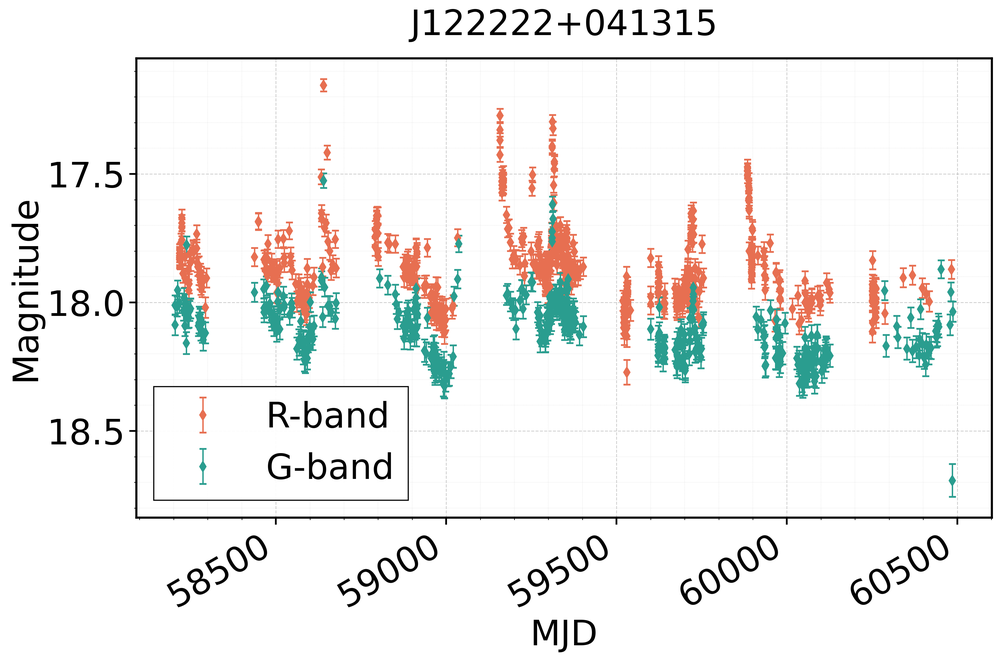}
    \end{minipage}
    \begin{minipage}{.3\textwidth}
        \centering
        \includegraphics[width=.99\linewidth]{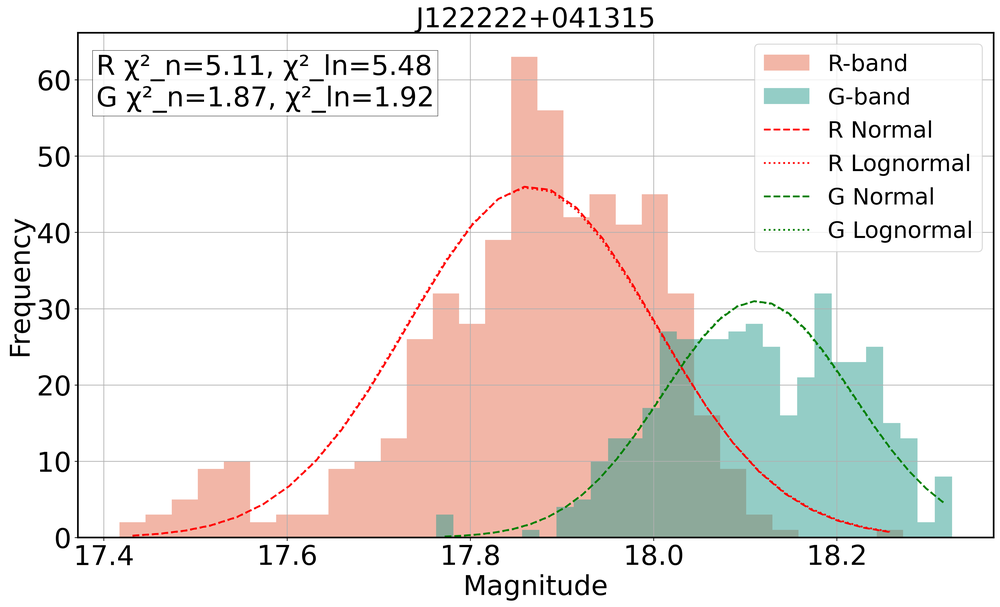}
    \end{minipage}
    \begin{minipage}{.3\textwidth}
        \centering
        \includegraphics[width=.99\linewidth]{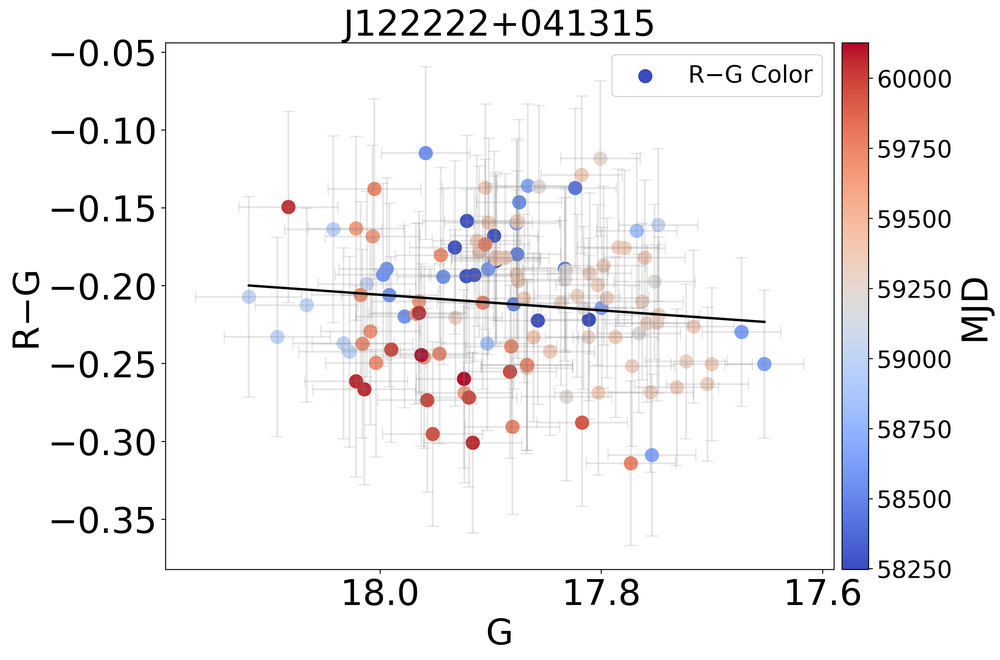}
    \end{minipage}
    \\
    \begin{minipage}{.3\textwidth}
        \centering
        \includegraphics[width=.99\linewidth]{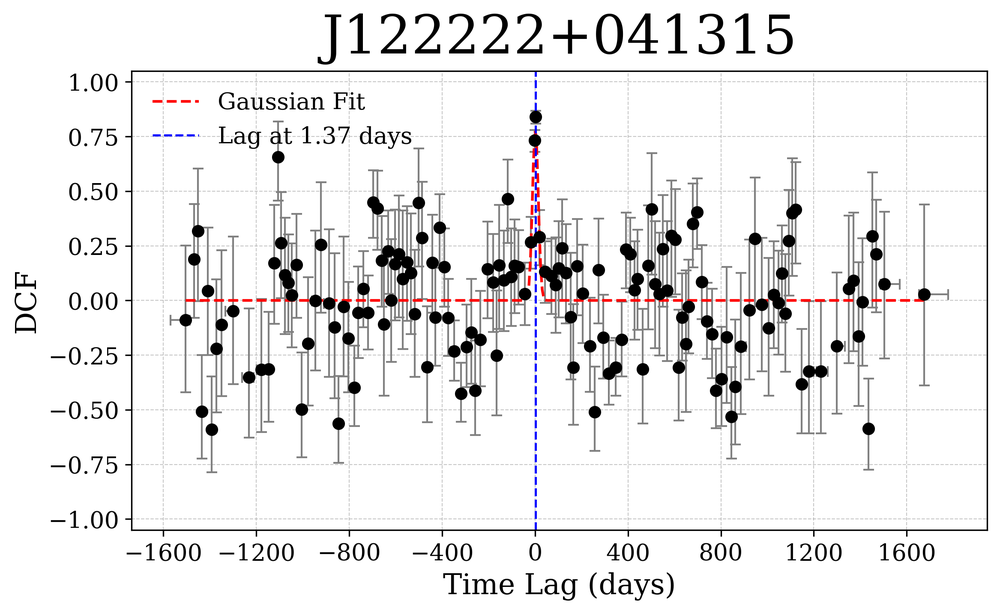}
    \end{minipage}
    \begin{minipage}{.3\textwidth}
        \centering
        \includegraphics[width=.99\linewidth]{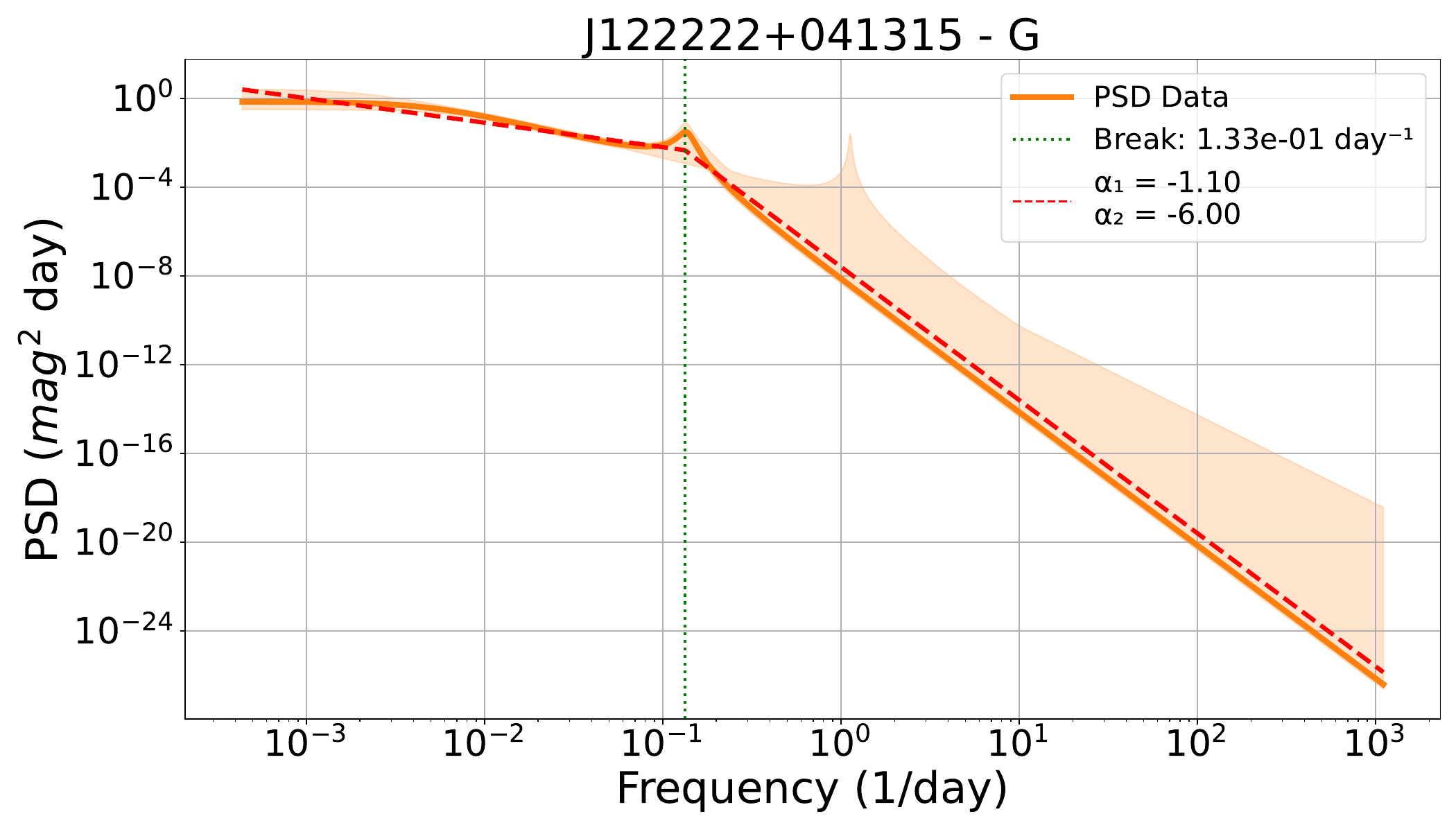}
    \end{minipage}
    \begin{minipage}{.3\textwidth}
        \centering
        \includegraphics[width=.99\linewidth]{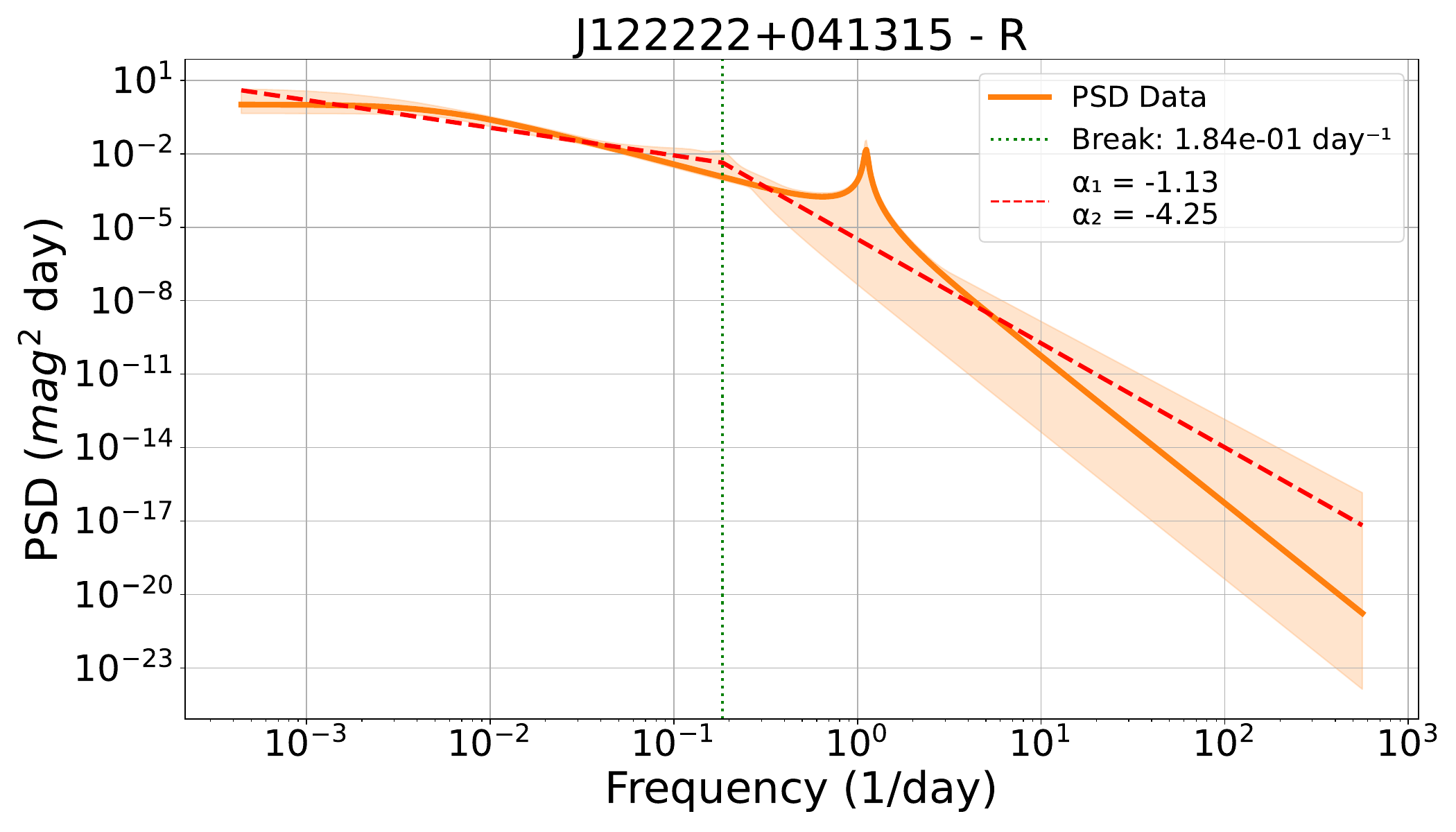}
    \end{minipage}
    \\
    \begin{minipage}{.3\textwidth}
        \centering
        \includegraphics[width=.99\linewidth]{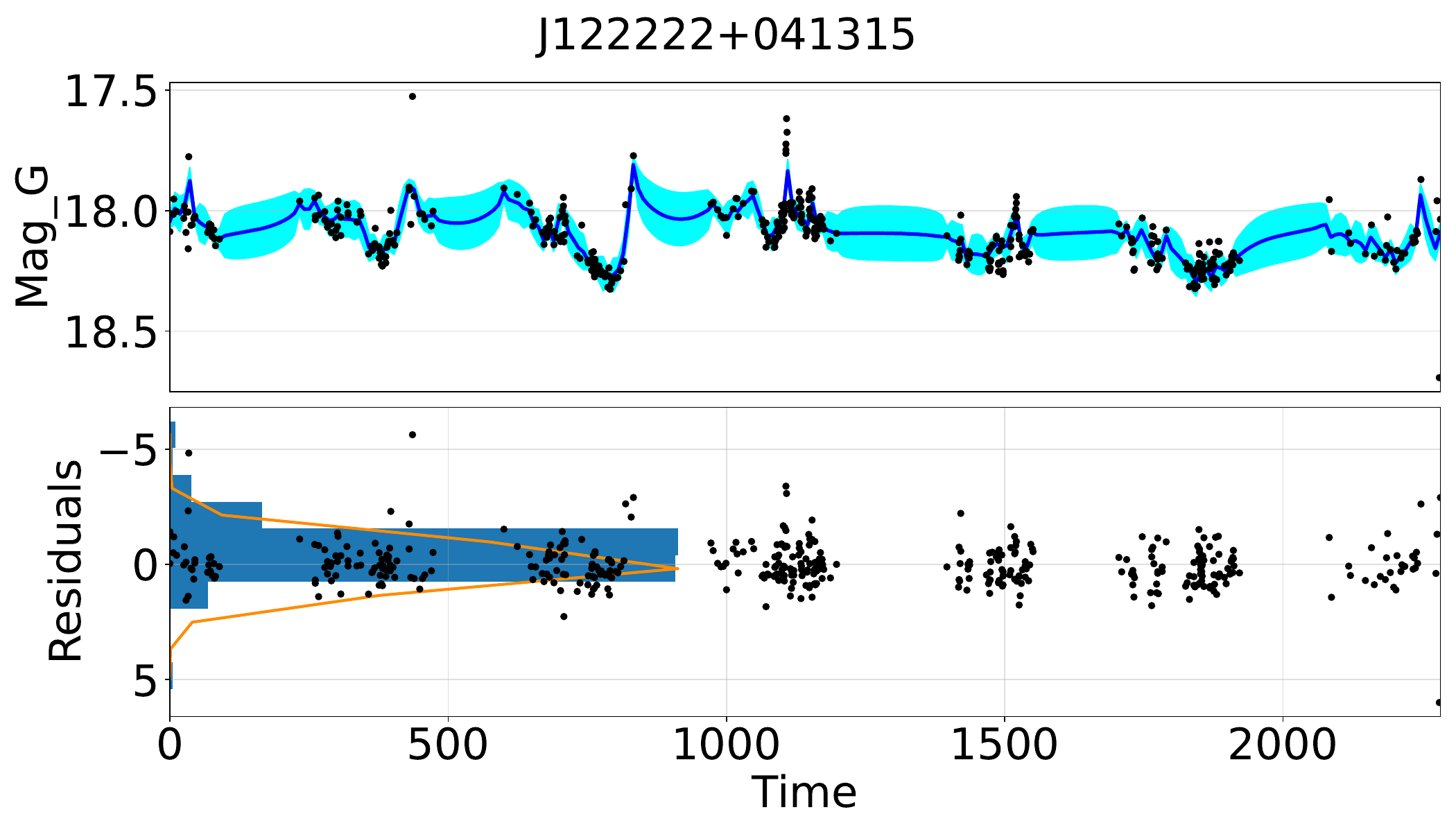}
    \end{minipage}
    \begin{minipage}{.3\textwidth}
        \centering
        \includegraphics[width=.99\linewidth]{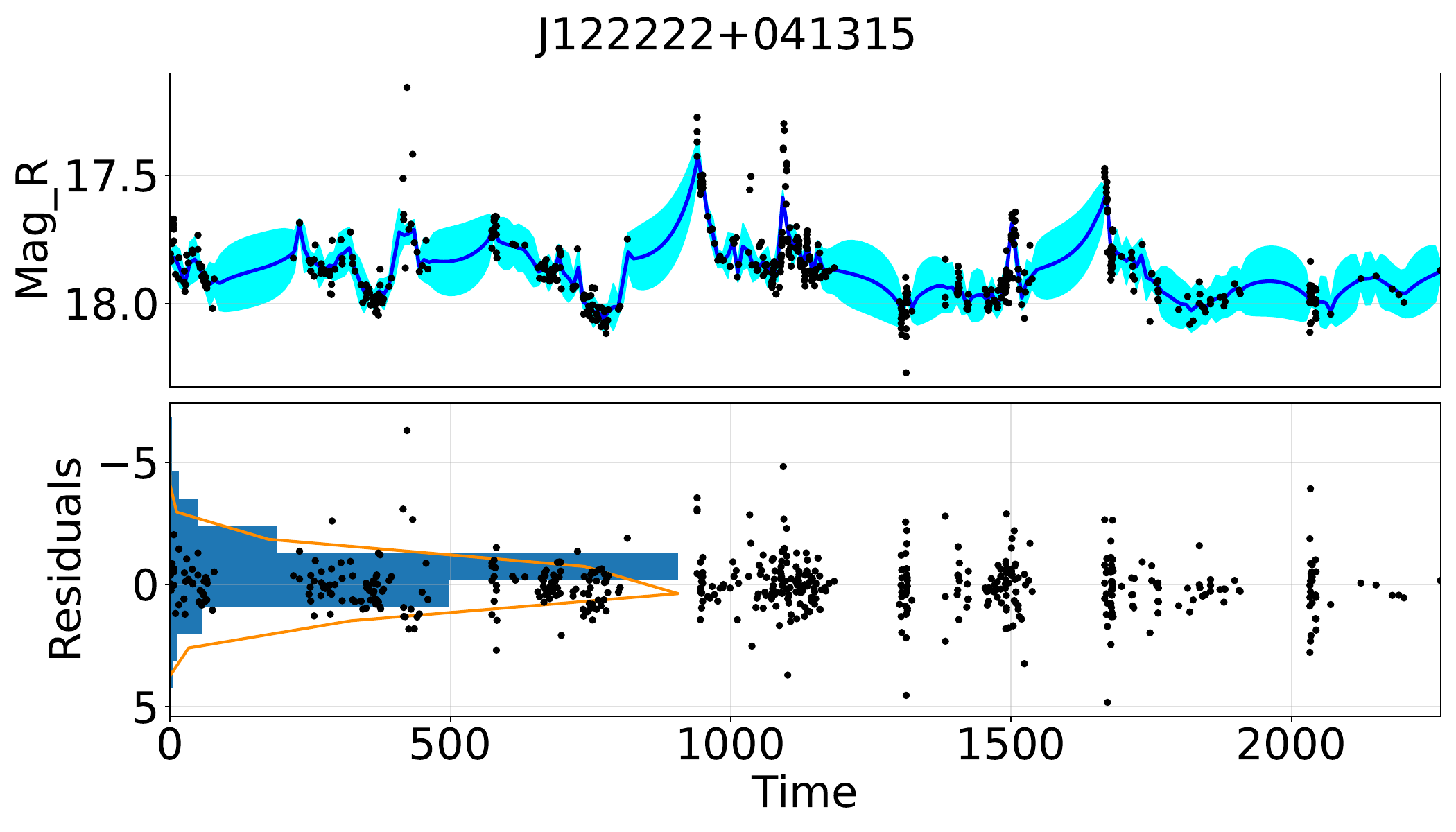}
    \end{minipage}
\end{figure*}

\begin{figure*}\label{J133108303032}
\centering
\caption{J133108+303032}
    \begin{minipage}{.3\textwidth}
        \centering
        \includegraphics[width=.99\linewidth]{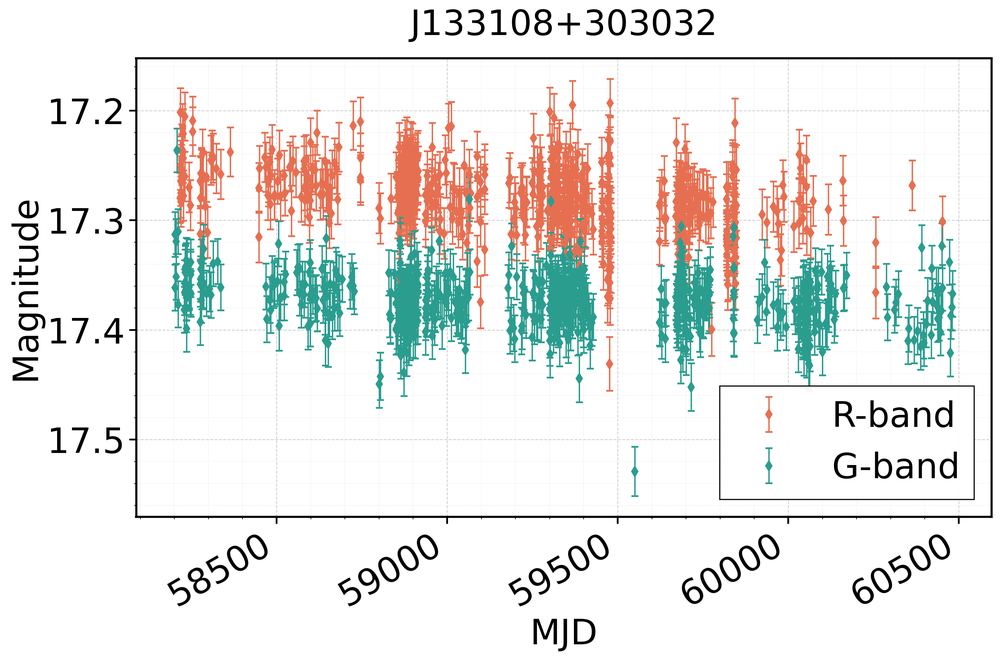}
    \end{minipage}
    \begin{minipage}{.3\textwidth}
        \centering
        \includegraphics[width=.99\linewidth]{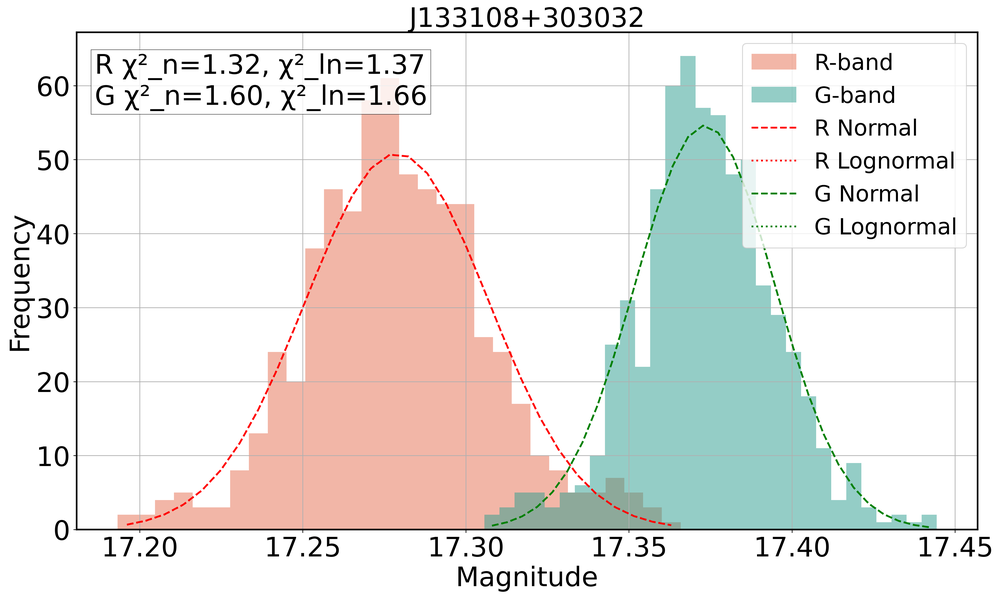}
    \end{minipage}
    \begin{minipage}{.3\textwidth}
        \centering
        \includegraphics[width=.99\linewidth]{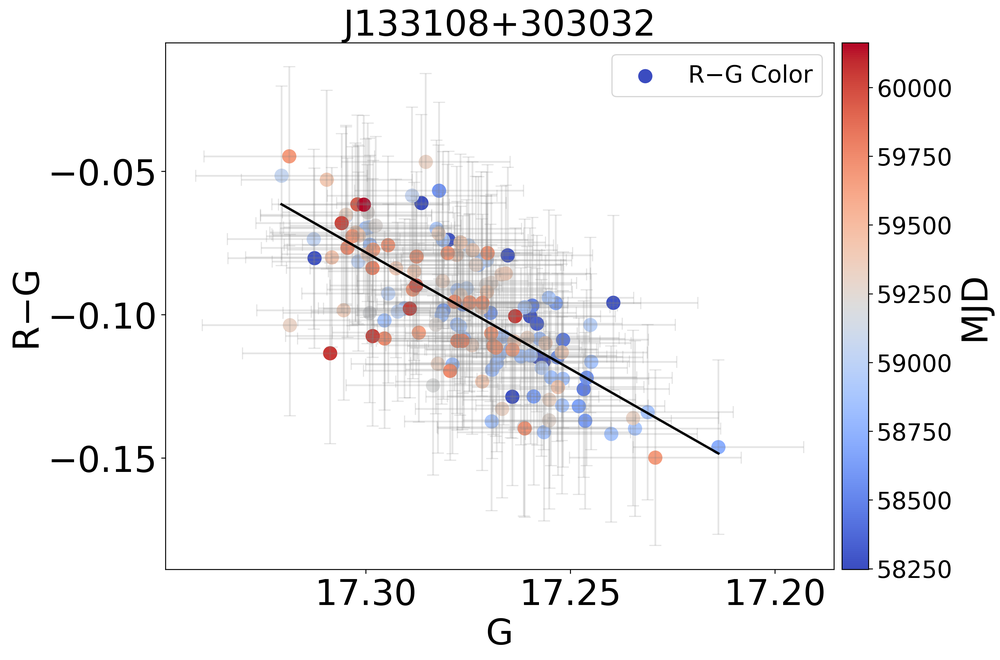}
    \end{minipage}
    \\
    \begin{minipage}{.3\textwidth}
        \centering
        \includegraphics[width=.99\linewidth]{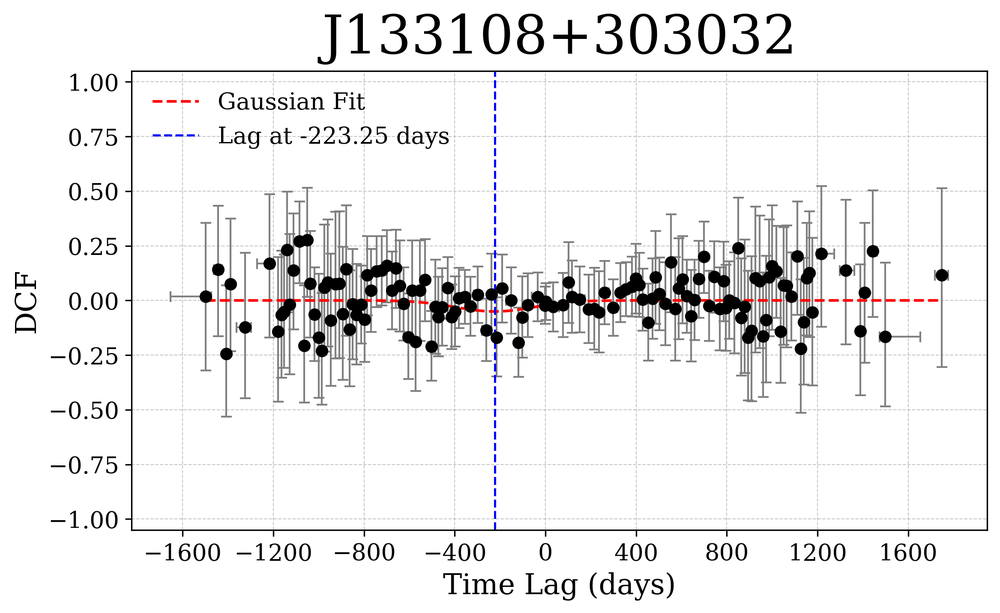}
    \end{minipage}
    \begin{minipage}{.3\textwidth}
        \centering
        \includegraphics[width=.99\linewidth]{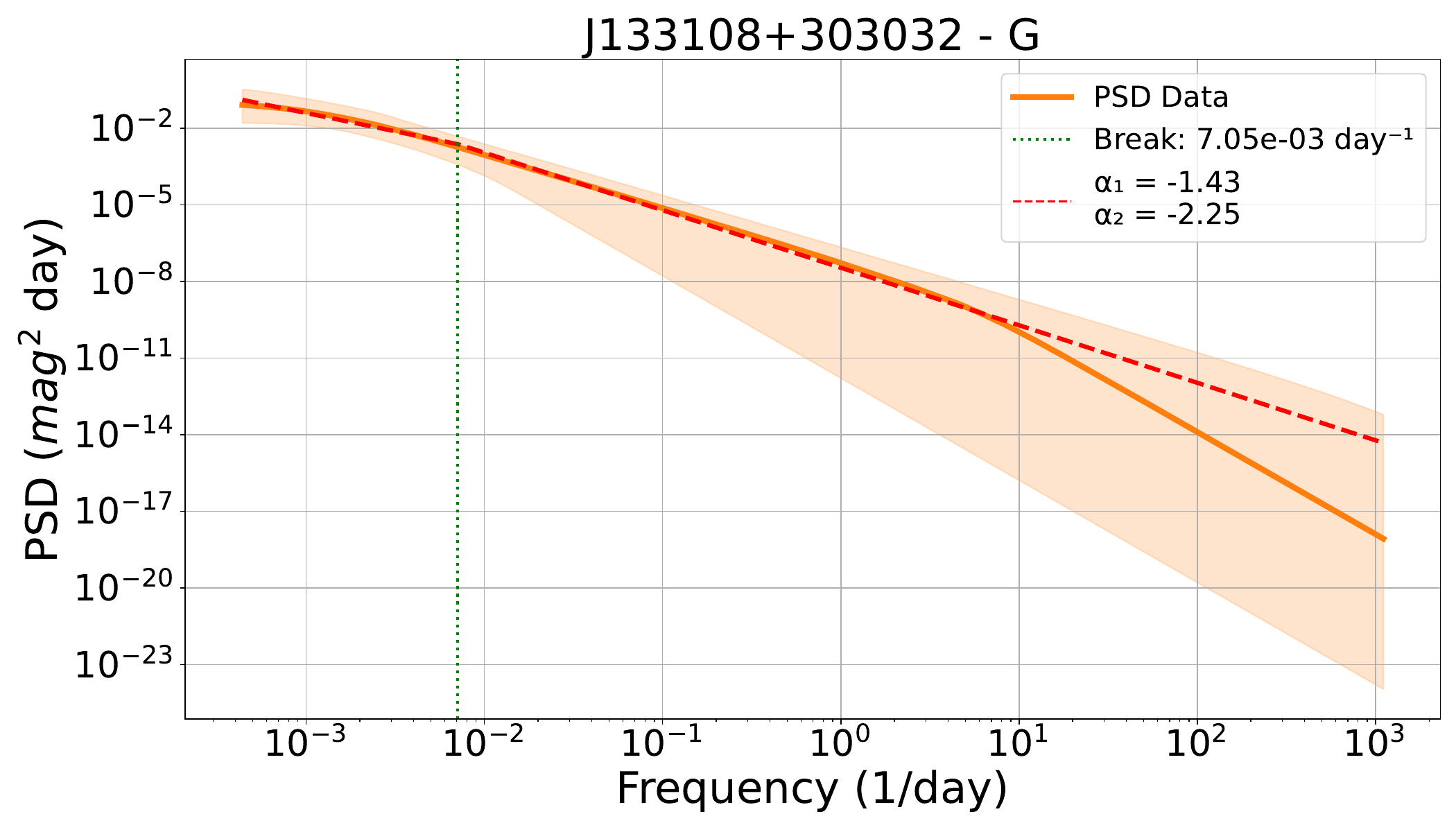}
    \end{minipage}
    \begin{minipage}{.3\textwidth}
        \centering
        \includegraphics[width=.99\linewidth]{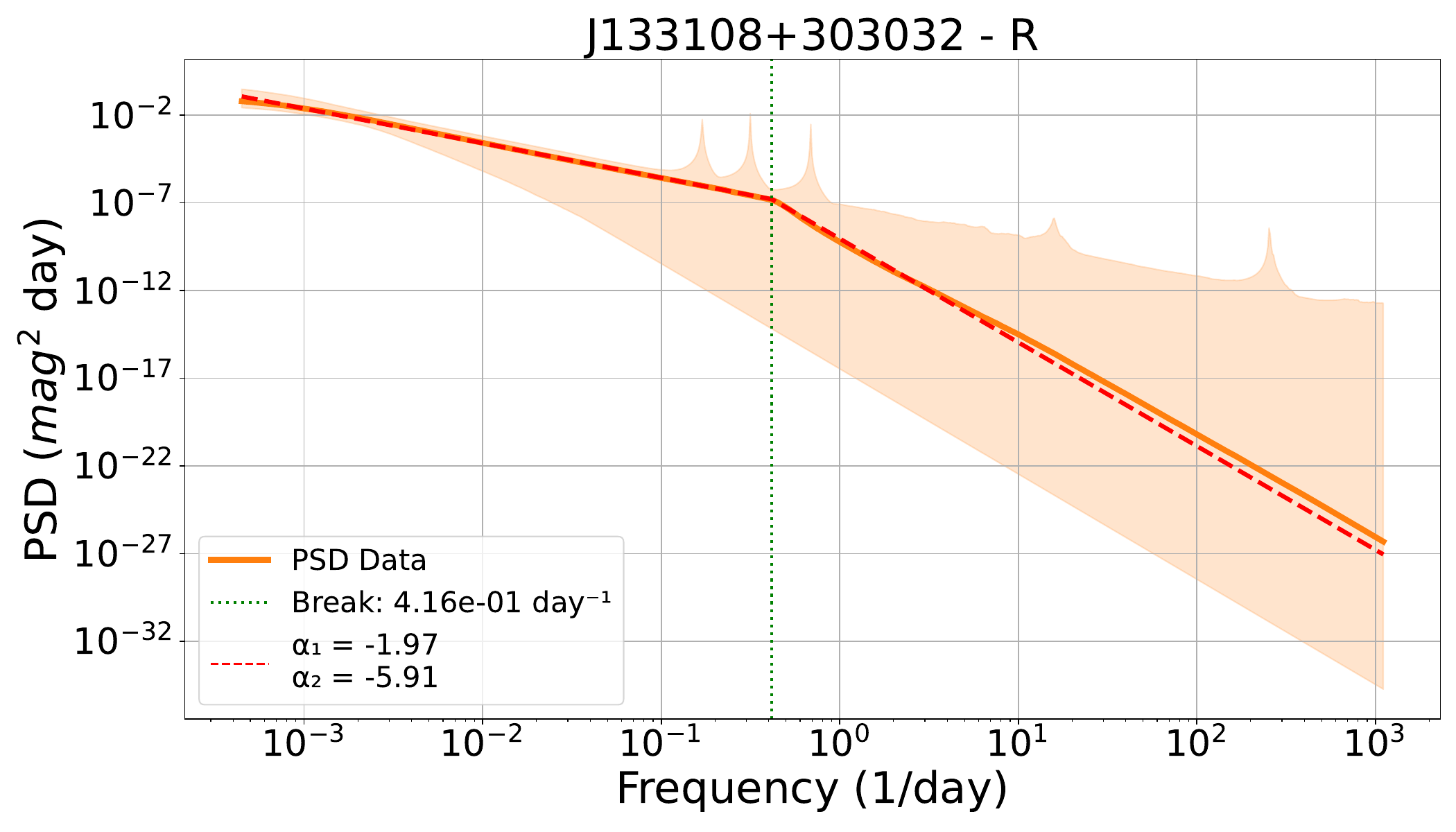}
    \end{minipage}
    \\
    \begin{minipage}{.3\textwidth}
        \centering
        \includegraphics[width=.99\linewidth]{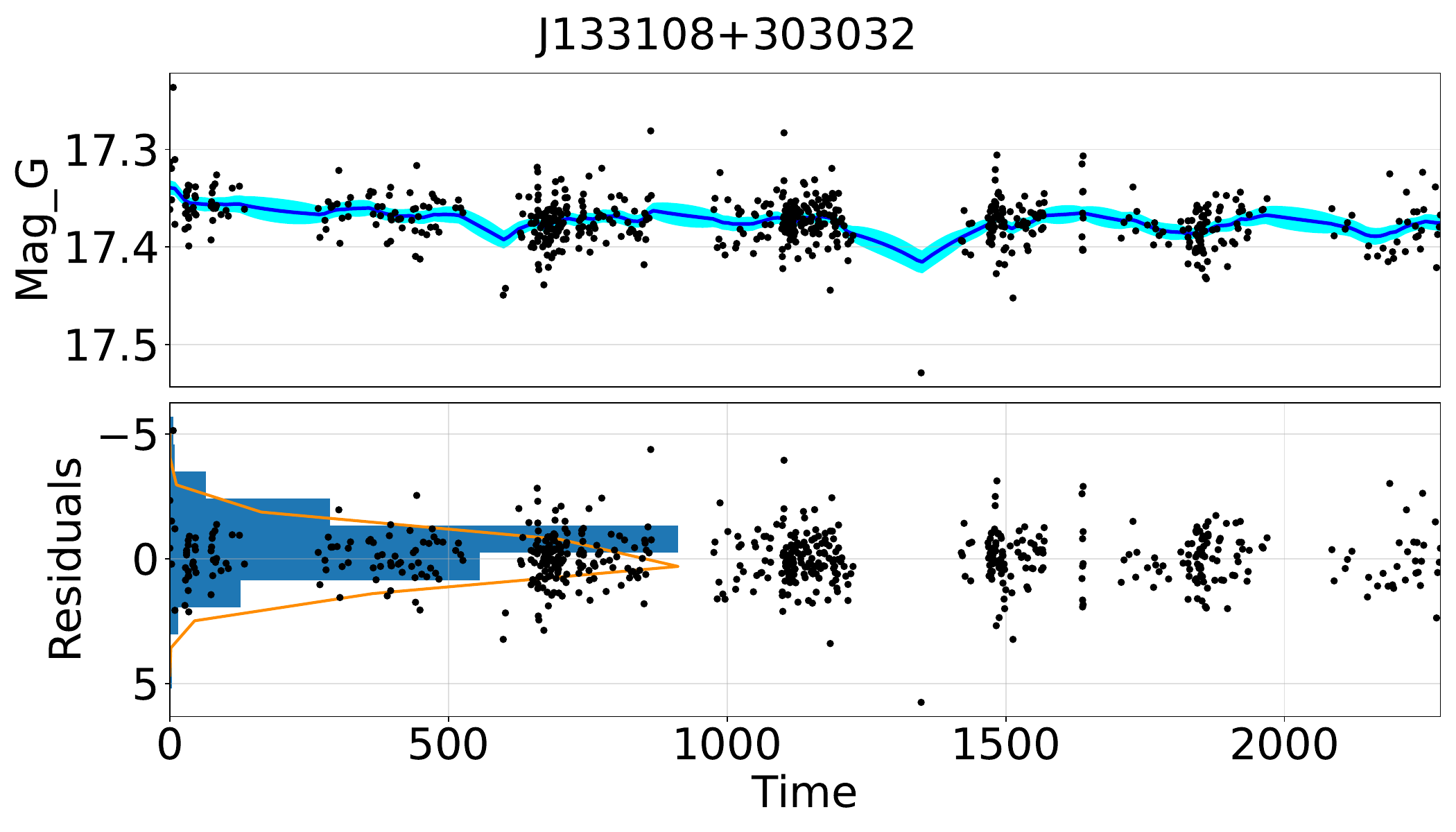}
    \end{minipage}
    \begin{minipage}{.3\textwidth}
        \centering
        \includegraphics[width=.99\linewidth]{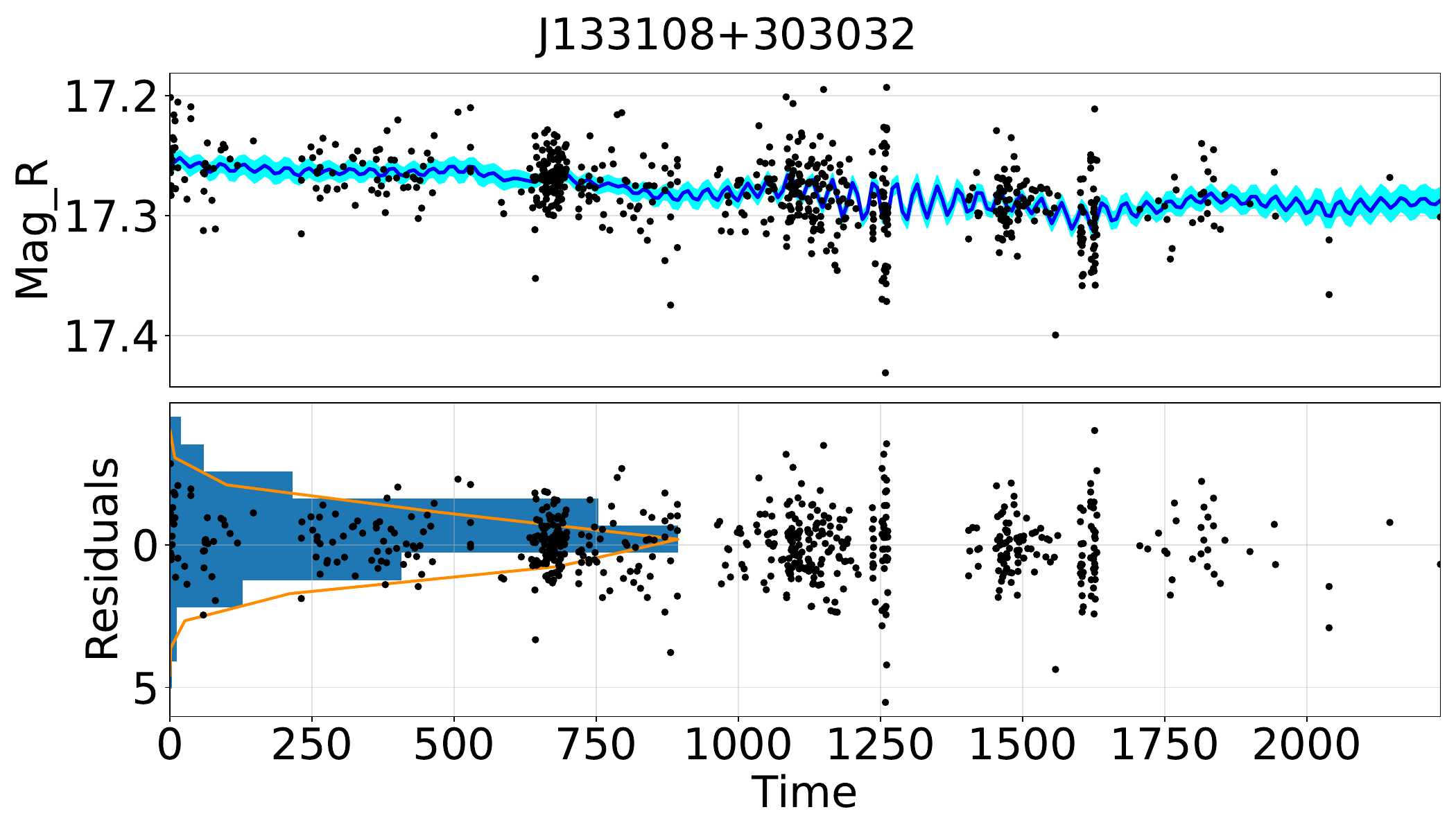}
    \end{minipage}
\end{figure*}

\begin{figure*}\label{J142105385522}
\centering
\caption{J142105+385522}
    \begin{minipage}{.3\textwidth}
        \centering
        \includegraphics[width=.99\linewidth]{plots/lightcurves/light_curve_J142105+385522.png}
    \end{minipage}
    \begin{minipage}{.3\textwidth}
        \centering
        \includegraphics[width=.99\linewidth]{plots/flux_distributions/flux_distribution_J142105+385522.png}
    \end{minipage}
    \begin{minipage}{.3\textwidth}
        \centering
        \includegraphics[width=.99\linewidth]{plots/cm/color_magnitude_J142105+385522.png}
    \end{minipage}
    \\
    \begin{minipage}{.3\textwidth}
        \centering
        \includegraphics[width=.99\linewidth]{plots/DCF/dcf_J142105+385522.png}
    \end{minipage}
    \begin{minipage}{.3\textwidth}
        \centering
        \includegraphics[width=.99\linewidth]{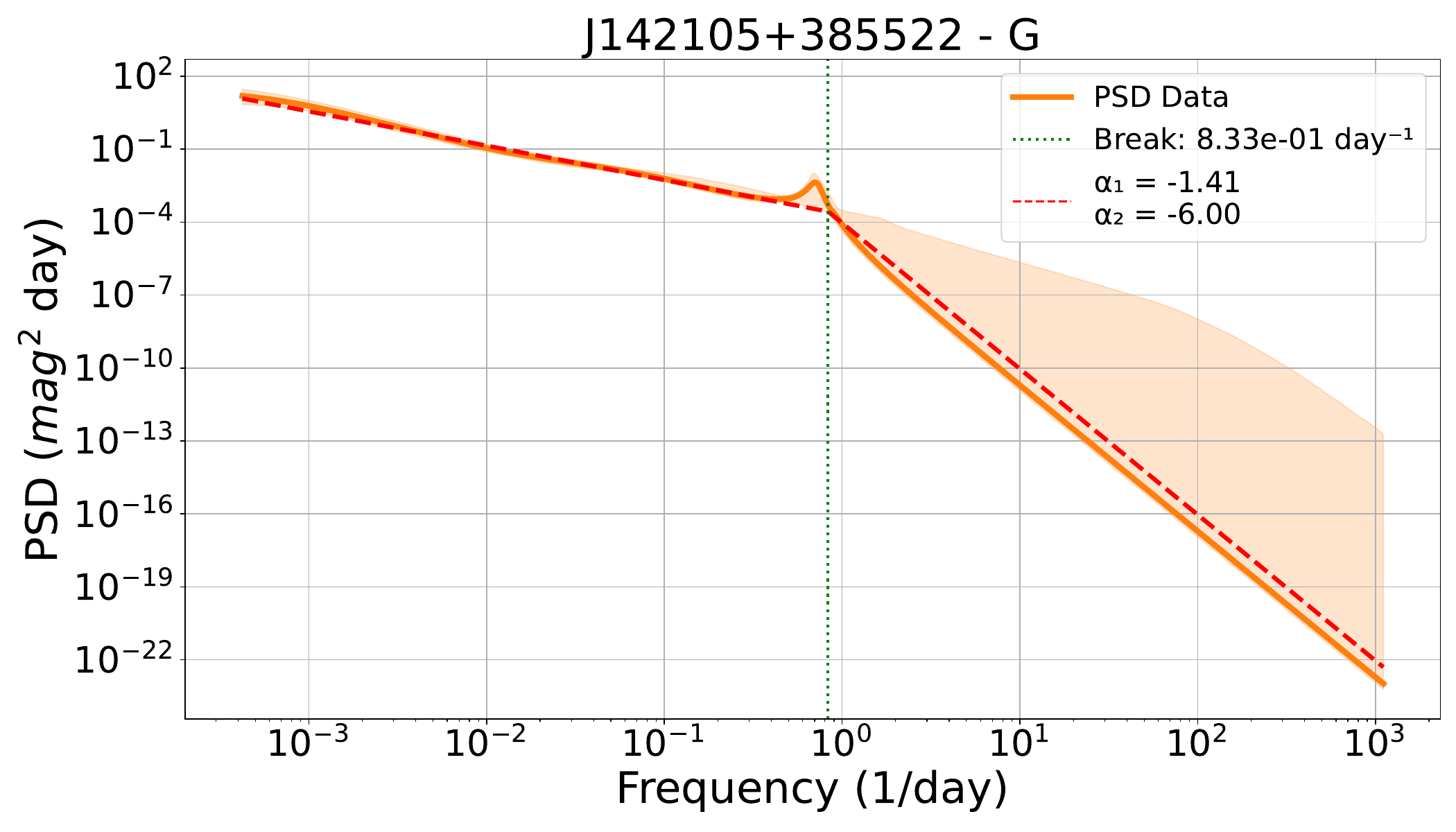}
    \end{minipage}
    \begin{minipage}{.3\textwidth}
        \centering
        \includegraphics[width=.99\linewidth]{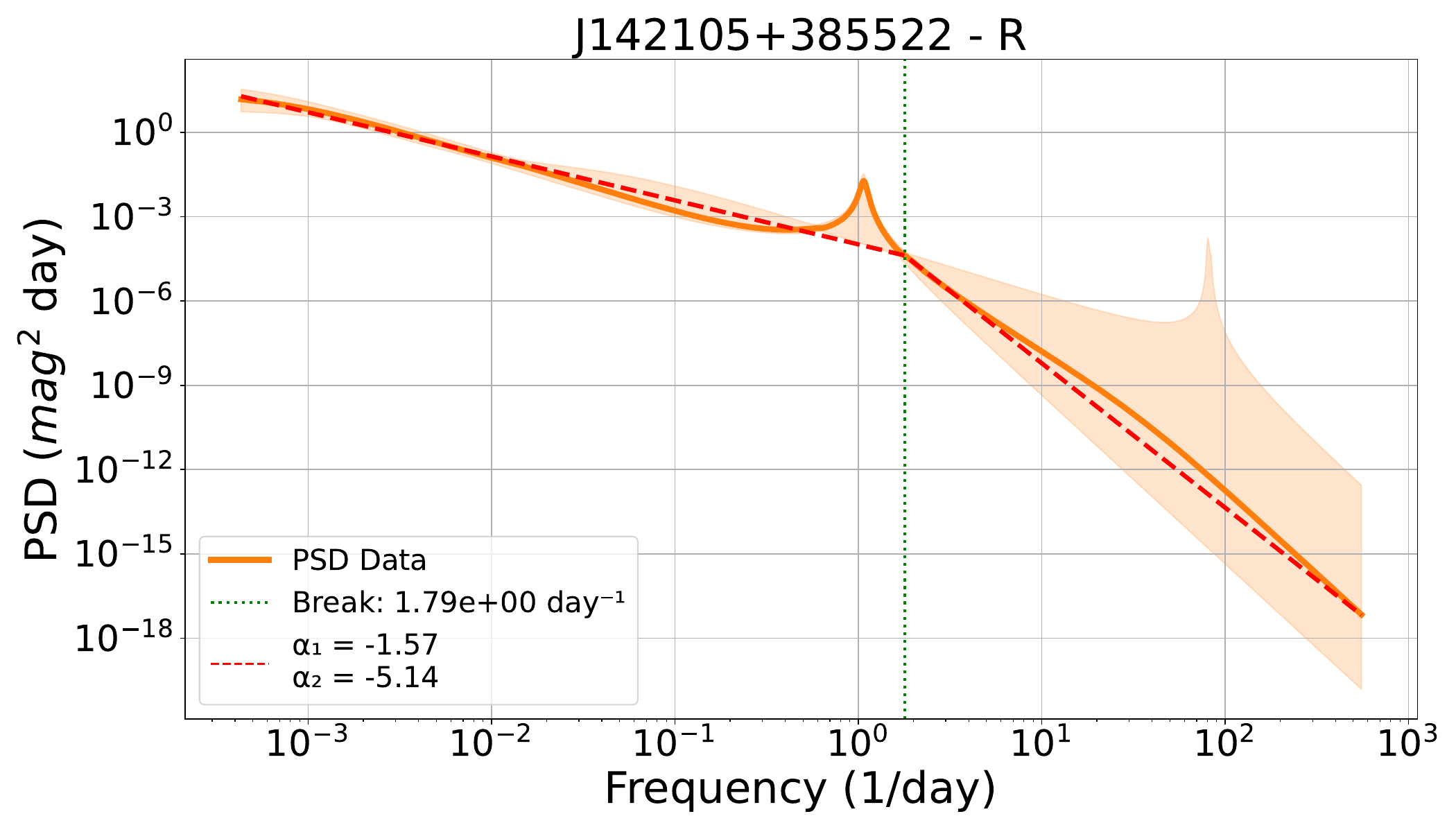}
    \end{minipage}
    \\
    \begin{minipage}{.3\textwidth}
        \centering
        \includegraphics[width=.99\linewidth]{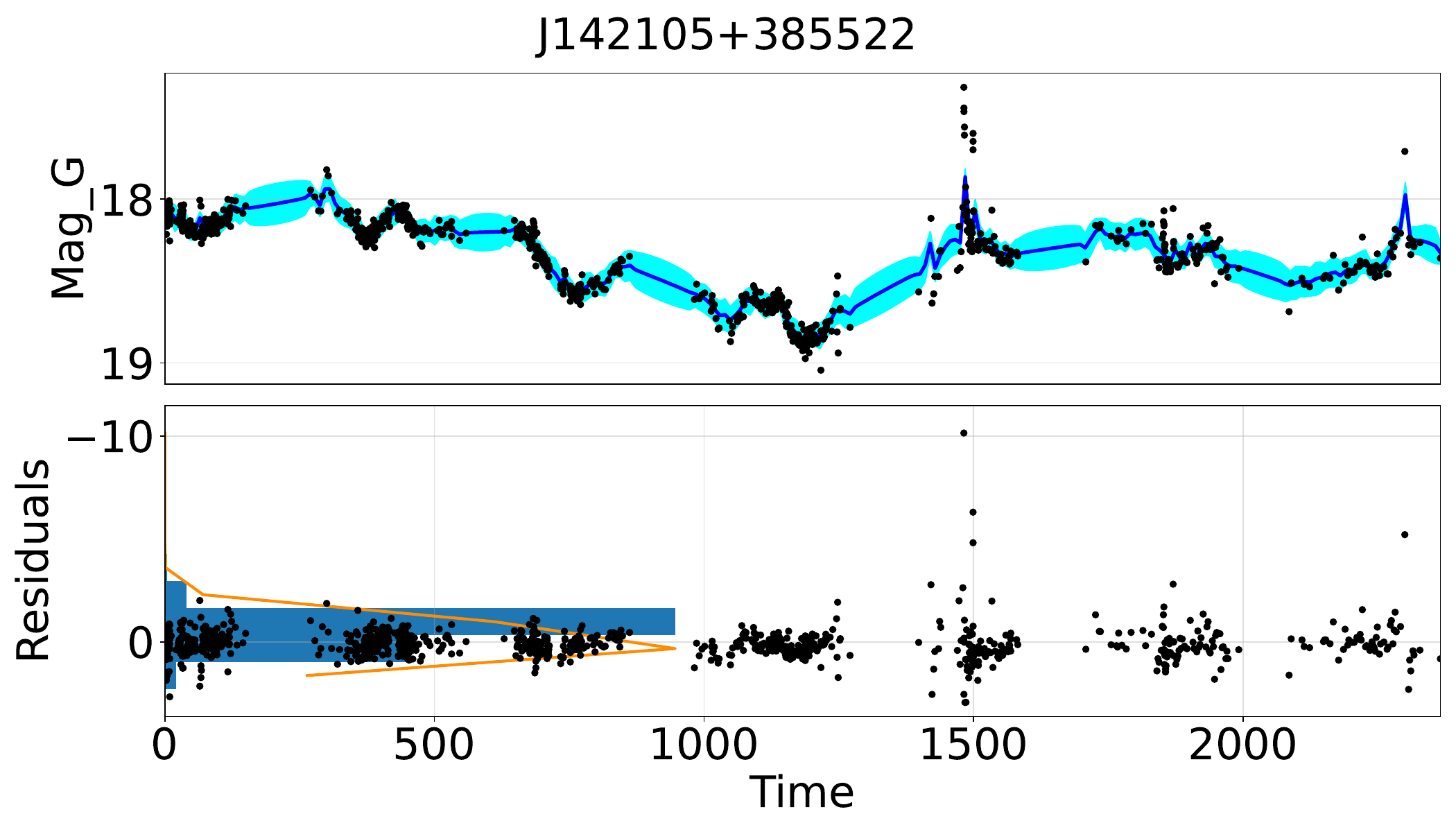}
    \end{minipage}
    \begin{minipage}{.3\textwidth}
        \centering
        \includegraphics[width=.99\linewidth]{carma/J142105+385522_R_lc.pdf}
    \end{minipage}
\end{figure*}

\begin{figure*}\label{J144318472556}
\centering
\caption{J144318+472556}
    \begin{minipage}{.3\textwidth}
        \centering
        \includegraphics[width=.99\linewidth]{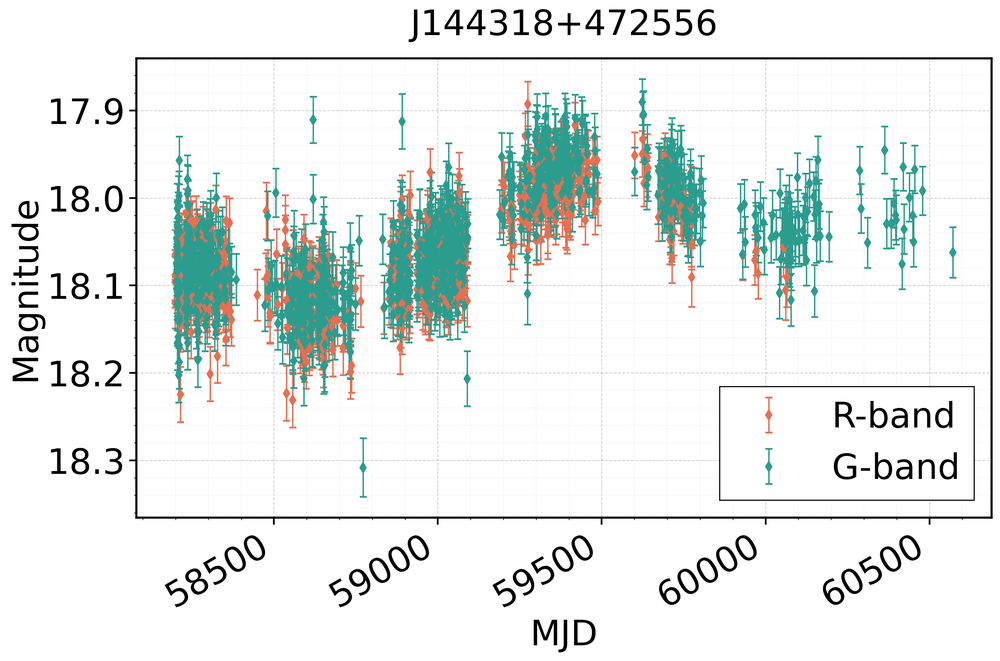}
    \end{minipage}
    \begin{minipage}{.3\textwidth}
        \centering
        \includegraphics[width=.99\linewidth]{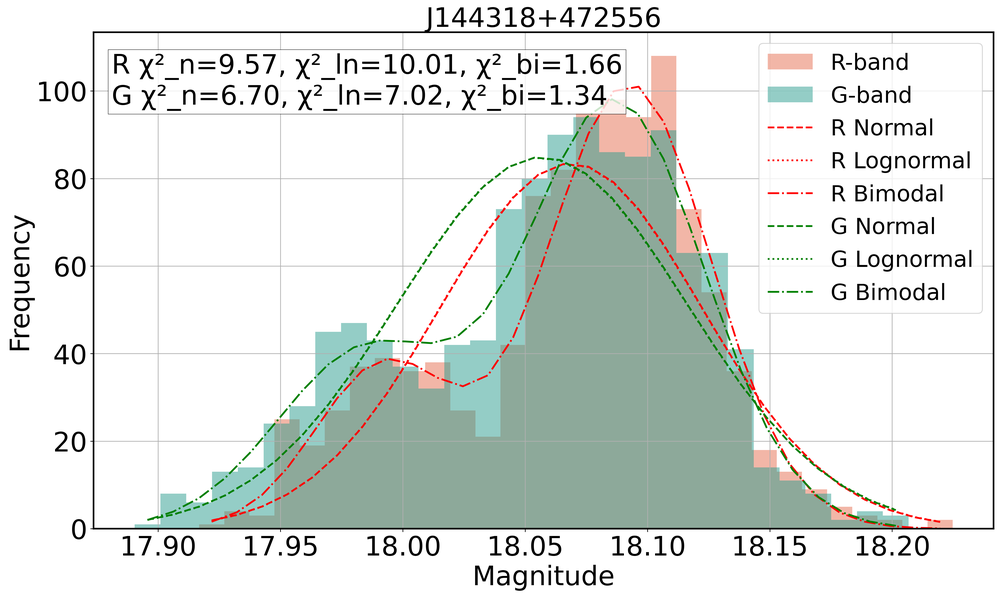}
    \end{minipage}
    \begin{minipage}{.3\textwidth}
        \centering
        \includegraphics[width=.99\linewidth]{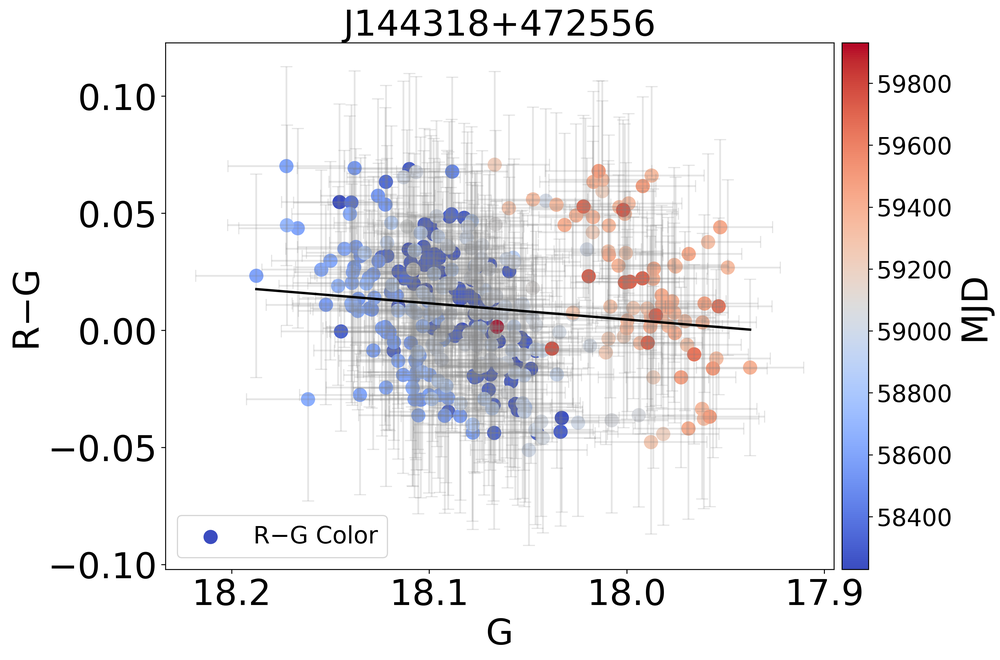}
    \end{minipage}
    \\
    \begin{minipage}{.3\textwidth}
        \centering
        \includegraphics[width=.99\linewidth]{plots/DCF/dcf_J144318+472556.png}
    \end{minipage}
    \begin{minipage}{.3\textwidth}
        \centering
        \includegraphics[width=.99\linewidth]{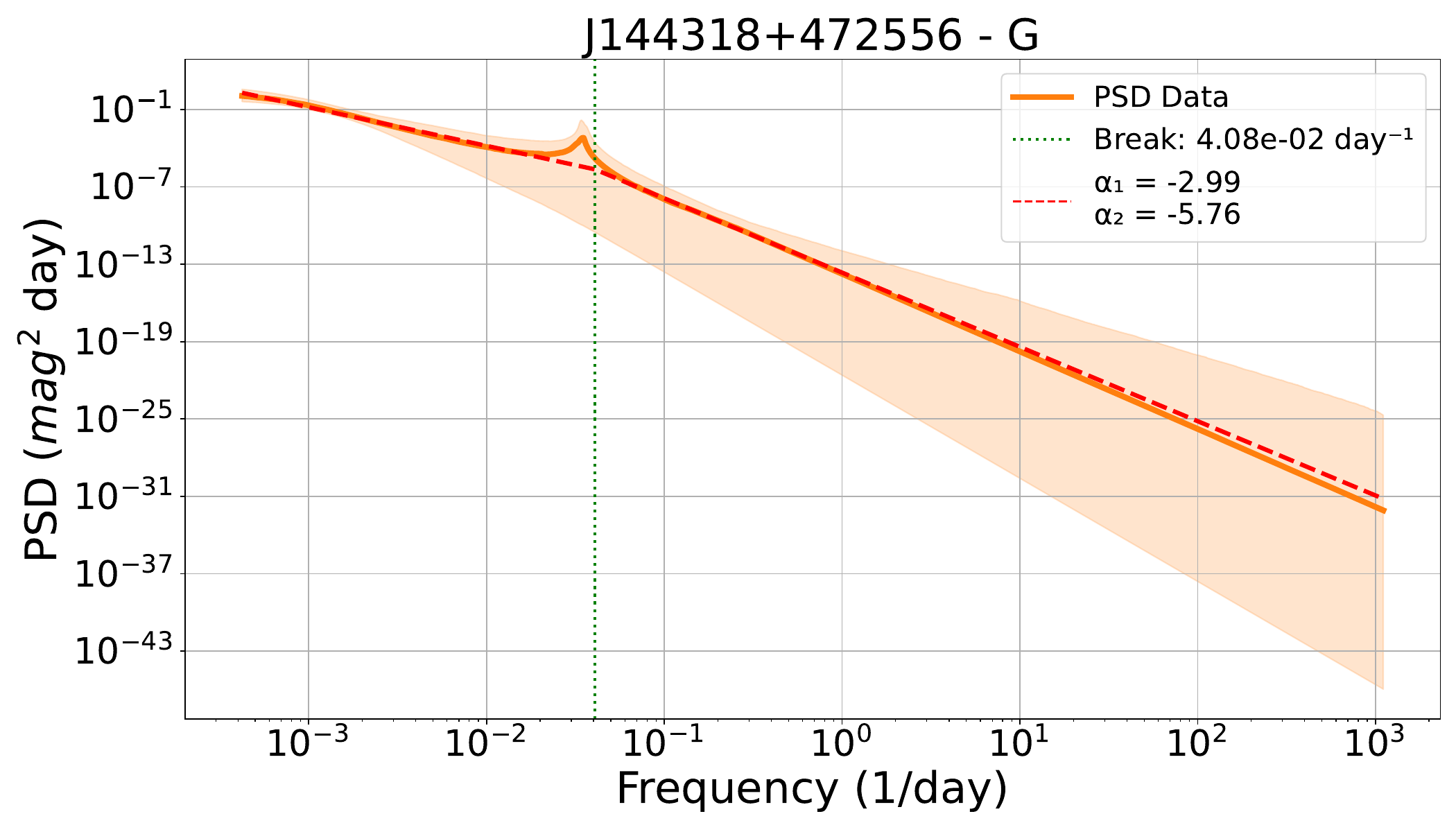}
    \end{minipage}
    \begin{minipage}{.3\textwidth}
        \centering
        \includegraphics[width=.99\linewidth]{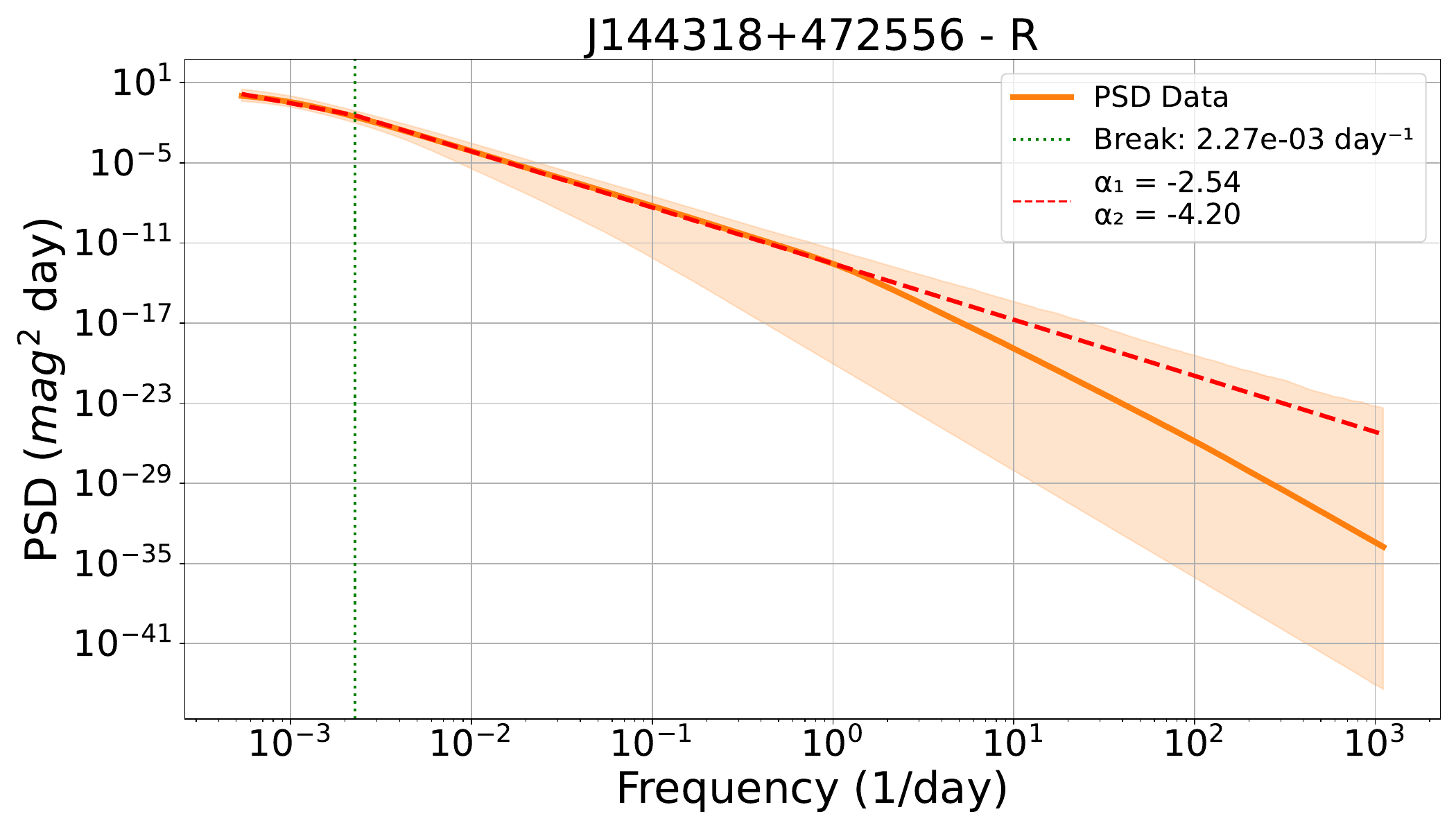}
    \end{minipage}
    \\
    \begin{minipage}{.3\textwidth}
        \centering
        \includegraphics[width=.99\linewidth]{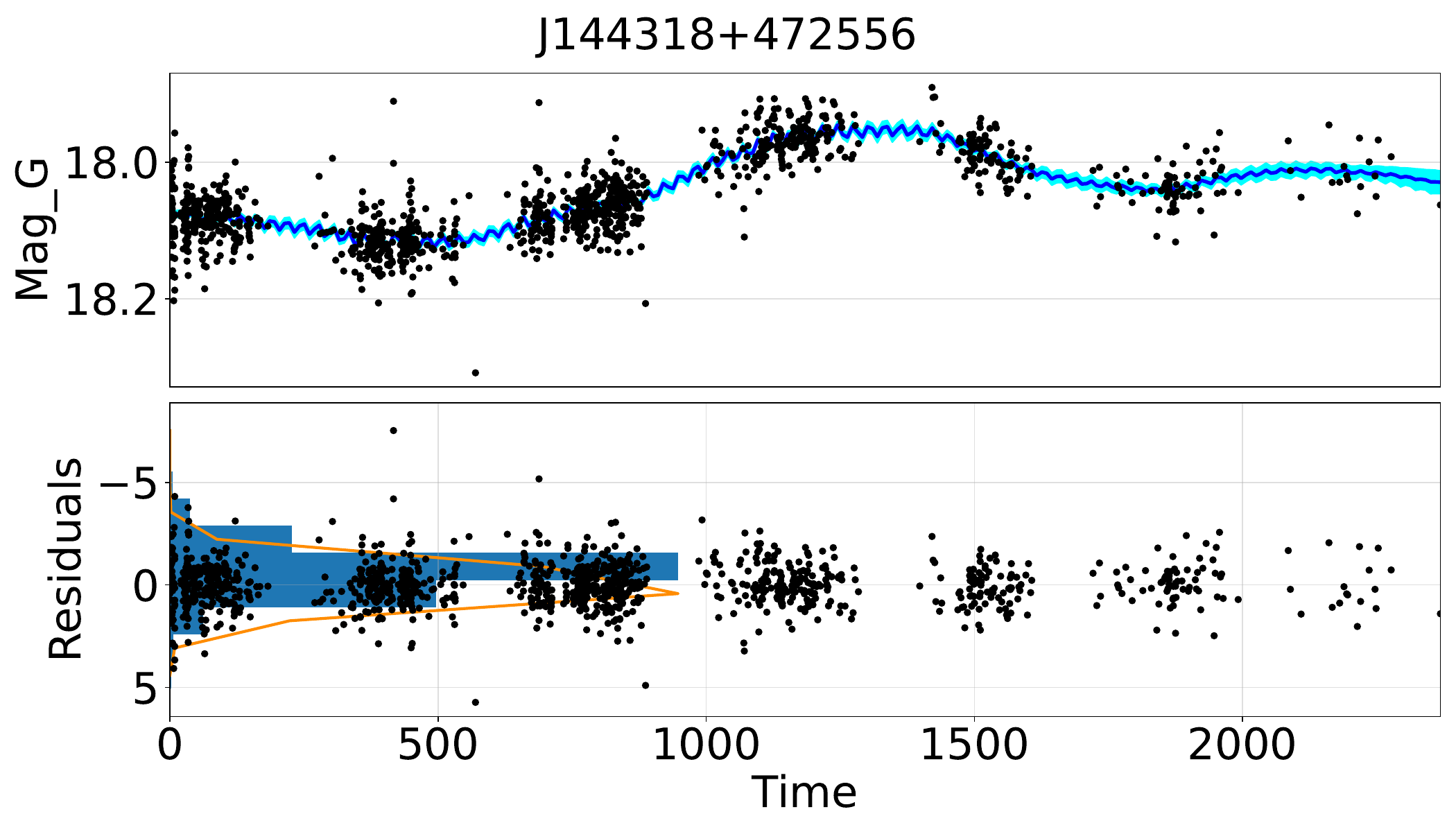}
    \end{minipage}
    \begin{minipage}{.3\textwidth}
        \centering
        \includegraphics[width=.99\linewidth]{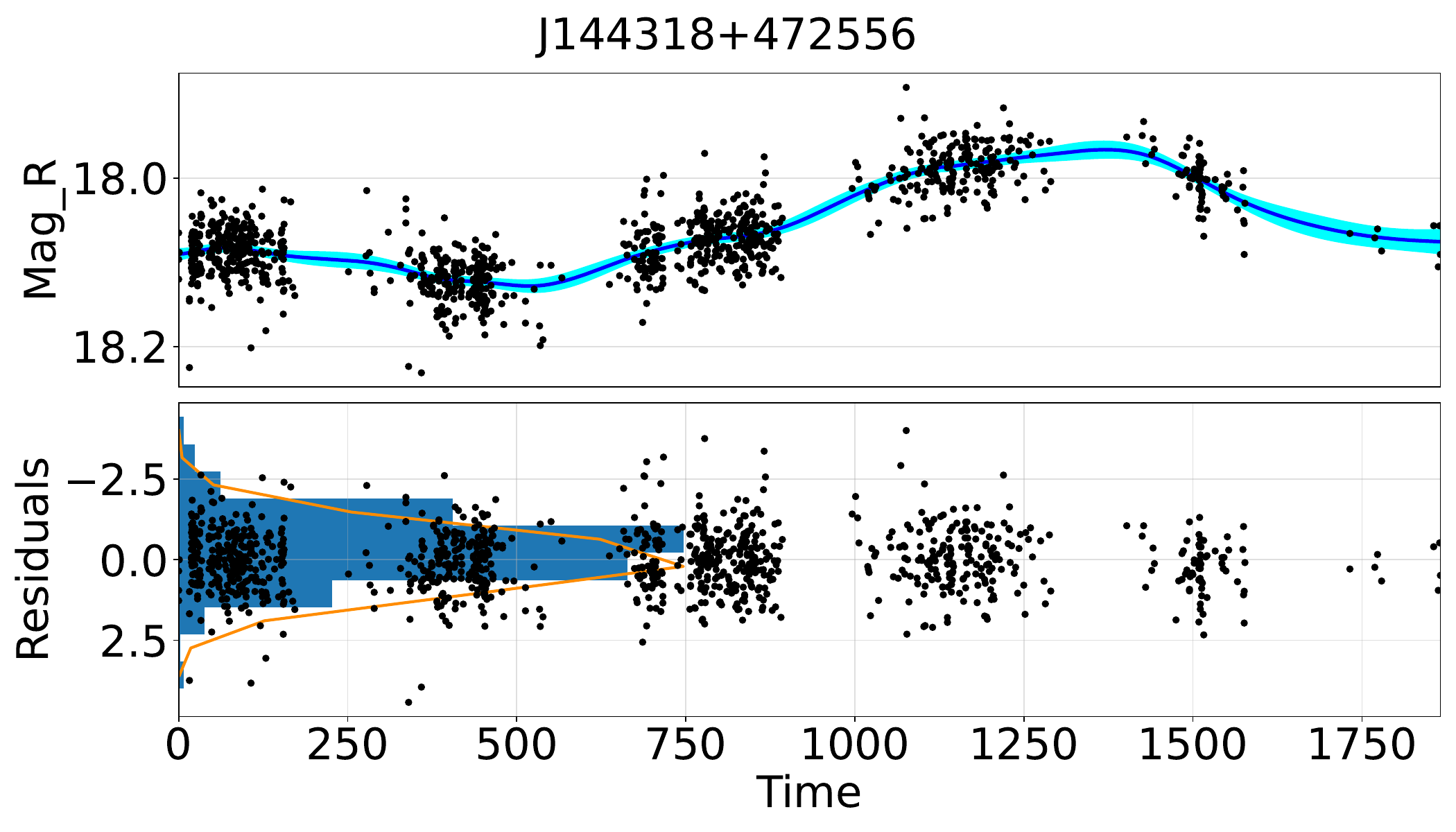}
    \end{minipage}
\end{figure*}

\begin{figure*}\label{J150506032631}
\centering
\caption{J150506+032631}
    \begin{minipage}{.3\textwidth}
        \centering
        \includegraphics[width=.99\linewidth]{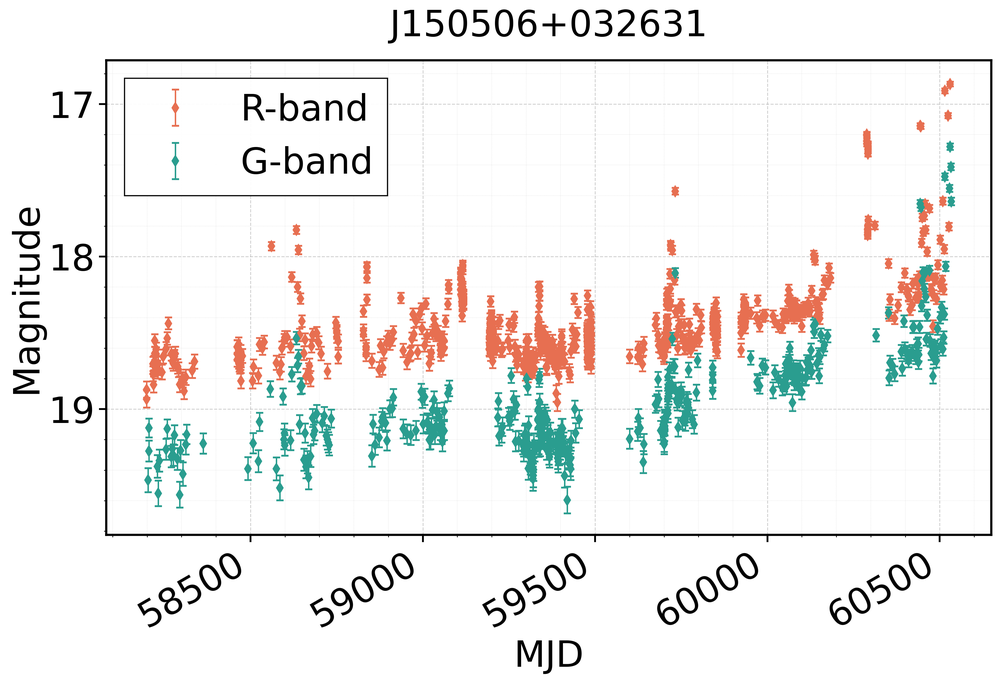}
    \end{minipage}
    \begin{minipage}{.3\textwidth}
        \centering
        \includegraphics[width=.99\linewidth]{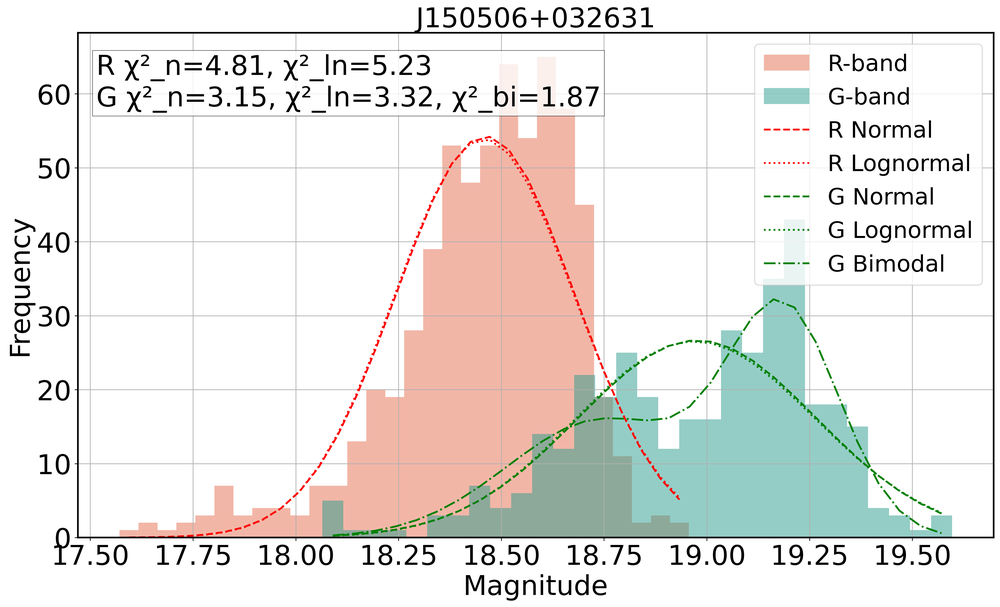}
    \end{minipage}
    \begin{minipage}{.3\textwidth}
        \centering
        \includegraphics[width=.99\linewidth]{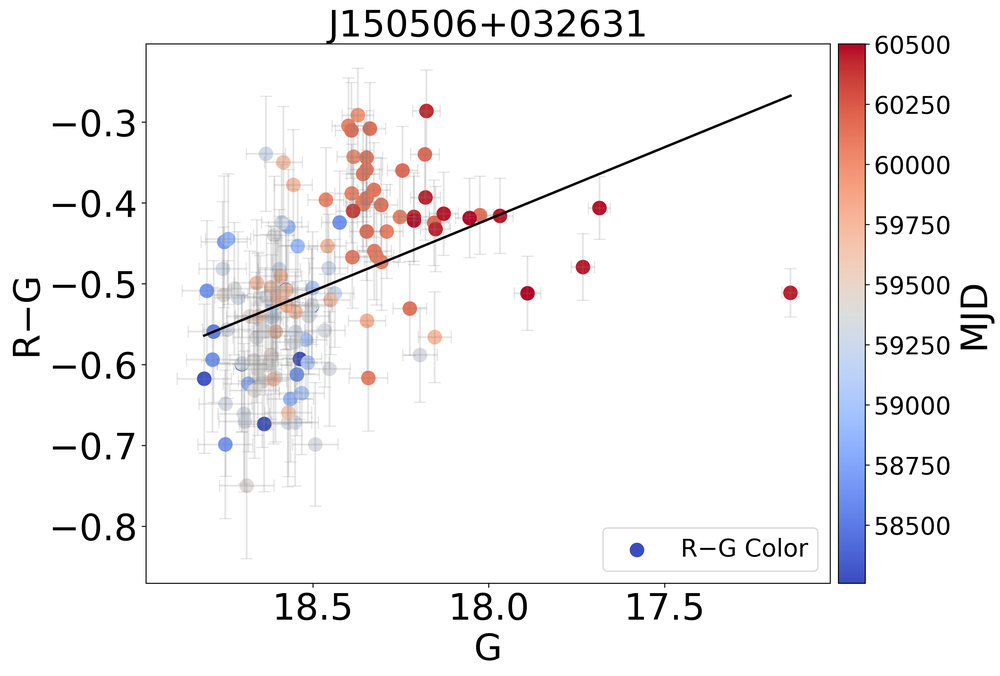}
    \end{minipage}
    \\
    \begin{minipage}{.3\textwidth}
        \centering
        \includegraphics[width=.99\linewidth]{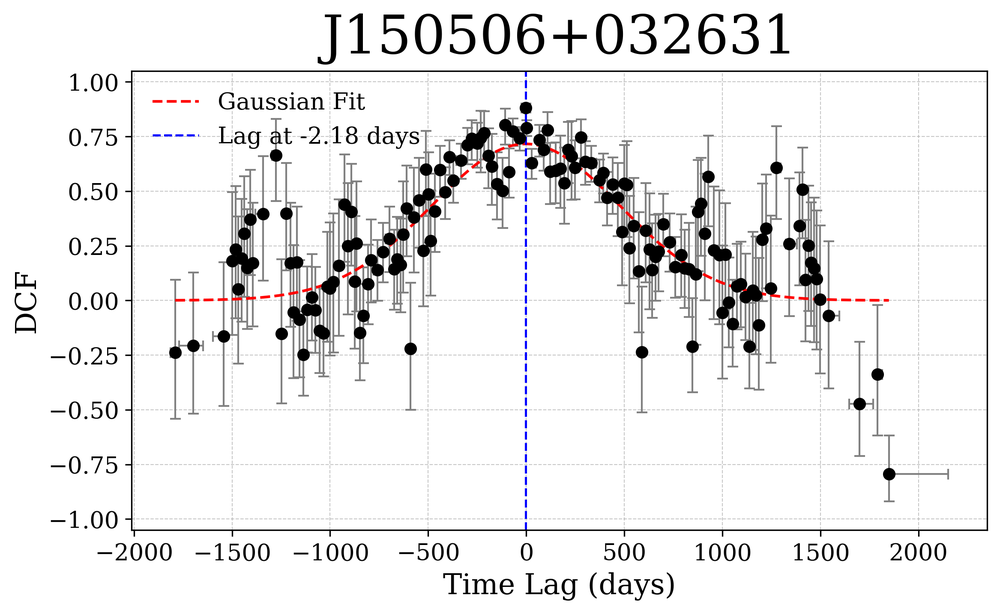}
    \end{minipage}
    \begin{minipage}{.3\textwidth}
        \centering
        \includegraphics[width=.99\linewidth]{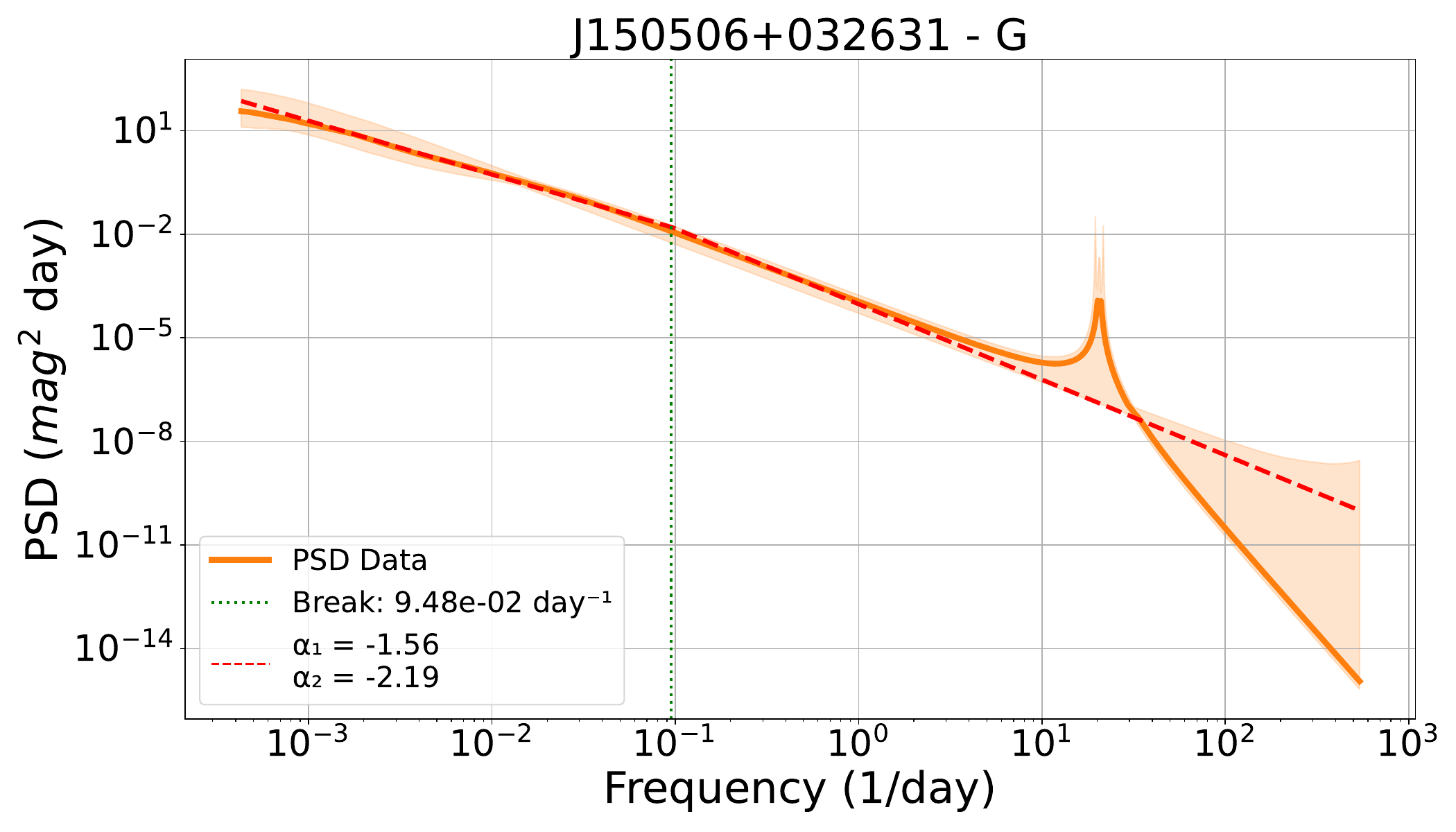}
    \end{minipage}
    \begin{minipage}{.3\textwidth}
        \centering
        \includegraphics[width=.99\linewidth]{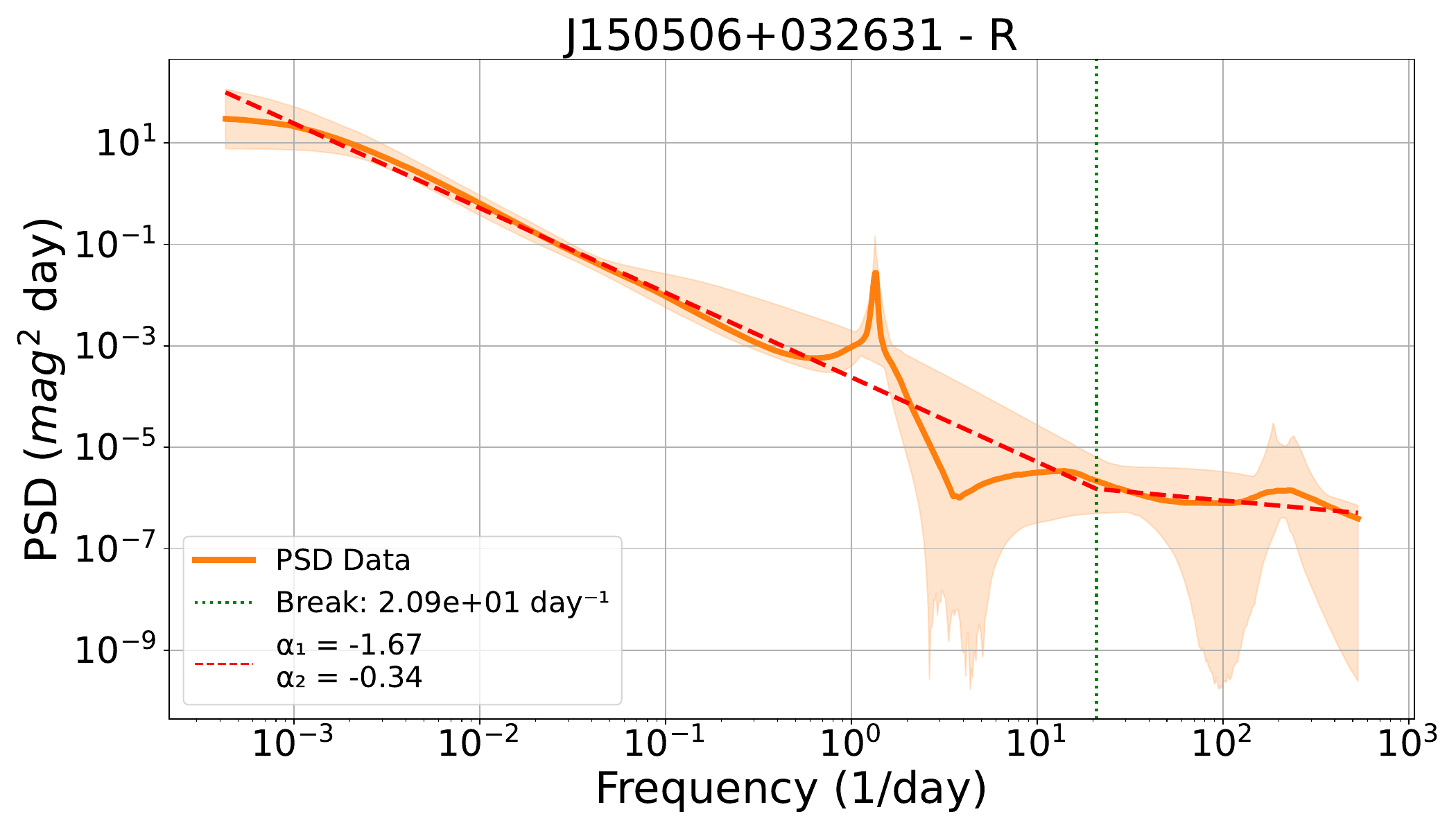}
    \end{minipage}
    \\
    \begin{minipage}{.3\textwidth}
        \centering
        \includegraphics[width=.99\linewidth]{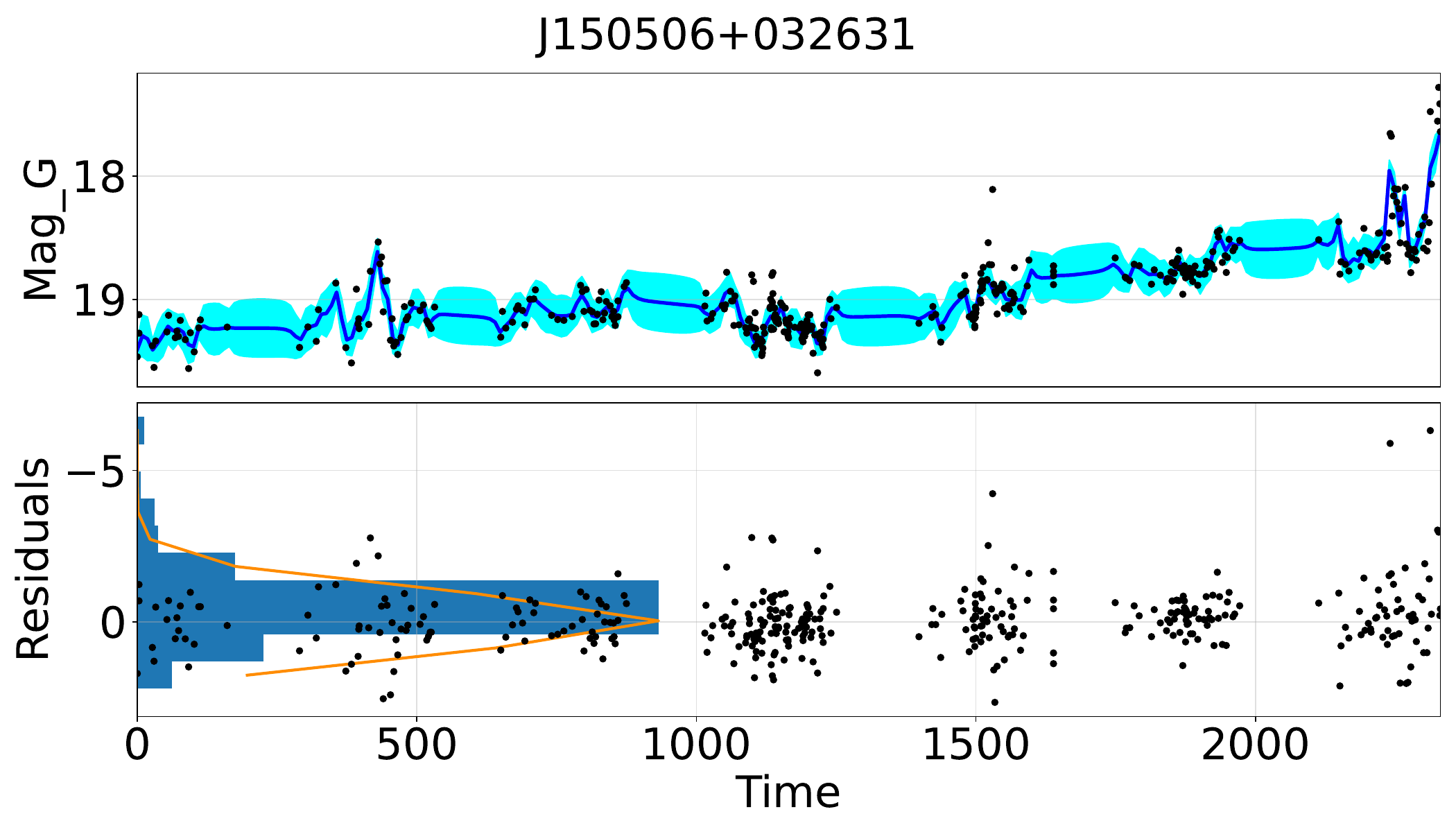}
    \end{minipage}
    \begin{minipage}{.3\textwidth}
        \centering
        \includegraphics[width=.99\linewidth]{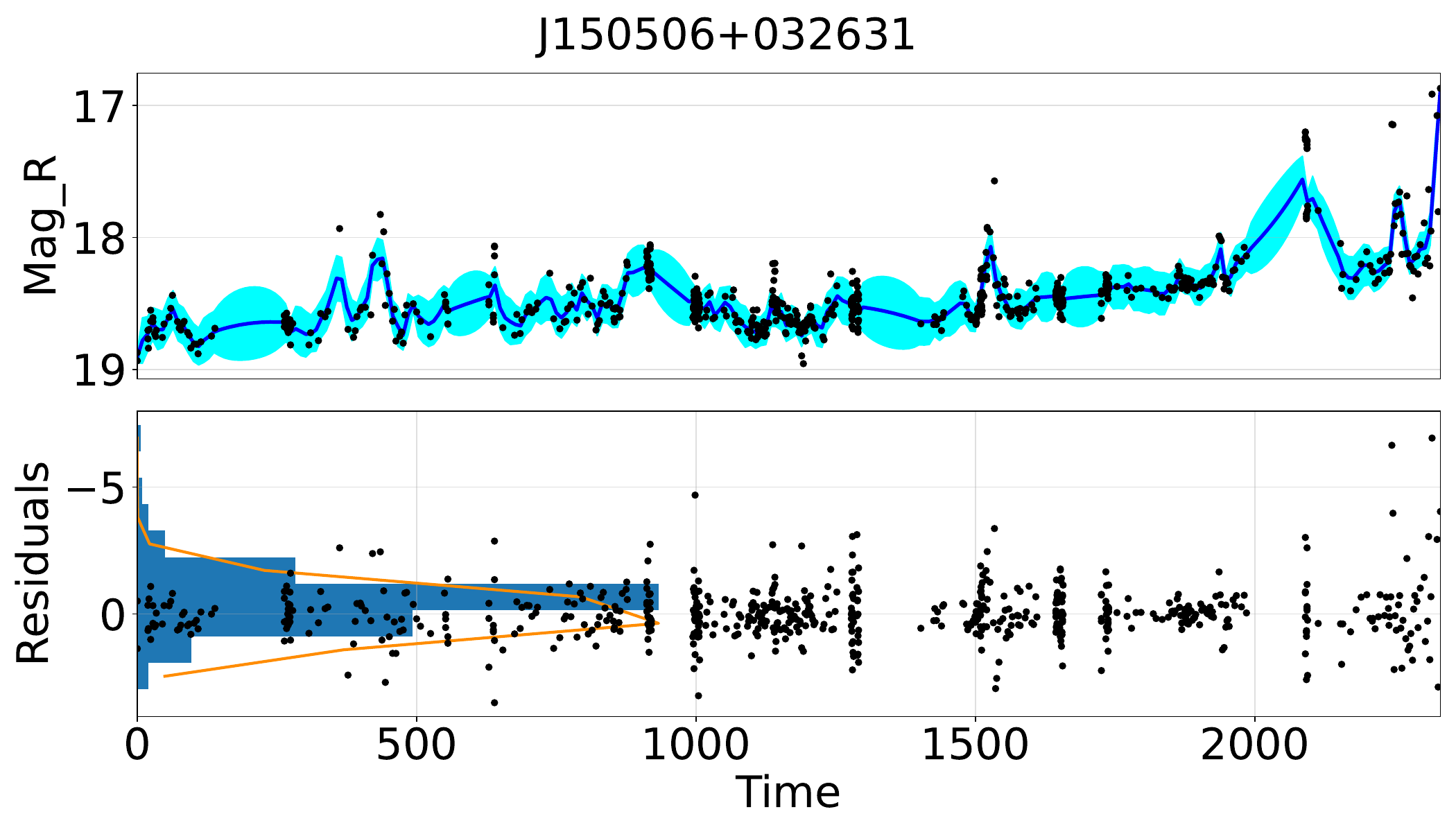}
    \end{minipage}
\end{figure*}

\begin{figure*}\label{J164100345453}
\centering
\caption{J164100+345453}
    \begin{minipage}{.3\textwidth}
        \centering
        \includegraphics[width=.99\linewidth]{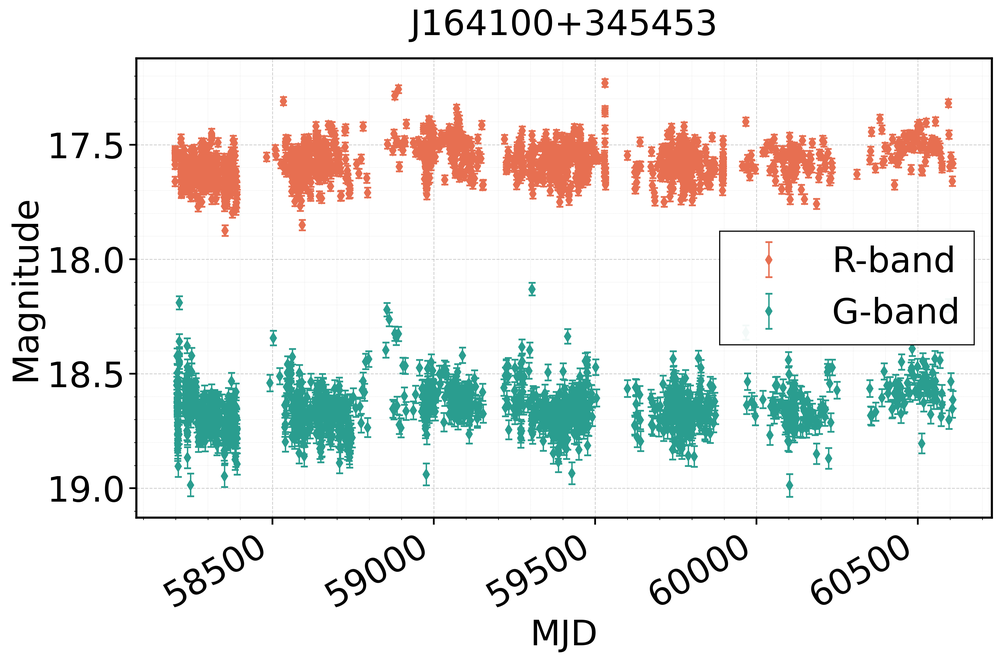}
    \end{minipage}
    \begin{minipage}{.3\textwidth}
        \centering
        \includegraphics[width=.99\linewidth]{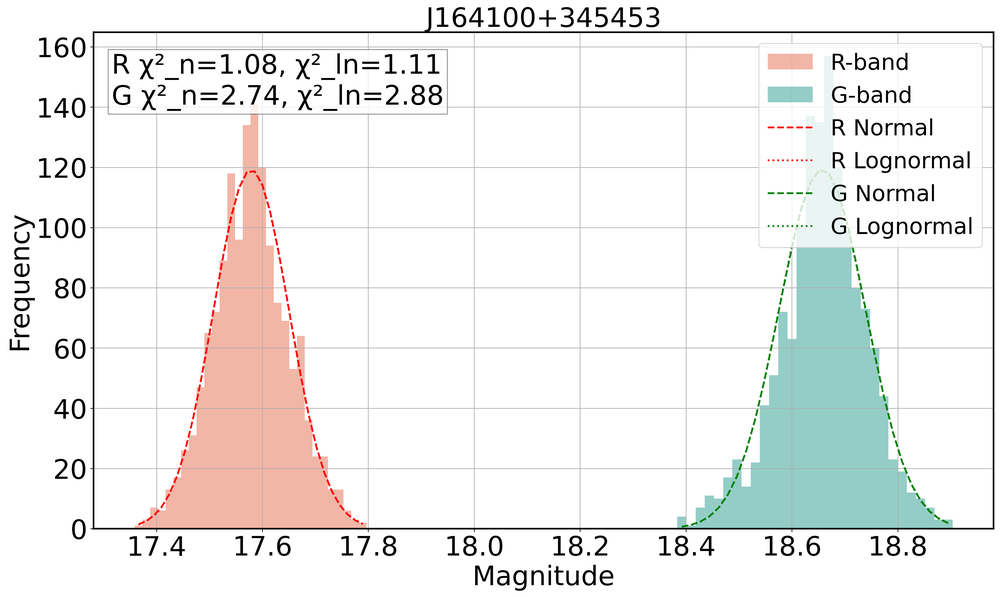}
    \end{minipage}
    \begin{minipage}{.3\textwidth}
        \centering
        \includegraphics[width=.99\linewidth]{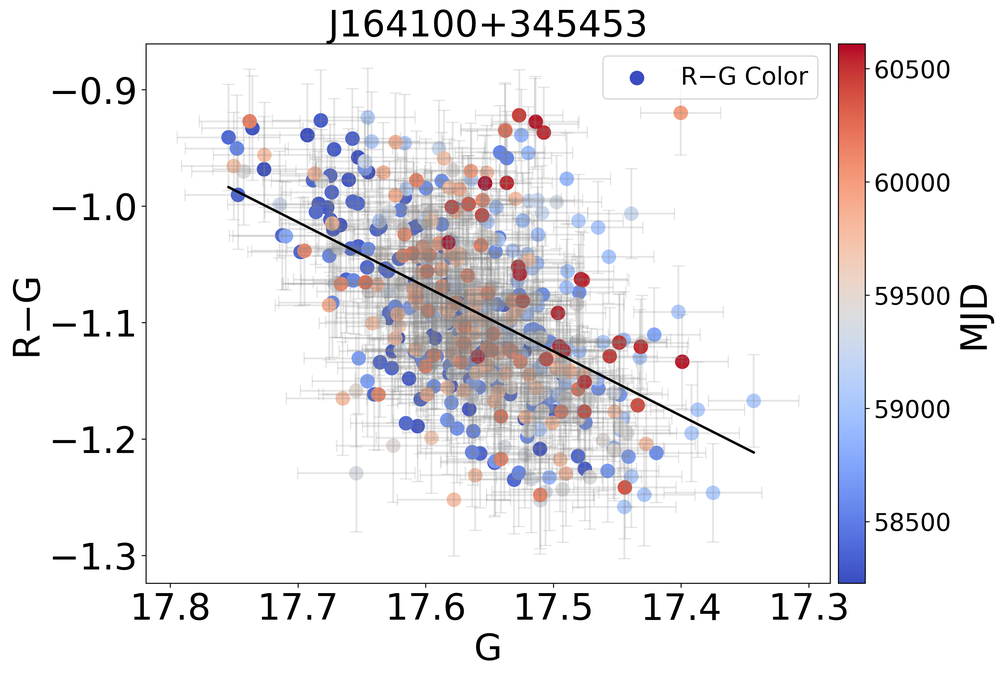}
    \end{minipage}
    \\
    \begin{minipage}{.3\textwidth}
        \centering
        \includegraphics[width=.99\linewidth]{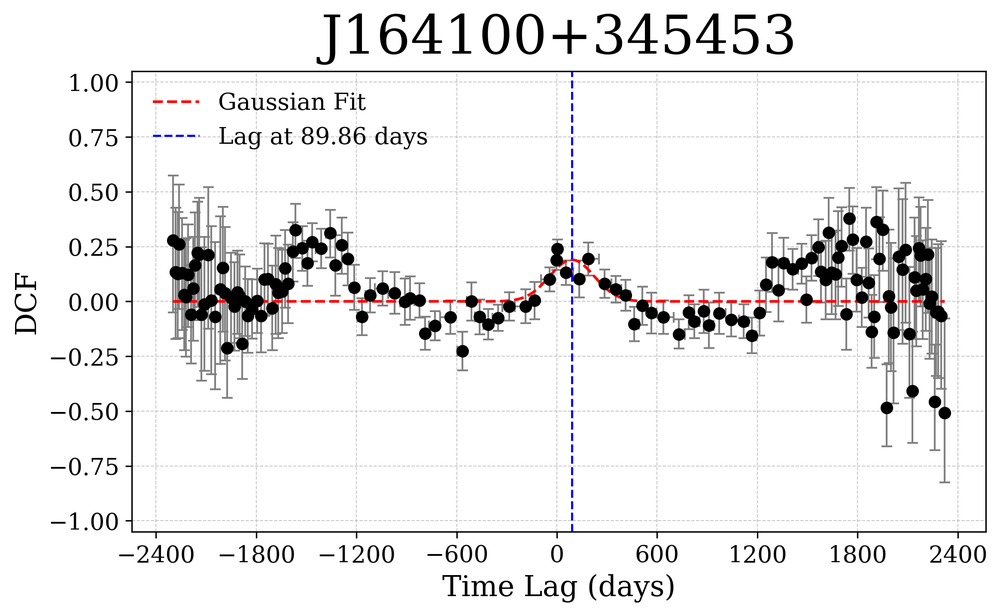}
    \end{minipage}
    \begin{minipage}{.3\textwidth}
        \centering
        \includegraphics[width=.99\linewidth]{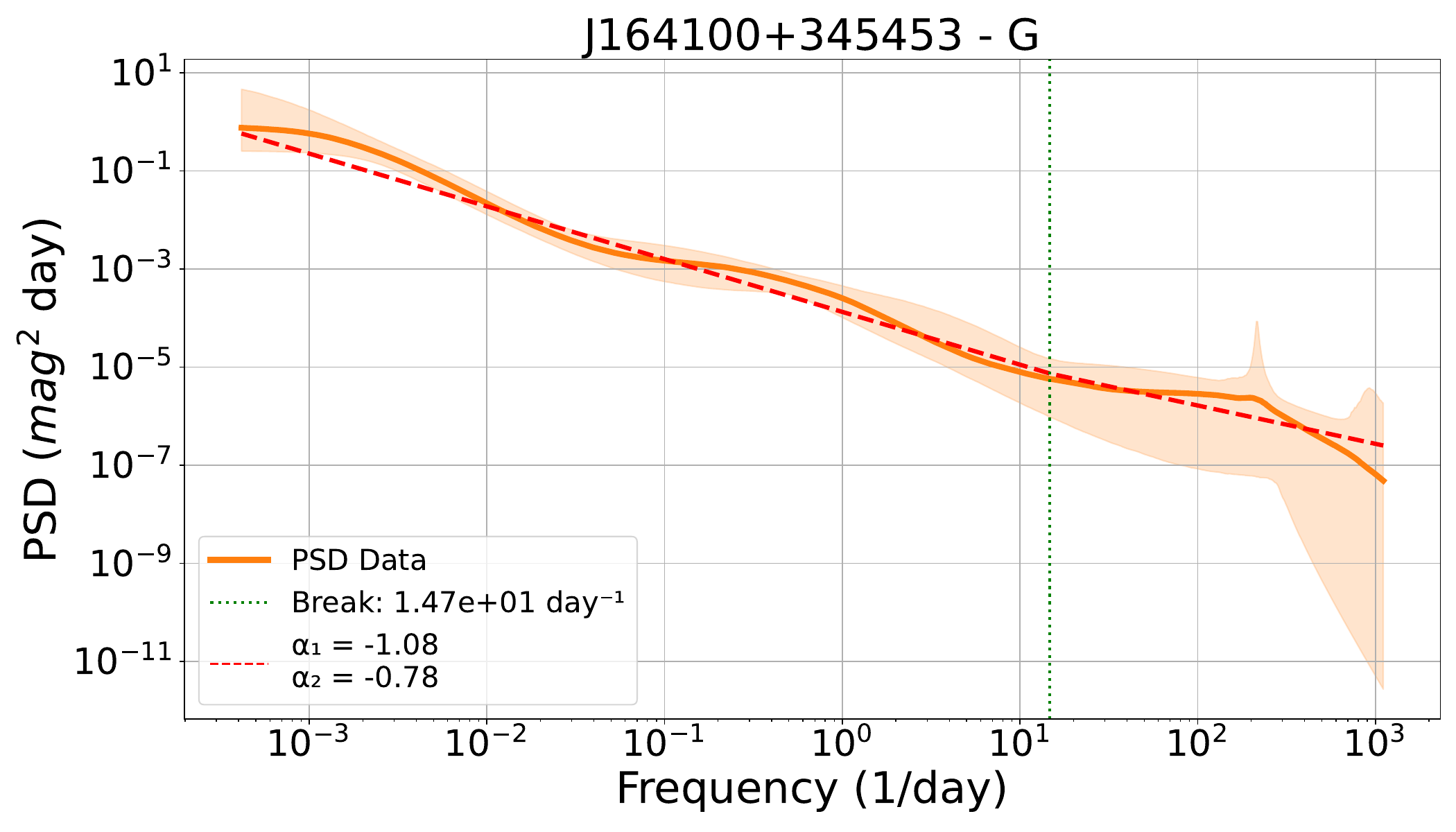}
    \end{minipage}
    \begin{minipage}{.3\textwidth}
        \centering
        \includegraphics[width=.99\linewidth]{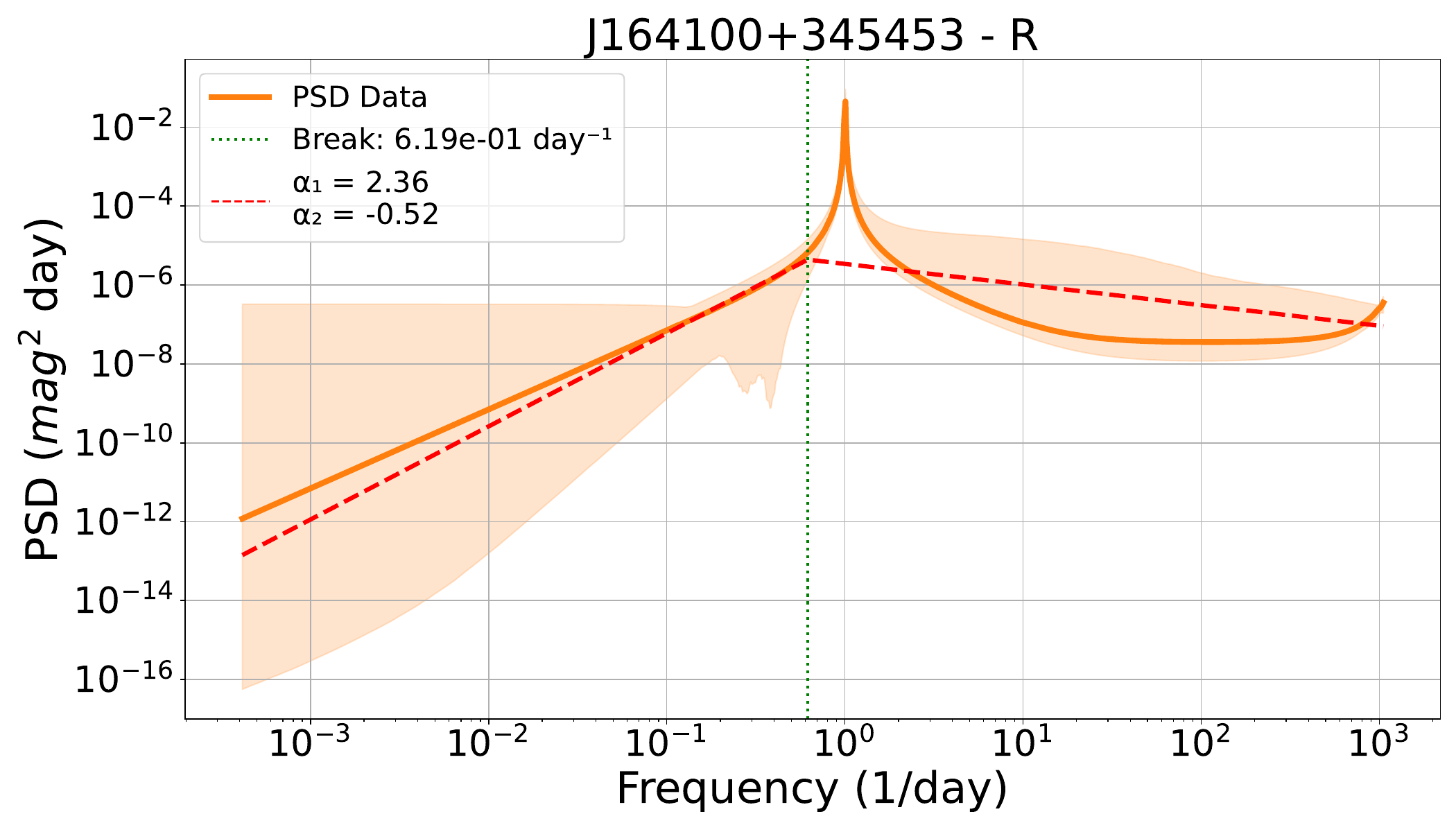}
    \end{minipage}
    \\
    \begin{minipage}{.3\textwidth}
        \centering
        \includegraphics[width=.99\linewidth]{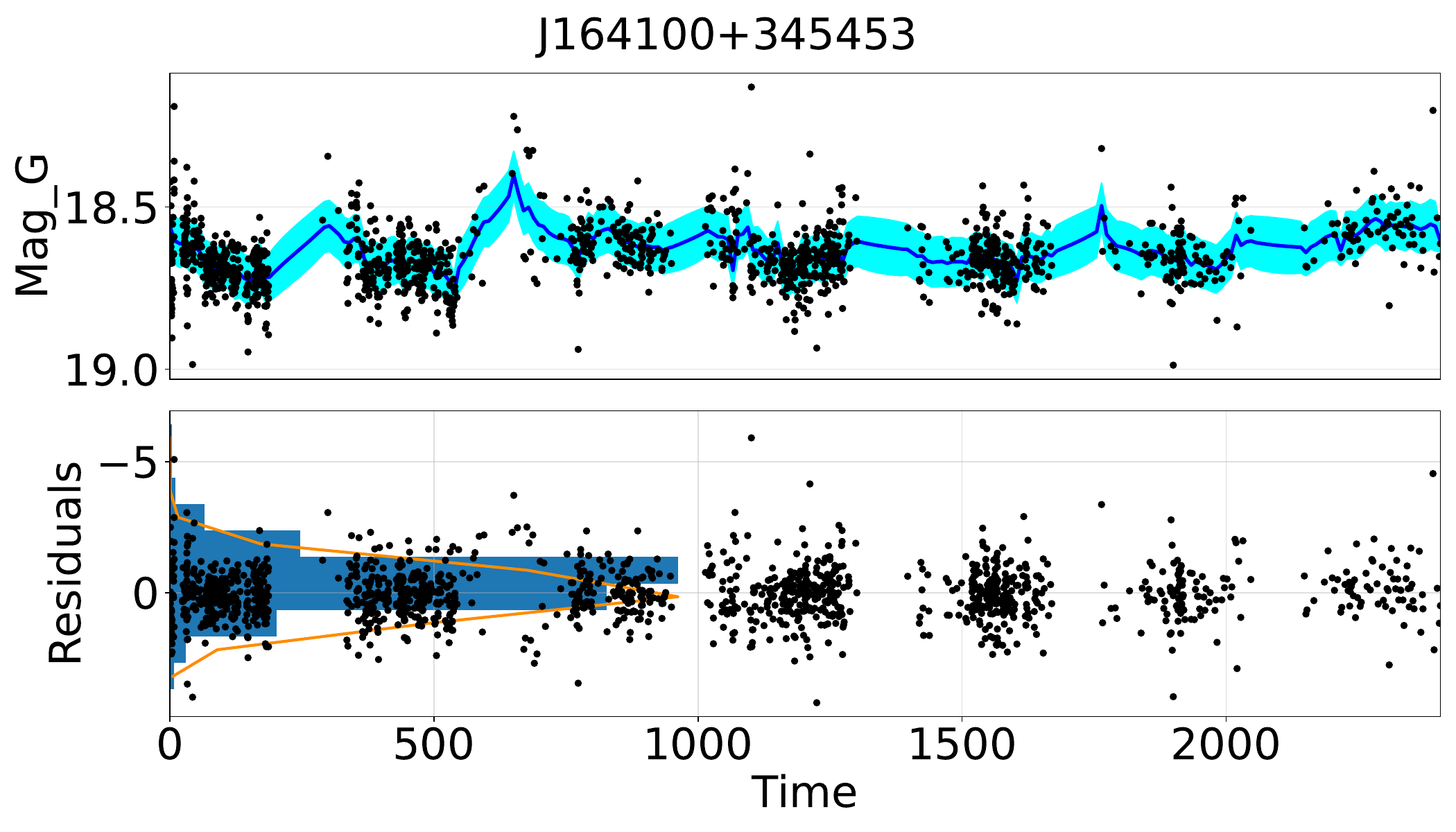}
    \end{minipage}
    \begin{minipage}{.3\textwidth}
        \centering
        \includegraphics[width=.99\linewidth]{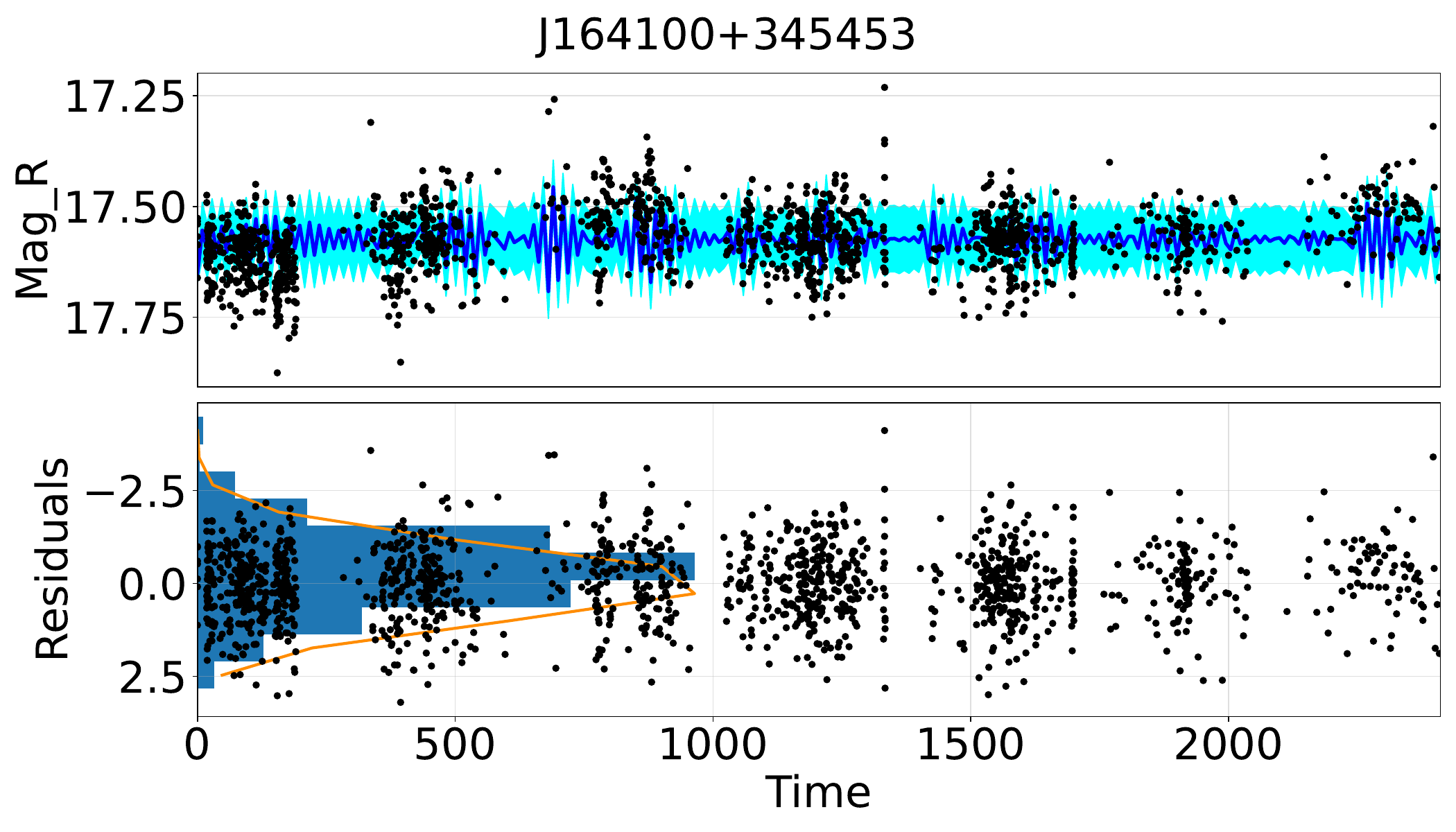}
    \end{minipage}
\end{figure*}

\begin{figure*}\label{J164442261913}
\centering
\caption{J164442+261913}
    \begin{minipage}{.3\textwidth}
        \centering
        \includegraphics[width=.99\linewidth]{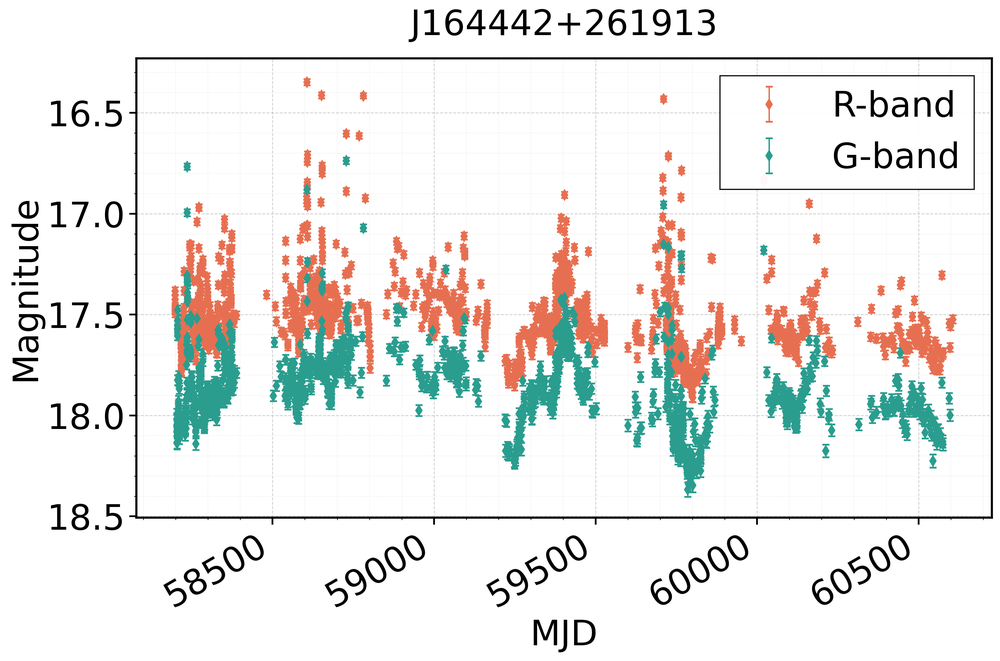}
    \end{minipage}
    \begin{minipage}{.3\textwidth}
        \centering
        \includegraphics[width=.99\linewidth]{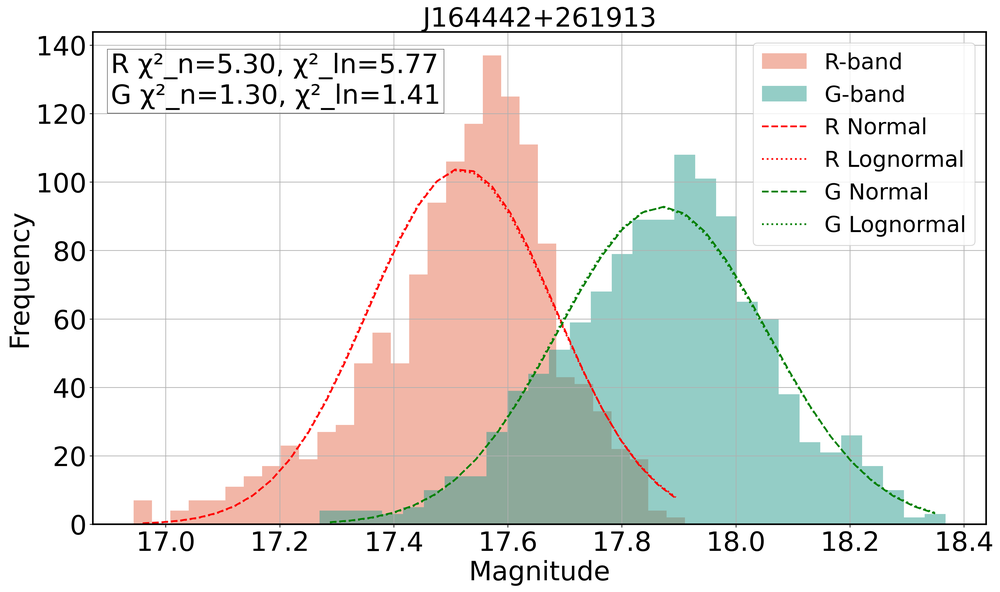}
    \end{minipage}
    \begin{minipage}{.3\textwidth}
        \centering
        \includegraphics[width=.99\linewidth]{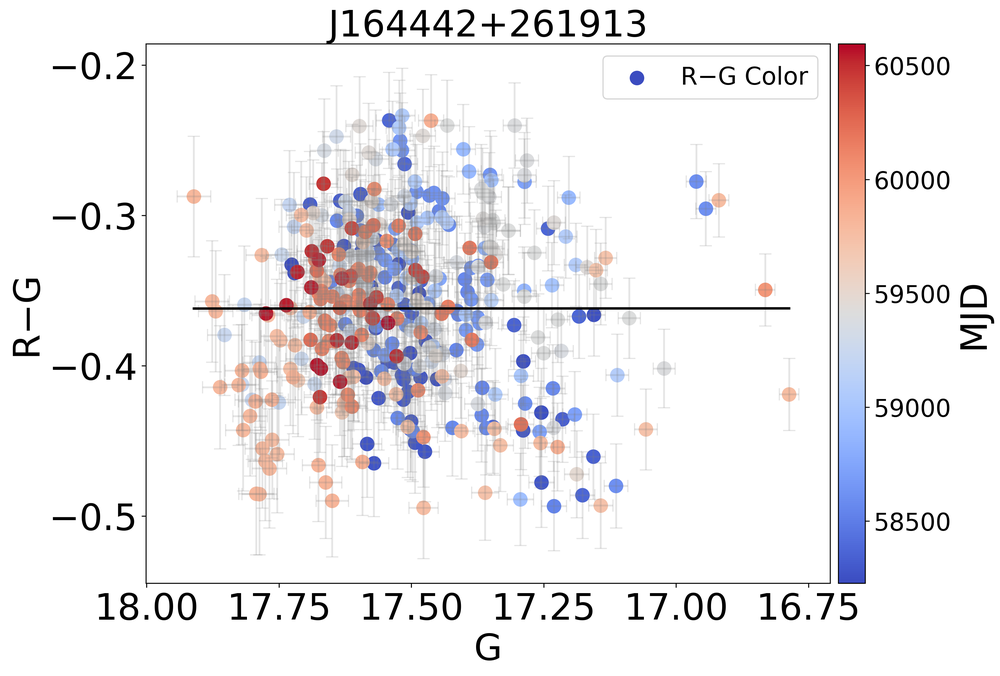}
    \end{minipage}
    \\
    \begin{minipage}{.3\textwidth}
        \centering
        \includegraphics[width=.99\linewidth]{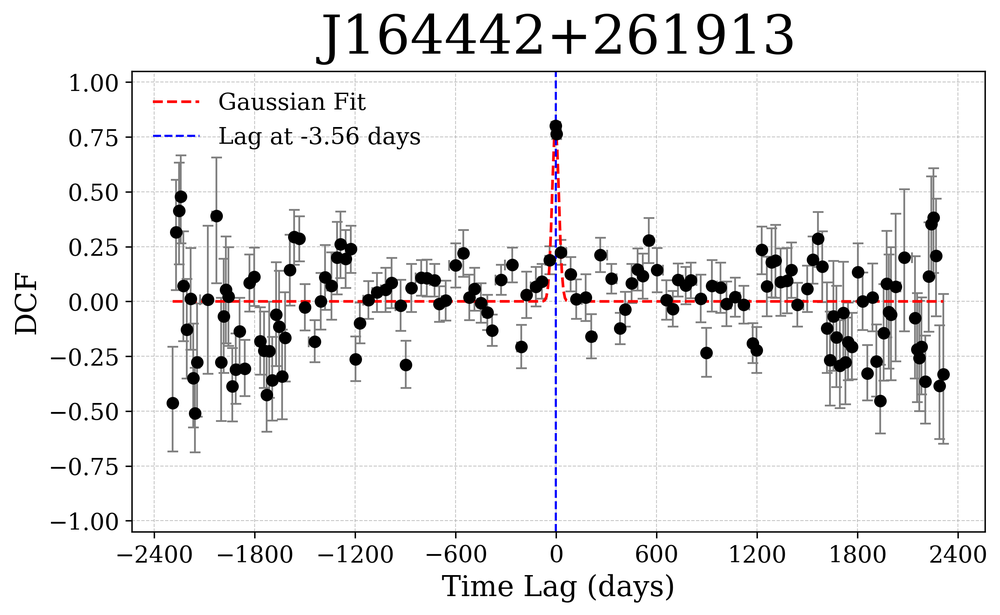}
    \end{minipage}
    \begin{minipage}{.3\textwidth}
        \centering
        \includegraphics[width=.99\linewidth]{carma/J164442+261913_G_psd.pdf}
    \end{minipage}
    \begin{minipage}{.3\textwidth}
        \centering
        \includegraphics[width=.99\linewidth]{carma/J164442+261913_R_psd.pdf}
    \end{minipage}
    \\
    \begin{minipage}{.3\textwidth}
        \centering
        \includegraphics[width=.99\linewidth]{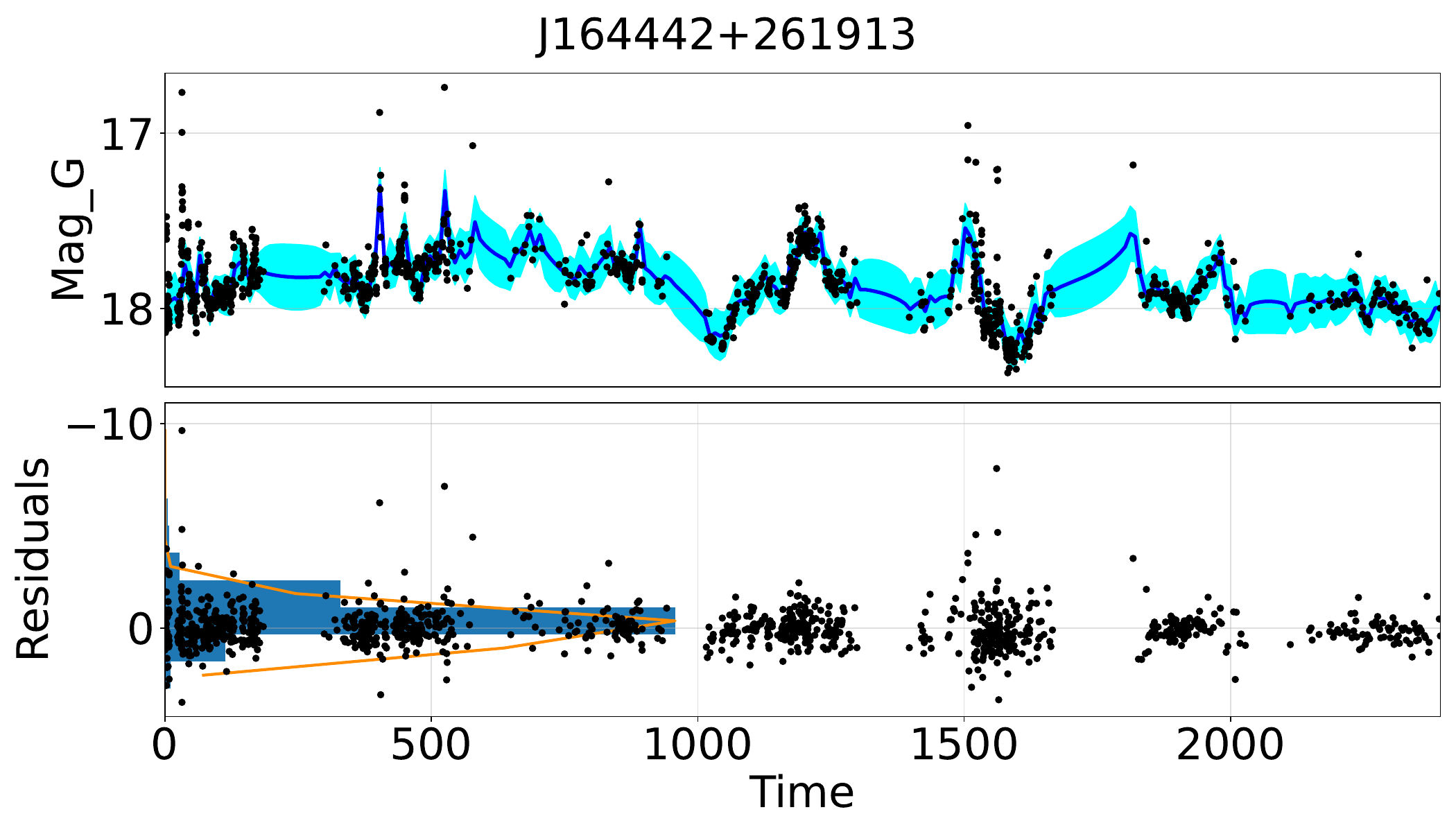}
    \end{minipage}
    \begin{minipage}{.3\textwidth}
        \centering
        \includegraphics[width=.99\linewidth]{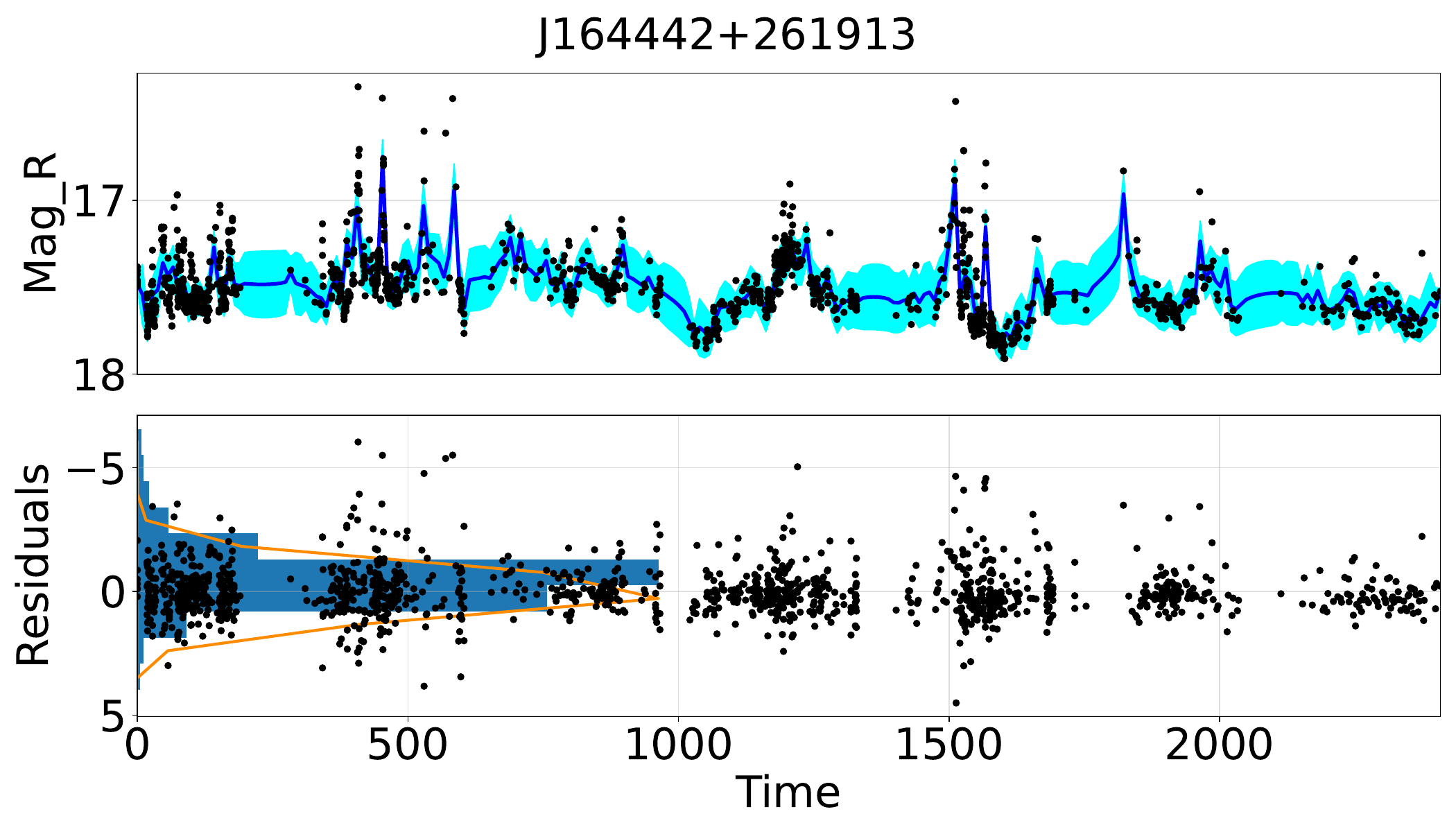}
    \end{minipage}
\end{figure*}

\begin{figure*}\label{J211852-073228}
\centering
\caption{J211852-073228}
    \begin{minipage}{.3\textwidth}
        \centering
        \includegraphics[width=.99\linewidth]{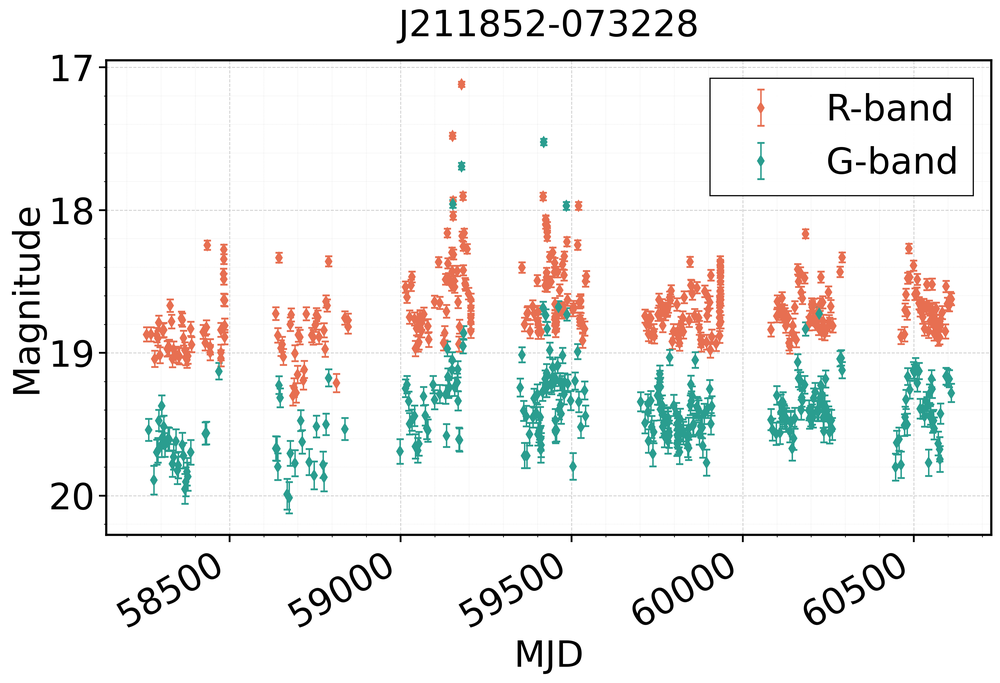}
    \end{minipage}
    \begin{minipage}{.3\textwidth}
        \centering
        \includegraphics[width=.99\linewidth]{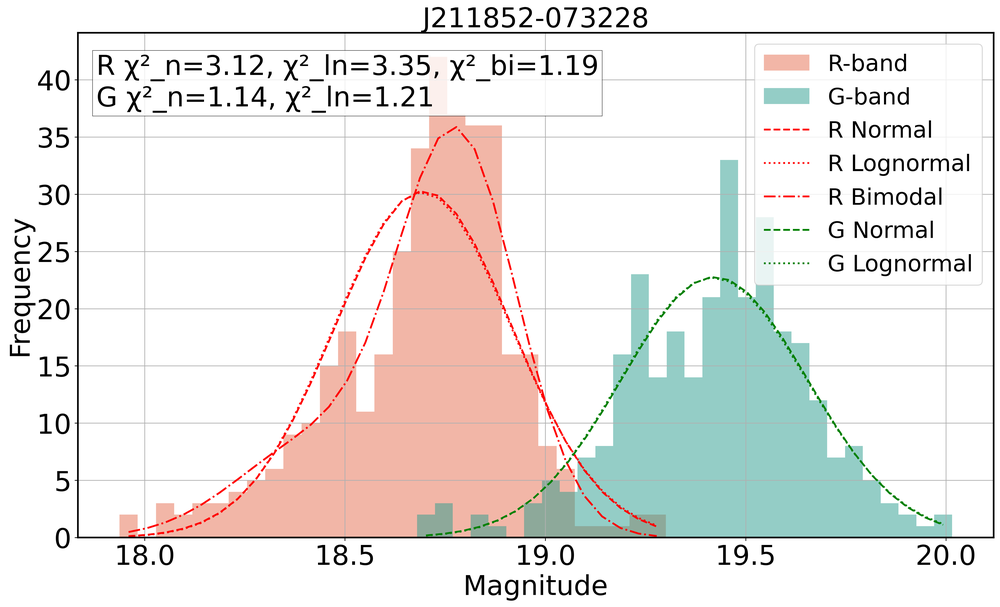}
    \end{minipage}
    \begin{minipage}{.3\textwidth}
        \centering
        \includegraphics[width=.99\linewidth]{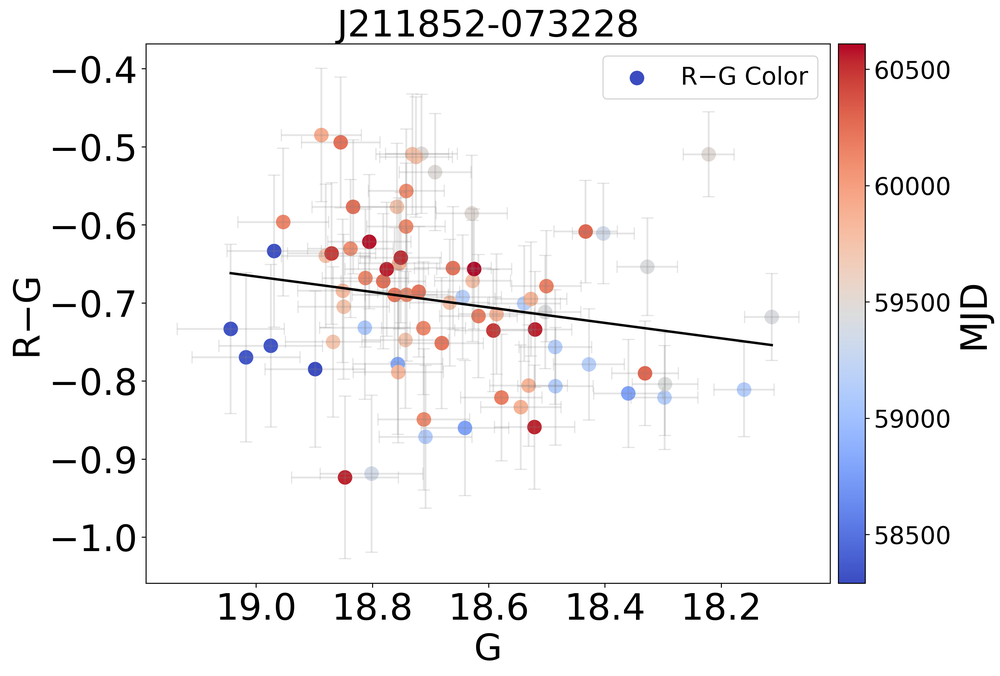}
    \end{minipage}
    \\
    \begin{minipage}{.3\textwidth}
        \centering
        \includegraphics[width=.99\linewidth]{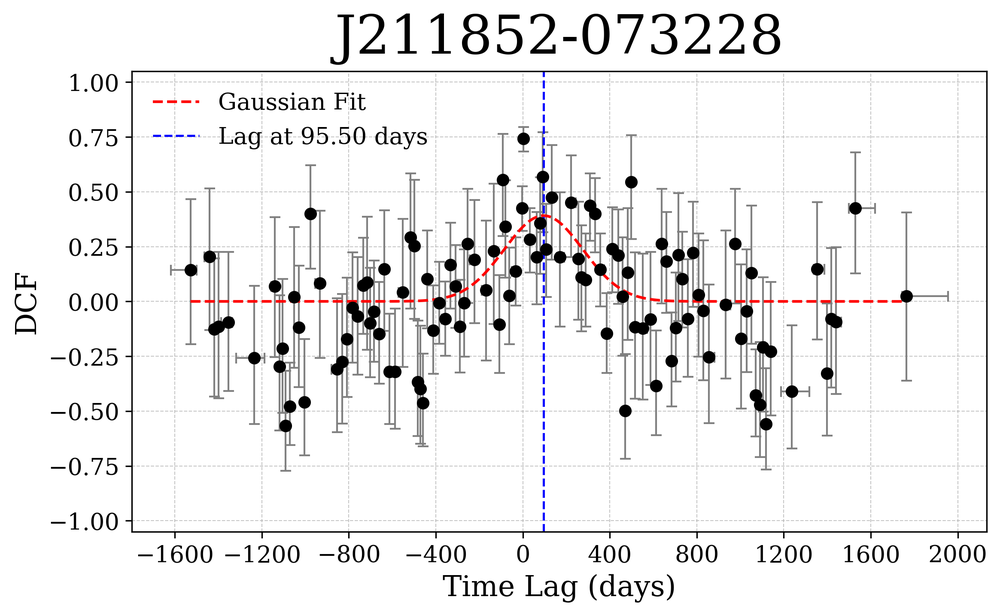}
    \end{minipage}
    \begin{minipage}{.3\textwidth}
        \centering
        \includegraphics[width=.99\linewidth]{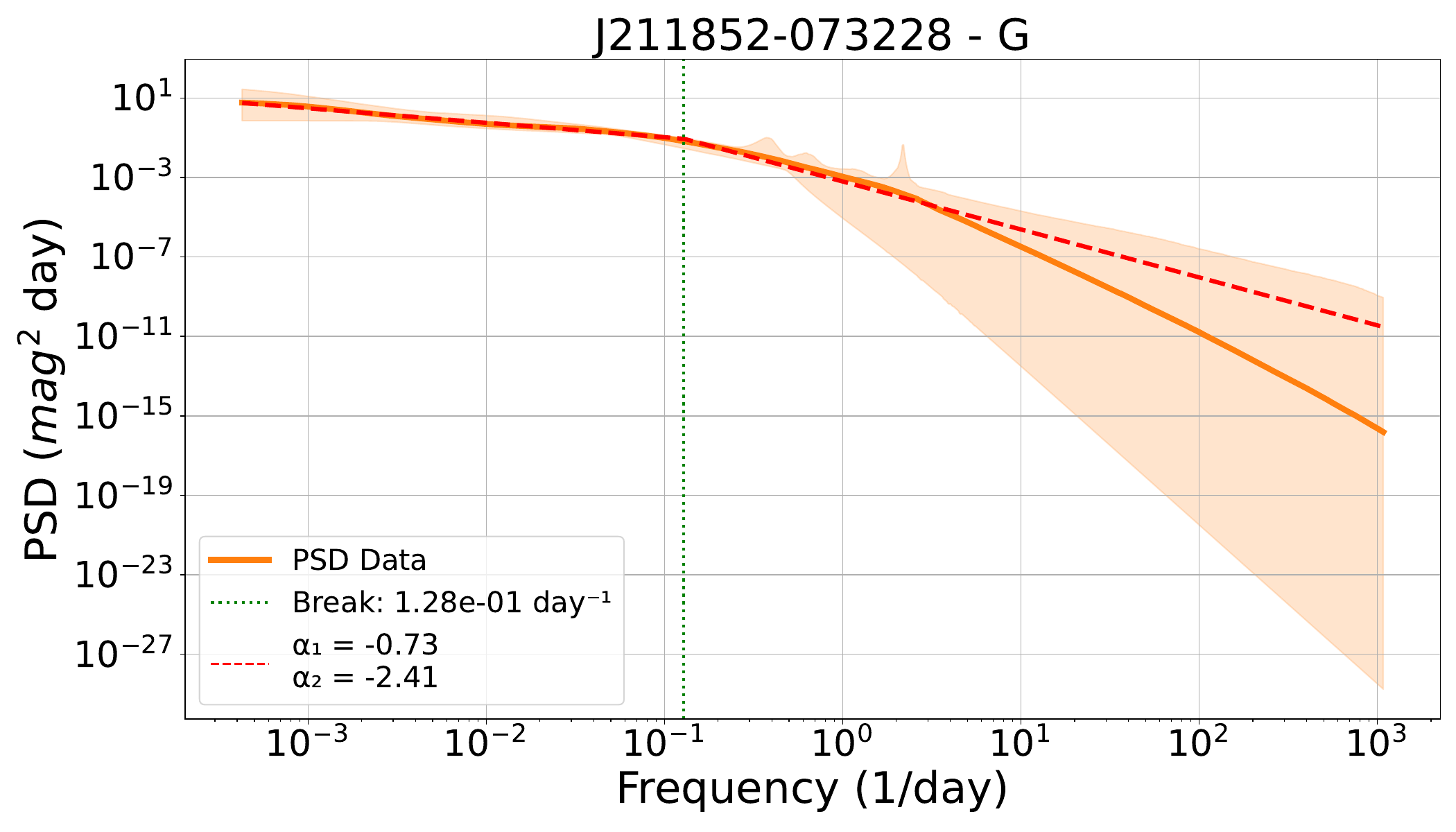}
    \end{minipage}
    \begin{minipage}{.3\textwidth}
        \centering
        \includegraphics[width=.99\linewidth]{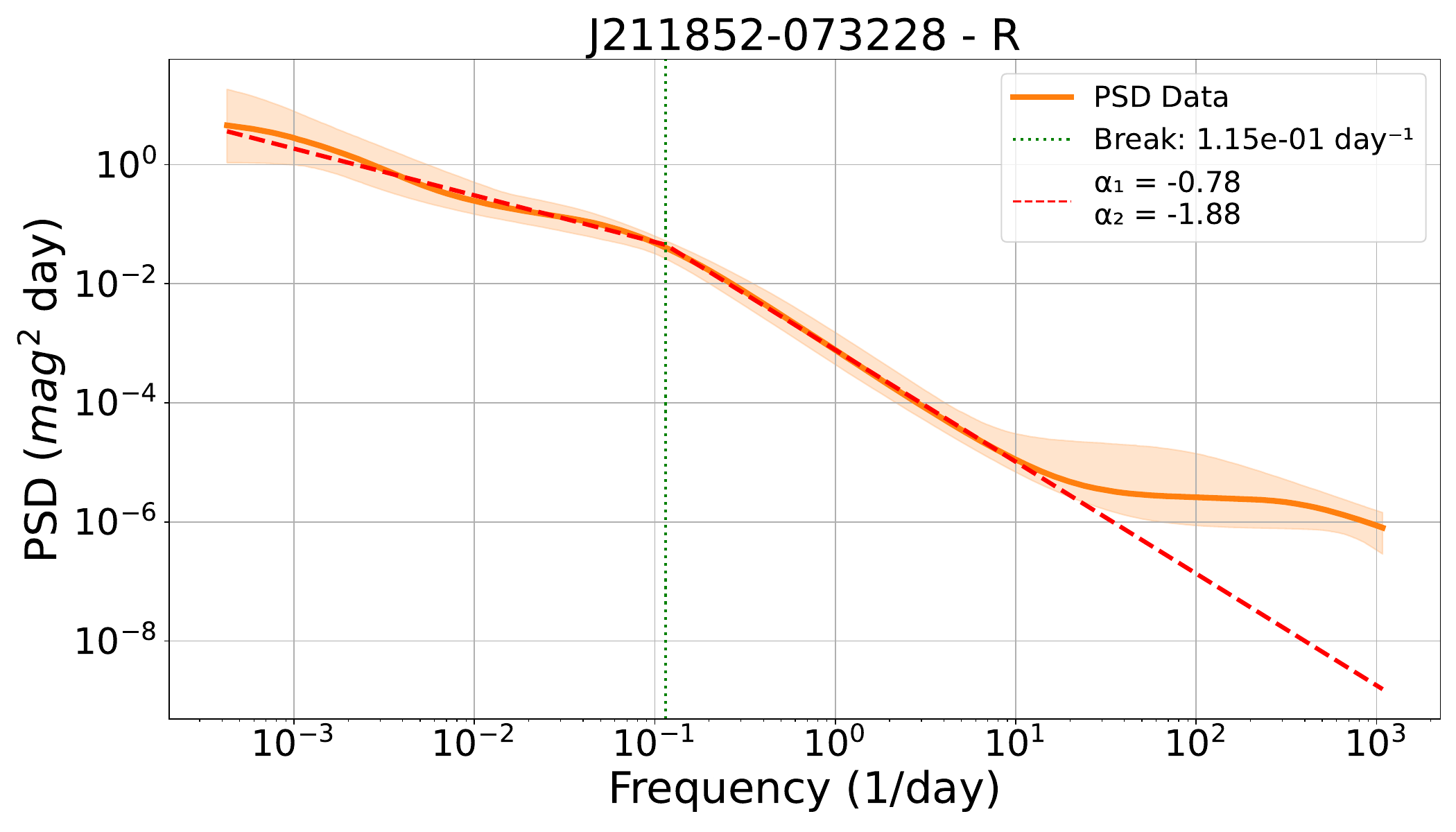}
    \end{minipage}
    \\
    \begin{minipage}{.3\textwidth}
        \centering
        \includegraphics[width=.99\linewidth]{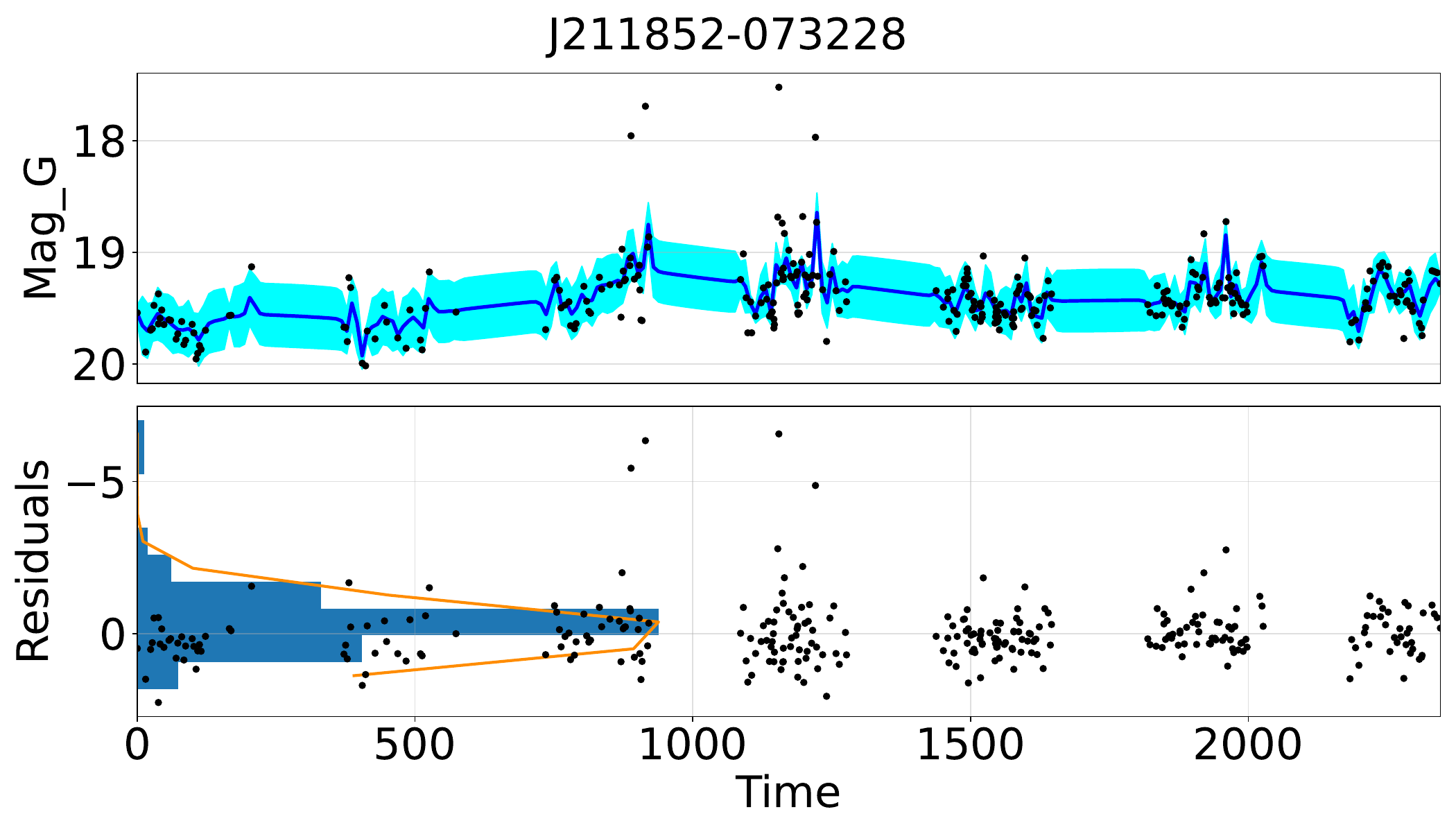}
    \end{minipage}
    \begin{minipage}{.3\textwidth}
        \centering
        \includegraphics[width=.99\linewidth]{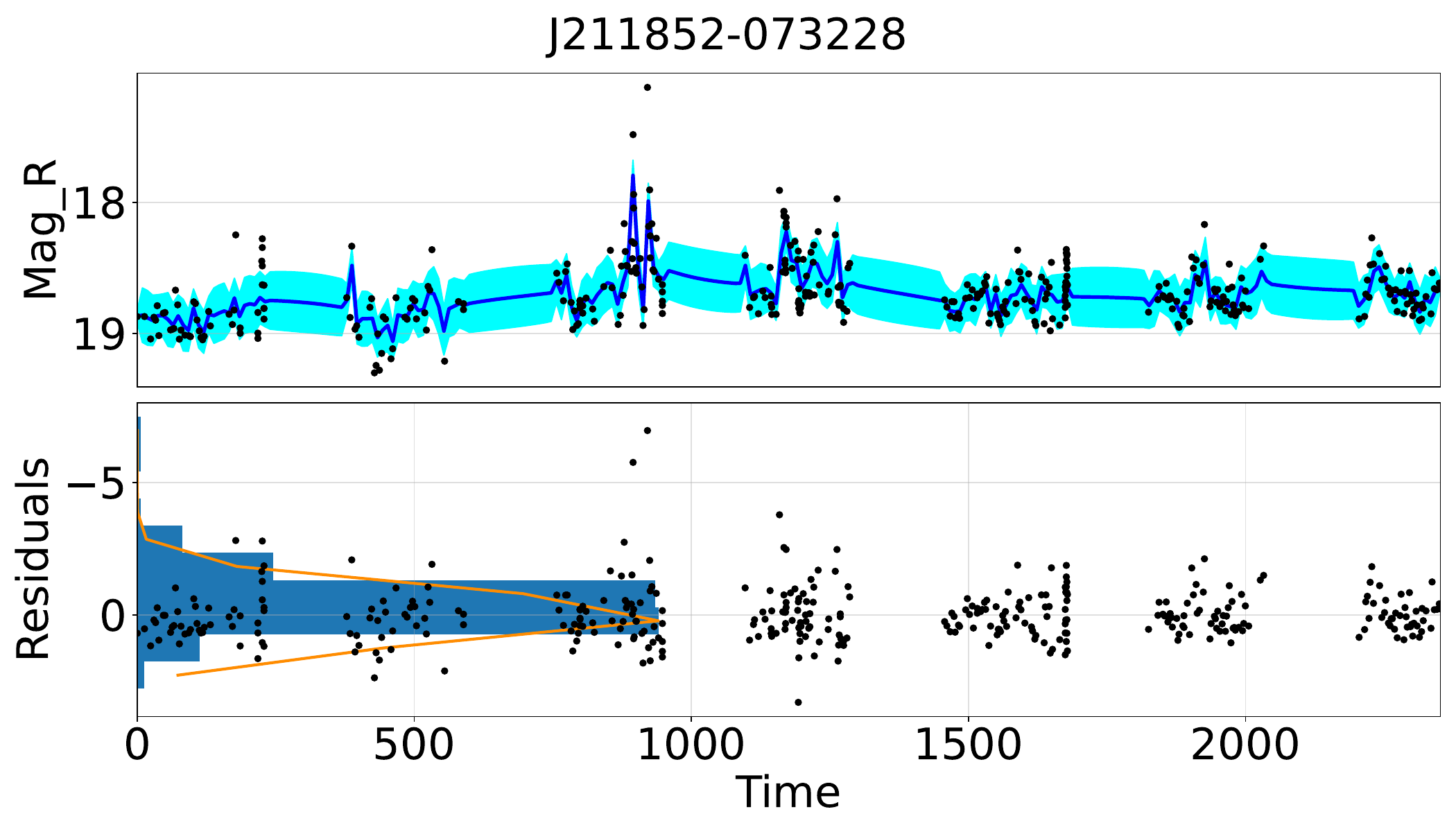}
    \end{minipage}
\end{figure*}

\label{lastpage}
\end{document}